\documentclass[12pt]{book}
\usepackage[utf8]{inputenc}
\usepackage[T1]{fontenc}
\usepackage[english]{babel}
\usepackage{a4wide}
\usepackage[linktocpage=true]{hyperref}
\usepackage{amsfonts}
\usepackage{amssymb}
\usepackage{graphics,amsmath}
\usepackage{graphicx}
\usepackage{cancel}
\usepackage{mathrsfs}
\usepackage[usenames,dvipsnames,svgnames,table]{xcolor}
\usepackage{fancyhdr}
\usepackage{cite}
\usepackage{slashed}
\usepackage{etoolbox}

\usepackage[font=small,labelfont=bf]{caption}

\def\be{\begin{equation}}
\def\ee{\end{equation}}
\def\ba{\begin{eqnarray}}
\def\ea{\end{eqnarray}}

\newcommand{\bra}[1]{\left\langle #1 \right|}
\newcommand{\ket}[1]{\left| #1 \right\rangle}
\newcommand{\scal}[2]{\left\langle #1 | #2 \right\rangle}
\newcommand{\Bigscal}[2]{\left\langle #1 \Big| #2 \right\rangle}

\newcommand{\mf}[1]{\mathfrak{#1}}
\newcommand{\mb}[1]{\mathbb{#1}}
\newcommand{\mc}[1]{\mathcal{#1}}
\newcommand{\order}[1]{\mathcal{O} \left( #1 \right)}

\newcommand{\esperado}[1]{\left\langle #1 \right\rangle}
\newcommand{\granesperado}[3]{\left\langle #1 \left| #2 \right| #3 \right\rangle}
\newcommand{\Biggranesperado}[3]{\left\langle #1 \Big| #2 \Big| #3 \right\rangle}
\newcommand{\dpartial}[3][]{\frac{\partial^{#1} #2}{\partial #3 ^{#1}}}
\def\l{\lambda}
\def\m{\mu}
\def\e{\epsilon}
\def\a{\alpha}
\def\O{\mathcal{O}}
\def\p{\pi} 
\def\k{\kappa}
\def\b{\beta}
\def\a{\alpha}
\def\g{\gamma}
\def\s{\sigma} 
\def\d{\delta}

\makeatletter
\newenvironment{chapquote}[2][3em]
  {\setlength{\@tempdima}{#1}%
   \def\chapquote@author{#2}%
   \parshape 1 \@tempdima \dimexpr\textwidth-2\@tempdima\relax%
   \itshape}
  {\par\normalfont\hfill--\ \chapquote@author\hspace*{\@tempdima}\par\bigskip}
\makeatother

\DeclareMathOperator\sech{sech}
\DeclareMathOperator\csch{csch}

\DeclareMathOperator\arcsinh{arcsinh}
\DeclareMathOperator\Tr{Tr}
\DeclareMathOperator\STr{STr}
\DeclareMathOperator\Res{Res}
\DeclareMathOperator\Ad{Ad}

\renewcommand{\theequation}{\thesection.\arabic{equation}}
\csname @addtoreset\endcsname{equation}{section}

\begin{document}

\pagestyle{plain}

\begin{titlepage}
\begin{center}

\vskip .5in

{\LARGE \bf Cuerdas en rotación y funciones de correlación en la correspondencia AdS/CFT }
\rule{.6667\linewidth}{.4pt}
\vskip 0.15in
{\LARGE \bf Spinning strings and correlation functions in the AdS/CFT correspondence }

\vskip 0.4in
por
\vskip 0.2in
          
{\Large \bf Juan Miguel Nieto Garc\'ia} 
\vskip 0.2in

bajo la supervisi\'on de

\vskip 0.2 in

{\Large Rafael Hern\'andez Redondo}

\vskip 0.3 in

\begin{figure}[h]
\begin{center}
	\includegraphics[width=7cm,keepaspectratio]{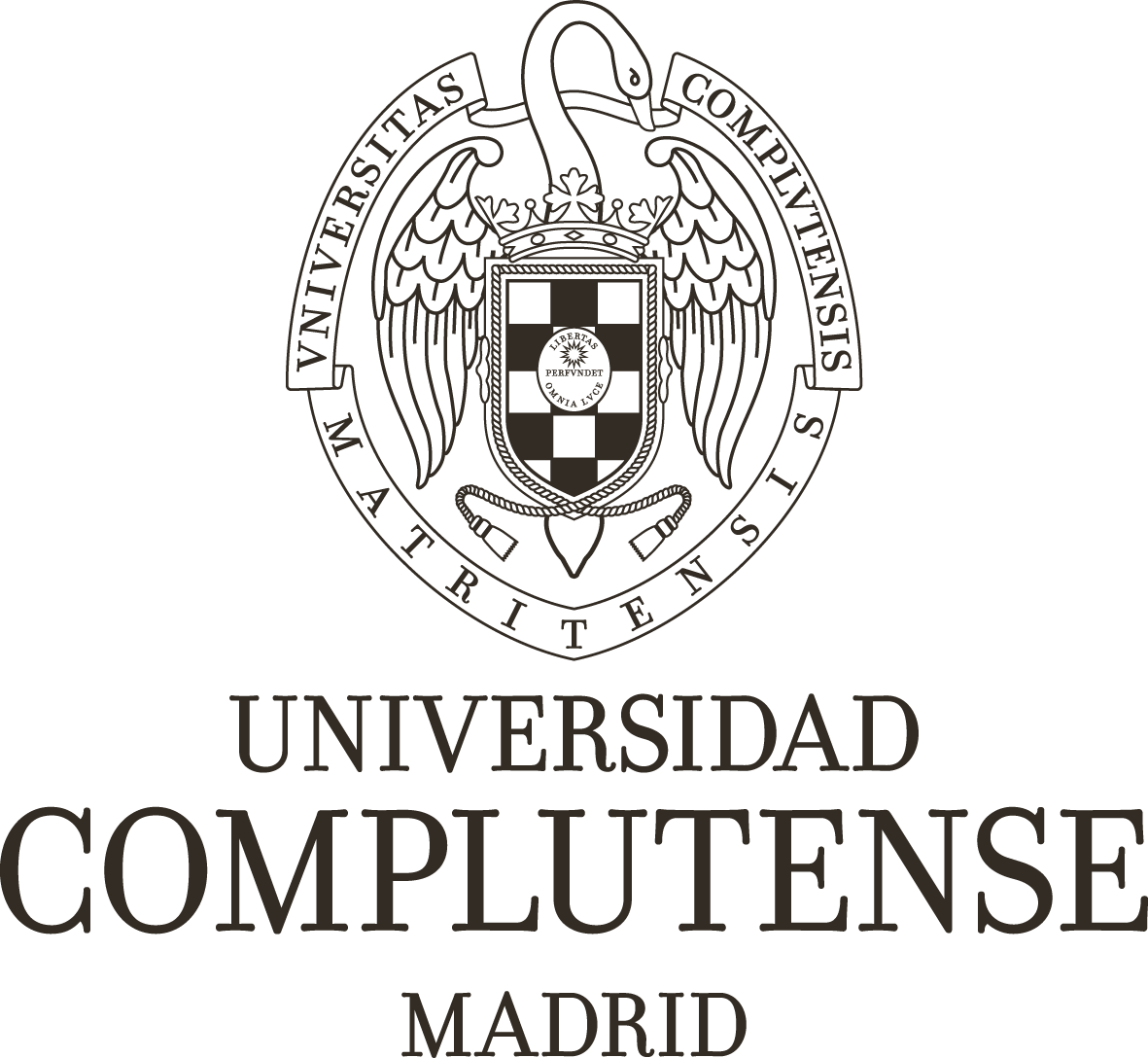}
\end{center}
\end{figure}

\vskip 0.3 in

{\small Tesis presentada en la\\
Universidad Complutense de Madrid \\
para el grado de Doctor en Física
\vskip 0.1in
Departamento de F\'{\i}sica Te\'orica I\\
Facultad de Ciencias F\'isicas\\
Abril 2017}

\end{center}

\vskip .4in
\noindent

\end{titlepage}

\vfill
\eject

\def\baselinestretch{1.2}


\baselineskip 20pt

\null 
\thispagestyle{empty}
\newpage
\setcounter{page}{0}
\pagenumbering{roman}

\setcounter{tocdepth}{1}
\tableofcontents
\addtocontents{toc}{\null \hfill \textbf{Page}\par}

\chapter*{Abstract}
\addcontentsline{toc}{chapter}{Abstract}

The AdS/CFT correspondence states that the strong-coupling limit of four dimensional Yang-Mills theory with $\mc{N}=4$ supersymmetry can be identified with the weak-coupling limit of type IIB supersymmetric string theory compactified in $AdS_5 \times S^5$ and vice-versa. This correspondence was later broadened to other compactifications like $AdS_4 \times \mathbb{CP}^3$ and $AdS_3 \times S^3 \times \mc{M}_4$. As it relates theories at weak and strong-coupling, it allow us to access the non-perturbative regime of gauge theories and string theories. However its proof requires to face extremely complex problems, e.g.  computing the spectrum of conformal dimensions of a gauge theory or quantizing type IIB string theory in a curved background.

Symmetries are a powerful way to simplify computations on those theories, the most typical example of which is the exact conformal symmetry of both the type IIB string theory in $AdS_5 \times S^5$ and the $\mc{N}=4$ supersymmetric Yang-Mills theory, which completely fixes the functional form of the two and three-point correlation functions. Another important simplification is the appearance of an integrable structure in the correspondence. The presence of these structures gave rise to an exhaustive exploration of the planar spectrum of anomalous dimensions of gauge invariant operators and the spectrum of energies of rotating strings in $AdS_5 \times S^5$. Techniques based on integrability reported important successes and even the possibility of performing interpolation between both regimes in some particular cases.

In this dissertation we will present some computations made in both sides of the AdS/CFT holographic correspondence, the string theory side and the field theory side, using the integrability of both theories as a starting point and a method to simplify these computations.

Regarding the string theory side, this dissertation is focused in the computation of the energy spectrum of closed spinning strings in some deformed $AdS_3 \times S^3$ backgrounds. In particular we are going to focus in the deformation provided by the mixing of R-R (Ramond-Ramond) and NS-NS (Neveu-Schwarz-Neveu-Schwarz) fluxes and the so-called $\eta$-deformation. These computations are made using the classical integrability of these two deformed string theories, which is provided by the presence of a set of conserved quantities called ``Uhlenbeck constants''. The existence of the Uhlenbeck constants is central for the method used to derive the dispersion relations.

Regarding the gauge theory side, we are interested in the computation of two and three-point correlation functions. At weak-coupling these correlation functions can be obtained in a perturbative computation, but it is also possible to use techniques derived from the integrable structure of the theory. For that purpose, instead of using directly the field theory formulation, we are going to use the isomorphism between gauge-invariant single-trace operators in $\mc{N}=4$ supersymmetric Yang-Mills theory and states of a $PSU(2,2|4)$ invariant spin chain. Concerning the two-point function a computation of correlation functions involving different operators and different number of excitations is performed using the Algebraic Bethe Ansatz and the Quantum Inverse Scattering Method. These results are compared with computations done with the Coordinate Bethe Ansatz and with Zamolodchikov-Faddeev operators. Concerning the three-point functions, we will explore the novel construction given by the hexagon framework. In particular we are going to start from the already proposed hexagon form factor and rewrite it in a language more resembling of the Algebraic Bethe Ansatz. For this intent we construct an invariant vertex using Zamolodchikov-Faddeev operators, which is checked for some simple cases.

\chapter*{Resumen}

\addcontentsline{toc}{chapter}{Resumen}

De acuerdo con la formulación original de la correspondencia AdS/CFT el límite de acoplamiento fuerte de una teoría de Yang-Mills en cuatro dimensiones con supersimetría $\mc{N}=4$ se puede identificar con el límite de acoplamiento débil de la teoría de cuerdas supersimétrica de tipo IIB compactificada en $AdS_5 \times S^5$, y viceversa. Dicha correspondencia fue posteriormente extendida a otras compactificaciones como $AdS_4 \times \mathbb{CP}^3$ y $AdS_3 \times S^3 \times \mc{M}_4$. Puesto que relaciona teorías a acoplamiento fuerte y débil, dicha correspondencia permite acceder al régimen no perturbativo tanto en una teoría gauge como en una teoría de gravedad. Su demostración requiere sin embargo afrontar problemas extremadamente complejos, entre los cuales est\'an encontrar el espectro completo de la teoría gauge, y cuantizar cuerdas de tipo IIB en un espacio curvo. 

Las simetrías son una manera muy importante de simplificar los cálculos en dichas teorías. El ejemplo por antonomasia de esto es la simetría conforme exacta de la teoría de cuerdas en $AdS_5 \times S^5$ y en la teoría de Yang-Mills supersimétrica $\mc{N}=4$, que especifica completamente la forma funcional de las funciones de correlación a dos y tres puntos. Otra simplificaci\'on importante proviene de la aparici\'on de estructuras integrables en la correspondencia. La presencia de estas estructuras dio lugar a una labor exhaustiva de exploraci\'on de las dimensiones an\'omalas de operadores invariantes gauge en el l\'imite planar y del espectro de energ\'ias de cuerdas en rotación en $AdS_5 \times S^5$. El uso de t\'ecnicas basadas en la integrabilidad proporcionaron importantes \'exitos e incluso la posibilidad de interpolar entre ambos l\'imites de la correspondencias en algunos casos concretos.

En esta tesis presentaremos algunos c\'alculos realizados en ambos lados de la correspondencia hologr\'afica AdS/CFT, el l\'imite de teor\'ia de cuerdas y el l\'imite de teor\'ia de campos, usando la integrabilidad de ambas teor\'ias como punto de partida y como herramienta para simplificar dichos c\'alculos.

En lo que respecta al lado de teor\'ia de cuerdas, esta tesis se centra en el c\'alculo del espectro de energ\'ias de cuerdas cerradas en rotación en deformaciones del espacio $AdS_3 \times S^3$. En concreto nos centraremos en las deformaciones producidas por tener una mezcla de flujos de tipo R-R (Ramond-Ramond) y de tipo NS-NS (Neveu-Schwarz-Neveu-Schwarz) y la llamada deformaci\'on $\eta$. Estos cálculos se han hecho usando la integrabilidad de ambas deformaciones a nivel clásico. Dicha integrabilidad está asociada a la presencia de un conjunto de cantidades conservadas llamadas ``constantes de Uhlenbek''. La existencia de dichas constantes es central en el método usado para derivar las relaciones de dispersión.

En lo que respecta a la teoría de gauge, nos interesaremos por el cálculo de funciones de correlación a dos y tres puntos. En acoplamiento débil estas funciones de correlación pueden obtenerse usando cálculos perturbativos, pero también es posible usar técnicas derivadas de la estructura integrable de la teoría. Para ello, en lugar de usar directamente la formulación de teoría de campos, usaremos el isomorfismo existente entre operadores invariantes gauge compuestos de una sola traza en una teoría de Yang-Mills con supersimetría $\mc{N}=4$ y una cadena de spines con simetría $PSU(2,2|4)$. En lo que respecta a funciones de correlación a dos puntos, hemos calculado funciones de correlación que involucran diferentes operadores y diferente número de excitaciones usando el Ansatz de Bethe Algebraico y el Método Cuántico de Dispersion Inversa. Estos resultados se comparan con los obtenidos usando el Ansatz de Bethe Coordenado y usando operadores de Zamolodchikov-Faddeev. En lo que respecta a funciones a tres puntos, exploramos la reciente construcción dada por el método del hexágono. En concreto, empezaremos estudiando factor de forma hexagonal actualmente propuesto y reescribiremos dicho factor de forma en un lenguaje más parecido al Ansatz de Bethe Algebraico. Para ello construiremos un vértice invariante usando operadores de Zamolodchikov-Faddeev, que comprobaremos para algunos casos sencillos.

\chapter*{Acknowledgements}

\addcontentsline{toc}{chapter}{Acknowledgements}

First of all, I would like to thank my advisor Rafael Hernández for guiding me during my doctorate. However I cannot thank him without thanking other three people also that also guided me: Agustín Sabio, Didina Serban and Ivan Kostov.

I want to thank my family for always helping, supporting and encouraging me during all my studies.

I also want to thank all the people I met and discussed with during this journey. But because my memory is not perfect so this will be comprehensive but not exhaustive list: from the Universidad Autonoma de Madrid (Arancha Gómez, Raquel Gómez, Irene Valenzuela....), from the Instituto de Física Teórica (Mario Herrero, Victor Martín, Miguel Montero, Mikel Berasaluce, Ander Retolaza, Gianluca Zoccarato, Josu Hernandez, Xabier Marcano, Georgios Korpas, Aitor Landete...), from the Universidad Complutende de Madrid (Miguel Aparicio, Santos J. Núñez, Héctor Villarrubia, Jose Manuel Sánchez, Arkaitz Rodas, Daniel Gutiérrez, José de Ramón...), from the Institut de Physique Théorique (Andrei Petrovskii, Romain Couvreur, Kemal Bidzhiev, Benoit Vallet, Lais Sarem Schunk, Raphaël Belliard...), from the GATIS network (Xinyi Chen-Lin, Christian Marboe, Joao Caetano, Stefano Negro, Rouven Frassek, István M. Szécsényi, Brenda Penante, Mikhail Alfimov, Alessandra Cagnazzo, Fedor Levkovich-Maslyuk...) and from workshops, schools and conferences (Daniel Medina-Rincón, Joakim Strömwall, Frank Coronado, Lucia Gomez Cordova...).

During my doctoral studies I have been financially supported by the European Community's Seventh Framework Programme FP7/2007-2013 under grand agreement No 317089 (GATIS), and by CT45/15 grant from the Universidad Complutense of Madrid.

Finally, a special acknowledgement to all the people that have read this thesis looking for typos: Laura Ortiz, Santiago Varona and Marta Azucena.

\newpage
\null
\newpage

\pagestyle{fancy}
\setcounter{page}{0}
\pagenumbering{arabic}
\part{Introduction}

\chapter{Integrability in the AdS/CFT correspondence} \chaptermark{Integrability in  the A\MakeLowercase{d}S/CFT correspondence}

\begin{chapquote}{William Shakespeare, \textit{Henry V}}
	For ’tis your thoughts that now must deck our kings,\\
Carry them here and there, jumping o'er times,\\
Turning th' accomplishment of many years\\
Into an hour-glass: for the which supply,\\
Admit me chorus to this history;\\
Who prologue-like your humble patience pray,\\
Gently to hear, kindly to judge, our play.
\end{chapquote}

\section{The AdS/CFT conjecture} \sectionmark{The A\MakeLowercase{d}S/CFT conjecture}

Since the AdS/CFT correspondence was conjectured \cite{original,Witten_1998,Gubser_1998} there has been a huge de\-velop\-ment in this field. The basic idea of this conjecture is that a (type IIB super)string theory where the strings propagate on an $AdS_5 \times S^5$ background is partnered with a particular Conformal Field Theory (CFT) defined at the (conformal) boundary (of the $AdS_5$ space), being this boundary flat $4$-dimensional spacetime while the $S^5$ part becomes a symmetry of the theory.

The particular CFT involved in this duality is $\mathcal{N}=4$ Super Yang-Mills (SYM) with color group $SU(N_c)$ in four dimensionsal flat space. This theory is a gauge theory whose field content is one vector multiplet in the adjoint representation of $SU(N_c)$. As we have $\mc{N}=4$ supersymmetry we can organize the vector multiplet into bosons and fermions that transform in different representations of an $SU(4)$ symmetry called \emph{R-symmetry}. Therefore the vector multiplet can be divided into a vector gauge field $A^a_\mu$ that transform as a scalar under said R-symmetry, four Weyl spinors $\lambda^{a,A}_\alpha$ that transform as a vector and six real scalars $\phi^{a,AB}$ that transform as the second rank complex self dual representation. For some computations it is more interesting to organize it in one vector multiplet and three chiral multiplets of $\mc{N}=1$ supersymmetry or one vector multiplet and one chiral multiplet of $\mc{N}=2$ supersymmetry \cite{Kovacks1999}. Supersymmetry also completely fixes the Lagrangian of the theory to
\begin{equation}
\begin{split}
	S_{\text{SYM}}=\int d^4 x \Tr \left\{  (D_\mu \phi^{AB}) (D^\mu \bar{\phi}_{AB})  -\frac{i}{2} ( \lambda^{\alpha A} \overleftrightarrow{\slashed{D})}_{\alpha \dot{\alpha}} \bar{\lambda}^{\dot{\alpha}}_A ) - \frac{1}{4} F^{\mu \nu} F_{\mu \nu}   \right.  \\
	\left. \vphantom{\frac{i}{2} ( \lambda^{\alpha A} \overleftrightarrow{\slashed{D})}_{\alpha \dot{\alpha}}} -g\lambda^{\alpha A} [ \lambda_\alpha^B , \bar{\phi}_{AB} ] -g\bar{\lambda}_{\dot{\alpha} A} [ \bar{\lambda}^{\dot{\alpha}}_B , \phi^{AB} ] + 2g^2 [\phi^{AB} , \phi^{CD} ] [\bar{\phi}_{AB} , \bar{\phi}_{CD} ] \right\} \ ,
\end{split}
\end{equation}
where the covariant derivative is defined as $D_\mu=\partial_\mu +ig [A_\mu , \cdot ]$ and Tr is a trace over the color indices. It can be checked that the matter content of this theory makes the one-loop $\beta$-funcion of the theory vanishes by using the well known formula for the beta functions of Yang-Mills theories \cite{betafunction}. This computation can be done for higher loops, giving a vanishing of the $\beta$-funcion for two and three-loop \cite{Jones_1977,Poggio_1977,Avdeev_1980}. Even more, using light-cone gauge it was argued that the $\beta$-function should be zero to all loops \cite{Mandelstam_1983,Brink_1983}, making it a quantum theory with exact conformal symmetry. The generating functional of correlation functions involving the local operator $\mc{O}$ is given by
\begin{equation}
	Z_{CFT,\mc{O}} [J]=\int{\mc{D} \text{[fields]} e^{-S_{\text{SYM}}+\int{d^4 x \mc{O} (x) J (x)}}} \ .
\end{equation}

Type IIB string theory is a chiral superstring theory in 10 dimensions with $\mc{N}=(2,0)$ supersymmetry. There is no manifestly Lorentz-invariant action for this theory \cite{Marcus_1982}, but one can write down the equations of motion, and the symmetries and transformation rules. Together with type IIA string, which is related with this one by T-duality, they are the two string theories with two supersymmetries in ten dimensions. In addition to the flat ten-dimensional Minkowski space, type IIB supergravity admits another maximally supersymmetric solution which is product of the five-dimensional Anti-de-Sitter space $AdS_5$ and the five-sphere $S^5$. This solution is supported by a self-dual Ramond-Ramond five-form flux, whose presence precludes the usage of the standard Neveu-Schwarz-Ramond approach in a straightforward way to build up the action. This will make us look for alternative approaches.

The AdS/CFT duality relates the string partition function with sources $\phi$ for the string vertex operators fixed to value $J$ at the boundary of the AdS space to the CFT partition function with sources $J$ for local operators (operators composed from the fundamental fields all residing at a common point in spacetime)
\be
Z_{str} \left[ \left. \phi \right|_{\partial AdS}=J \right] =Z_{CFT} [J] \text{ .}
\ee


One of the cornerstones of the AdS/CFT correspondence is the so-called \emph{Planar Limit}, that is, the limit where the range of the gauge group $N_c$ goes to infinity. This limit was first described in \cite{Planar} and so it is also sometimes called \emph{'t Hooft Limit}. The idea behind it is that we can classify Feynmann diagrams of an $SU(N_c)$ gauge theory by their topology, so the dependence of the diagrams on the Yang-Mills coupling constant $g_{YM}$ and the number of colors $N_c$ is given by $g_{YM}^{2(P-V)} N^{F-L}$ where $P$ is the number of propagators, $V$ the number of vertices, $L$ the number of quark loops and $F$ the number of faces, which is equal to the sum of quark\footnote{By quark here we mean matter in the fundamental representation. Matter in other representations has to be treated in a different way.} loops $L$ and index loops $I$. Using Euler's Theorem $F-P+V=\chi=2-2H$, where $\chi$ is the Euler characteristic and $H$ counts the number of ``holes'' of the surfaces ($H=0$ for the sphere, $H=1$ for the torus, etc), the prefactor can be rewritten as $(g_{YM}^2 N_c)^{P-V} N_c^{2-2H-L}$. As a consequence, diagrams with no quark loops and planar topology, that is diagrams with $H=0$ or without self-intersections, dominate at large $N_c$. The immediate corolary is that a quarkless theory, which means $L=0$ for every diagram, with a large number of colors it is more naturally described as a perturbation theory in the combination $\lambda=g_{YM}^2 N_c$, called \emph{'t Hooft coupling constant}, than in the usual Yang-Mills coupling constant $g_{YM}$. This can be seen as the gauge theory turning into surfaces, which remind of the genus expansion in string theory generated by joining and splitting of string. This suggest a possible relationship between both, in which gauge theory diagrams would triangulate the worldsheet of an effective string \cite{Ooguri_2002}. The AdS/CFT correspondence is a concrete realization of this connection.

We can classify different versions of the AdS/CFT duality depending on the range of validity of it:
\begin{itemize}
	\item The weakest version: the duality is only valid in the planar limit with $\lambda=g_{YM}^2 N_c \gg 1$ in the CFT side and in the supergravity approximation (the limit where the string coupling constant $g_{\text{str}} \rightarrow 0$) of the string theory with $g_{\text{str}} N_c \gg 1$ restriction\footnote{This restriction is a consequence of the coupling constants being related as $\frac{g_{YM}^2}{g_{\text{str}}}=$ const., where the precise constant depends on the normalization of the actions. The two main choices are 1, which is used for example in \cite{Boer_2015}, and $4\pi$, which can be found in \cite{Polchinski_2011}.}.
	
	\item A stronger version will be to move away from the low energy limit, that is, to include $\alpha '$ corrections to the string theory, but remaining in the $g_{\text{str}} \rightarrow 0$ limit. In the gauge theory this corresponds to taking into account all $\lambda$ corrections but still in the planar limit.
	
	\item The strongest version would be a duality between the two full theories, that is, for any values of $g_{\text{str}}$ and $N_c$ (or $g_{YM}$ and $N_c$).
\end{itemize}

\begin{figure}[t]
\begin{center}
	\includegraphics[width=10cm,keepaspectratio]{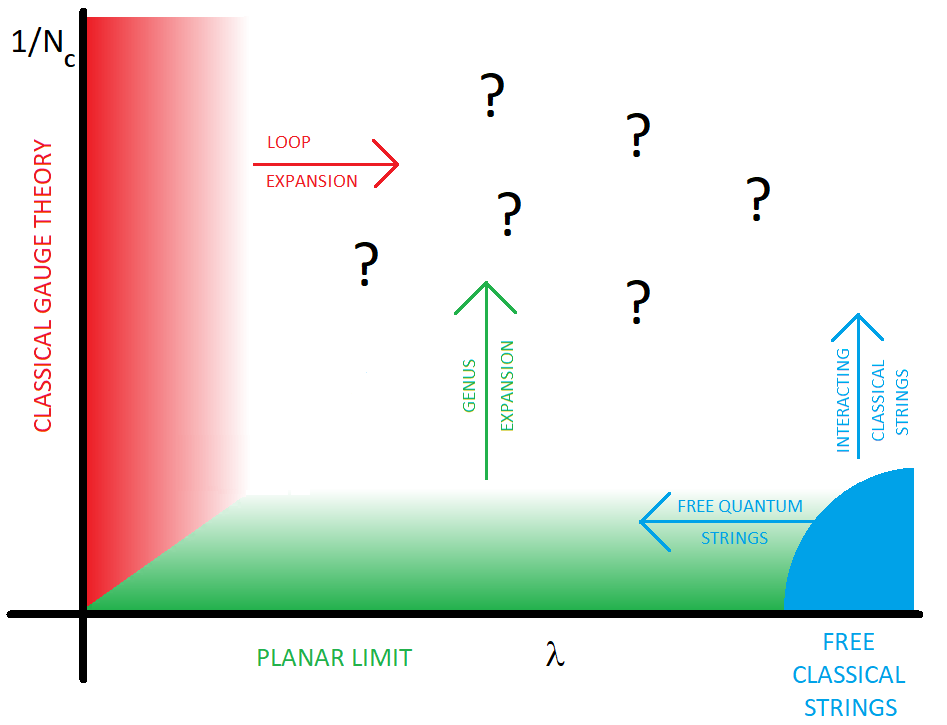}
\caption{Map of the parameter space of $\mathcal{N}=4$ SYM.}
\label{parameters}
\end{center}
\end{figure}

As $\lambda$ is the natural expansion parameter of the SYM theory, the weakest AdS/CFT correspondence will relate the weak-coupling limit of the string theory with the strongly coupled limit of the SYM theory. In the same way, we can examine the weak-coupling limit of the SYM theory, but we will find that the dual string theory will be a strongly coupled one. This makes it a \emph{strong/weak duality}. The different regimes of the theories are represented in Figure~\ref{parameters}: \emph{Classical gauge theory}, where we work at weak 't Hooft coupling (or strong $\alpha'$) $\lambda=g_{YM}^2 N_c=\frac{R^4}{\alpha ^{'2}}\ll 1$ and general value of the number of colors $N_c$; the \emph{planar limit} where we  perform calculations in the limit $N_c\rightarrow \infty$; and the \emph{free classical strings limit}, where we work around the point $\lambda=\infty$ and $g_{str}=\frac{\alpha ^{\prime 2}}{R^4 N_c}\ll 1$. In the classical gauge theory limit we can obtain more accurate estimations by the usual procedure of pertubative gauge theory. In the planar limit we can do the same by taking non-planar Feynman graphs into account. In the classical strings limit we can perform two different expansions, an expansion in $\lambda$ by adding quantum corrections to the worldsheet sigma model, or an expansion in $g_{str}$ by adding handles to the string worldsheet.

Solving the theory requires to compute all its observables. For a gauge theory there are the scaling dimensions (the sum of the constituents dimensions plus quantum corrections from interactions between them), scattering amplitudes, structure constants, expectations values of Wilson loops, etc. One of the most powerful tools to perform these computations is the use of symmetries of the theory. For example, conformal invariance heavily restricts the functional form of correlation functions. In the case of the two-point functions it is restricted to
\begin{equation}
	\esperado{\O_1 (x_1) \O_2 (x_2)}=\frac{\sqrt{\mc{N}_1 \mc{N}_2} \delta_{\Delta_1,\Delta_2}}{|x_{1}-x_{2}|^{\Delta_1+\Delta_2}} \text{ ,}
\end{equation}
which is completely fixed up to two constants: the scaling dimension $\Delta_i$, which is characterized by how the operator transforms under a dilatation, and the normalization $\mc{N}_i$, which can be set to one in general. In a similar way the three-point function can be constrained
\begin{equation}
	\esperado{\O_1 (x_1) \O_2 (x_2) \O_3 (x_3)} =\frac{\sqrt{\mc{N}_1 \mc{N}_2 \mc{N}_3}C_{123}}{|x_{12}|^{\Delta_1+\Delta_2-\Delta_3} \, |x_{13}|^{\Delta_1-\Delta_2+\Delta_3} \, |x_{23}|^{-\Delta_1+\Delta_2+\Delta_3}} \text{ ,}  \label{conformalcorrelators}
\end{equation}
where $x_{ij}=x_i-x_j$ and $C_{123}$ are called \emph{structure constants}. Again the correlation function is completely constrained up to the structure constants and the scaling dimensions. However conformal invariance is not enough to completely constrain the functional form of the four-point functions
\begin{displaymath}
	\esperado{\O_1 (x_1) \O_2 (x_2) \O_3 (x_3) \O_4 (x_4) } =\mathcal{F}_{1234} \left( \frac{|x_{12}| \, |x_{34}|}{|x_{13}| \, |x_{24}|} \, , \, \frac{|x_{14}| \, |x_{23}|}{|x_{13}| \, |x_{24}|} \right) \prod_{1\leq i<j \leq 4} \frac{1}{|x_{ij}|^{\Delta_i+\Delta_j-\frac{\sum_k{\Delta_k}}{3}}} \text{ ,} \notag
\end{displaymath}
where $\mathcal{F}$ is an arbitrary function. The quantities $\frac{|x_{12}| \, |x_{34}|}{|x_{13}| \, |x_{24}|}$ and $\frac{|x_{14}| \, |x_{23}|}{|x_{13}| \, |x_{24}|}$ are usually refereed in the literature as \emph{conformal cross-ratios} (sometimes they are also referred as double ratios, anharmonic ratios or simply as ratios).

Another way to simplify our computation of observables is to use the Operator Product Expansion (OPE). The idea behind this expansion is that, in the limit of the positions of two operator insertions approaching one another, the product of those two operators can be approximated as a series of local operators\footnote{Actually the operator in the right hand side can be inserted in any point on the line $z_\lambda =\lambda x +(1-\lambda ) y$, $\lambda \in \mathbb{R}$. Although different points result in a different choices of the structure constants $C_{ij}^k (x-y)$, they depend only on the difference $x-y$ regardless of this choice. The choice of this insertion point depends on the author. For example \cite{CardyCFT} chooses $\lambda=\frac{1}{2}$ while \cite{9780521672276} chooses $\lambda=0$, being this last one the most common choice in string theory textbooks and articles.}
\begin{equation}
	\O_i (x) \O_j (y) = \sum_k C_{ij}^k (x-y) \O_k (y) \ . \label{OPE}
\end{equation}
The $C_{ij}^k (x-y)$ functions are also called \emph{structure constants} and, although they are not the same as the $C_{123}$ constants from the three-point functions, they are related. Note that this is an operator statement, meaning that it only holds inside a general expectation value as long as the distance between $x$ and $y$ is small compared with the distance to any other operator. This expansion can be done to arbitrary accuracy and within a conformal field theory, like $\mc{N}=4$ SYM, the expression is not asymptotic but exact at finite separation. Performing several expansions we can write all correlation functions involving $n$ local operators in terms of correlation functions involving only two or three operators\footnote{This is not completely true if we are working on the planar limit of the theory, as higher-point functions of single trace operators in general need information about multi-trace operators even in this limit. However there are certain limits where these contributions are suppressed and the OPE expansion can be performed only with single trace operators. We refer to \cite{2016arXiv161105577F} for a complete discussion.}.

Another very powerful method that has been widely used in the recent years and allow us to perform exact computations of some particular quantities in supersymmetric field theories is the \emph{supersymmetric localization}. This method was already known in the context of cohomology theory and topological field theories (see the Duistermaat-Heckman \cite{Duistermaat_1982} and Atiyah-Bott-Berline-Vergne \cite{Atiyah_1984,BerlineVergne} formulae) but it was the computation of partition functions on $\mc{N}=2$ supersymmetric theories made Pestun \cite{Pestun_2012} what brought the current attention to this method. The main idea is the following: If our action is invariant under a fermionic symmetry generated by the supercharge $Q$, we can try to construct a functional $V$ such that $QV$ is $Q$ invariant ($Q(QV)=0$) and has a definite positive bosonic part. If we add this extra $QV$ term to the partition function in the following way
\begin{equation}
	Z(t)=\int \mc{D} \Phi e^{-S[\Phi]-t QV} \ ,
\end{equation}
we can prove that such partition function is independent of the parameter $t$
\begin{equation}
 \frac{d Z(t)}{dt}=\int \mc{D} \Phi e^{-S[\Phi]-tQV} (-QV)=-\int \mc{D} \Phi Q(e^{-S[\Phi]-tQV} V)=0 \ ,
\end{equation}
if we assume that the measure are $Q$-invariant, that is, the fermionic symmetric is not anomalous. This implies that vacuum expectation values of operators depend only on the $Q$-cohomology of the operator and they are independent of the inclusion of the extra $QV$ term. Therefore, we can perform all our computations in the large $t$ limit, where the path integral is given by the saddle point approximation $\left. Q V\right|_{bos}[\Phi_0 ]=0$, as other configurations are going to be suppressed because $QV$ is definite positive. Keeping up to quadratic expansion around this fixed point we get
\begin{equation}
	Z(t\rightarrow \infty )=\int_{\left. Q V\right|_{bos}[\Phi_0 ]=0}{\mc{D}\Phi_0  e^{-S[\Phi_0]} \frac{1}{\text{SDet} (QV[\Phi_0]_{\text{quad}})}} \ .
\end{equation}
With proper treatment if bosonic zero modes are present. And, because the partition function is independent of $t$, this saddle point approximation has to become \emph{exact}.

In this thesis we mostly will make use of a third tool at our disposal, the integrability of both the classical string theory and the $\mc{N}=4$ SYM gauge theory, chose implications are going to be explained in the following two sections.

\section{Integrability on the CFT side}
\label{MinahanZaremboargument}

In \cite{MinahanZarembo} Minahan and Zarembo showed that the $\mc{N}=4$ SYM one-loop dilatation operator for scalar operators in the planar limit is isomorphic to the Hamiltonian of an $SO(6)$ spin chain. The main idea behind this isomorphism comes from writing the operators in a basis of single trace operators made of products of scalar fields
\begin{equation}
	\O (\psi )=\psi^{i_1 , \dots , i_L} \Tr \{ \Phi_{i_1} \dots \Phi_{i_L} \} \ ,
\end{equation}
where $\Phi_{i}$ is a generic scalar $i=1,\dots ,6$. Therefore each operator $\O (\psi )$ is associated with an $SO(6)$ tensor with $L$ indices $\psi^{i_1 , \dots , i_L}$. These tensors form a linear space $\mathscr{H}=\bigotimes_{l=1}^L V_l$ with $V_l=\mb{R}^6$, which can be understood as a lattice with $L$ sites whose ends are identified and each lattice site host a six-dimensional real vector. Therefore it can be regarded as the Hilbert space of a spin system.

Composite operators have to be renormalized due to the emergence of UV divergences in the loop integrals of Feynman diagrams. Renormalized operators in general are linear combinations of bare operators, so we can write
\begin{equation}
	\O^A_{\text{ren}}=Z^A_{\text{\phantom{A}}B} (\lambda, \Lambda ) \O^B \ ,
\end{equation}
where the renormalization factor $Z^A_{\text{\phantom{A}}B}$ depends on the UV cutoff $\Lambda$ and on the ’t Hooft coupling $\lambda$ in the large $N_c$ limit. By standard arguments, the matrix of anomalous dimensions can be computed as
\begin{equation}
	\Gamma =\frac{d \ln Z}{d\ln \Lambda} \ ,
\end{equation}
whose eigenvalues determine the anomalous dimensions of the (multiplicatively renormalizable) operators on our theory. After computing and adding the only three possible kinds of diagrams that contribute at one-loop in the planar limit\footnote{As we commented above, the non-planar graphs are suppressed by a factor of $1/N^2$. However we also need the length of the operator $L$ to be $L\ll N$ to suppress the non-planar contributions, as there are $L!$ tree level diagrams of which only $L$ are planar.} (gluon exchange, $\Phi^4$ interaction and self-energy corrections) the renormalization factor reads
\begin{align}
	Z_{\dots i_{l} i_{l+1} \dots}^{\dots j_{l} j_{l+1} \dots} &=\mb{I}+ \frac{\lambda}{16\pi^2 }\ln \Lambda  \left( \delta_{i_{l} i_{l+1}} \delta^{j_{l} j_{l+1}} +2\delta^{j_{l}}_{i_{l}} \delta^{j_{l+1}}_{i_{l+1}} -2\delta^{j_{l+1}}_{i_{l}} \delta^{j_{l}}_{i_{l+1}} \right) \\
	&=\mb{I}+ \frac{\lambda}{16\pi^2 }\ln \Lambda  \left( \mc{K} +2\mb{I} -2\mc{P} \right) \ ,
\end{align}
for each link of the lattice. The total $Z$ factor is the product over all links. The operator $\mc{K}$ is called \emph{trace operator} and the operator $\mc{P}$ is called \emph{permutation operator}. They act as
\begin{align}
	\mc{K} (a\otimes b) &=(a \cdot b) \sum_i \hat{e}^i \otimes \hat{e}^i \mc{K} \ , \\
	\mc{P} (a\otimes b) &= (b\otimes a) \mc{P} \ ,
\end{align}
where $\hat{e}^i$ are orthonormal unit vectors in $\mb{R}^6$. With the expression for the renormalization factor we can compute the matrix of anomalous dimensions
\begin{equation}
	\Gamma =\frac{\lambda}{16 \pi^2} \sum_{i=1}^L (\mc{K}_{l,l+1} +2 -2\mc{P}_{l,l+1} ) \ , \label{MZHamiltonian}
\end{equation}
which can be interpreted as the local Hamiltonian of an $SO(6)$ spin chain. Hence the isomorphism of states can be generalized to an isomorphism between the anomalous dimensions and the spectrum. It is important to point out that this Hamiltonian was known to be integrable \cite{Reshetikhin_1983,Reshetikhin_1985}, so it can be diagonalized using the Bethe Ansatz technique (either the coordinate or the algebraic version, both ansätze will be reviewed in chapter \ref{Betheansatze}).

The isomorphism between the dilatation operator and a Hamiltonian was later expanded to the full $PSU(2,2|4)$ superconformal symmetry at one-loop \cite{BeisertStaudacher_2003,Beisert_2004}. It was also proven to be generalizable to two-loops in the planar limit \cite{Beisert_2003}, although non-planar corrections were proven to be non-integrable. This last development prompted people to think about the existence of a Hamiltonian that captures the full non-perturbative planar structure of the dilatation operator. A proposal for the $SU(2)$ sector appeared \cite{BDS} based on the integrability of the theory, field theory considerations and comparisons with string theory results. This proposal together with similar arguments for the $SU(1|1)$ and $SL(2)$ sectors lead to a more general hypothesis for the full $PSU(2,2|4)$ spin chain describing planar $\mc{N}=4$ Super Yang-Mills Theory (asymptotically) to arbitrary loop order was proposed in \cite{Beisert_2005}. However this description breaks down at some point due to finite size corrections\footnote{The spin chain Hamiltonian associated to $\mc{N}=4$ SYM at $K$-loops has interactions involving, at most, up to the $K^{\text{th}}$ nearest neighbour. Hence, the description breaks down when the loop order is greater than the length of the spin chain, and finite size or \emph{wrapping} corrections have to be incorporated. Note that this also imply that the Hamiltonian associated to the dilatation operator beyond one-loop has long-range interactions.} so more sophisticated machinery were developed. The Y-system \cite{Gromov_2009,Cavagli__2011}, the Thermodynamical Bethe ansatz \cite{Bombardelli_2009,Arutyunov_2009,Gromov_2012}, and the Quantum Spectral Curve \cite{Gromov_2015} are examples of upgrades of the usual Bethe ansätze that are able to deal with long-range interactions, wrapping effects, or both. In part III of this thesis we are going to make an extensive use of these isomorphisms.

\section{Integrability on the string theory side}\sectionmark{Integrability on the $A\MakeLowercase{d}S$ side}

As the action for the $AdS_5 \times S^5$ superstring has a complicated structure, people have tried to bypass this difficulty by considering special limits involving other parameters apart from the ’t Hooft coupling $\lambda$ to simplify computations. The two main approaches are
\begin{itemize}
	\item The BMN limit: in \cite{BMN} Berenstein, Maldacena and Nastase considered the case of near-BPS states, which are related to near point-like strings rotating along the great circle of $S^5$ with angular momentum $J\gg 1$. In the limit $J\rightarrow \infty$ but $J/\sqrt{\lambda}$ finite one obtain a string theory in the pp-wave background and it is possible to compute the dispersion relation
	\begin{equation}
		E-J=\sqrt{1+\frac{2\pi g N_c n^2}{J^2}} \ , \label{BMNdispersion}
	\end{equation}
	where $R^4 =2\pi g N_c \alpha^{\prime 2}$ is the AdS radius.
	
	The most important part of this limit is the possibility to identify the unique state with zero light-cone Hamiltonian with the chiral primary operator of $\mc{N}=4$ SYM,
	\begin{equation}
		\frac{1}{\sqrt{J} N_c^{J/2}} \Tr [Z^J ] \longleftrightarrow \ket{0,p_+ }_{l.c.}
	\end{equation}
	where $Z$ is a complex scalar. The energy of the state can be identified with the conformal dimension of the operator, as both are protected from corrections.
	
	Luckily the identification does not end with the ground state. If we move to the case of modes with $\Delta-J=E-J=1$, on the string theory side new states can be constructed by applying the zero momentum oscillators $a^i_0$ and $S^i_0$, with $i=1,\dots ,8$, to the light-cone vacuum (because the total light-cone energy is equal to the total number of oscillator that are acting, as they are massive modes), while on the field theory side we have the possibility of adding four possible scalars (the scalars that do not form the complex scalar $Z$), the four possible $\mb{R}^4$ derivatives and the eight components with $J=\frac{1}{2}$ of the sixteen component gaugino. As we have the same number of possibilities in both sides it is very tempting to identify both of them, but to do that we have to prove that adding more fields is an operation equivalent to adding excitations and that the dispersion relation is equal to the conformal dimensions.
	
	The first part is easily proven, as each time we act with a rotation of $S^5$ that does not commute with the $SO(2)$ symmetry singled out by the $Z$ operator we modify the chiral primary into
	\begin{equation}
		\frac{1}{\sqrt{J}} \sum_l \frac{\Tr [ Z^l \phi^r Z^{J-l} ]}{\sqrt{J} N_c^{(J+1)/2}} =\frac{\Tr [\phi^r Z^{J}]}{N_c^{(J+1)/2}} \ ,
	\end{equation}
	where $\phi^r$ is a scalar that does not form part of $Z$ and the equality comes from the cyclicity of the trace. If we apply another time a rotation of the same kind we can change the $Z$ operators that were left unchanged by the first rotation or we can apply it to the same operator as the first rotation. However, this second case is subleading in a $\frac{1}{J}$ expansion and can be neglected in the large $J$ limit we are considering. A similar scenario can be proven for other transformations. 
	
	The second part is a little bit more tricky to prove. We can compute the one-loop correction to the dilatation operator (as described in the previous section) and compare it with the expansion of the BMN dispersion relation $E-J\approx 1+\frac{2\pi g N_c n^2}{J^2} +\dots $, which they agree when we write $gN_c=\sqrt{\lambda}$ and identify $\frac{n}{J}$ with the momentum of the excitations. The details of the computation can be found in \cite{BMN}, together with a very incomplete Hamiltonian realization of the dilatation operator that, although it reproduces the full BMN dispersion relation, leaves out many diagrams and operators.
	
	After stating these two-points we can say that the ``string of $Z$ operators'' becomes the physical string and a correspondence between the energies of the string states and scaling dimensions of $\mc{N}=4$ SYM can be established in the near-BPS limit.

	\item The GKP string: the opposite case is to investigate strings far from the BPS states. A very important result was found in \cite{GKP} by Gubser, Klebanov and Polyakov, where they considered strings with large spin in $AdS_5$. On one hand, they found that the dispersion relation of these kind of string can be written as
	\begin{equation}
		E-S=f(\lambda ) \ln S +\dots
	\end{equation}
	where $f(\lambda )=b_0 \sqrt{\lambda} +b_1 +\frac{b_2}{\sqrt{\lambda}}+\dots$. On the other hand, they found that the conformal dimension of twist two operators (operators formed by a high number of derivatives and two scalars) have the same expression but with $f(\lambda )=a_1 \lambda +a_2 \lambda^2 +\dots$. As the expression are the same up to the $f(\lambda )$ function, they might represent different asymptotics of the same function. This means that we can identify the GKP strings with twist two operators.
	
\end{itemize}

These results helped to present a map between classical string solutions and Riemann surfaces (finite gap equations) \cite{Kazakov2004}. The structure of cuts connecting the sheets composing these Riemann surfaces bear some resemblance with the strings of solutions of the Bethe equations in the scaling limit, which lead to the proposal of a quantum string Bethe ansatz in \cite{ArutyunovStringBethe}. However it was shown a discrepancy between the BMN scaling and four-loop perturbation theory in the field theory side (while integrability persists) \cite{Fischbacher2004}. This discrepancy was later explained by using the freedom of adding an scalar phase (dressing phase) to the S-matrix.

	Similar expansions can be expected for other semiclassical string states. However we will focus our attention into the following case: some particular multispin string states (with at least one large angular momentum in $S^5$) have regular expansion in $\frac{\lambda}{J^2}$ while the quantum superstring sigma model corrections are suppressed in the limit $J\rightarrow  \infty$ with $\frac{\lambda}{J^2}$ constant. The first step to test the non-BPS sector of AdS/CFT was made with the construction of solutions with one spin $S$ and one angular momentum $J$ \cite{FrolovSemiclassical}, which can be obtained by boosting the center of mass of the string rotating in $AdS_5$ along a circle of $S^5$. Solutions with two and three angular momenta were constructed in \cite{RussoTwoSpins} and \cite{Multi-spin} respectively. The possibility to test the non-BPS sector of AdS/CFT by comparing the $\frac{\lambda}{J^2}\ll 1$ limit of the dispersion relation/conformal dimension was successfully accomplished in \cite{stringingspins-spinningstrings,Rotatingstrings,Arutyunov_2003} among others. In part II of this thesis we are going to focus mainly in these kind of string states.

In addition, the world-sheet sigma model on $AdS_5 \times S^5$ background supported by a self-dual Ramond-Ramond five-form flux was proven to be classically integrable by the explicit construction of its Lax connection \cite{Mandal2002,Bena_2004,AldayIntegrability}. Later the exact $S$-matrix for the world-sheet excitations of this theory was found in \cite{Arutyunov_2007} using that, in the uniform light-cone gauge, the symmetry breaks to $PSU(2|2)^2$ and can be centrally extended by relaxing the level-matching condition\footnote{We have to relax the level-matching condition because, although a multiparticle state would fulfil it, arbitrary pairs of particles forming this state do not necessarily obeys it, therefore the two particle S-matrix needs to be computed without imposing this condition.}. This $S$-matrix was shown to be equivalent to the $\mc{N}=4$ SYM $S$-matrix up to some twists \cite{Arutyunov_ZFalgebra}.

\section{Outline of the thesis}

The rest of this thesis is divided in three well differentiated parts. 
The first part of this thesis is devoted to the string theory side of the duality and it contains chapters two to four. In the second chapter we will present an introduction to string theory where we explain the construction of bosonic string theories, both from the Polyakov action and from the Principal Chiral Model over a symmetric coset, using the last as an example to introduce the concept of classical integrability and the toolbox it provides; and the construction of supersymmetric string theories, mostly using Wess-Zumino-Witten models over semi-symmetric cosets. In the third chapter we will use one of the tools provided by integrability to study a deformation of the $AdS_3 \times S^3$ background by the presence of mixed R-R and NS-NS fluxes. In particular that tool will be the rewriting of the string Lagrangian for spinning strings in $AdS_5 \times S^5$ as a Neumann-Rosochatius model, developed in \cite{Arutyunov_2004}. This will allow us to write analytical expressions for the dispersion relations for spinning strings as a series in inverse powers of the total angular momentum. In the fourth chapter we will apply the rewriting of the string Lagrangian in $AdS_5 \times S^5$ as a Neumann-Rosochatius model to study spinning strings in $\eta$-deformed $AdS_3 \times S^3$ space.

The second part of this thesis is centered around the spin chain interpretation of the field theory side of the duality and it constains chapters five to seven. In the fifth chapter we will present the main computational tools we are going to use: the Coordinate Bethe Ansatz and the Algebraic Bethe Ansatz. Some problems arising from the different normalization of the states obtained from both methods are also discussed. We will dedicate a section to the Beisert-Dippel-Staudacher (BDS) spin chain \cite{BDS}, an all-loops ansatz for the spin chain picture of $\mc{N}=4$ SYM. We will end this chapter by presenting the bootstrap program and discussing the Smirnov's form factors axioms \cite{Smirnov}. The sixth chapter is focused on the computation of two-point correlation functions. In particular we will present computations of form factors using the Algebraic Bethe Ansatz for the Heisenberg spin chain and for the BDS spin chain. The seventh chapter is instead focused on three-point functions. We will describe first the Tailoring method \cite{EscobedoTailoring,TailoringII,TailoringIII,TailoringIV,Tailoringnoncompact} and a proposal for an all-loop generalization, called the hexagon form factor\cite{BKVhexagon}. We end this chapter presenting an ``algebraic version'' of the hexagon proposal that gives an explanation to some of its characteristics.

We conclude this thesis with the third and last part, presenting a summary and conclusions and two appendices with some details on our computations.

\part{Integrability on the string theory side}

\chapter{Strings in coset spaces}

\begin{chapquote}{Lubos Motls, \textit{Formal string theory is physics, not mathematics}}
	The motivation of formal string theory is to understand the truly fundamental ideas in string theory which is assumed by the practitioners to be the theory explaining or predicting everything in the Universe that may be explained or predicted. How may someone say that the motivation is similar to that of mathematicians?
\end{chapquote}

In this chapter we will present a review of semi-classical string theory and classical integrability, together with the tools we are going to use for the rest of this part. In the first section we introduce the basis of classical bosonic string theory, mostly following \cite{Theisen-blumenhagen}, and an alternative construction using the Principal Chiral Model. This second construction is explained following \cite{Francesco} for the description of the models involved and \cite{ZaremboLesHouches} for the construction of the bosonic string theory. After explaining the Polyakov action, we will present the Principal Chiral Model Lagrangian and use it as an example to introduce the concept of classical integrability and the tools it provides. After it, the Wess-Zumino-Witten model is presented both as a way to introduce a $B$-field for a bosonic string theory and as knowledge needed for the extension of the method to supersymmetric theories. We end this section by constructing a bosonic string theory as a Principal Chiral Model over a symmetric coset. The second section is completely devoted to the construction of supersymmetric string theories, mostly following \cite{foundationsAdSstring} and \cite{HeinzeThesis,Cagnazzo-Zarembo} for some specific parts. First we will review the concepts of Neveu-Schwarz-Ramond superstring, Green-Schwarz superstring and the rewriting of the last one as a coset model. Before building the coset models explicitly we will present the superconformal algebra we will use for that intent and the idea of semi-symmetric spaces. After that we will construct a Wess-Zumino-Witten model on a semi-symmetric space, presenting some of the characteristics like the construction of the Lax and the $\kappa$-symmetric. Following \cite{foundationsAdSstring} and \cite{Cagnazzo-Zarembo} we will do it for the two cases we are interested in: the $AdS_5 \times S^5$ and the $AdS_3 \times S^3 \times T^4$ backgrounds.

\section{Principal Chiral Model}

We are going to start this section describing the basic concepts of a bosonic string theory. However, our end goal is to construct a supersymmetric string theory, so it will be more useful to write the theory as a coset sigma model. We will accomplish it at the end of the section. Before doing that we are going to introduce a $\sigma$-model called \emph{Principal Chiral Model} (PCM) to stablish some characteristics and to see how integrability arises. 

\subsection{Basic concepts. The Polyakov action}

As the action of a relativistic particle can be written as the length of its trajectory or \emph{world-line}, we can directly generalize it to write the action of a relativistic string as the area of the \emph{world-sheet} swept out by the string. 
This action, written using an auxiliary field $h_{\alpha \beta}$ that can be interpreted as a metric for the world-sheet, is called the \emph{Polyakov action},
\begin{equation}
	S_P=-\frac{T}{2} \int_S{d^2\sigma \sqrt{-h} h^{\alpha \beta} G_{\mu \nu}(X) \partial_\alpha X^\mu \partial_\beta X^\nu } \ ,
\end{equation}
where $\sigma^\alpha=(\tau , \sigma )$ are the two coordinates of the world-sheet (chosen such that $\tau_1 <\tau <\tau_2$ and $0\leq \sigma <l$), $X^\mu (\tau , \sigma )$ are the embeddings functions, $h=\det h_{\alpha \beta}$, $G_{\mu \nu}$ is the metric of the space-time and $T$ is a constant called \emph{string tension}. This action is not only invariant under re-parametrization of the world-sheet but also Weyl invariant (invariant under re-scaling of the auxiliary field). These two invariances allow us to fix the auxiliary field to $h_{\alpha \beta}=\eta_{\alpha \beta}=\text{diag} (-1,1)$, a choice called \emph{conformal gauge}, but with the equations of motion for this metric as constraints
. Both constrains are related with the energy-momentum tensor of the world-sheet theory, defined treating the auxiliary field as a metric
\begin{equation}
	T_{\alpha \beta}=\frac{4\pi}{\sqrt{-h}} \, \frac{\delta S_P}{\delta h^{\alpha \beta}}=- \frac{G_{\mu \nu}(X)}{\alpha '} \left( \partial_\alpha X^\mu \partial_\beta X^\nu -\frac{1}{2} h_{\alpha \beta} h^{\gamma \delta} \partial_\gamma X^\mu \partial_\delta X^\nu \right) \ ,
\end{equation}
where $\alpha ' =\frac{1}{2\pi T}$ is called \emph{Regge slope}. In particular, the diffeomorphism invariance imply that the energy-momentum tensor is conserved, the Weyl invariance requires it tracelessness $h^{\alpha \beta} T_{\alpha \beta}=0$ and re-parametrization invariance requires to fix it to zero $T_{\alpha \beta}=0$. These last conditions are usually called \emph{Virasoro constrains} because, together with energy-momentum conservation, give rise to an infinite number of conserved charges which form a Virasoro algebra.

\subsection{PCM lagrangian and integrability}
\label{PCMintegrability}

Let us consider a field $g(\sigma ,\tau )$ periodic in $\sigma$ that takes values over a Lie group. We define the Lagrangian density of the PCM as
\begin{equation}
	L=\frac{1}{4a^2} \sqrt{-h} h^{\alpha \beta} \text{Tr} \Big\{ (g^{-1}\partial_\alpha g) (g^{-1} \partial_\beta g) \Big\} \ ,
\end{equation}
where $h_{\alpha \beta}$ is a metric and $a$ is a coupling constant. There are some similarities between this Lagrangian and Polyakov action regarding the treatment of $h_{\alpha \beta}$. This allow us to impose conformal gauge $\sqrt{-h} h^{\alpha \beta}=\eta^{\alpha \beta}$ without further discussion. The domain of the coordinates is chosen as $0 <\tau <T$ and $0\leq \sigma <l$, so $g(\sigma +l,\tau )=g(\sigma ,\tau )$.

The Lagrangian is more conveniently written in term of left-invariant and right-invariant currents\footnote{The definition of left and right currents depend on the authors. This is because some authors define left-invariant and right-invariant currents (for example \cite{Lorenzothesis}) and other authors define Noether currents corresponding to multiplications of $g$ by a constant element of the group from the left and from the right (for example \cite{Gromov_finitegap}). This gives opposite definitions of these currents and some confusions. Sometimes the definition of the left-invariant current has an extra global minus sign (for example \cite{foundationsAdSstring}).}
\begin{align}
	j^L_\alpha &= g^{-1} \partial_\alpha g \ , & j^R_\alpha &= (\partial_\alpha g )  g^{-1}	\ ,
\end{align}
which are, respectively, the Noether currents associated to the transformation of the field by the right and left multiplication by a constant element of the group. Note that because $g$ is defined over a Lie group, the currents are defined over the corresponding Lie algebra. The Lagrangian can be written now as
\begin{equation}
	L=\frac{1}{4a^2} \text{Tr} \Big\{ j^L_\alpha j^{L ,\alpha} \Big\}=\frac{1}{4a^2} \text{Tr} \Big\{ j^R_\alpha j^{R,\alpha} \Big\} \ ,
\end{equation}
where we have used the cyclicity of the trace to get the second Lagrangian. The equations of motion associated to the first Lagrangian are
\begin{equation}
	\partial_\alpha j^{L, \alpha}=0 \ .
\end{equation}
These equations can be supported with an equation reflecting the fact that the current is exact
\begin{equation}
	\partial_\alpha j^L_\beta + \partial_\beta j^L_\alpha +[j^L_\alpha , j^L_\beta ]=0 \ ,
\end{equation}
with similar equations for the right current.

One can define now a new current, which we are going to call \emph{Lax connection}, that contains both equations at the same time
\begin{equation}
	L_\alpha (z) =\frac{j_\alpha + z \epsilon_{\alpha \beta} \sqrt{-h} h^{\beta \gamma} j_\gamma}{1-z^2} \ ,
\end{equation}
where $z$ is a real parameter called \emph{spectral parameter} and the current used can be either the left or the right current\footnote{The Lax connection of a system is not unique. Given an arbitrary matrix $f(\tau , \sigma ,z)$, the flatness condition is invariant under the gauge transformation
\begin{equation}
	L_\alpha \rightarrow L'_\alpha=f L_\alpha f^{-1} + (\partial_\alpha f )f^{-1} \ .
\end{equation}
If we choose the particular case $f(\tau , \sigma ,z)\propto g(\tau , \sigma)$ we can relate the Lax connection written in the left and in the right currents.}. This new current can actually be viewed as a flat connection $\partial_\alpha L_\beta + \partial_\beta L_\alpha +[L_\alpha , L_\beta ]=F_{\alpha \beta}=0$ and this flatness for all values of the spectral parameter implies both the equation of motion and the flatness of the original current $j_\alpha$.

This flatness equation, usually called \emph{zero curvature equation} in the literature, can be seen as the compatibility condition of the linear problem
\begin{equation}
	\left\{ \begin{array}{c}
	D_\sigma \Psi (\tau , \sigma ,z) =(\partial_\sigma -L_\sigma ) \Psi (\tau , \sigma ,z)=0 \\ 
	D_\tau \Psi (\tau , \sigma ,z) =(\partial_\tau -L_\tau ) \Psi (\tau , \sigma ,z)=0
	\end{array} \right. \ . \label{classicalwaveequation} 
\end{equation}
Solving this linear system will give us information about the solution of the Lagrangian\cite{BabelonTalon,ashokdas}. The function $\Psi (\tau , \sigma ,z)$, called \emph{classical wave function}, is determined up to a constant, usually fixed to $\Psi (0,0 ,z)=1$.

The existence of this connection implies the (classical) integrability of the Lagrangian. Its flatness allow us to define a well behaved parallel transport
\begin{equation}
	U_\gamma (\tau_2 ,\sigma_2 ; \tau_1 ,\sigma_1 )=\text{P exp } \left[ \int_\gamma {dx^\alpha L_\alpha (\tau ,\sigma , z)} \right] \ ,
\end{equation}
where $\gamma$ is a path from $( \tau_1 ,\sigma_1 )$ to $(\tau_2 ,\sigma_2 )$ and P stands for an ordering of the points along the path of integration such that the points closer to $(\tau_2 ,\sigma_2 )$ stand to the left of those closer to $( \tau_1 ,\sigma_1 )$. Using the Baker-Cambell-Hausdorff formula and Stoke's theorem we can prove that
\begin{equation}
	U_\delta (\tau_1 ,\sigma_1 ; \tau_2 ,\sigma_2 )U_\gamma (\tau_2 ,\sigma_2 ; \tau_1 ,\sigma_1 )=\exp \left[ -\frac{1}{2} \int_{\gamma + \delta}{d\sigma^{\alpha \beta} F_{\alpha \beta}} \right]\ ,
\end{equation}
therefore the vanishing of the curvature implies that the parallel transport defined by this connection is independent of the path. The parallel transport can be used to compute the wave function as $\Psi (\tau , \sigma ,z)=U_\gamma (\tau ,\sigma ; 0 ,0 )$. 

We will be particularly interested in the path given by constant $\tau$ and $\sigma$ varying from $0$ to $l$. The parallel transport for this particular path,
\begin{equation}
	T(\tau ,z)=\text{P exp } \int_{0}^l{d\sigma L_\sigma (\tau ,\sigma , z)} \label{monodromymatrix} \ ,
\end{equation}
is called \emph{monodromy matrix}. The $\tau$ evolution of the monodromy matrix can be computed in the following way
\begin{align}
	\partial_\tau T(\tau ,z) &= \int_{0}^l{d\Sigma \left[ \text{P exp } \int_{\Sigma}^l{d\sigma L_\sigma (\tau ,\sigma , z)} \right] \partial_\tau L_\sigma (\tau , \Sigma , z)} \left[ \text{P exp } \int_{0}^\Sigma{d\sigma L_\sigma (\tau ,\sigma , z)} \right] \notag \\
	&=\int_{0}^l{d\Sigma \, [...] \big( \partial_\sigma L_\tau +[L_\sigma , L_\tau ] \big) [...]}= \int_{0}^l{d\Sigma \,\partial_\sigma \big( [...] L_\tau [...] \big)} \notag \\
	&= \left[ L_\tau (\tau ,0 ,z), T(\tau ,z) \right] \ ,
\end{align}
where we have used the vanishing of the curvature and periodicity of the Lax connection in the $\sigma$ coordinate. From here it is obvious that the trace of the monodromy matrix $\mathcal{T}=$Tr~$(T)$, called \emph{transfer matrix}, is independent of the $\tau$ coordinate on-shell. Therefore if we expand this trace on inverse powers of the spectral parameter we get an infinite set of charges that are conserved (we will prove later that they Poisson-commute), proving the integrability of the Lagrangian. This formalism can be generalized to non-periodic solutions with $\sigma \in (-\infty , \infty )$. However the definition of the monodromy matrix case present some subtleties when we take the limit in which the endpoints go to infinity \cite{ashokdas,Caudrelier_2003}.

To end this section we are going to define the concept of \emph{R-matrix.} Although it is a useful concept in classical integrability, it is not as useful as in quantum integrability, where it will become the central element of the formalism. At the level of the Poisson Brackets of the Lax operators, we assume the existence of a matrix such that
\begin{equation}
	\left\{ L_{\sigma ,1} (\tau ,\sigma , z) \underset{,}{\otimes} L_{\sigma ,2} (\tau ,\sigma ' , z') \right\}= \left[ r_{12} (z,z') , L_{\sigma ,1} (\tau ,\sigma , z)+L_{\sigma ,2} (\tau ,\sigma ' , z') \right] \delta (\sigma -\sigma ') \ ,\label{rmatrixdefinition}
\end{equation}
this is an equation on $\mf{g}\otimes \mf{g}$, with $\mf{g}$ a Lie algebra, where the subindices $1$ and $2$ labels the algebra in which the operators act. This is the definition of the classical R-matrix, which is a $\mf{g}\otimes \mf{g}$-valued function. The Jacobi identity of the Poisson brackets implies the following property
\begin{equation}
	\left[ r_{12} (u) , r_{13} (u+v) \right] +\left[ r_{12} (u) , r_{23} (v) \right] +\left[ r_{13} (u+v) , r_{23} (v) \right] =0 \ , \label{classicalyangbaxter}
\end{equation}
called \emph{Classical Yang-Baxter Equation} (CYBE). This equation has been studied in detail \cite{Belavin_1983} and two important property are: 1. If $r(u)$ is a nondenerate solution of the CYBE meromorphic around $u=0$, then it can be extended meromorphically to the whole complex plane having only simple poles. 2. The set of these poles is a discrete subgroup of $\mathbb{C}$ relative to the addition and it allows to classify the R-matrices into three different categories: rational R-matrices (when the rank of this subgroup is zero), trigonometric R-matrices (when its rank is $1$) and elliptic R-matrices (when it has rank $2$, which only exists for $\mf{g}=\mf{sl}(n)$).

From the definition of the R-matrix we can find the Poisson-Lie brackets between transport matrices, called \emph{fundamental Sklyanin relation} \cite{BabelonTalon}
\begin{multline}
	\left\{ U_1 (\tau_1 ,\sigma_1 ; \tau_2 ,\sigma_2; z ) \underset{,}{\otimes} U_2(\tau_1 ,\sigma_1 ; \tau_2 ,\sigma_2 ; z') \right\}=\\=\left[ r_{12} (z,z') , U_1 (\tau_1 ,\sigma_1 ; \tau_2 ,\sigma_2; z ) U_2(\tau_1 ,\sigma_1 ; \tau_2 ,\sigma_2 ; z') \right] \ , \label{sklyaninrelation}
\end{multline}
whose consequence is that the traces of powers of the monodromy matrix generate Poisson-commuting quantities\footnote{To prove that we used the property $\Tr _{12} \{ A\otimes B\}=\Tr (A) \Tr (B)$.}, a statement equivalent to the one obtained from the expansion of the transfer matrix.

\subsection{WZW models}

The PCM is enough for writing bosonic string theories as sigma models. But to write supersymmetric string theories or to write a bosonic string theory with a B-field we need a further extension of the PCM called Wess-Zumino-Witten models.

To understand what we want to accomplish first we have to write the world-sheet in euclidean light-cone coordinates $z=\tau +i\sigma$ (usually called holomorphic coordinate) and $\bar{z}=\tau -i\sigma$ (usually called anti-holomorphic coordinate). In this new coordinates, conservation of left and write currents can be written as
\begin{equation}
	\partial_z j_{\bar{z}}+\partial_{\bar{z}} j_z= \partial \bar{j}+\bar{\partial} j=0 \ .
\end{equation}
The idea behind the extension is to enhance the symmetry of the model to make both components separately conserved $\partial \bar{j}=\bar{\partial} j=0$. This enhancement is done in a non-obvious way by adding a Wess-Zumino term
\begin{equation}
	S=S_0 -\frac{ki}{24\pi} \int_{\mc{B}}{d^3y \, \epsilon^{\alpha \beta \gamma} \text{Tr} \Big\{ (\tilde{g}^{-1}\partial_\alpha \tilde{g}) (\tilde{g}^{-1} \partial_\beta \tilde{g}) (\tilde{g}^{-1} \partial_\gamma \tilde{g}) \Big\} } \ ,
\end{equation}
where $\mc{B}$ is a 3-dimensional manifold whose boundary is the compactification of our original space and $\tilde{g}$ is the extension of our original field $g(z,\bar{z})$ to this manifold. However this extension is not unique, hence a potential ambiguity in the definition of this term arise. But the difference between two choices gives $k$ times a topological quantity, defined modulo $2\pi i$ ($\pi i$ if our group is $SO(3)$). So a well defined path integral needs $k$ to be integer (or an even integer for $SO(3)$).

If we derive now the equations of motion for this action we get
\begin{equation}
	\left( 1+\frac{a^2 k}{4\pi} \right) \partial \bar{j} + \left( 1-\frac{a^2 k}{4\pi} \right) \bar{\partial} j=0 \ ,
\end{equation}
thus the choice $a^2=4\pi /k$ selects the anti-holomorphicity of the anti-holomorphic current as the equations of motion. That is, it reduces the equations of motion to $\partial \bar{j}=0$. Furthermore, as $j_\alpha$ is a conserved current, it also imposes $\bar{\partial} j=0$. The equations of motion can then be solved by choosing $g(z,\bar{z})=f(z) \bar{f} (\bar{z})$ for arbitrary functions $f(z)$ and $\bar{f} (\bar{z})$, reminding a free-field theory.

It is very important to notices that the original $G_{\text{left}}\times G_{\text{right}}$ symmetry of the PCM is now enhanced to a local $G(z) \times G (\bar{z})$ symmetry\footnote{Which one acts on the left and on the right depends on if we are working with the left or the right invariant current. In particular, for the left invariant current we have $G_{\text{left}}(z) \times G_{\text{right}} (\bar{z})$.}.

\subsection{Bosonic string theory as a coset model} \label{bosoniccosetmodel}

We are going to move now to the construction of a PCM-like Lagrangian on a coset $G/H$. 
To do that we are going to impose the equivalence relation $g(\tau ,\sigma ) \equiv g (\tau , \sigma) h(\tau , \sigma)$, where $h(\tau , \sigma)\in H$, which will treat the fields $h(\tau , \sigma)$ as gauge fields. As in section~\ref{PCMintegrability}, it is better to write everything in term of the currents instead of the fields. The equivalence relation between fields maps into the decomposition of the currents into a direct sum of the two algebras $g^{-1} \partial_\alpha g\in \mathfrak{g}=\mathfrak{h} \oplus \mathfrak{f}$ where $\mathfrak{h}$ is the algebra associated to the group $H$ and $\mathfrak{f}$ is the orthogonal complement to $\mathfrak{h}$ in $\mathfrak{g}$. Hence dividing the current into $A_\alpha \in \mathfrak{h}$ and $K_\alpha \in \mathfrak{f}$, they transform under the equivalence relation as
\begin{equation}
	g(\sigma^\alpha) \equiv g(\sigma^\alpha) h(\sigma^\alpha) \Longrightarrow \left\{\begin{array}{l}
	A_\alpha \equiv h^{-1} A_\alpha h+h^{-1} \partial_\alpha h \\
	K_\alpha \equiv h^{-1} K_\alpha h
\end{array}	  \right. \ , 
\end{equation}
where we have assumed that the algebra we are using is simple, so $h^{-1} \partial_\alpha h\in \mathfrak{h}$. This transformation property imply that the $A_\alpha$ field can be understood as a gauge field as it has the same transformation properties as one, while $K_\alpha$ only undergoes a similarity transformation. Thus a Lagrangian of the Principal Chiral Model on the coset $G/H$ can be written as a gauge invariant Lagrangian
\begin{equation}
	L=-\frac{1}{4} \text{Tr} \Big\{ K_\alpha K^\alpha \Big\} \ ,
\end{equation}
whose equations of motion are $D_\alpha K^\alpha=0$, where we define the covariant derivative as $D_\alpha =\partial_\alpha  + [A_\alpha ,\cdot]$.

In principle, having a simple algebra ensures that $[\mathfrak{h} , \mathfrak{h}] \subset \mathfrak{h}$ and $[\mathfrak{h} , \mathfrak{f}] \subset \mathfrak{f}$, so the equations of motion are well defined. Nevertheless it is interesting to also impose $[\mathfrak{f},\mathfrak{f}]\subset \mathfrak{h}$. This extra restriction allows us to introduce a $\mathbb{Z}_2$ symmetry that acts as $\Omega (\mathfrak{h})=\mathfrak{h}$ and $\Omega (\mathfrak{f})=-\mathfrak{f}$. These kinds of cosets are called \emph{symmetric cosets}. The reason to impose the last condition is related to the flatness condition of $K_\alpha$ and, by extension as we saw in the last section, to the integrability of the Lagrangian.

Because $G/H$ is a symmetric coset, the flatness condition can be broken into
\begin{equation}
	F_{\alpha \beta}+D_\alpha K_\beta -D_\beta K_\alpha +[K_\alpha , K_\beta]=0 \Longrightarrow \left\{ \begin{array}{l}
	F_{\alpha \beta} +[K_\alpha , K_\beta]=0 \\
	D_\alpha K_\beta -D_\beta K_\alpha=0
	\end{array} \right. \ , 
\end{equation}
where $F_{\alpha \beta}$ is the field strength associated to $A_\alpha$, allowing us to define the following Lax connection
\begin{equation}
	L_\alpha =A_\alpha +\frac{z^2+1}{z^2-1} K_\alpha +\frac{2z}{z^2-1} \epsilon_{\alpha \beta} K^\beta \ ,
\end{equation}
whose existence implies the integrability of the system.

\section{Supersymmetric string as coset model}

In this section we will discuss how to construct a supersymmetric string theory. First we present the two different approaches to construct it, the Neveu-Schwarz-Ramond superstring and the Green-Schwarz superstring, of which we will present their most important characteristics. Of those approaches we are going to choose the second one as our approach to supersymmetric strings, as this one can be constructed using coset models in a similar way as the bosonic string in the previous subsection \ref{bosoniccosetmodel}. Before doing it, we will study the supersymmetric algebra we are going to use as the group over which we are going to quotient. Finally, we will take all the elements together to present the construction of the GS action.

\subsection{NSR and GS string theories}

There are two ways to introduce supersymmetry in a string theory: the Neveu-Schwarz-Ramond superstring (NSR) and the Green-Schwarz superstring (GS). Both theories should give the same physical results and have supersymmetry both in the world-sheet and in the space-time. The only difference between them is which of those two supersymmetries is explicit.

\begin{itemize}
	\item NSR superstring introduces the supersymmetry in the world-sheet. It does that by introducing Grassmann superspace coordinates $\theta^A_\alpha$, where $A=1,2$ count two different components. These coordinates should be spinors under the usual Clifford algebra in the conformal gauge, so we can define the matrices
	\begin{equation}
		\{ \rho^\alpha , \rho^\beta \}=-2\eta^{\alpha \beta} \ .
	\end{equation}
	We can generalize now our space-time coordinates to superfields, assuming that its fermionic components are Majorana (real) fermions
	\begin{equation}
		Y^\mu (\tau, \sigma , \theta )=X^\mu (\tau ,\sigma ) +\bar{\theta} \psi^\mu (\tau ,\sigma ) +\frac{1}{2} \theta \bar{\theta} B^\mu (\tau ,\sigma) \ ,
	\end{equation}
	where spinor indices are omitted. Because our action should be invariant under the supersymmetric charges $Q_A = \frac{\partial}{\partial \bar{\theta}^A} + i (\rho^\alpha \theta)_A \partial_\alpha$, we have not only to generalize our coordinates but also our derivatives to a covariant derivative. We can check that $D_A=\frac{\partial}{\partial \bar{\theta}^A} - i (\rho^\alpha \theta)_A \partial_\alpha$ indeed satisfy $\{ D_A , Q_B \}=0$. Therefore we can choose our action as
	\begin{equation}
		S_{RNS}=\frac{iT}{4} \int{d^2 \sigma \, d^2 \theta \, \bar{D} Y^\mu D Y_\mu} \ ,
	\end{equation}
	where 
	\begin{align*}
	DY^\mu = \psi^\mu +\theta B^\mu -i \rho^\alpha \theta \partial_\alpha X^\mu +\frac{i}{2} \bar{\theta}\theta \rho^\alpha \partial_\alpha \psi^\mu \ , \\
	\bar{D}Y^\mu = \bar{\psi}^\mu + B^\mu \bar{\theta} +i \partial_\alpha X^\mu \bar{\theta} \rho^\alpha -\frac{i}{2} \bar{\theta}\theta \rho^\alpha \partial_\alpha \psi^\mu \ .
	\end{align*}
	Expanding the derivatives and the fields and using that $\int{d^2 \theta \, \bar{\theta} \theta}=-2i$ is the only non-vanishing integral of Grassmann coordinates, gives us
	\begin{equation}
		S_{RNS}=\frac{-T}{4} \int{d^2 \sigma \, \left( \partial_\alpha X^\mu \partial^\alpha X_\mu -i\bar{\psi} \rho^\alpha \partial_\alpha \psi_\mu -B^\mu B_\mu \right)} \ ,
	\end{equation}
	where we can see that $B^\mu$ is an auxiliary field whose field equations imply $B^\mu=0$\footnote{In reality, things are more complex because the auxiliary field that is the world-sheet metric $h_{\alpha \beta}$ also needs to be supersymetrized. Hence we don't work with it, but with the \emph{zweibein} $e^\alpha _a$ defined as $e^\alpha_a e^\beta_b \eta^{ab}=h^{\alpha \beta}$ and their supersymmetric partner, a Majorana spinor-vector called \emph{gravitino}. However, at the end we can gauge away these fields and get a \emph{superconformal gauge} where we put the gravitino to zero and the zweibein to identity.}.
	
	This formalism has some problems: first of all, the two components of fermions $\psi_+$ and $\psi_-$ can be made independently periodic (called Ramond boundary conditions) or antiperiodic (called Neveu-Schwarz boundary conditions) in the sigma coordinate, but we cannot have all four combinations at the same time. We have to project out some of them, this process is called the GSO (Gliozzi-Scherk-Olive) projection. Second we are interested in backgrounds like $AdS_5 \times S^5$, which has a five-form (Ramond) flux, but the Ramond-Ramond vertex needed to construct them is non-local in terms of world-sheet fields \cite{Friedan_1986}, so it is unclear how to couple it to the string world-sheet. Because of these problems we are going to work with the other formalism.
	
	\item GS superstring instead introduces the supersymmetry directly in the space-time. Constructing the GS action for arbitrary superstring solutions is difficult because the full structure of the superfields has to be determined from the bosonic solution, something not generally known. However, inspired from formulations like \cite{Henneaux_1985}, where the authors constructed the flat space GS action as a WZW-type non-linear sigma model on the Poincaré group modded by $SO(1,9)$, other constructions have been developed. In particular
	\begin{align*}
		AdS_5 \times S^5 & \cong \frac{PSU(2,2|4)}{SO(4,1) \times SO(5)} \ ,\\
		AdS_4 \times \mathbb{CP}^3 &  \cong \frac{OSP(2,2|6)}{SO(3,1)\times U(3)} \ , \\
		AdS_3 \times S^3 \times S^3 \times S^1 &\cong \frac{D(2,1;\alpha )^2}{SL(2) \times SU(2)^2} \times U(1) \ , \\
		AdS_3 \times S^3 \times T^4 &\cong \frac{PSU(1,1|2)^2}{SL(2) \times SU(2)}\times U(1)^4 \ .
	\end{align*}
	In this thesis we are going to be mostly interested in the first and the last one.
\end{itemize}

\subsection{Superconformal algebra}

Before writing the coset for the supersymmetric theories we are interested in, it is better to understand first the $\mf{psu}(n,n|2n)$ algebras with $n\in \mathbb{N}$. This includes the superalgebras $\mf{psu} (1,1|2)$and $\mf{psu} (2,2|4)$, which we will use for the construction of the $AdS_3 \times S^3\times T^4$ and $AdS_5 \times S^5$ backgrounds respectively.

The matrix representation of the superalgebra $\mf{sl}(n,n|2n)$ is spanned by $4n\times 4n$ matrices with vanishing supertrace. If we write them in term of $2n\times 2n$ blocks
\begin{equation}
	M=\left( \begin{array}{cc}
	a & \theta \\
	\eta & b
	\end{array} \right) \ ,
\end{equation}
then we have to impose that $\text{STr } M=\text{Tr } a - \text{Tr }b=0$. There is an obvious $\mathbb{Z}_2$ grading where $a$ and $b$ are even and $\theta$ and $\eta$ are odd. This symmetry will mix with another $\mathbb{Z}_2$ symmetry when defining the coset. By the fundamental theorem of finitely generated abelian groups they will give either $\mathbb{Z}_2 \otimes \mathbb{Z}_2$ or $\mathbb{Z}_4$. As we will see, the later one is the correct.

We can reduce the superalgebra to $\mf{su}(n,n|2n)$ by imposing the additional constrain
\begin{equation}
	M^\dagger H + H M=0 \ ,
\end{equation}
where
\begin{equation}
	H=\left( \begin{array}{cc}
	\Sigma & 0 \\
	0 & \mathbb{I}_{2n}
	\end{array} \right) \, \text{ and } \, \Sigma =\left( \begin{array}{cc}
	\mathbb{I}_n & 0 \\
	0 & -\mathbb{I}_n
	\end{array} \right) \ .
\end{equation}
Note that apart from the obvious $\mf{su}(n,n)\oplus\mf{su}(2n)$ bosonic subalgebras there is an additional $\mf{u}(1)$ coming from the identity, which is supertraceless in this case. We can quotient our algebra by this generator\footnote{Sadly this quotient has no realization in terms of $4n\times 4n$ matrices.} to get the $\mf{psu}(n,n|2n)$ superalgebra.

However the explicit construction of the $\mathbb{Z}_4$ action is going to be different for $n=1$ and $n=2$. Let us first consider the case of the $\mf{su}(2,2|4)$ algebra. The outer automorphism group Out$(\mf{sl}(2,2|4))$ contains a finite subgroup isomorph to the Klein group $\mathbb{Z}_2\times \mathbb{Z}_2$ with generators
\begin{align*}
	M\rightarrow \left(\begin{array}{cc}
	b & \eta \\
	\theta & a
	\end{array} \right) \ , \quad M\rightarrow -M^{st}=\left(\begin{array}{cc}
	-a^t & \eta^t \\
	-\theta^t & -b^t
	\end{array} \right) \ ,
\end{align*}
where $st$ denote the supertranspose. Although the second one seems to be of order four, in fact it is of order two in the group of outer automorphism as it squares to the grading transformation we have defined before, which is an inner automorphism. Actually the $M\rightarrow -M^{st}$ transformation is the one that generates the $\mathbb{Z}_4$ structure we talked about, but instead of using it, we are going to work with the automorphism
\begin{equation}
	M\rightarrow \Omega (M)=-\pmb{\Sigma}M^{st}\pmb{\Sigma}^{-1}=-\left(\begin{array}{cc}
	\Sigma & 0 \\
	0 & \Sigma
	\end{array}\right) M^{st} \left(\begin{array}{cc}
	\Sigma & 0 \\
	0 & \Sigma
	\end{array} \right) \ ,
\end{equation}
which conserves the commutator but not the product nor the hermitian conjugation
\begin{align}
	\Omega([M_1 ,M_2])&=[\Omega (M_1) , \Omega (M_2) ] \ , \\
	\Omega(M_1 M_2)&=-\Omega (M_2) \Omega (M_1) \ , \\
	\Omega (M)^\dagger &=\Upsilon \Omega (M^\dagger ) \Upsilon^{-1}=-(\Upsilon H) \Omega (M) (\Upsilon H)^{-1} \ ,
\end{align}
where $\Upsilon=\left(\begin{array}{cc}
	\mathbb{I}_4 & 0 \\
	0 & -\mathbb{I}_4
	\end{array}\right)$ is called hypercharge. This allows us to split the algebra into four separated graded spaces
\begin{align}
	\mf{g} &=\mf{g}^{(0)} \oplus \mf{g}^{(1)} \oplus \mf{g}^{(2)} \oplus \mf{g}^{(3)} \ , & \mf{g}^{(k)}&=\left\{ M\in g : \Omega (M) =i^k M\right\} \ ,
\end{align}
with the property
\begin{equation}
	[g^{(k)},g^{(l)}] \subset g^{(k+l)\text{ mod } 4} \ .
\end{equation}
The coset $G/G^{(0)}$, obtained when we mod out the group associated to the bosonic algebra $\mf{g}^{(0)}$, is then called a \emph{semi-symmetric} superspace. The semi-symmetric spaces can be considered a supersymmetric generalization of the symmetric spaces we have used to construct the bosonic string theories. Therefore it is natural to try to write the supersymmetric string theories with them as a building block.

In the case of the $AdS_3 \times S^3\times T^4$ background we are going to make use of having two copies of $\mf{psu}(1,1|2)$ to construct the automorphism as an operation that interchanges the two copies and whose square is $(-\mathbb{I})^{F}$ where $F$ is the grading (+1 for odd generators and 0 for even ones). In particular we can choose
\begin{align}
	\Omega (\mf{g}_{L,R}^B) =\mf{g}_{R,L}^B \ , \quad \Omega (\mf{g}_{L,R}^F) =\mp \mf{g}_{R,L}^F \ ,
\end{align}
as the $\mb{Z}_4$ transformation. Similarly this transformation allow us to define an splitting of the algebra and to construct a semi-symmetric coset.

\subsection{Green-Schwarz string as a coset model}
\label{GScoset}

In this section we are going to see how to construct the Green-Schwarz action for $AdS_5 \times S^5$ and $AdS_3 \times S^3\times T^4$ backgrounds using the superalgebras presented in the previous section.

\subsubsection{$AdS_5 \times S^5$ GS action}

As we have seen just discussed, we can write the $\mf{psu}(2,2|4)$ superalgebra as a direct sum of four subalgebras defined by how they transform under the $\mathbb{Z}_4$ symmetry. This property can be directly applied to the construction of invariant currents
\begin{equation}
	j_\alpha=g^{-1} \partial g=j^{(0)}_\alpha +j^{(1)}_\alpha +j^{(2)}_\alpha +j^{(3)}_\alpha \ ,
\end{equation}
but we have still to quotient this algebra by the subgroup generated by the subalegra invariant by the $\Omega$ automorphism. Taking this quotient is equivalent to treating the current $j^{(0)}$ as a gauge degree of freedom in the same way as we have done above. This is because applying a transformation by one element of this subgroup transforms the current $j^{(0)}$ as a gauge field and acts as a similarity transformation on the rest of the currents.

The coset action we are going to work with is an extension of the already studied one\footnote{We have left a general $\gamma^{\alpha \beta}=\sqrt{-h} h^{\alpha \beta}$ metric instead of directly choosing the conformal gauge  $\gamma^{\alpha \beta}=\eta^{\alpha \beta}$  because later we are going to consider a symmetry transformation that has $\delta \gamma \neq 0$.}
\begin{equation}
	S=-\frac{\sqrt{\lambda}}{4\pi} \int{d^2 \sigma \left[ \gamma^{\alpha \beta} \text{ STr} \left( j_\alpha ^{(2)} j_\beta^{(2)} \right) + \kappa \epsilon^{\alpha \beta} \text{ STr} \left( j_\alpha^{(1)} j_\beta^{(3)} \right) \right]} \ ,
\end{equation}
formed by a kinetic term that is equivalent to the bosonic coset model we presented above and a Wess-Zumino term formed only with fermionic currents. Obviously it has the right symmetry and is $\Omega$ invariant. Although it seems that the action depend on the full $SU(2,2|4)$ group because the identity element is inside $j^{(2)}$, the supertraces of the identity and of the rest of components of $j^{(2)}$ vanish. Therefore the action will only depend on the square of $\tilde{j}^{(2)}$, the part of $j^{(2)}$ that does not contain the identity, and so it do not depend on this $\mf{u}(1)$ subalgebra.

Note also that, naïvely, the second term of the action does not seem like a WZ term. This is because the invariant three-form
\begin{equation}
	\Theta_3 =\text{STr} ( j^{(2)} \wedge j^{(3)} \wedge j^{(3)} - j^{(2)} \wedge j^{(1)} \wedge j^{(1)} )= \frac{1}{2} d \text{ STr} (j^{(1)} \wedge j^{(3)} ) \ ,
\end{equation}
is an exact form, proving that the second term in the action is indeed a WZ term.

If we define the current
\begin{equation}
	\Lambda^\alpha = g \left[ \gamma^{\alpha \beta} j^{(2)}_\beta -\frac{\kappa}{2} \epsilon^{\alpha \beta} \left( j^{(1)}_\beta - j^{(3)}_\beta \right) \right] \ ,
\end{equation}
we can express the conserved currents associated to the global $PSU(2,2|4)$ symmetry and the equations of motion in a compact way,
\begin{align}
	& J^\alpha = \frac{\sqrt{\lambda}}{2\pi} g \Lambda^\alpha g^{-1} \Longrightarrow \partial_\alpha J^\alpha=0 \ , \\
	&\partial_\alpha \Lambda^\alpha -[j_\alpha , \Lambda^\alpha ]=0 \ .
\end{align}
The three components of the last equation (as the $(0)$-th component vanishes identically) can be separated into
\begin{align}
	&\gamma^{\alpha \beta} \partial_\alpha j_\beta^{(2)} - \gamma^{\alpha \beta} [j_\alpha^{(0)} , j_\beta^{(2)}] + \frac{\kappa}{2} \epsilon^{\alpha \beta} \left( [j_\alpha^{(1)} , j_\beta^{(1)}] - [j_\alpha^{(3)} , j_\beta^{(3)}] \right)=0 \ ,\\
	&\gamma^{\alpha \beta} [j_\alpha^{(3)} ,j_\beta^{(2)}] +\kappa \epsilon^{\alpha \beta} [j_\alpha^{(2)} , j_\beta^{(3)}]=-2P_-^{\alpha \beta} [j_\alpha^{(2)} , j_\beta^{(3)}]=0 \ , \\
	&\gamma^{\alpha \beta} [j_\alpha^{(1)} ,j_\beta^{(2)}] -\kappa \epsilon^{\alpha \beta} [j_\alpha^{(2)} , j_\beta^{(1)}]=-2P_+^{\alpha \beta} [j_\alpha^{(2)} , j_\beta^{(1)}]=0 \ ,
\end{align}
where in the last two equations we have introduced the ``projections'' $P^{\alpha \beta}_\pm=\frac{\gamma^{\alpha \beta} \pm \kappa \epsilon^{\alpha \beta}}{2}$. We put the word projections between quotations because those operators are orthogonal projectors only when $\kappa=\pm 1$.

The Lax connection for this Lagrangian is given by
\begin{equation}
	L_\alpha =j^{(0)}_\alpha +\frac{1}{2} \left( z^2 +\frac{1}{z^2} \right) j^{(2)}_\alpha -\frac{1}{2\kappa} \left( z^2 -\frac{1}{z^2} \right) \gamma_{\alpha \beta} \epsilon^{\beta \mu} j^{(2)}_\mu  + z j^{(1)}_\alpha  +\frac{1}{z} j^{(3)}_\alpha \ ,
\end{equation}
proving that it is indeed an integrable Lagrangian.

Finally, the equations of motion for the world-sheet metric give the stress energy tensor
\begin{equation}
	T_{\alpha \beta}=\text{STr} \{ j_\alpha^{(2)} j_\beta^{(2)} \}- \gamma_{\alpha \beta} \gamma^{\gamma \delta} \text{STr} \{ j_\gamma^{(2)} j_\delta^{(2)} \} \ ,
\end{equation}
which has to be set to zero to recover the Virasoro constraints.

\subsubsection{Kappa symmetry in $AdS_5 \times S^5$}

We have built the Lagrangian with an explicit invariance under global left $PSU(2,2|4)$ transformation. In this paragraph we are going to see a remaining symmetry generated by right multiplication by a local fermionic element $\epsilon (\tau , \sigma )\in \mf{psu} (2,2|4)$ called $\kappa$-symmetry. Because we are interested on the coset model we will allow to have a compensating element $h\in SO(1,4) \times SO(5)$, that is
\begin{equation}
	g e^{\epsilon (\tau , \sigma )}=g' h \ .
\end{equation}
However the superstring action in general is not invariant under this transformation. Hence some restriction has to be imposed on $\epsilon$. We can divide $\epsilon$ into the part that transforms in $\mf{g}^{(1)}$ and $\mf{g}^{(3)}$ under the $\Omega$ transformation, that is, $\epsilon=\epsilon^{(1)} +\epsilon^{(3)}$. Then the transformation of the current, that is, $\delta_\epsilon j=-d\epsilon +[j,\epsilon ]$ can be expanded into the different four components of the algebra as
\begin{align}
	\delta_\epsilon j^{(1)} &=d\epsilon^{(1)} +[j^{(0)} , \epsilon^{(1)} ]+[j^{(2)} , \epsilon^{(3)} ] \ ,\\
	\delta_\epsilon j^{(3)} &=d\epsilon^{(3)} +[j^{(2)} , \epsilon^{(1)} ]+[j^{(0)} , \epsilon^{(3)} ] \ ,\\
	\delta_\epsilon j^{(2)} &=[j^{(1)} , \epsilon^{(1)} ]+[j^{(3)} , \epsilon^{(3)} ] \ , \\
	\delta_\epsilon j^{(0)} &=[j^{(3)} , \epsilon^{(1)} ]+[j^{(1)} , \epsilon^{(3)} ] \ ,
\end{align}
and the same can be done to the variation of the Lagrangian density,
\begin{align}
	-\frac{2}{g} \delta_\epsilon \mc{L} &= \delta \gamma^{\alpha \beta} \STr \left( j^{(2)}_\alpha j^{(2)}_\beta \right) -4 \STr \left( P^{\alpha \beta}_+ [j^{(1)}_\beta , j^{(2)}_\alpha ] \epsilon^{(1)} +P^{\alpha \beta}_- [j^{(3)}_\beta , j^{(2)}_\alpha ] \epsilon^{(3)} \right) \ ,
\end{align}
where we have used the flatness condition of $j^{(1)}$ and $j^{(3)}$ to simplify it. To reduce this expression even further in order to find the variation of the metric, we will choose an ansatz for the $\epsilon$ functions
\begin{align}
	\epsilon^{(1)} &=j^{(2)}_{\alpha , -} \kappa^{(1),\alpha}_+ + \kappa^{(1),\alpha}_+ j^{(2)}_{\alpha , -} \ , \notag \\
	\epsilon^{(3)} &=j^{(2)}_{\alpha , -} \kappa^{(3),\alpha}_+ + \kappa^{(3),\alpha}_+ j^{(2)}_{\alpha , -} \ , \label{epsilonansatz}
\end{align}
where we have written $K^\alpha_\pm = P^{\alpha \beta}_\pm K_\beta$. These expressions have indeed the right transformation properties
\begin{multline}
	\Omega (\epsilon^{(n)} )= \Omega (j^{(2)}_{\alpha , -} \kappa^{(n),\alpha}_+ + \kappa^{(n),\alpha}_+ j^{(2)}_{\alpha , -})=\\
	=-\Omega (\kappa^{(n),\alpha}_+) \Omega (j^{(2)}_{\alpha , -}) - \Omega (j^{(2)}_{\alpha , -}) \Omega (\kappa^{(n),\alpha}_+)=\\
	=- i^n \kappa^{(n),\alpha}_+ (-j^{(2)}_{\alpha , -}) - (-j^{(2)}_{\alpha , -}) i^n \kappa^{(n),\alpha}_+=i^n \epsilon^{(n)} \ ,
\end{multline}
and, after some computations the variation of the metric can be written as\cite{foundationsAdSstring}
\begin{equation}
	\delta \gamma^{\alpha \beta} =\frac{1}{2} \Tr \left( [\kappa^{(1),\alpha}_+ , j^{(1),\beta}_{-}] +[\kappa^{(3),\alpha}_+ , j^{(3),\beta}_{-}] \right) \ .
\end{equation}
We must note that to get this result we had to assume that $P^{\alpha \beta}_\pm$ are projectors, therefore this $\kappa$-symmetry only exists if $\kappa=\pm 1$. Now the important question is how many fermionic degrees of freedom can we ``gauged away'' using it. It can be checked that the answer is 16 real fermions \cite{foundationsAdSstring}, leaving 16 physical fermionic degrees of freedom.

\subsubsection{$AdS_3 \times S^3\times T^4$ GS action}

The construction of the action for the $AdS_3 \times S^3\times T^4$ background is very similar to the one we have already described for $AdS_5 \times S^5$. Consequently we will only point out the differences between both constructions. The most important difference is the definition of the currents, which boils down to the different definition of the $\Omega$ transformation and the $\mb{Z}_4$ grading. In this case we can define the currents on the different symmetry subalgebras as
\begin{align}
	 j^{(0)} &=\frac{j^{\text{even}}_L +j^{\text{even}}_R}{2} \ ,& j^{(2)} &=\frac{j^{\text{even}}_L -j^{\text{even}}_R}{2} \ , \\
	 j^{(1)} &=\frac{j^{\text{odd}}_L +j^{\text{odd}}_R}{2} \ , & j^{(3)} &=\frac{j^{\text{odd}}_L -j^{\text{odd}}_R}{2} \ ,
\end{align}
where $L$ and $R$ refers to each of the copies of $PSU(1,1|2)$, not to left and right invariant currents. The Lagrangian, the equations of motion, and the Virasoro constrains remain the same. However the construction of the $\kappa$-symmetry needs some modifications. For example the ansatz \ref{epsilonansatz} for $\epsilon^{(1)}$ and $\epsilon^{(3)}$ has to be modified to
\begin{equation}
	\epsilon^{(1)} =j^{(2)}_{\alpha , -} j^{(2)}_{\beta , -} \kappa^{\alpha \beta} +j^{(2)}_{\alpha , -} \kappa^{\alpha \beta} j^{(2)}_{\beta , -} +\kappa^{\alpha \beta} j^{(2)}_{\alpha , -} j^{(2)}_{\beta , -} -\frac{1}{8} \STr ( \Sigma j^{(2)}_{\alpha , -} j^{(2)}_{\beta , -}) \kappa^{\alpha \beta} \ .
\end{equation}

\subsubsection{$AdS_3 \times S^3\times T^4$ GS action with mixed flux}

For the case of $AdS_3 \times S^3\times T^4$ background we can add an extra WZ term to the coset action. In particular the new action reads\footnote{All the expressions of this subsection are written in the language of differential form to make the expressions more readable and compact.}
\begin{align}
	S &=\frac{1}{2} \int_{\mc{M}=\partial \mc{B}}{\text{STr} \left\{ j_2 \wedge *j_2 +\kappa j_1 \wedge j_3 \right\}} +\notag \\
	 &+q\int_{\cal B}{ \left( \frac{2}{3} j_2 \wedge j_2 \wedge j_2 + j_1 \wedge j_3 \wedge j_2 + j_3 \wedge j_1 \wedge j_2 \right)} \ ,
\end{align}
where $\kappa$ and $q$ are coupling constants that have to be fixed. The requirements to fix these constants are to preserve conformal invariance, integrability and $\kappa$ symmetry of the theory.

Inspired from the other Lax connections we have written, we take the ansatz
\begin{equation}
	L=j_0 + \alpha_1 j_2 +\alpha_2 *j_2+\beta_1 j_1 + \beta_2 j_3 \ ,
\end{equation}
and impose its flatness $dL+L\wedge L=0$. The equations obtained in this way have no solutions unless the coupling constants are related as
\begin{equation}
	\kappa^2 = 1-q^2 \ ,
\end{equation}
but we cannot fix all the constants of the ansatz. This is because one of them adopts the role of the spectral parameter. After choosing a convenient parametrization, the Lax connection can be written as
\begin{multline}
	L (z) =j_0 + \kappa \frac{z^2 +1}{z^2-1} j_2 +\left( q -\frac{2\kappa z}{z^2 -1} \right) *j_2+ \\
	+\left( z + \frac{\kappa}{1-q} \right) \sqrt{\frac{\kappa (1-q)}{z^2 -1}} j_1 +\left( z - \frac{\kappa}{1+q} \right) \sqrt{\frac{\kappa (1+q)}{z^2 -1}} j_3 \ .
\end{multline}
The conformal invariance and the kappa symmetry of the action can be proved expanding around a fixed classical background \cite{Cagnazzo-Zarembo}.

\chapter{Flux-deformed Neumann-Rosochatius system} \chaptermark{Flux-deformed N-R system}

	\begin{chapquote}{M. Kac, \textit{Some Stochastic Problems in Physics and Mathematics} \cite{1258668041}}
 		One should try to formulate even familiar things in as many different ways as possible.
	\end{chapquote}

While the use of integrability is widely developed on $\mc{N}=4$ SYM theory and $AdS_5 \times S^5$, integrability methods are potentially applicable to other $AdS_{n}$ backgrounds with RR flux and consequently to their dual CFTs. For instance one of the manifestations of the integrability in the case of the $AdS_5/CFT_4$ correspondence that can be generalized is the identification of the Lagrangian describing closed strings rotating in $AdS_5 \times S^5$ with the Neumann and the Neumann-Rosochatius integrable systems. Furthermore it allowed to write an analytical expression (at least for the large spin limit) for the dispersion relations and their one-loop correction, a quantity easily comparable with the anomalous scaling dimensions of operator to perform a check of the duality.

This chapter is divided into four sections. In the first section we will present the Neumann system and the Neumann-Rosochatius systems. We will prove the integrability of these systems and how the string Lagrangian in $AdS_5 \times S^5$ reduces to them when the closed spinning string ansatz is chosen. This section will be mostly based on references \cite{Arutyunov_2003,Arutyunov_2004}. In the second section we will introduce part of the case of interest for us, as we will study the Lagrangian for strings moving in $\mb{R} \times S^3$ with R-R and NS-NS three-form flux. We will prove that this Lagrangian can be identified with a deformation of the Neumann-Rosochatius Lagrangian when we substitute the closed spinning string anstaz. We will also prove that this system is integrable by computing the deformed version of the integrals of motion, called Uhlenbeck constants. Once these tools are presented, we will compute the solutions of the equations of motion for the case of constant radii (a proper definition of these radii will be presented later). For this case we can construct the dispersion relation as a series in inverse powers of the total angular momentum. We will also be able to build the exact dispersion relation for some particular simple limits. After that we will make use of the deformed Uhlenbeck constants to compute more general solutions, which we are going to call elliptic strings, given that they are described by Jacobi Elliptic functions. Finally we will examine in detail the limit of pure NS-NS flux where the system simplifies, as the coset model becomes a pure WZW model. In this case the Elliptic functions degenerate to trigonometric ones, and an exact dispersion relation can be constructed. In the third section the case of strings moving only in $AdS_3$ space with R-R and NS-NS three-form flux will be studied. Most of the methods developed in the previous section can be applied to this problem as the $AdS$ space can be formally treated as an analytical continuation of the sphere. In section four, we will address the full problem of strings moving in $AdS_3 \times S^3$ with R-R and NS-NS three-form flux. In this section we take the tools developed in the previous two sections and put them together to describe the full Lagrangian. While we mostly follow the same steps as the two previous sections, we are not going to present the general elliptic solution, only their pure NS-NS limit. This is because the full solution is not very enlightening as some of the relations involving winding numbers and angular momenta are very difficult to invert in order to obtain the dispersion relation. We end this chapter changing our focus from the spinning string ansatz to its natural counterpart, the pulsating string ansatz, defined in the same way but with the roles of $\tau$ and $\sigma$ exchanged. The results presented in sections 2, 3 and 4 are extracted from references \cite{Spinningstrings1,Spinningstrings2}. The results presented in the last section are novel and will be published soon.

\section{Neumann-Rosochatius systems in $AdS_5\times S^5$} \sectionmark{N-R systems in $A\MakeLowercase{d}S_5\times S^5$}

Let us consider the bosonic part of the classical closed string propagating in the $AdS_5 \times S^5$ space-time. In the conformal gauge the Polyakov action can be written as
\begin{align}
	S&=-\frac{\sqrt{\lambda}}{4\pi} \int{d^2 \sigma \left\{ \left[ -\frac{1}{2} \partial_\alpha X_M \partial^a X_M +\frac{1}{2} \Lambda (X_M X_M -1) \right] + \right. }\notag \\
	&\left. +\left[ -\frac{1}{2} \eta^{MN} \partial_\alpha Y_M \partial^\alpha Y_N +\frac{1}{2} \tilde{\Lambda} (\eta^{MN} Y_M Y_N +1) \right] \right\} \ , \label{PolyakovNR}
\end{align}
where the first term correspond to the $S^5$ part and the second one to the $AdS_5$ part with metric $\eta^{MN}=\text{diag}(-1,+1,+1,+1,+1,-1)$. Throughout the rest of this chapter we are going to use three different parametrizations of the $AdS_5\times S^5$ background
\begin{itemize}
	\item The first one is the $\mb{R}^6 \times \mb{R}^{(2,4)}$ embedding we have already used to formulate the Lagrangian: $X_M$, $M=1,\dots ,6$ and $Y_M$, $M=0,\dots ,5$ with the constrains $X_M X_M=1$ and $\eta^{MN} Y_M Y_N =-1$. These constrains are implemented using Lagrange multipliers.
	
	\item The second one is the angle parametrization, which can be related to the embedding coordinates as
	\begin{align*}
		X_1 +iX_2 &= \sin \gamma \cos \psi \, e^{i\varphi_1} \ , & X_3 +iX_4 &= \sin \gamma \sin \psi \, e^{i\varphi_2} \ , & X_5 +iX_6 &= \cos \gamma \, e^{i\varphi_3} \ , \\
		Y_1 +iY_2 &= \sinh \rho \cos \theta \, e^{i\phi_1} \ , & Y_3 +iY_4 &= \sin \rho \sin \theta \, e^{i\phi_2} \ , & Y_5 +iY_0 &= \cos \rho \, e^{it} \ .
	\end{align*}
	Which does not need to impose any constrain, so a Lagrange multiplier is not needed.
	
	\item Throughout this chapter will mostly use the third parametrization, as it is adapted to the spinning string ansatz we will consider below
	\begin{align}
		X_1 +iX_2 &= r_1 e^{i\varphi_1} \ , & X_3 +iX_4 &= r_2 e^{i\varphi_2} \ , & X_5 +iX_6 &= r_3 e^{i\varphi_3} \ , \label{parametrizationspinning}\\
		Y_1 +iY_2 &= z_1 e^{i\phi_1} \ , & Y_3 +iY_4 &= z_2 e^{i\phi_2} \ , & Y_5 +iY_0 &= z_0 e^{it} \ ,
	\end{align}
	with the constrains $r_1^2+r_2^2+r_3^2=1$ and $z_1^2+z_2^2-z_0^2=-1$ imposed through a Lagrange multiplier.
\end{itemize} 
Both metrics have three commuting translational isometries which give rise to six global commuting integrals of motion: three spins $J_i$, $i=1,2,3$ from the sphere and two spins $S_j$, $j=1,2$ and the energy $E$ from the AdS space. In the embedding coordinates those are defined as
\begin{align}
	J_i &=J_{2i-1,2i}=\sqrt{\lambda} \int_0^{2\pi}{ \frac{d\sigma}{2\pi} \left( X_{2i-1} \dot{X}_{2i} - X_{2i} \dot{X}_{2i-1} \right) } \ , \\ 
	S_j &=S_{2j-1,2j}=\sqrt{\lambda} \int_0^{2\pi}{ \frac{d\sigma}{2\pi} \left( Y_{2j-1} \dot{Y}_{2j} - Y_{2j} \dot{Y}_{2j-1} 	\right) } \ ,  \\ 
	E &=S_{5,0}=\sqrt{\lambda} \int_0^{2\pi}{ \frac{d\sigma}{2\pi} \left( Y_{5} \dot{Y}_{0} - Y_{0} \dot{Y}_{5} \right) } \ , 
\end{align}
where the dot means derivative with respect to $\tau$ and, for convenience, we have set the domain of the $\sigma$ coordinate to $[0,2\pi )$. Index $i$ runs from 1 to 3 and index $j$ can take values 1 and 2.

Let us focus for the moment only on the sphere and set the AdS to a trivial solution $Y_1=Y_2=Y_3=Y_4=0$, $Y_5+iY_0=e^{i\kappa \tau}$. We are interested in the periodic motion with three non-vanishing spins, so it is natural to choose the ansatz
\begin{equation}
	X_1 +iX_2=r_1 (\sigma ) e^{i\varphi_1(\sigma , \tau)} \ , \,  X_3 +iX_4=r_2 (\sigma ) e^{i\varphi_2(\sigma , \tau)} \ , \, X_5 +iX_6=r_3 (\sigma ) e^{i\varphi_3(\sigma , \tau)} \ .
\end{equation}
While the constrain on the radii remains unchanged
\begin{equation}
	\sum_{i=1}^3 r_i^2 (\sigma )=1 \ ,
\end{equation}
the angular momenta can be simplified to
\begin{equation}
	J_i=\sqrt{\lambda} \int_0^{2\pi} \frac{d\sigma}{2\pi} r_i^2 (\sigma) \dot{\varphi_i} (\sigma , \tau ) \ ,
\end{equation}
and the energy can be written as
\begin{equation}
	E^2=\lambda \kappa^2=\lambda \left[ (\partial_\tau X_M)^2 +(\partial_\sigma X_M)^2 \right] =\lambda \sum_{i=1}^3 \left( r_i^{\prime 2} +r_i^2 \dot{\varphi}_i^{2} + r_i^2 \varphi_i^{\prime 2} \right) \ ,
\end{equation}
where the prime means derivative with respect to $\sigma$ and the second equality comes from one of the Virasoro constrains. The other Virasoro constrain is given by
\begin{equation}
	0=\partial_\tau X_M \partial_\sigma X_M=\dot{X}_M X'_M=2\sum_{i=1}^3{r_i^2 \dot{\varphi}_i \varphi '_i} \ .
\end{equation}

We cannot go much further without giving an ansatz for the $\varphi$ variables. In the rest of this section three different ansätze are examined, giving rise to three different integrable models.

\subsection{Neumann system}

The most simple ansatz we can choose for $\varphi$ is to assume independence of the $\sigma$ variable, making the periodicity condition of the embedding coordinates easier to implement, and linear in the $\tau$ variable. This means
\begin{equation}
	\varphi_i (\tau , \sigma )=\omega_i \tau \label{simpletsansatz} \ .
\end{equation}
Thus the periodicity constrain reads now $r_i (\sigma +2\pi)=r_i (\sigma)$. This ansatz also simplifies the expression of the spins, being now
\begin{equation}
	J_i=\sqrt{\lambda} \omega_i \int_0^{2\pi} \frac{d\sigma}{2\pi} r_i^2 (\sigma) \ .
\end{equation}
Therefore the constraints on the radii can also be rewritten as a condition over the spins and the $\omega$'s,
\begin{equation}
	\frac{J_1}{\omega_1} +\frac{J_2}{\omega_2} +\frac{J_3}{\omega_3} =\sqrt{\lambda} \ .
\end{equation}
Substituting this ansatz into the Lagrangian (\ref{PolyakovNR}) we get
\begin{equation}
	L=\frac{1}{2} \sum_{i=1}^3 (r_i^{\prime 2} -\omega_i^2 r_i^2) + \frac{1}{2} \Lambda \left( \sum_{i=1}^3 r_i^2 -1 \right) \ .
\end{equation}
This is the Lagrangian of a $n=3$ dimensional harmonic oscillator constrained to remain on an unit 2-sphere. This is a special case of the Neumann system \cite{Neumann1859}, which is known to be integrable \cite{Uhlenbeck_1982}. The integrals of motion are given by the so called \emph{Uhlenbeck constants},
\begin{equation}
	I^N_i=r_i^2 +\sum_{j\neq i} \frac{(r_i r'_j -r'_i r_j)^2}{\omega_i^2 -\omega_j^2} \ , \label{Uhlenbeckconst}
\end{equation}
but not all of them are independent, being the constraint $I_1^N+I_2^N+I_3^N=1$. The Hamiltonian can be written as a function of these conserved quantities as
\begin{equation}
	H=\frac{1}{2} \sum_{i=1}^3 \omega_i^2 I^N_i \ .
\end{equation}

This Lagrangian can be explicitly solved, but we are not going to do that here. Instead we are going to find the solutions for a more general ansatz which includes this one as a particular case. More results about the Neumann system can be found in \cite{Arutyunov_2003,Babelon_1992}.

\subsection{Rosochatius system}

The second ansatz is, in some way, opposite to the previous one: we are going to assume independence of the $\tau$ variable but general behaviour on the $\sigma$ coordinate. This means
\begin{equation}
	\varphi_i (\tau , \sigma )=\alpha_i (\sigma) \ ,
\end{equation}
so the whole Lagrangian is independent of the $\tau$ variable. However the periodicity constraint now is more involved because we have to impose not only $r_i (\sigma +2\pi)=r_i (\sigma)$ but also
\begin{equation}
	\int_0^{2\pi}{ \frac{d\sigma}{2\pi} \alpha '_i (\sigma)}=\alpha_i (2\pi ) - \alpha_i (0)=2\pi m_i \ ,
\end{equation}
where the $m_i$ can be interpreted as winding numbers. Substituting this ansatz into the Lagrangian we get
\begin{equation}
	L=\frac{1}{2} \sum_{i=1}^3 (r_i^{\prime 2} +r_i^2 \alpha_i^{\prime 2} ) + \frac{1}{2} \Lambda \left( \sum_{i=1}^3 r_i^2 -1 \right) \ ,
\end{equation}
which depends on $\alpha_i (\sigma )$ only through its derivative. This means that there are a con\-served quantities associated to shifts of these functions and we can get rid of them in the Lagrangian. The equations of motion for these functions are
\begin{equation}
	\left( r_i^2 \alpha '_i \right)'=0 \Longrightarrow \alpha '_i=\frac{v_i}{r_i^2} \ ,
\end{equation}
where $v_i$ are the three integrals of motion from the shifts of the angles. We can substitute these equations of motion back into the Lagrangian
\begin{equation}
	L=\frac{1}{2} \sum_{i=1}^3 \left( r_i^{\prime 2} -\frac{v_i^2}{r_i^2} \right) + \frac{1}{2} \Lambda \left( \sum_{i=1}^3 r_i^2 -1 \right) \label{Rlagrangian} \ ,
\end{equation}
which is a particular case of the Rosochatius system in a 2-sphere. This system was shown to be integrable by Rosochatius \cite{Rosochatius}, while the Lax pair was computed in \cite{Moser}. We can write its integrals of motion as
\begin{equation}
	I^R_i=\sum_{j\neq i} \left[ (r_i r'_j -r'_i r_j)^2+\frac{v_i^2}{r_i^2} r_j^2 +\frac{v_j^2}{r_j^2}{r_i^2} \right] \ ,
\end{equation}
which is very reminiscent of the Uhlenbeck integrals.

There are some similarities between the Rosochatius system and the Neumann system, as pointed out by \cite{Ratiu_1982}. Thus we may think about an ansatz that incorporated both systems. That is what we are going to present next.

\subsection{Neumann-Rosochatius system}

The last ansatz we are going to study is the mixture of both previous ansätze;
\begin{equation}
	\varphi_i (\tau , \sigma )=\omega_i \tau +\alpha_i (\sigma ) \label{mostgeneralansatz} \ .
\end{equation}
In consequence we have to impose the same periodicity condition as the Rosochatius system
\begin{align*}
	\int_0^{2\pi}{ \frac{d\sigma}{2\pi} \alpha '_i (\sigma)}&=2\pi m_i \ , & r_i (\sigma +2\pi )&=r_i (\sigma) \ .
\end{align*}
Substituting this ansatz into the Lagrangian we get
\begin{equation}
	L=\frac{1}{2} \sum_{i=1}^3 (r_i^{\prime 2} +r_i^2 \alpha_i^{\prime 2}-\omega_i^2 r_i^2) + \frac{1}{2} \Lambda \left( \sum_{i=1}^3 r_i^2 -1 \right) \ ,
\end{equation}
which again depends on $\alpha_i (\sigma )$ only through its derivative. Repeating the same steps as before we can rewrite it as
\begin{equation}
	L=\frac{1}{2} \sum_{i=1}^3 \left( r_i^{\prime 2} -\omega_i^2 r_i^2 -\frac{v_i^2}{r_i^2} \right) + \frac{1}{2} \Lambda \left( \sum_{i=1}^3 r_i^2 -1 \right) \ , \label{NRlagrangian}
\end{equation}
which describes the so called Neumann-Rosochatius integrable system, studied, among others, in \cite{Bartocci_2004}. This system is also integrable, being the integrals of motion a deformation of the original Uhlenbeck constants
\begin{equation}
	I^{NR}_i =r_i^2 +\sum_{j\neq i} \frac{1}{\omega_i^2 -\omega_j^2} \left[(r_i r'_j -r'_i r_j)^2+\frac{v_i^2}{r_i^2} r_j^2 +\frac{v_j^2}{r_j^2}{r_i^2} \right] \ ,
\end{equation}
with the same constraint as the original ones $I_1^{NR}+I_2^{NR}+I_3^{NR}=1$. Furthermore the Hamiltonian of the Neumann-Rosochatius system,
\be
H = \frac {1}{2} \sum_{i=1}^2 \big[ r_i^{\prime 2} +r_i^2 \alpha_i^{\prime 2} + r_i^2 \omega_i^2 \big] \ ,
\ee
can be written in terms of the Uhlenbeck constants and the integrals of motion $v_i$,  
\be
H = \frac {1}{2} \sum_{i=1}^2 \big[ \omega_i^2 I_i^{NR} + v_i^2 \big] \ .
\ee

Now we are going to analyse the solutions of this Lagrangian with constant radii. To do that it is better to look at the equations of motion that follow from~(\ref{NRlagrangian}) instead of the Uhlenbeck constants. These equations of motion read
\begin{equation}
	r''_i=-\omega_i^ 2 r_i +\frac{v_i^2}{r_i^3}-\Lambda r_i \ ,
\end{equation}
while the constraints from the Lagrange multiplier impose
\begin{align}
	\Lambda &=\sum_{j=1}^3{r_j^{\prime 2}-\omega_j^2 r_j^2+\frac{v_j^2}{r_j^2}} \ , & \sum_{j=1}^3 r_j^2=1 \ .
\end{align}
We can see that indeed $r_i=a_i=$ const. and $\Lambda=$ const. are a solution. This imposes that the derivatives of the angles $\alpha_i$ also become constant and thus
\be
\alpha_i = m_i \sigma + \a_{0i} \ ,
\ee
where the integration constants $\a_{0i}$ can be set to zero through a rotation, and the constants $m_i$ must be integers in order to satisfy the closed string periodicity condition. After some algebra we can write the relations
\begin{align}
	 &\omega_i^2 =m_i^2 -\Lambda \ , & &v_i=a_i^2 m_i \ , \\
	 &\sum_{j=1}^3 a_j^2 \omega_j m_j =0 \ , & &\kappa^2=2\left( \sum_{j=1}^3 a_j^2 \omega_j^2 \right) +\Lambda \ ,
\end{align}
which, in terms of the spins and the energy, read
\begin{align}
	E^2 &=2\sqrt{\lambda} \left( \sum_{j=1}^3{J_j \sqrt{m_j^2 -\Lambda^2}} \right) +\lambda \Lambda \ , \\
	1 &=\sum_{j=1}^3 \frac{J_j}{\sqrt{m_j^2 -\Lambda}} \ , \\
	0 &= \sum_{j=1}^3 m_j J_j \ .
\end{align}
But the solution to these equations cannot be written down explicitly for generic values of the spins. However we can expand them for large total spin $J=\sum_{j=1}^3 J_j$ obtaining
\begin{align}
	\Lambda &=-\frac{J^2}{\lambda} +\sum_{j=1}^3 m_j^2 \frac{J_j}{J} +\dots \ , & E^2 &=J^2 +\sum_{j=1}^3 \lambda m_j^2 \frac{J_j}{J} +\dots \ , \label{generalenergyNeumannsystem}
\end{align}
where $\sum_{j=1}^3 m_j J_j=0$ has still to be imposed. Luckily some particular cases can be solved analytically. Those are the case where two of the spin vanish, which gives the trivial point-like string rotating in $S^1$ with 
\begin{equation}
E=J_1=\sqrt{\lambda} \omega_1 \ ,
\end{equation}
and the case of one vanishing spin with equal and opposite windings $m_1=-m_2=m$, which imposes $J_1=J_2=\frac{J}{2}$, and has an energy
\begin{equation}
	E=\sqrt{J^2+\lambda m^2} \label{oppositewindingenergy} \ .
\end{equation}

These results can be generalized to solutions dynamical in the full $AdS_5\times S^5$\cite{Arutyunov_2004}.

\section[The flux-deformed Neumann-Rosochatius. Spinning strings in $\mathbb{R} \times S^3$]{The flux-deformed Neumann-Rosochatius. Spinning strings in $\mathbb{R}\times S^3$\sectionmark{Spinning strings in $\mathbb{R}\times S^3$}} \sectionmark{Spinning strings in $\mathbb{R}\times S^3$}

Now we will move our attention to closed spinning string solutions in $AdS_3 \times S^3 \times T^4$ with NS-NS three-form flux. The solutions that we will study will have no dynamics along the torus and thus we will not include these directions in what follows. As we said in the previous chapter, the $AdS_3 \times S^3$ background admits an NS-NS B-field. This B-field is of the form
\begin{equation}
	b_{t \phi} = q \sinh^2 \rho \ , \quad b_{\phi_1 \phi_2} = - q \cos^2 \theta \ , 
\end{equation}
where $0 \leq q \leq 1$. 
The presence of this B-field can be related to the additional WZ term that, as explained in section~\ref{GScoset}, can be included in the action for the $AdS_3 \times S^3 \times T^4$ background. The value $q=0$ would correspond to not adding this extra WZ term, so the theory only has R-R flux and can be formulated in terms of a pure Green-Schwarz action as for the case of the $AdS_5 \times S^5$ background. In this limit of vanishing NS-NS flux the sigma model for closed strings rotating in $AdS_3 \times S^3$ becomes the Neumann-Rosochatius integrable system presented in the previous section. On the other hand the value $q=1$ would correspond to pure NS-NS flux, as the usual WZ term has a prefactor $\sqrt{1-q^2}$ to maintain the conformal invariance and integrability of the action, and so it cancels. This limit can be described purely as a supersymmetric WZW model, so some simplifications are expected.

For the moment we are going to restrict our computations again to the case of rotation on $S^3$, so that we will take $Y_1=Y_2=0$ and $Y_3 + i Y_0 = e^{i w_0 \tau}$ for the $AdS$ coordinates. For the coordinates along $S^3$ we will choose again the ansatz where $\varphi$ has dynamics both in $\sigma$ and $\tau$,
\begin{equation}
X_1 + i X_2 = r_1 (\sigma) \, e^{i \omega_1 \tau + i\alpha_1(\sigma)} \ , \quad X_3 + i X_4 = r_2 (\sigma) \, e^{i \omega_2 \tau + i\alpha_2(\sigma)} \ ,
\label{ansatz}
\end{equation}
where the functions $r_i(\sigma)$ must satisfy
\be
	r_1^2+r_2^2=1 \ , \label{sphere}
\ee
and the periodicity constraints are the same as in the undeformed N-R system,
\be
	\quad r_i(\sigma + 2 \pi) = r_i(\sigma) \ , \quad \alpha_i(\sigma + 2 \pi) = \alpha_i(\sigma) + 2 \pi \bar{m}_i \ .
\ee
The bar over the winding numbers is added to distinguish them from the undeformed ones, as they will have an extra contribution coming from the presence of the flux. When we enter this ansatz in the Polyakov action with a B-field term
\be
S = \frac {\sqrt{\lambda}}{4 \pi} \int d^2 \sigma \big[ \sqrt{-h} h^{ab} G_{MN} \partial_a X^M \partial_b X^N 
- \e^{ab} B_{MN} \partial_a X^M \partial_b X^N \big] \ ,
\ee
we find the lagrangian
\be
L_{S^3} = \frac {\sqrt{\lambda}}{2 \pi} \Big\{  \sum_{i=1}^2 \frac {1}{2} \big[ (r_i')^2 + r_i^2 (\a_i')^2 - r_i^2 \omega_i^2 \big] 
- \frac {\Lambda}{2} ( r_1^2 + r_2^2 - 1) + q r_2^2 \, ( \omega_1 \alpha_2' - \omega_2 \alpha_1' ) \Big\} \ ,
\label{NRq}
\ee
where we have chosen the conformal gauge. The first piece in (\ref{NRq}) is the Neumann-Rosochatius integrable system from eq.~(\ref{NRlagrangian}). The presence 
of the non-vanishing flux introduces the last term in the Lagrangian\footnote{Note that in the 
WZW model limit $q=1$ the Lagrangian simplifies greatly because we can complete squares. 
We will find further evidence on this simplification below.}.

The equations of motion for the $\alpha_i$ variables can be written compactly using the cyclicity of the Lagrangian on them
\be
\alpha_i' = \frac {v_i + q r_2^2 \epsilon_{ij} \omega_j}{r_i^2} \ , \quad i = 1,2 \ , \label{alphaprime}
\ee
where $\epsilon_{12}=+1$ (we assume summation on $j$). The variation of the Lagrangian with respect to the radial coordinates gives us
\begin{align}
r_1''&=-r_1 \omega_1^2 +r_1 \alpha_1^{\prime 2}-\Lambda r_1 \ , \label{r1prime} \\
r_2''&=-r_2 \omega_2^2 +r_2 \alpha_2^{\prime 2}-\Lambda r_2 + 2 q r_2 ( \omega_1 \alpha_2' - \omega_2 \alpha_1' )\ .
\label{r2prime}
\end{align}
To these equations we have to add the Virasoro constraints,
\begin{align}
&\sum_{i=1}^2 \big( r_i^{\prime 2} +r_i^2 ( \alpha_i^{\prime 2}+ \omega_i^2 ) \big)=w_0^2 \ , \\
&\sum_{i=1}^2 r_i^2 \omega_i \alpha_i'=0 \ , \label{virasoro1} 
\end{align}
In terms of the integrals $v_i$ the second Virasoro constraint can be rewritten as
\begin{equation}
\omega_1 v_1+\omega_2 v_2=0 \ .
\end{equation}
The energy and the angular momenta of the string are given by
\begin{align}
E & = \sqrt{\lambda} \, w_0 \ , \label{E} \\
J_1 & = \sqrt{\lambda} \int_0^{2\pi} {\frac {d\sigma}{2 \pi} \left( r_1^2 \omega_1 - q r_2^2 \alpha_2' \right)} \ , \label{J1} \\
J_2 & = \sqrt{\lambda} \int_0^{2 \pi} {\frac {d\sigma}{2 \pi} \left( r_2^2 \omega_2 + q r_2^2 \alpha_1' \right)} \label{J2} \ .
\end{align}

\subsection{Constant radii solutions}

A simple solution to the equations of motion can be obtained if we take the radii $r_i$ to be constant, $r_i=a_i$. As in the undeformed case, the derivatives of the angles $\alpha_i$ become constant and thus
\be
\alpha_i = \bar{m}_i \sigma + \a_{0i} \ ,
\ee
while the windings get deformed to
\be
\bar{m}_i \equiv \frac {v_i + q a_2^2 \epsilon_{ij} \omega_j}{a_i^2} \ .
\ee
The integration constants $\a_{0i}$ can be set to zero through a rotation, and the constants $\bar{m}_i$ must be integers in order to satisfy the closed string periodicity condition. The equations of motion for $r_i$ reduce now to
\begin{align}
& \omega_1^2 - \bar{m}_1^2 + \Lambda = 0 \label{o1} \ , \\
& \omega_2^2 -\bar{m}_2^{2} - 2 q ( \omega_1 \bar{m}_2 - \omega_2 \bar{m}_1 )+ \Lambda = 0 \label{o2} \ ,
\end{align}
and thus we conclude that the Lagrange multiplier $\Lambda$ is constant on this solution. The Virasoro constraints can then be written as
\begin{align}	
& \sum_{i=1}^2 a_i^2 \left( \bar{m}_i^{2}+\omega_i^2 \right) = \kappa^2 \ , \label{energy} \\
& \bar{m}_1 J_1 + \bar{m}_2 J_2 = 0 \ . \label{virasoro2}
\end{align}
We will now find the energy as a function of the angular momenta and the integer winding numbers $\bar{m}_i$. In order to do this we will first use 
equations (\ref{sphere}) and (\ref{virasoro1}) to write the radii as functions of $\omega_i$ and $\bar{m}_i$,
\be
a_1^2 = \frac{\omega_2 \bar{m}_2}{\omega_2 \bar{m}_2-\omega_1 \bar{m}_1} \ , \quad 
a_2^2 = \frac{\omega_1 \bar{m}_1}{\omega_1 \bar{m}_1 - \omega_2 \bar{m}_2} \ .
\ee
With these relations at hand and the definitions (\ref{E})-(\ref{J2}), together with (\ref{energy}), we find
\be
E^2 = \frac{(J_1+ \sqrt{\lambda} q a_2^2 \bar{m}_2)^2}{a_1^2} +\frac{(J_2- \sqrt{\lambda} q a_2^2 \bar{m}_1)^2}{a_2^2} 
+ \lambda \left( a_1^2 \bar{m}_1^2 + a_2^2 \bar{m}_2^2 \right) \ ,
\ee
or after some immediate algebra,
\begin{align}
E^2 & = (J_1+J_2)^2 +J_1 J_2 \frac{(1 - \hbox{w})^2}{\hbox{w}} -2 \sqrt{\lambda} q \bar{m}_1 ( J_1 \hbox{w} + J_2 ) \notag \\
& + \lambda \left( \bar{m}_ 1 \bar{m}_2 - q^2 \bar{m}_1^2 \hbox{w} \right) \frac{\bar{m}_1 - \bar{m}_2 \hbox{w}}{\bar{m}_2 - \bar{m}_1 \hbox{w}} \ ,
\label{EJ1J2}
\end{align}
where we have made use of (\ref{virasoro2}) and we have introduced $\hbox{w} \equiv \omega_1/\omega_2$. 
Now we need to write the ratio $\hbox{w}$ as a function of the windings $\bar{m}_i$ and the angular momenta $J_i$. 
This can be done by adding equations (\ref{J1}) and (\ref{J2}), subtracting equation (\ref{o2}) from (\ref{o1}), 
and solving the resulting system of equations,
\begin{align}
& \big[ \bar{m}_1 J- \sqrt{\lambda} q \bar{m}_1 (\bar{m}_1 -\bar{m}_2) \big] \hbox{w} - \bar{m}_2  J - \sqrt{\lambda} (\bar{m}_1 -\bar{m}_2)\omega_1 = 0 \ , 
\label{J1J2} \\
& \omega_1^2 -\bar{m}_1^2-\frac{\omega_1^2}{\hbox{w}^2} +\bar{m}_2^2 +2 q \bar{m}_2 \omega_1 -2 q \bar{m}_1 \frac{\omega_1}{\hbox{w}} = 0 \ ,
\end{align}
where $J \equiv J_1+J_2$ is the total angular momentum. When we eliminate $\omega_1$ in these expressions we are left 
with a quartic equation in w
\begin{align}
& (\bar{m}_1 \hbox{w} - \bar{m}_2 )^2 \Big[ 1 - \Big( 1 - \frac {\sqrt{\lambda}}{J} q (\bar{m}_1 -\bar{m}_2 ) \Big)^2 
\hbox{w}^2 \Big] \nonumber \\
& + \frac {\lambda}{J^2} \hbox{w}^2 (\bar{m}_1 + \bar{m}_2 ) (\bar{m}_1 - \bar{m}_2 )^3 (1- q^2) = 0 \ .
\label{quartic}
\end{align}
Rather than trying to solve this equation explicitly, we can write the solution as a power series expansion in large $J/\sqrt{\lambda}$. Out of the four different solutions to (\ref{quartic}), the only one with a well-defined expansion\footnote{Two of the expansions will give us $\omega_i \sim \mc{O} (1)$ and, because also $a_i^2  \sim \mc{O} (1)$, this gives us that $J_1+J_2  \sim \mc{O} (1)$ too instead of the expected $J$. The third and fourth expansions will be given by w$=1+\dots$ and w$=-1+\dots$, the first one will be well defined when $J_1,J_2 \geq 0$, which also imply $m_1\geq 0\geq m_2$ or $m_2\geq 0\geq m_1$, and the second one will be well defined when $J_1$ and $J_2$ are one positive and the other negative.}  is 
\be
\hbox{w} =1 + \frac {\sqrt{\lambda}}{J} q ( \bar{m}_1-\bar{m}_2)  
+ \frac {\lambda}{2J^2} ( \bar{m}_1-\bar{m}_2) \big( \bar{m}_1+\bar{m}_2 +q^2 ( \bar{m}_1 - 3 \bar{m}_2) \big)  + \cdots 
\label{w}
\ee
which implies that
\begin{align}
\omega_1 & = \frac {J}{\sqrt{\lambda}} + \frac {\sqrt{\lambda}}{2J} \bar{m}_1 (\bar{m}_1 +\bar{m}_2) (1-q^2)  
\Big[ 1 - \frac {\sqrt{\lambda}}{J} q \bar{m}_2 + \dots \Big] \ , \label{w1} \\
\omega_2 & = \frac {J}{\sqrt{\lambda}} - q (\bar{m}_1 -\bar{m}_2) + \frac {\sqrt{\lambda}}{2J} \bar{m}_2 (\bar{m}_1 + \bar{m}_2) (1-q^2) 
\Big[ 1 - \frac {\sqrt{\lambda}}{J} q (\bar{m}_1 + \bar{m}_2) + \dots \Big] \ . \label{w2}
\end{align}
Note that the ${\cal O}(\sqrt{\lambda}/J)$ terms and the subsequent corrections in (\ref{w1}) and (\ref{w2}) are dressed 
with a common factor of $\bar{m}_1 + \bar{m}_2$ that vanishes for equal angular momenta. 
We can easily prove the existence of this factor if we set $\bar{m}_1 = - \bar{m}_2 \equiv m$ in equation (\ref{quartic}), which reduces to
\be
(1 + \hbox{w})^2 \big[ (J - 2 \sqrt{\lambda} q m )^2 \hbox{w}^2 - J^2 \big] = 0 \ ,
\ee
whose only well-defined solution is
\be
\hbox{w} = \frac {J}{J - 2 \sqrt{\lambda} q m} \ ,
\label{wexact}
\ee
and therefore we can calculate the frequencies $\omega_1$ and $\omega_2$ exactly, 
\be
\omega_1 = \frac {J}{\sqrt{\lambda}} \ , \quad 
\omega_2 = \frac {J}{\sqrt{\lambda}} - 2 q m \ . 
\ee
Substituting these values into relation (\ref{EJ1J2}) we find
\be
E^2 = J^2 - 2 \sqrt{\lambda} q m J + \lambda m^2 \ ,
\label{HSTdispersion}
\ee
which is an exact expression as the ratio $\hbox{w}$ is exact. This dispersion relation is a generalization of  the expression of circular string solutions with two equal angular momenta we have already seen (\ref{oppositewindingenergy}). It can be compared with the one obtained in \cite{HST} via the deformation the original bosonic currents, coinciding both.

An identical reasoning can be employed to prove the existence of the global factor $1-q^2$. If we substitute the value $q=1$, which corresponds to the pure NS-NS flux, in the equation (\ref{quartic}) we can solve it to get
\begin{align*}
\omega_1 & = \frac {J}{\sqrt{\lambda}} + \frac {\sqrt{\lambda} \bar{m}_1 (\bar{m}_1 + \bar{m}_2) J}{(J + \sqrt{\lambda} \bar{m}_2)^2} (1-q) +  \dots \ , \\
\omega_2 & = \frac {J}{\sqrt{\lambda}} - (\bar{m}_1 -\bar{m}_2) 
+ \frac {(J^2 - \lambda \bar{m}_1 \bar{m}_2) (\bar{m}_1 - \bar{m}_2) + \sqrt{\lambda} \bar{m}_2 J (3 \bar{m}_1 - \bar{m}_2)}{(J + \sqrt{\lambda} \bar{m}_2)^2} (1-q) + \dots  \ .
\end{align*}
Substituting into equation~(\ref{EJ1J2}) 
and performing some algebra, the expression for the energy reduces to 
\be
E = J - \sqrt{\l} \bar{m}_1 \ ,
\ee
when we write the angular momenta $J_1$ and $J_2$ in terms of the total momentum $J$.

If we substitute the general value of $\hbox{w}$ from equation (\ref{w}) in relation (\ref{EJ1J2}) we find
\be
E^2 = J^2  - 2 \sqrt{\lambda} q \bar{m}_1 J + \frac {\lambda}{J} \big[ (\bar{m}_1^2 J_1+\bar{m}_2^2 J_2) (1-q^2) 
+ q^2 \bar{m}_1^2 J \big] + \cdots \ .
\label{qdispersion}
\ee
When the flux vanishes this expression becomes the expansion for the energy in the Neumann-Rosochatius system 
describing closed string solutions rotating with two different angular momenta (\ref{generalenergyNeumannsystem}). 
We must note that the subleading terms not included in (\ref{qdispersion}) contain a common factor of $\bar{m}_1 + \bar{m}_2$. We can check that indeed at the particular case of $\bar{m}_1= - \bar{m}_2$ relation (\ref{qdispersion}) simplifies to the correct one. Similarly happens with $(1-q^2)$ factors.

\subsection{Non-constant radii computations. Elliptic strings}
\label{nonconstantsphere}

To study the solutions with non-constant radii it will prove more convenient not to work with the equations of motion but with the integrals of motion. We have already seen that integrability of the Neumann-Rosochatius system follows from the existence of a set of integrals of motion in involution, the Uhlenbeck constants. In the case of a closed string rotating in $S^3$ there are only two integrals $I_1^{NR}$ and $I_2^{NR}$, but as they must satisfy the constraint $I_1^{NR} + I_2^{NR} = 1$, we are left with a single independent constant. As we saw before this constant is given by 
\be
I_1^{NR} = r_1^2 + \frac{1}{\omega_1^2 - \omega_2^2} \left[ (r_1 r'_2 - r'_1 r_2)^2 
+ \frac{v_1^2}{r_1^2} r_2^2 + \frac{v_2^2}{r_2^2} r_1^2 \right] \ . 
\ee

When the NS-NS three-form is turned on the Uhlenbeck constants should be deformed in some way. In order to find this deformation we will 
assume that the extended constant can be written as
\be
\bar{I}_1=r_1^2 +\frac{1}{\omega_1^2 -\omega_2^2} \left[ (r_1 r'_2-r'_1 r_2)^2 +\frac{v_1^2}{r_1^2} r_2^2 +\frac{v_2^2}{r_2^2} r_1^2 +2 f \right] \ ,  
\label{I1tilde}
\ee
where $f=f(r_1,r_2,q)$. This function can be determined if we impose that $\bar{I}_1'=0$. After some immediate algebra we find that
\be
f' + \frac{(q^2 \omega_2^2 +2q\omega_2 v_1) r'_1}{r_1^3} +q^2 (\omega_1^2 -\omega_2^2) r_1 r'_1 =0 \ ,
\label{fprime}
\ee
where we have used the constraint (\ref{sphere}) together with 
\be
r_1 r'_1 + r_2 r'_2=0 \ , \quad r_1 r''_1+(r'_1)^2 + r_2 r''_2 +(r'_2)^2 = 0 \ , 
\ee
and the equations of motion (\ref{alphaprime}), (\ref{r1prime}) and (\ref{r2prime}). As all three terms in relation (\ref{fprime}) are total derivatives, 
integration is immediate and we readily conclude that the deformation of the Uhlenbeck constant is given by
\be
\bar{I}_1=r_1^2 (1-q^2) +\frac{1}{\omega_1^2 -\omega_2^2} \left[ (r_1 r'_2-r'_1 r_2)^2 +\frac{(v_1+q\omega_2)^2}{r_1^2} r_2^2 +\frac{v_2^2}{r_2^2} r_1^2 \right] \ .
\label{deformedUhlenbeck}
\ee
The Hamiltonian including the contribution from the NS-NS flux can also be written now using the deformed Uhlenbeck constants and the integrals of motion $v_i$,
\be
H_{S^3} = \frac {1}{2} \sum_{i=1}^2 \big[ \omega_i^2 \bar{I}_i + v_i^2 \big] + \frac {1}{2} q^2 (\omega_1^2 - \omega_2^2 ) - q \omega_1 v_2 \ .
\label{H}
\ee

A convenient way to proceed is to change variables to an ellipsoidal coordinate \cite{Arutyunov_2003,Babelon_1992}.
The ellipsoidal coordinate $\zeta$ is defined as the root of the equation
\be
\frac{r_1^ 2}{\zeta -\omega_1^ 2} + \frac{r_2^2}{\zeta -\omega_2^2} = 0 \ .
\ee
If we choose the angular frequencies such that $\omega_1 < \omega_2$ the range of the ellipsoidal coordinate is $\omega_1^2 \leq \zeta \leq \omega_2^2$. Using that
\be
(r_1 r'_2 -r_2 r'_1)^2=\frac{\zeta^{\prime 2}}{4 (\omega_1^2 -\zeta) (\zeta -\omega_2^2)} \ ,
\ee
and solving for $\zeta'^2$ in the deformed Uhlenbeck constant (\ref{deformedUhlenbeck}) we conclude that
\be
\zeta'^2 = - 4 P_3(\zeta) \ ,
\label{diffeqzeta}
\ee
where $P_3(\zeta)$ is the third order polynomial 
\begin{align}
P_3(\zeta)
& = (1-q^2) (\zeta -\omega_1^2 )^2 (\zeta -\omega_2^2) + (\zeta- \omega_1^2) (\zeta- \omega_2^2) (\omega_1^2 -\omega_2^2) \bar{I}_1 \notag \\
& + (\zeta - \omega_1^2)^2  v_2^2 + (\zeta - \omega_2^2)^2 (v_1 + q \omega_2)^2 = (1-q^2 ) \prod_{i=1}^3 (\zeta -\zeta_i)  \ .
\end{align}
This polynomial defines an elliptic curve $s^2 + P_3(\zeta)=0$. In fact if we change variables to
\be
\zeta = \zeta_3 + (\zeta_2 - \zeta_3) \eta^2 \ ,
\ee
equation (\ref{diffeqzeta}) becomes the differential equation for the Jacobian elliptic sine\footnote{All the Jacobi elliptic functions and elliptic integrals in this thesis are written following the convention from \cite{9780486612720}. This implies, for example, that $\text{dn}^2 (x,\kappa) +\kappa \text{ sn}^2 (x,\kappa)=1$.},
\be
\eta^{\prime 2}=(1-q^2) (\zeta_3 - \zeta_1) (1 - \eta ^2) (1- \kappa \eta^2) \ ,
\ee
where the elliptic modulus is given by $\kappa = (\zeta_2 - \zeta_3)/(\zeta_1 - \zeta_3)$. The solution is thus
\be
\eta(\sigma) = \hbox {sn} \big( \sigma \sqrt{(1-q^2) (\zeta_3 -\zeta_1)} +\sigma_0, \kappa \big) \ ,
\label{etasn}
\ee
with $\sigma_0$ an integration constant that can be set to zero by performing a rotation. Therefore we conclude that
\be
r_1^2(\sigma) = \frac{\zeta_3 - \omega_1^2}{\omega_2^2 - \omega_1^2} + \frac{\zeta_2 - \zeta_3}{\omega_2^2 - \omega_1^2} \, \hbox{sn}^2
\big( \sigma \sqrt{(1-q^2) (\zeta_3 -\zeta_1)} , \kappa \big) \ .
\label{r1elliptic}
\ee
We must note that we need to order the roots in such a way that $\zeta_1 < \zeta_3$ to make sure that the argument of the elliptic sine is real. We also need 
$\zeta_2 < \zeta_3$ to have $\kappa > 0$, together with $\zeta_1 < \zeta_2 $ to keep $\kappa < 1$.\footnote{We made this two choices because, although the Jacobi elliptic functions are defined for any real valued elliptic parameter $\kappa$, they can be re-expressed as Jacobi elliptic functions with elliptic parameter in the interval $[0,1]$, which we are going to denominate as their fundamental domain.} Furthermore, 
imposing that (\ref{r1elliptic}) must have codomain between $0$ and $1$ demands 
$\omega_1^2 \leq \zeta_{2,3} \leq \omega_2^2$. Note that this last restriction does not apply to $\zeta_1$. The periodicity condition on the radial coordinates implies that
\be
\pi \sqrt{(1-q^2) (\zeta_3 -\zeta_1)} = n \, \hbox{K} (\kappa ) \ , 
\ee
where we have used that $2\hbox{K}(\kappa)$ is the period of the square of the Jacobi sine, with 
$\hbox{K}(\kappa)$ being the complete elliptic integral of first kind and $n$ an integer number\footnote{There are four cases in which we have to alter this periodicity condition.
When either $v_1+ q\omega_2 = 0$ or $v_2 = 0$ the condition becomes $\frac{\pi}{2} \sqrt{(1-q^2) (\zeta_3 -\zeta_1)} = n \, \hbox{K} (\kappa )$ 
because of a change of branch in the square root in (\ref{r1elliptic}) that increase the perodicity to $4\hbox{K}(\kappa)$, the periodicity of the Jacobi sine. The two remaining cases correspond to the limit $\zeta_3 \rightarrow \zeta_2$, 
which is the constant radii case, and to the limit $\kappa \rightarrow 1$, where the periodicity of the elliptic sine becomes infinite. In both cases 
there is no periodicity condition. We will discuss these two limits later in this section.}. We also want to comment that our solution is of circular type. An exception could happen in the absence of R-R flux and setting the $v_i$ integrals to zero. This choice of parameters 
corresponds to solutions of circular type when $I_1$ is taken as negative, or solutions of folded\footnote{We call a solution \emph{folded} if $X_M(\sigma , \tau )=X_M (2\pi -\sigma , \tau )$, that is, the string is the same if it is traversed forward and backward.} type when $I_1$ is positive~\cite{Arutyunov_2003}. 

We must note that there are important cases where this solution can be reduced to a simpler one. 
They correspond to the choices of parameters that make the discriminant of $P_3(\zeta)$ equal to zero. Our hierarchy of roots implies that there are only three cases able to fulfil 
this condition. The first corresponds to solutions with constant radii, where $\zeta_2=\zeta_3$. These solutions were first constructed in~\cite{HST} 
and later on recovered by deriving the corresponding finite-gap equations in~\cite{Babichenko} or by solving the equations of motion for the flux-deformed 
Neumann-Rosochatius system in~\cite{Spinningstrings1}. The second case corresponds to the limit $\kappa=1$, which is obtained when $\zeta_1=\zeta_2$. 
These are the giant magnons analyzed in~\cite{BPP} for the $v_2=0$ case and in~\cite{ABozhilov} for general values of $v_2$ (giant magnon solutions were also constructed in~\cite{HST,Babichenko}). The third case corresponds to setting $\zeta_1=\zeta_2=\zeta_3$ and cannot be obtained unless we have equal angular frequencies, $\omega_1=\omega_2$.

Going back to the general case, we can use now equation (\ref{r1elliptic}) to write the winding numbers 
$\bar{m}_i$ in terms of the integration constants $v_i$ and the angular frequencies $\omega_i$. From the periodicity condition on $\alpha_1$,
\be
2 \pi \bar{m}_1 = \int_0^{2\pi}{\alpha'_1 d\sigma}=\int_0^{2\pi}{\left( \frac{v_1}{r_1^2} + q \omega_2 \frac{r_2^2}{r_1^2} \right) d\sigma} \ ,
\ee
we can write
\be
\frac{\bar{m}_1 + q \omega_2}{v_1 + q \omega_2}= \int_0^{2\pi} \frac{d\sigma}{2\pi} \frac{1}{r_1^2} \ .
\ee
Inserting (\ref{r1elliptic}) in this expression and performing the integration we find
\be
\bar{m}_1 + q \omega_2 = \frac{(v_1 + q \omega_2) (\omega_2^2 - \omega_1^2)}{(\zeta_3 - \omega_1^2)
\hbox{K} (\kappa)} \Pi \left( \frac{\zeta_3 - \zeta_2}{\zeta_3 - \omega_1^2} , \kappa \right) \ ,
\ee
where $\Pi(a,b)$ is the complete elliptic integral of third kind. In a similar way, from the periodicity condition for $\alpha_2$, 
\be
2 \pi \bar{m}_2 = \int_0^{2\pi}{\alpha'_2 d\sigma} = \int_0^{2\pi}{\left( \frac{v_2}{r_2^2} - q \omega_1 \right) d \sigma} \ ,
\ee
we find that
\be
\frac{\bar{m}_2 + q \omega_1}{v_2}= \int_0^{2\pi} \frac{d\sigma}{2\pi} \frac{1}{r_2^2} \ ,
\ee
that we can integrate to get
\be
\bar{m}_2 + q \omega_1 = \frac{v_2 (\omega_2^2 - \omega_1^2 )}{(\omega_2^2 - \zeta_1) \hbox{K} (\kappa)} \Pi 
\left( - \ \frac{\zeta_3 - \zeta_2}{\omega_2^2 - \zeta_3} ,\kappa \right) \ .
\ee
We can perform an identical computation to obtain the angular momenta. From equation (\ref{J1}) we get
\be
\frac{J_1} {\sqrt{\lambda}} + q v_2 - q^2 \omega_1 = \omega_ 1 (1-q^2) \int_0^{2\pi}{\frac{d\sigma}{2\pi} r_1^2} \ ,
\ee
and therefore
\be
\frac{J_1}{\sqrt{\lambda}} = \frac{\omega_1 (1-q^2)}{\omega_2^2 -\omega_1^2} \left[ \zeta_3 - \omega_1^2 - (\zeta_3 - \zeta_1) \
\left( 1-\frac{ \hbox{E} (\kappa)}{\hbox{K} (\kappa)} \right) \right] - q v_2 +q^2 \omega_1 \ .
\ee
with $\hbox{E}(\kappa)$ the complete elliptic integral of the second kind. As before, (\ref{J2}) implies
\be
\frac{J_2}{\sqrt{\lambda}} + q v_1 - q \bar{m}_1= \omega_2 (1-q^2) \int_0^{2\pi}{\frac{d\sigma}{2\pi} r_2^2} \ ,
\ee
and thus after integration we conclude that
\be
\frac {J_2}{\sqrt{\lambda}} = \frac{\omega_2 (1-q^2)}{\omega_2^2 -\omega_1^2} \left[ \omega_2^2 -\zeta_3 + (\zeta_3 - \zeta_1) \
\left( 1-\frac{\hbox{E}(\kappa)}{\hbox{K}(\kappa)} \right) \right] - q v_1 +q \bar{m}_1 \ .
\ee
These expressions for the angular momenta can be used to rewrite the first Virasoro constraint (\ref{virasoro1}) as
\be
\omega_2 J_1 +\omega_1 J_2 = \sqrt{\lambda} \left( \omega_1 \omega_2 +q\omega_1 \bar{m}_1 \right) \ .
\ee
We could now employ these relations to write the energy in terms of the winding numbers $\bar{m}_i$ and the angular momenta $J_i$. However the resulting expression is rather lengthy and cumbersome. Instead in the following subsection we will focus on the analysis of the above solutions in the limit of pure NS-NS flux. 

We must stress that the expressions  we have obtained have to be modified in the giant magnon solution and other cases where the periodicity condition cannot be imposed or the string does not close. Therefore factors $\sqrt{(1-q^2) (\zeta_3 -\zeta_1)} / n \hbox{K} (\kappa )$, which had been cancelled 
in the expressions we have obtained for the angular momenta and windings, do not cancel anymore.


\subsubsection{Solutions with pure NS-NS flux}

The cubic term in the polynomial $P_3(\zeta)$ is dressed with a factor $1-q^2$. Therefore in the case of pure NS-NS three-form flux 
the degree of the polynomial reduces to two and the solution can be written using trigonometric functions. In this limit\footnote{We can also take the limit directly in equation (\ref{etasn}) if we note that 
$\zeta_1$ goes to minus infinity when we set $q = 1$. In this limit the elliptic modulus vanishes but the factor $(1-q^2)\zeta_1$ 
in the argument of the elliptic sine remains finite and we just need to recall that $\hbox{sn} (x,0) = \sin x$.}
\be
\zeta^{\prime 2} = - 4 P_2 (\zeta) \ ,
\label{Qequation}
\ee
with $P_2(\zeta)$ the second order polynomial
\begin{align}
P_2 (\zeta) & = (\zeta - \omega_1^2) (\zeta - \omega_2^2) (\omega_1^2 - \omega_2^2) \bar{I}_1 + (\zeta - \omega_1^2)^2  v_2^2 \nonumber \\ 
& + (\zeta - \omega_2^2)^2 (v_1 + \omega_2)^2 = \omega^2 (\zeta - \tilde{\zeta}_1) (\zeta - \tilde{\zeta}_2) \ ,
\end{align}
where $\omega^2$ is 
\be
\omega^2 = (\omega_1^2 -\omega_2^2) \bar{I}_1 + (v_1 + \omega_2)^2 + v_2^2 \ .
\label{omega}
\ee
The solution to equation (\ref{Qequation}) is given by
\be
\zeta(\sigma) = \tilde{\zeta}_2 +( \tilde{\zeta}_1 - \tilde{\zeta}_2 ) \sin ^2 ( \omega \sigma ) \ ,
\ee
where we have set to zero an integration constant by performing a rotation. Therefore
\be
r_1^2(\sigma) = \frac{\tilde{\zeta}_2 - \omega_1^2}{\omega_2^2 - \omega_1^2} + \frac{\tilde{\zeta}_1 - \tilde{\zeta}_2}{\omega_2^2 - \omega_1^2} \, \sin^2
(\omega \sigma) \ .
\label{r1q1}
\ee
Periodicity of the radial coordinates implies that $\omega$ must be a half-integer number\footnote{An important exception to this condition happens when $\omega = J + \bar{m}_2$, which is a solution even if it is not a half-integer. This value corresponds to the case of constant radii we have already seen in the previous subsection,
\[
\bar{I}_1 = -\left| \frac{2(v_1+\omega_2) v_2}{\omega_2^2-\omega_1^2} \right| \ , \quad 
J_1 = \frac{\bar{m}_2 J}{\bar{m}_2 - \bar{m}_1} \ , \quad J_2 = \frac{\bar{m}_1 J}{\bar{m}_1 - \bar{m}_2} \ , \quad E  = J - \sqrt{\lambda} \bar{m}_1 \ .
\]
}. The relation between the winding numbers $\bar{m}_i$ and the constants $v_i$ and the frequencies $\omega_i$ is now rather simple. 
The periodicity condition for the angles implies
\begin{align}
\bar{m}_1 + \omega_2 & = \frac{(v_1 + \omega_2)(\omega_1^2 - \omega_2^2)}{\sqrt{(\omega_1^ 2-\tilde{\zeta}_1)(\omega_1^ 2 - \tilde{\zeta}_2)}}
= \frac{ \omega (v_1 + \omega_2)(\omega_1^2 - \omega_2^2)}{\sqrt{P_2(\omega_1^2)}} = \omega \, \hbox{sgn}(v_1 + \omega_2) \ ,\label{p1} \\
\bar{m}_2 + \omega_1& = \frac{v_2(\omega_1^2-\omega_2^2)}{\sqrt{(\omega_2^ 2-\tilde{\zeta}_1)(\omega_2^ 2 - \tilde{\zeta}_2)}}
= \frac{\omega v_2(\omega_1^2-\omega_2^2)}{\sqrt{P_2(\omega_2^2)}} = \omega \, \hbox{sgn} (v_2) \ . \label{p2}
\end{align}
From the definition of the angular momenta we find
\be
\frac{J_1}{\sqrt{\lambda}} = \omega_1-v_2 \ , \quad \frac{J_2}{\sqrt{\lambda}} = \bar{m}_1 - v_1 \ .
\label{p3}
\ee
We can now write the energy as a function of the winding numbers and the angular momenta. 
A convenient way to do this is recalling the relation between the energy and the Uhlenbeck constant. 
If we assume that both $v_1+\omega_2$ and $v_2$ are positive (the extension to the other possible signs of $v_1+\omega_2$ and $v_2$ is immediate)  
and we combine equations (\ref{H}) and (\ref{omega}) we can write
\be
E^2 = \lambda \big(  \omega^2 + \omega_1^2 - \omega_2^2 - 2v_1 \omega_2 - 2 v_2 \omega_1 \big) \ ,
\ee
and thus using relations (\ref{p1})-(\ref{p3}) we can write it in terms of windings and angular momenta
\be
E^2 = \lambda \bar{m}_1^2 + \big( 2 \sqrt{\lambda} J_1 - \lambda ( \omega - \bar{m}_2 ) \big) (\omega - \bar{m}_2) 
+ 2 \sqrt{\lambda} J_2 (\omega - \bar{m}_1) \ .
\label{secondEJ1J2}
\ee
Now we can use the Virasoro constraint (\ref{virasoro1}) to express it only in terms of the total angular momentum $J=J_1 +J_2$, as it allow us to write
\be
J_1 = \frac{(J - \sqrt{\lambda} \omega)(\omega - \bar{m}_2)}{\bar{m}_1 - \bar{m}_2} \ , \quad J_2 = \frac{ J (\bar{m}_1 - \omega) 
+ \sqrt{\lambda} \omega ( \omega - \bar{m}_2 )}{\bar{m}_1 - \bar{m}_2} \ .
\ee
Replacing these expressions in (\ref{secondEJ1J2}) we obtain the energy 
as a function of the winding numbers and the total momentum, 
\begin{equation}
E^2 = \lambda \big( \bar{m}_1^2 - \bar{m}_2^2 + 4 \omega \bar{m}_2- 3 \omega^2 \big) - 2\sqrt{\lambda} J (\bar{m}_1 + \bar{m}_2 -2\omega) \ .
\end{equation}


\section{Spinning strings in $AdS_3$} \sectionmark{Spinning strings in  $A\MakeLowercase{d}S_3$}

Before moving to the full $AdS_3  \times S^3$ we are going to take a look to the case with no dynamics in the sphere. We can describe these configurations with the ansatz
\be
Y_3 + i Y_0 = z_0 (\sigma) \, e^{i w_0 \tau + i\beta_0 (\sigma)} \ , \quad Y_1 + i Y_2 = z_1 (\sigma) \, e^{i w_1 \tau + i \beta_1 (\sigma)} \ ,
\label{ansatzAdS}
\ee
together with the periodicity conditions
\be
z_a(\sigma + 2 \pi) = z_a (\sigma) \ , \quad \beta_a(\sigma + 2 \pi) = \beta_a(\sigma) + 2 \pi \bar{k}_a \ ,
\ee
with $a=0,1$. Note however that the time direction has to be single-valued so we need to exclude windings 
along the time coordinate. Therefore we must take $\bar{k}_0=0$. When we substitute this ansatz in the Polyakov action 
in the conformal gauge we obtain
\begin{equation}
L_{AdS_3} = \frac{\sqrt{\lambda}}{4\pi} \Big[ g^{ab} \left( z'_a z'_b + z_a z_a \beta_b'^2 - z_a z_a w_b^2 \right) 
-\frac {\tilde{\Lambda}}{2} \left( g^{ab} z_a z_b +1 \right) - 2 q z_1^2 ( w_0 \beta '_1 - w_1 \beta '_0 ) \Big] \ , \label{AdSNRq}
\end{equation}
where we have chosen $g=\hbox{diag}(-1,1)$ and $\tilde{\Lambda}$ is a Lagrange multiplier. The equations of motion for $z_a$ are
\begin{align}
z''_0 &=z_0 \beta^{'2}_0 -z_0 w_0^2 -\tilde{\Lambda} z_0 \ , \label{z0prime} \\ 
z''_1 &=z_1 \beta^{'2}_1 -z_1 w_1^2 -\tilde{\Lambda} z_1 -2 q z_1 ( w_0 \beta'_1 - w_1 \beta'_0 ) \ , \label{z1prime}
\end{align}
and the equations for the angular functions are 
\be
\beta '_a = \frac{u_a + q z_1^2 \epsilon_{ab} w_b}{g^{aa} z_a^2} \ , \label{betaprime}
\ee
where $u_a$ are some integration constants analogous to the $v_i$ we have defined for the spherical case. To these equations we need to add the $AdS$ constraint 
\be
- z_0^2 + z_1^2 = - 1 \ ,
\label{AdSconstraint}
\ee
together with the Virasoro constraints
\begin{align}
& z^{'2}_0 + z_0^2 (\beta ^{'2}_0 + w_0^2 ) = z^{'2}_1 + z_1^2 (\beta ^{'2}_1 + w_1^2) \label{Virasoro1AdS} \ , \\
& z_1^2 w_1 \beta '_1 = z_0^2 w_0 \beta '_0 \label{Virasoro2AdS} \ .
\end{align}
The spin and the energy in this case are given by
\begin{align}
E & = \sqrt{\lambda} \int_0^{2\pi} {\frac{d\sigma}{2\pi} (z_0^2 w_0 + q z_1^2 \beta'_1)} \ , \label{EAdS} \\
S & = \sqrt{\lambda} \int_0^{2\pi} {\frac{d\sigma}{2\pi} 
(z_1^2 w_1 -q z_1^2 \beta'_0)} \ . \label{SAdS}
\end{align}

\subsection{Constant radii solutions}

As before we can start by looking at solutions where  the string radii are taken as constant, 
$z_a=b_a$. In this case the periodicity condition on $\beta_0$ and the fact that the time coordinate is single-valued implies 
\be
\beta'_0=0 \ ,
\ee
and thus the equations of motion reduce to
\begin{align}
& w_1^2 -\bar{k}_1^2 - w_0^2 + 2 q w_0 \bar{k}_1 = 0 \ .
\end{align}
where we have used that $\beta'_1=\bar{k}_1$. The Virasoro constraints become then
\begin{align}
& b_1^2 (w_1^2 +\bar{k}_1^{2}) = b_0^2 w_0^2 \ , \\
& \bar{k}_1 S= 0 \ .
\end{align}
Therefore there are only two kinds of solutions: those with no spin and those with no winding. However both of them are inconsistent. The first one gives a pure imaginary value of the energy and the second one imposes the constraint $b_0=b_1$, which is inconsistent with the AdS constraint $b_1^2+1=b_0^2$.

\subsection{Non-constant radii computations. Elliptic strings}
\label{nonconstantads}

Despite the fact that there are no consistent constant radii solutions, there exist consistent elliptic solutions. As in the previous section, in order to construct general solutions for strings rotating in $AdS_3$ it will be convenient 
to introduce an analytical continuation of the ellipsoidal coordinates. The definition of this coordinate $\mu$ can be directly borrowed 
from the definition for the sphere with a change of sign,
\be
\frac{z_1^ 2}{\mu -w_1^ 2} - \frac{z_0^2}{\mu-w_0^2} = 0 \ .
\ee
If we order the frequencies such that $w_1 > w_0$, the range of the ellipsoidal coordinate will be $w_1^2 \leq \mu$. 
Now we can again make use of the Uhlenbeck constants to obtain a first order differential equation for this coordinate. 
In the case of the (analytically-continued to $AdS_3$) Neumann-Rosochatius system the (anallytically-continued) Uhlenbeck integrals 
satisfy the constraint $F_1 - F_0 = -1$, and thus we are again left with a single independent constant. 
To obtain the deformation of, say, $F_1$ by the NS-NS flux we can proceed in the same way as in the previous section. 
After some immediate algebra we conclude that
\be
\bar{F}_1=z_1^2 (1-q^2) +\frac{1}{w_1^2 -w_0^2} \left[ (z_1 z'_0-z'_1 z_0)^2 +\frac{(u_0+q w_1)^2}{z_0^2} z_1^2 +\frac{u_1^2}{z_1^2} z_0^2 \right] \ .
\ee
The Hamiltonian can also be written now using the deformed Uhlenbeck constants and the integrals of motion $u_a$,
\be
H = \frac {1}{2} \sum_{a=0}^1 \big[ g_{aa} w_a^2 \bar{F}_a - u_a^2 \big] + q u_1 w_0 \ .
\label{HAdS}
\ee
Now we need to note that
\be
(z_1 z'_0 -z_0 z'_1)^2=\frac{\mu^{\prime 2}}{4 (\mu -w_1^2) (\mu-w_0^2)} \ . 
\ee
When we solve for $\mu'^2$ in the deformed integral of motion we find that
\be
\mu'^2 = -4 Q_3 (\mu) \ ,
\label{diffeqzmu}
\ee
where $Q_3(\mu)$ is the third order polynomial, 
\begin{align}
Q_3 (\mu)
& = (1-q^2) (\mu -w_1^2 )^2 (\mu -w_0^2) + (\mu -w_1^2 ) (\mu -w_0^2) (w_0^2 -w_1^2) \bar{F}_1 \notag \\
& + (\mu - w_1^2 )^2  (u_0+q w_1)^2 + (\mu -w_0^2)^2 u_1^2 = (1-q^2 ) \prod_{i=1}^3 (\mu -\mu_i)  \ .
\end{align}
This equation nearly identical to the spherical one, so we can write
\be
z_0^2(\sigma) = \frac{\mu_3 - w_0^2}{w_1^2 -w_0^2} +\frac{\mu_2 -\mu_3}{w_1^2 -w_0^2} \, \hbox{sn}^2 
\big( \sigma \sqrt{(1-q^2) (\mu_3 -\mu_1)} , \nu \big) \ ,
\label{z0elliptic}
\ee
where the elliptic modulus is $\nu = (\mu_3 - \mu_2)/(\mu_3 - \mu_1)$. 
As in the case of strings rotating in $S^3$, we must perform now an analysis of the roots of the polynomial. We need to choose $\mu_3 > \mu_1$ 
to make sure that the argument of the elliptic sine is real, and $\mu_3 > \mu_2$ to have $\nu>0$, together with $\mu_2>\mu_1$ to keep $\nu<1$. 
Furthermore we have to impose $z_0^2 \geq 1$ which constrains $\mu_2$ and $\mu_3$ to be greater or equal than $w_1^2$. This restriction does 
not apply to $\mu_1$. Note that this hierarchy of roots implies that not all possible combinations of the parameters $u_i$, $w_i$ and $\bar{F}_1$ are allowed.

As in the previous section, there are three possible cases where this general solution is simplified as a consequence of the vanishing discriminant of $Q_3(\mu)$. The first one is the constant radii case, where $\mu_2=\mu_3$. However this limit is not well defined because, as we have seen, there is no consistent constant radii solution. The second case corresponds to the limit $\kappa=1$ and it is obtained when $\mu_1=\mu_2$. In this case there is no periodicity condition because the elliptic sine has infinite period and thus the string does not close. The third case corresponds to $\mu_1=\mu_2=\mu_3$ and requires setting $w_1=w_0$.

The periodicity condition on the radial coordinates now implies that
\be
\pi \sqrt{(1-q^2) (\mu_3 -\mu_1)} = n' \hbox{K} (\nu) \ ,
\ee
with $n'$ an integer number\footnote{Again there are four different cases where this condition 
must be modified. When $u_0+q w_1=0$ or $u_1=0$ the periodicity condition becomes 
$\frac{\pi}{2} \sqrt{(1-q^2) (\mu_3 -\mu_1)} = n' \, \hbox{K} (\nu)$ because of a change of branch in the square root in (\ref{z0elliptic}). The other two cases 
are the degenerate limits where there is no periodicity.}. From the periodicity condition on $\beta_1$,
\be
2 \pi \bar{k}_1 = \int_0^{2\pi}{\beta'_1 d\sigma}=\int_0^{2\pi}{\left( \frac{u_1}{z_1^2} + q w_0  \right) d\sigma} \ ,
\ee
we can write
\be
\frac{\bar{k}_1 - q w_0}{u_1}= \int_0^{2\pi} \frac{d\sigma}{2\pi} \frac{1}{z_1^2} \ .
\ee
Performing the integration we find
\be
\bar{k}_1 - q w_0= \frac{u_1 (w_1^2 -w_0^2)}{(\mu_3 - w_1^2) 
\hbox{K}  (\nu) } \Pi \left( \frac{\mu_3 -\mu_2}{\mu_3 - w_1^2}, \nu \right) \ .
\ee
The periodicity condition for $\beta_0$ implies that
\be
2 \pi \bar{k}_0 = \int_0^{2\pi}{\beta'_0 d\sigma} = \int_0^{2\pi}{\left( -\frac{u_0}{z_0^2} + q w_1 \frac{z_1^2}{z_0^2} \right) d \sigma} \ .
\label{beta0periodicity}
\ee
Now we must remember that we are working in $AdS_3$ instead of its universal covering. The time coordinate should therefore be single-valued, 
and thus we have to exclude windings along the time direction. When we set $\bar{k}_0=0$ equation (\ref{beta0periodicity}) becomes
\be
\frac{q w_1}{u_0 +q w_1}= \int_0^{2\pi} \frac{d\sigma}{2\pi} \frac{1}{z_0^2} \ ,
\ee
that we can integrate to get
\be
q w_1= \frac{(u_0 +q w_1) (w_1^2-w_0^2 )}{(\mu_3 - w_0^2) \hbox{K} (\nu)} \Pi \left( \frac{\mu_3 -\mu_2}{\mu_3 - w_0^2} , \nu \right)  \ .
\ee
In the same way we can perform an identical computation to obtain the energy and the spin. From equation (\ref{EAdS}) we get
\be
\frac{E} {\sqrt{\lambda}} + q u_1 - q^2 w_0= (1-q^2) w_0 \int_0^{2\pi}{\frac{d\sigma}{2\pi} z_0^2} \ ,
\ee
and thus
\be
\frac{E} {\sqrt{\lambda}} =q^2 w_0- q u_1 +\frac{(1-q^2) w_0}{w_1^2 -w_0^2} \left[ \mu_3 - w_0^2 -(\mu_3 - \mu_1) 
\left( 1-\frac{ \hbox{E} (\nu)}{\hbox{K} (\nu)} \right) \right] \ .
\ee
Repeating the same steps with (\ref{SAdS}) we obtain an expression for the spin,
\be
\frac{S} {\sqrt{\lambda}} - q u_0 = (1-q^2) w_1 \int_0^{2\pi}{\frac{d\sigma}{2\pi} z_1^2} \ ,
\ee
and thus
\be
\frac{S} {\sqrt{\lambda}} =q u_0 +\frac{(1-q^2) w_1}{w_1^2 -w_0^2} \left[ \mu_3 - w_1^2 - (\mu_3 - \mu_1) 
\left( 1-\frac{ \hbox{E} (\nu)}{\hbox{K} (\nu)} \right) \right] \ .
\ee
These expressions for the energy and the spin can be used to rewrite the first Virasoro constraint (\ref{Virasoro2AdS}) as
\be
w_1 E - w_0 S =\sqrt{\lambda} w_0 w_1 \ .
\ee
which is already a very closed expression. However we need a relation involving only $E$, $S$ and~$\bar{k}_1$. This relation can be readily found from the above equations but it is again a lengthy and complicated expression and we will not present it here. We will consider instead in the following subsection the limit of pure NS-NS flux of these solutions. 

As happened in the spherical case, factors $\sqrt{(1-q^2) (\mu_3 -\mu_1)}/n \hbox{K} (\nu )$ do not cancel anymore in the expressions for the energy, the spin and the winding number for the giant magnon and non-periodic solutions.


\subsubsection{Solutions with pure NS-NS flux}

As in the case of strings rotating in $S^3$ in the limit of pure NS-NS three-form flux, the above solutions also can be written in terms of trigonometric functions in the same limit. Now (\ref{diffeqzmu}) reduces to
\be
\mu^{\prime 2} = - 4 Q_2 (\mu) \ ,
\label{Q2equation}
\ee
with $Q_2(\mu)$ the second order polynomial
\begin{align}
Q_2 (\mu) & = (\mu - w_1^2) (\mu - w_0^2) (w_0^2 - w_1^2) \bar{F}_1 + (\mu - w_0^2)^2  u_1^2 \nonumber \\ 
& + (\mu - w_1^2)^2 (u_0 + w_1)^2 = \omega'^2 (\mu - \tilde{\mu}_1) (\mu - \tilde{\mu}_2) \ ,
\end{align}
where $\omega'^2$ is 
\be
\omega'^2 = (w_0^2-w_1^2) \bar{F}_1 +(u_0+w_1)^2 +u_1^2 \ .
\label{omegaprime} 
\ee
Thus we conclude that
\be
z_0^2(\sigma) = \frac{\tilde{\mu}_2 - w_0^2}{w_1^2 - w_0^2} + \frac{\tilde{\mu}_1 - \tilde{\mu}_2}{w_1^2 - w_0^2} \, \sin ^2 (\omega' \sigma ) \ .
\label{z0trig}
\ee
The periodicity condition on the radial coordinates implies now that $\omega'$ should be a half-integer number. 
The frequencies $w_a$ and the integration constants $u_a$ are related to the energy, the spin and the winding number $\bar{k}_1$ by 
\begin{align}
w_1 & = \omega' \, \hbox{sgn} \, (u_0+w_1)\ , \quad  \omega' = (\bar{k}_1-w_0) \, \hbox{sgn}(u_1) \label{q1} \ , \\
S & = \sqrt{\lambda} u_0 \ , \quad E = \sqrt{\lambda} (w_0 -u_1)=\frac{w_0}{w_1} S+\sqrt{\lambda} w_0 \label{q2} \ .
\end{align}
Recalling now the Virasoro condition (\ref{Virasoro1AdS}) the spin can be written as
\be
S =  \sqrt{\lambda} \, \frac{(\bar{k}_1-\omega')^2 \omega'}{2 \bar{k}_1 (2 \omega' - \bar{k}_1)} \ , 
\ee
while the energy is given by
\be
E = \sqrt{\lambda} \, \frac{\bar{k}_1^3 - 3 \bar{k}_1^2 \omega' + \bar{k}_1 \omega^{\prime 2} 
+ \omega^{\prime 3}}{2 \bar{k}_1 (\bar{k}_1 - 2 \omega')} \ .
\ee
We must note that we still have to impose a restriction on the parameters.  
This restriction comes from imposing that the discriminant of $Q_2 (\mu)$ must be positive 
and taking the region in the parameter space with the correct hierarchy of roots. 
This condition can be written as
\be
 | 2 (u_0 + w_1) u_1 | \leq \left| \bar{F}_1 (w_1^2 -w_0^2)\right| =\left| \omega ^{\prime 2} -(u_0 +w_1)^2-u_1^2 \right| \ .
\ee
The inequality is saturated in the cases that would correspond to constant radii. However in this point our equations become not well defined as a consequence of these solutions being inconsistent.


\section{Spinning strings in $AdS_3  \times S^3$} \sectionmark{Spinning strings in  $A\MakeLowercase{d}S_3\times S^3$}

We will now extend the previous analysis to the case where the string can rotate both in $AdS_3$ and $S^3$, again with no dynamics along $T^4$. Therefore the string solutions that we are going to consider will have one spin $S$ in $AdS_3$ and two angular momenta $J_1$ and $J_2$ in $S^3$. We can describe these configurations by the two ansätze we have already seen (\ref{ansatz}) and (\ref{ansatzAdS}), with the same periodicity conditions. When we substitute this ansatz in the Polyakov action in the conformal gauge we obtain the Lagrangian
\begin{equation}
	L=L_{S^3}+L_{AdS_3}
\end{equation}
where $L_{S^3}$ is the Lagrangian (\ref{NRq}) and $L_{AdS_3}$ is the Lagrangian (\ref{AdSNRq}), that is, the pieces of the Lagrangian describing motion along $AdS_3$ and $S^3$ are decoupled. This implies that the equations of motion are not modified and are given directly by expressions (\ref{alphaprime})--(\ref{r2prime}) and (\ref{z0prime})--(\ref{betaprime}). The Virasoro constraints do get modified, and therefore are responsible for the coupling between the $AdS_3$ and the $S^3$ systems, 
\begin{align}
& z^{'2}_0 + z_0^2 (\beta ^{'2}_0 + w_0^2 ) = z^{'2}_1 + z_1^2 (\beta ^{'2}_1 + w_1^2)
+ \sum_{i=1}^2 \big( r^{'2}_i + r_i^2 (\alpha^{'2}_i + \omega^2_i ) \big)  \ , \\
& z_1^2 w_1 \beta '_1 + \sum_{i=1}^2 r_i^2 \omega _i \alpha '_i  = z_0^2 w_0 \beta '_0 \ .
\end{align}
The angular momenta are defined again as in equations (\ref{J1}) and (\ref{J2}), the spin is defined as in equation (\ref{SAdS}) and the energy is defined as in equation (\ref{EAdS}).

\subsection{Constant radii solutions}

As before a simple solution to these equations can be found when the string radii are taken as constant, 
$r_i=a_i$ and $z_a=b_a$. In this case the periodicity condition on $\beta_0$ and the fact that the time coordinate is single-valued implies 
\be
\beta'_0=0 \ .
\ee
Furthermore the angles can be easily integrated again,
\be
\beta'_1= \bar{k}_1 \ , \quad \alpha'_i=\bar{m}_i \ , \quad i=1,2 \ ,
\ee
and thus the equations of motion reduce to
\begin{align}
& w_1^2 -\bar{k}_1^2 - w_0^2 + 2 q w_0 \bar{k}_1 = 0 \ , \label{eom1} \\ 
& (\omega_2^2 -\omega_1^2)-(\bar{m}_2^2 -\bar{m}_1^2) - 2q ( \omega_1 \bar{m}_2 - \omega_2 \bar{m}_1 ) = 0 \label{eom2} \ .
\end{align}
The Virasoro constraints become then
\begin{align}
& b_1^2 (w_1^2 +\bar{k}_1^{2}) + \sum_{i=1}^2 a_i^2 (\omega^2_i + \bar{m}^{2}_i) = b_0^2 w_0^2 \ , \\
& \bar{k}_1 S + \bar{m}_1 J_1 + \bar{m}_2 J_2 = 0 \ .
\end{align}
Using the definitions of the energy and the spin, equations (\ref{EAdS}) and (\ref{SAdS}), together with the constraint (\ref{AdSconstraint}), 
we can write
\be
E_{\pm} =  \sqrt{\lambda} \, w_0 \pm \frac{S (w_0 - q \bar{k}_1)}{\sqrt{w_0^2 + \bar{k}^2 - 2 q \bar{k} w_0 }} \ .
\label{ESJ1J2}
\ee
The plus sign corresponds to the case where $w_0$ and $w_1$ are chosen to have equal signs, while the minus sign corresponds 
to the choice of opposite signs. 
We can use now this expression to write the energy as a function of the spin, the two angular momenta 
and the winding numbers $\bar{k}_1$ and $\bar{m}_i$. As in the previous section we can take the second 
Virasoro constraint together with the condition that $a_1^2+a_2^2=1$ to find that 
\be
a_1^2 = \frac{ \bar{k}_1 S + \sqrt{\lambda} \omega_2 \bar{m}_2}{\sqrt{\lambda} (\omega_2 \bar{m}_2 - \omega_1 \bar{m}_1)} \ , \quad 
a_2^2 = \frac{ \bar{k}_1 S +\sqrt{\lambda} \omega_1 \bar{m}_1}{\sqrt{\lambda} (\omega_1 \bar{m}_1 - \omega_2 \bar{m}_2)} \ . 
\label{a1a2}
\ee
Taking these relations into account when adding the angular momenta (\ref{J1}) and (\ref{J2}) we find
a relation between the frequencies $\omega_1$ and $\omega_2$,
\begin{align}
& \left[ \bar{k}_1 S + \bar{m}_1 J -\sqrt{\lambda} q \bar{m}_1 (\bar{m}_1 - \bar{m}_2) \right] 
\frac{\omega_1}{\omega_2} - ( \bar{k}_1 S + \bar{m}_2 J ) \nonumber \\ 
& - \sqrt{\lambda} ( \bar{m}_1 - \bar{m}_2) \, \omega_1 - \frac{q \bar{k}_1 S (\bar{m}_1 - \bar{m}_2)}{\omega_2} = 0 \ ,
\label{w1w2}
\end{align}
which extends expression (\ref{J1J2}) to the case of spin in $AdS_3$. Combining now equation (\ref{eom2}) with (\ref{w1w2}) 
we can solve for $\omega_1$. The result is again a quartic equation,
\begin{align}
& \Big[(\omega_1 +q \bar{m}_2)^2 -(\bar{m}_1^2-\bar{m}_2^2) (1-q^2) \Big] \Big[ \lambda (\bar{m}_1-\bar{m}_2)\omega_1^2
+ 2\sqrt{\lambda} (\bar{m}_2 J+ \bar{k}_1 S) \omega_1 \notag \\
& - \big( ( \bar{m}_1 + \bar{m}_2 ) J + 2 \bar{k}_1 S \big) J \Big]
- ( \bar{m}_1 + \bar{m}_2 ) ( \bar{m}_1 J + \bar{k}_1 S )^2 (1-q^2) = 0 \ .
\label{quarticAdS}
\end{align}
Once we have found the solution to this equation, we can read $\omega_2$ from (\ref{w1w2}) and 
use then the first Virasoro constraint to calculate $w_0$. But before writing the resulting equation let us first take into account that
\be
b_1^2 w_1^2 +b_1^2 \bar{k}_1^2 -b_0^2 w_0^2= b_1^2 (2\bar{k}_1^2 -2q w_0 \bar{k}_1)- w_0^2 
= \frac{ 2\bar{k}_1 S (\bar{k}_1-q w_0)}{\sqrt{\lambda (w_0^2 +\bar{k}_1^2 -2q \bar{k}_1 w_0 )}} - w_0^2 \ ,
\ee
where we have made use of (\ref{eom1}). The Virasoro constraint becomes thus a sixth-grade equation for $w_0$,
\begin{equation}
\frac{4\bar{k}_1^2 S^2 (\bar{k}_1-qw_0)^2}{\lambda (w_0^2 -2q\bar{k}_1 w_0 + \bar{k}_1^2)}=\big( w_0^2 - a_1^2 (\omega_1^2 + \bar{m}_1^2 ) 
- a_2^2 ( \omega_2^2 + \bar{m}_2^2) \big)^2 \ .
\label{kappaequation}
\end{equation}
The solution to this equation provides $w_0$, and thus the energy, as a function of the spin, the angular momenta, 
and the winding numbers $\bar{k}_1$ and $\bar{m}_i$. However equations (\ref{quarticAdS}) and (\ref{kappaequation}) 
are difficult to solve exactly. As in previous sections, instead of trying to find an exact solution we can write the solution 
in the limit $J_i/\sqrt{\lambda}  \sim S/\sqrt{\lambda} \gg 1$. Out of the four different solutions to (\ref{quarticAdS}), 
the only one with a well-defined limit is 
\be
\omega_1 = \frac {J}{\sqrt{\lambda}} + \frac{\sqrt{\lambda}}{2J^2} (\bar{m}_1 + \bar{m}_2) (\bar{m}_1 J + \bar{k}_1 S ) (1-q^2) 
\Big[ 1 - \frac {\sqrt{\lambda}}{J} q \bar{m}_2 + \dots \Big] \ .
\label{w1AdS}
\ee
Using now relation (\ref{w1w2}) we find\footnote{Note that, as in the case of rotation just in the sphere, the ${\cal O}(\sqrt{\lambda}/J)$ terms and the subsequent corrections 
in the expansions for $\omega_1$ and $\omega_2$ are again proportional to $\bar{m}_1 + \bar{m}_2$. We can prove the existence 
of this factor as in the previous section by setting $\bar{m}_1 = - \bar{m}_2$ in equation (\ref{quarticAdS}).}
\begin{align}
\omega_2 & = \frac {J}{\sqrt{\lambda}} - q ( \bar{m}_1 - \bar{m}_2) + \frac {\sqrt{\lambda}}{2J^2} (\bar{m}_1 + \bar{m}_2) (1-q^2) \nonumber \\
& \times \Big[ \bar{m}_2 J + \bar{k}_1 S 
- \frac {\sqrt{\lambda}}{J} q \bar{m}_2 (\bar{m}_1 J + \bar{m}_2 J + 2 \bar{k}_1 S ) + \dots \Big] \ .
\label{w2AdS}
\end{align}
Next we can calculate the radii $a_1$ and $a_2$ using (\ref{a1a2}), and solve equation (\ref{kappaequation}) to get
\begin{align}
w_{0,+} & = \frac {J}{\sqrt{\lambda}} - q \left( \bar{m}_1 + 2 \frac{ \bar{k}_1 S}{J} \right) 
+ \frac{\sqrt{\lambda}}{2 J^2}(\bar{m}_1^2 J_1+\bar{m}_2^2 J_2 + 2 \bar{k}_1^2 S) (1-q^2) 
\notag \\
& - \frac{\sqrt{\lambda}}{J^3} 2 q^2  \bar{k}_1 S ( \bar{m}_1 J + \bar{k}_1 S ) + \dots \ ,\\
w_{0,-} & = \frac {J}{\sqrt{\lambda}} - q \bar{m}_1 
+ \frac{\sqrt{\lambda}}{2 J^2}(\bar{m}_1^2 J_1+\bar{m}_2^2 J_2 - 2 \bar{k}_1^2 S) (1-q^2)  + \dots \ ,
\end{align}
where as in equation (\ref{ESJ1J2}) the plus or minus subindices refer respectively to the cases where $w_0$ and $w_1$ are chosen 
with identical or opposite signs. These expressions can now be substituted in relation (\ref{ESJ1J2}) to obtain
\begin{align}
E_+ & = J + S - \sqrt{\lambda} q \left( \bar{m}_1 + 2 \frac{ \bar{k}_1 S}{J} \right) + \frac{\lambda}{2 J^2}(\bar{m}_1^2 J_1+\bar{m}_2^2 J_2 + \bar{k}_1^2 S) (1-q^2)  \notag \\
& - \frac{\lambda}{J^3} 2 q^2 \bar{k}_1 S (\bar{m} J+\bar{k}_1 S) + \dots \ ,\\
E_- & = J - S - \sqrt{\lambda} q \bar{m}_1 + \frac{\lambda}{2 J^2}(\bar{m}_1^2 J_1+\bar{m}_2^2 J_2 - \bar{k}_1^2 S) (1-q^2) + \dots \ .
\end{align}
In the absence of flux the expression for $E_+$ reduces to the expansion for the energy in the Neumann-Rosochatius system for a closed circular 
string of constant radius rotating with one spin in $AdS_3$ and two different angular momenta in $S^3$ \cite{Arutyunov_2004}.

As in the previous section, we can now consider the limit of pure NS-NS flux. In this case the above expressions simplify greatly, and we get
\begin{align}
E_+ & = S + \sqrt{(J-\sqrt{\lambda} \bar{m}_1)^2 - 4 \sqrt{\lambda} \bar{k}_1 S} \ , \\
E_- & = J - S - \sqrt{\lambda} \bar{m}_1 \ .
\end{align}

\subsection{Non-constant radii computations with pure NS-NS flux}

Now we will consider the case where the string is allowed to rotate both in $AdS_3$ and $S^3$ with non-constant radii. We will restrict the analysis 
to the limit of pure NS-NS flux\footnote{Solutions in $AdS_3 \times S^3$ were analyzed for the case where $u_2=v_2=0$ in the limit $\kappa=\nu=1$ in \cite{BPP}.}. In this case the second Virasoro constraint can be rewritten as
\be
\omega_2 J_1 +\omega_1 J_2 +w_1 E -w_0 S= \sqrt{\lambda} \left( \omega_1 \omega_2 + w_0 w_1 +q \omega_1 \bar{m_1} \right) \ .
\ee

As the pieces in the Lagrangian describing motion in $AdS_3$ and $S^3$ are decoupled the equations (\ref{p1}), (\ref{p2}), (\ref{q1}) and (\ref{q2}) are still applicable. We only have to substitute them in the more general Virasoro constraints, 
\begin{align}
& z^{\prime 2}_0 + z_0^2 (\beta ^{\prime 2}_0 + w_0^2 ) = z^{\prime 2}_1 + z_1^2 (\beta ^{\prime 2}_1 + w_1^2)
+ \sum_{i=1}^2 \big( r^{\prime 2}_i + r_i^2 (\alpha^{\prime 2}_i + \omega^2_i ) \big)  \ , \\
& z_1^2 w_1 \beta '_1 + \sum_{i=1}^2 r_i^2 \omega _i \alpha '_i  = z_0^2 w_0 \beta '_0 \ .
\end{align}
With this relation and the equations of motion (\ref{p1}), (\ref{p2}), (\ref{q1}) and (\ref{q2}) it is immediate to write the angular momenta and the energy as functions of $\omega$, $\omega'$, the winding numbers $\bar{m}_1$, $\bar{m}_2$, and $\bar{k}_1$, and the spin $S$ and the total angular momentum $J$. In the case where $w_0+\bar{k}_1=-w_1=-\omega'$ we conclude that
\begin{align}
J_1  & = \big[ -\bar{k}_1^2 (\sqrt{\lambda} \omega ' + 2 S) + 2 \bar{k}_1\big( \sqrt{\lambda} \omega^{\prime 2} + 2 \omega' S 
+ (\bar{m}_2 - \omega) (\sqrt{\lambda} \omega - J) \big) \\
& + \omega' \big( \sqrt{\lambda} (\bar{m}_1^2 - \bar{m}_2^2 - \omega^{\prime 2} + \omega^2 )
- 2 (\bar{m}_1 - \bar{m}_2) J \big) \big] / 
\big( 2 (\bar{m}_1 - \bar{m}_2) (\bar{k}-2\omega') \big) \ , \nonumber \\
J_2 & = \big[ \, \bar{k}_1^2 (\sqrt{\lambda} \omega'+2S) - 2 \bar{k}_1\big( \sqrt{\lambda} \omega^{\prime 2} + 2 \omega' S - \bar{m}_1 J 
+ \omega (\sqrt{\lambda} \bar{m}_2 - \sqrt{\lambda} \omega + J) \big) \\
& - \omega' \big( \sqrt{\lambda} (\bar{m}_1^2 - \bar{m}_2^2 - \omega^{\prime 2} + \omega^2) 
+ 2 (\bar{m}_1 - \bar{m}_2) J \big) \big] / \big( 2 (\bar{m}_1 - \bar{m}_2) 
(\bar{k}_1-2\omega') \big) \ , \nonumber \\
E  & = \big[ \sqrt{\lambda} \big( \bar{k}_1^2 + \bar{m}_1^2 - (\bar{m}_2 -3\omega)(\bar{m}_2 -\omega) \big)-2 \bar{k}_1 (2 \sqrt{\lambda} \omega' +S) \\
& + \omega' (3 \sqrt{\lambda} \omega' +4S) -2 J (\bar{m}_1 + \bar{m}_2 -2\omega) \big] / \big( 2 (\bar{k}_1-2 \omega') \big) \ . \nonumber
\end{align}
If we choose $w_0 + \bar{k}_1 = w_1 = \omega'$ we find
\begin{align}
J_1  & =  \big[ -\bar{k}_1^2 (\sqrt{\lambda} \omega ' + 2 S) + 2 \bar{k}_1 (-\sqrt{\lambda} \omega^{\prime 2} -2 \omega' S 
+ (\bar{m}_2 - \omega) (\sqrt{\lambda} \omega - J)) \\ 
& +  \omega' \big( \sqrt{\lambda} (\bar{m}_1^2 -\bar{m}_2^2 - \omega^{\prime 2} +4 \bar{m}_2 \omega -3 \omega^2 )
- 2 (\bar{m}_1 + \bar{m}_2 -2 \omega \big) J) \big] / 
\big( 2 \bar{k}_1 (\bar{m}_1 - \bar{m}_2) \big) \ , \nonumber \\
J_2 & = \big[ \, \bar{k}_1^2 (\sqrt{\lambda} \omega'+2S) + 2 \bar{k}_1 \big( \sqrt{\lambda} \omega^{\prime 2} +2 \omega' S + \bar{m}_1 J 
- \omega (\bar{m}_2 - \omega + J) \big) \\
& -\omega' \big( \sqrt{\lambda} (\bar{m}_1^2 - \bar{m}_2^2 - \omega^{\prime 2} + 4 \bar{m}_2 \omega -3 \omega^2)
- 2 (\bar{m}_1 + \bar{m}_2 -2\omega ) J\big) \big] / 
\big( 2 \bar{k}_1 (\bar{m}_1 - \bar{m}_2) \big) \ , \nonumber \displaybreak \\
E  & = \big[ \sqrt{\lambda} \big( \bar{k}_1^2 + \bar{m}_1^2-(\bar{m}_2 -3\omega)(\bar{m}_2 -\omega) - \omega'^2 \big) -2 \bar{k}_1 S \\
& - 2 J (\bar{m}_1 + \bar{m}_2 -2\omega) \big] / \big( 2 \bar{k}_1 \big) \ . \nonumber
\end{align}
When we take the limit $\bar{k}_1 \rightarrow 0$, $S \rightarrow 0$ and $\sqrt{\lambda} \omega' \rightarrow E$ we recover the expressions from subsection~\ref{nonconstantsphere} in both cases. In a similar way when we set to zero the angular momenta, the winding numbers $\bar{m}_i$ and $\omega$ we recover the analysis in subsection~\ref{nonconstantads}. We can also reproduce the solutions of constant radii from the previous subsection to check if the expressions are consistent. In this case, when $w_0+\bar{k}_1=-w_1$ the angular momenta are given by
\be
J_1 =\frac{\bar{k}_1 S + \bar{m}_2 J}{\bar{m}_2 - \bar{m}_1} \ , \quad J_2 = \frac{\bar{k}_1 S + \bar{m}_1 J}{\bar{m}_1 - \bar{m}_2} \ , 
\ee
and the energy reduces to 
\be
E = - S \pm (J- \sqrt{\lambda} \bar{m}_1) \ .
\ee
In the case where $w_0+\bar{k}_1=w_1$ the angular momenta are
\be
J_1 = \frac{\bar{k}_1 S + \bar{m}_2 J}{\bar{m}_2 - \bar{m}_1} \ , \quad J_2 = \frac{\bar{k}_1 S + \bar{m}_1 J}{\bar{m}_1 - \bar{m}_2} \ , 
\ee
and the energy becomes 
\be
E = S \pm \sqrt{(J - \lambda \bar{m}_1)^2-4 \sqrt{\lambda} \bar{k}_1 S} \ .
\ee

\section{Pulsating strings ansatz}

Throughout this chapter we have been working with the spinning string ansatz. Another interesting and useful ansatz is the so called pulsating string ansatz \cite{Minahan_2003}. This ansatz is exactly the same as the spinning string one but with the roles of $\tau$ and $\sigma$ reversed, that is
\begin{align}
	&Y_3 + i Y_0 = z_0 (\tau) e^{i \beta_0 (\tau)} \ , & &Y_1 +i Y_2 =z_1 (\tau) e^{i (\beta_1 (\tau) +k_1 \sigma)} \ , \\
	&X_1 + i X_2 = r_1 (\tau) \, e^{i (\alpha_1 (\tau) + m_1 \sigma)} \ , & &X_3 + i X_4 = r_2 (\tau) \, e^{i (\alpha_2 (\tau) + m_2 \sigma)}  \ .
\label{ansatzpulsating}
\end{align}
Note that there is no $\sigma$ dependence on $Y_3 + i Y_0$ as the time has to be single valued. Additional solutions have been studied before \cite{Moo,BanerjeePanigrahiPulsatingOscillating}.

We can take advantage of the similarities between both ansätze and follow the same steps as before. When we enter this ansatz in the world-sheet action in the conformal gauge the Lagrangian for the sphere reads\footnote{Note the changes of sign of the flux terms, as we are working here with $\tau$ derivatives instead of $\sigma$ derivatives.}
\be
L_{S^3} = \frac {\sqrt{\lambda}}{2 \pi} \Big[  \sum_{i=1}^2 \frac {1}{2} \big[ \dot{r}_i^2 + r_i^2 (\dot{\alpha}_i)^2 - r_i^2 m_i^2 \big] 
+ \frac {\Lambda}{2} ( r_1^2 + r_2^2 - 1) - q r_2^2 \, ( m_1 \dot{\alpha}_2 - m_2 \dot{\alpha}_1 ) \Big] \ ,
\label{NRqpulsating}
\ee
while for the $AdS$ part
\be
L_{AdS_3} = \frac {\sqrt{\lambda}}{2 \pi} \Big[  \sum_{i=1}^2 \frac {g^{ii}}{2} \big[ \dot{z}_i^2 + z_i^2 (\dot{\beta}_i)^2 - z_i^2 k_i^2 \big] 
+ \frac {\tilde{\Lambda}}{2} ( z_0^2 - z_1^2 - 1) - q z_1^2 k_1 \dot{\beta}_0 \Big] \ .
\label{AdSNRqpulsating}
\ee
Supported with the Virasoro constraints
\begin{align}
	&z_1^2 k_1 \dot{\beta}_1 +\sum_i{r_i^2 \dot{\alpha}_i m_i}=z_0^2 k_0 \dot{\beta}_0 \ , & H_{S^3}+H_{AdS_3}=0 \ .
\end{align}

In the same spirit we can define deformed Uhlembeck constants to reduce the equations of motion of this Lagrangian to first order differential equations. The computation can be done by taking the undeformed Uhlembeck constant and adding deformation terms $f(m_i,v_i,q,r_1)$ and $g(k_1, u_i,q,z_1)$, respectively. After some algebra we obtain
\begin{align}
	\tilde{I}_1 &=r_1^2 (1-q^2) +\frac{1}{m_1^2 -m_2^2} \left[ (r_1 \dot{r}_2 -\dot{r}_1 r_2 )^2 +(v_1 -q m_2)^2 \frac{r_2^2}{r_1^2} +v_2^2 \frac{r_1^2}{r_2^2} \right] \ , \\
	\tilde{F}_1 &=z_1^2 (1-q^2) +\frac{1}{k_1^2} \left[ (z_1 \dot{z}_0 -\dot{z}_1 z_0 )^2 +(u_0 -q k_1)^2 \frac{z_1^2}{z_0^2} +u_1^2 \frac{z_0^2}{z_1^2} \right] \ ,
\end{align}
where $u_i=\beta_i z_i^2$ and $v_i=\alpha_i r_i^2$. To simplify our equations we can change to the ellipsoidal coordinates
\begin{align*}
	\frac{r_1^2}{\zeta -m_1^2} +\frac{r_2^2}{\zeta -m_2^2} &=0 \ , & \frac{z_1^2}{\mu -k_1^2}-\frac{z_0^2}{\mu} &=0 \ ,
\end{align*}
which gives us
\begin{align}
	\frac{-\dot{\zeta}^2}{4} &=P_3 (\zeta) = (1-q^2) (\zeta -m_1^2)^2 (\zeta -m_2^2) -\tilde{I}_1 (m_2^2 -m_1^2) (\zeta -m_1^2) (\zeta -m_2^2) \notag \\
	&+(v_1 -q m_2)^2 (\zeta -m_2^2)^2 +v_2^2 (\zeta -m_1^2 )^2=(1-q^2) \prod_{i=1}^3{(\zeta -\zeta_i )} \ , \label{Sequation} \\
	\frac{-\dot{\mu}^2}{4} &=Q_3 (\mu) =(1-q^2) \mu (\mu-k_1^2)^2 -k_1^2 \tilde{F}_1 \mu (\mu-k_1^2) +(u_0 -q k_1)^2 (\mu -k_1^2)^2 +u_1^2 \mu^2 \notag \\
	&=(1-q^2) \prod_{i=1}^3{(\mu -\mu_i )}\ . \label{AdSequation}
\end{align}

By looking at the results we have obtained it is obvious that the pulsating string ansatz can be treated as the spinning string ansatz just by changing $\sigma \leftrightarrow \tau$, $k_i \leftrightarrow w_i$, $q\rightarrow -q$ and $m_i \leftrightarrow \omega_i$. However here is where the similarities end, because the periodicity in the variable $\sigma$ constrains in different ways in the two cases. While in the spinning string these constraints have to be imposed, in the pulsating string they are directly fulfilled (provided that $m_1$, $m_2$ and $k_1$ are integers). Furthermore the radial functions $r_i$ and $z_i$ are trivially periodic, so the restrictions equivalent to $\zeta_3 -\zeta_1 >0$ and $\mu_3 -\mu_1>0$ for spinning strings, described in sections \ref{nonconstantsphere} and \ref{nonconstantads}, no longer apply. This difference gives rise to a richer set of solutions.

Another important difference with the spinning string is that the conserved quantities now take the very simple form
\begin{align}
	E &=-\sqrt{\lambda} u_0 \ , & S &=\sqrt{\lambda} u_1 \ , & J_1 &=\sqrt{\lambda} v_1 \ ,& J_2 &=\sqrt{\lambda} v_2 \ .
\end{align}

However, we are not going to explore the most general solution but only two particular cases: $AdS_3 \times S^1$, that is, a case where nearly all the dynamics of the sphere is frozen; and $S^3 \times \mathbb{R}$, the opposite case, where nearly all the dynamics of the $AdS_3$ is frozen.

\subsection{Pulsating string in $AdS^3 \times S^1$}

The first of the particular cases we are going to study is the $AdS_3 \times S^1$ space. To restrict our ansatz to this space we are going to fix $r_1=\alpha_1=m_1=m_2=0$, $r_2=1$ and $\alpha_2=\omega \tau$ while the degrees of freedom of the $AdS_3$ will remain unconstrained. Madacena and Ooguri \cite{Moo} have already studied the $AdS_3 \times \mc{M}$ background, with $\mc{M}$ a compact space, in a setting that corresponds to our limit $q\rightarrow 1$. Their idea was to study this background, using the methods provided by understanding of the $SL(2)$ WZW model underlying it, as it enables computations in the AdS/CFT correspondence beyond the gravity approximation and provides an understanding of string theory on a curved space-time with non-trivial $g_{00}$ component.

Under the restrictions we are considering, the Virasoro constraints read
\begin{align}
	&z_1^2 k_1 \dot{\beta}_1 +\sum_i{r_i^2 \dot{\alpha}_i m_i}=z_0^2 k_0 \dot{\beta}_0 \Longrightarrow u_1=0 \ , \\
	&H_{S^3}+H_{AdS_3}=\frac{\sqrt{\lambda}}{4 \pi} \left[ (\omega^2) +\sum_i{ g_{ii} (\dot{z}_i^2 +z_i^2 \dot{\beta}_i^2 +z_i^2 k_i^2)} \right]=0 \Longrightarrow k_1^2 \tilde{F}_1=u_0^2- \omega^2 \ .
\end{align}
The first of the Virasoro constraints imposes that $u_1=0$ while the second one will be useful to find the dispersion relation of our solution.

Because $u_1=0$ the cubic equation $Q_3 (\mu)=0$ is easier to solve as one of the roots can be found by direct inspection ($\mu=k_1^2$). The other roots then are easily obtained
\begin{displaymath}
	\mu=\frac{(1-q^2) f_1^2 -(u_0-q k_1)^2 \pm \sqrt{((1-q^2) f_1^2 -(u_0-q k_1)^2)^2 +4(1-q^2)(u_0-q k_1)^2 k_1^2}}{2(1-q^2)} \ ,
\end{displaymath}
where
\begin{equation}
f_1^2=k_1^2 +\frac{k_1^2 \tilde{F}_1}{1-q^2} \ .
\end{equation}
Solving the  differential equation~(\ref{AdSequation}) is exactly the same as in the spinning string case\footnote{Our solution is a periodic solution in $\tau$, usually called ``short string''. This solution can be shifted in $\tau$ and continued analytically by $n=\sqrt{(1-q^2) (\mu_3 -\mu_1)}\rightarrow in$ into a solution that comes from the boundary, contracts to zero and expands again, with no periodicity. These are called ``long strings''. To unify our notation with the notation from \cite{Moo} $\frac{\mu_3 - k_1^2}{k_1^2}=\cosh \rho_0$ for short strings and to $\sinh \rho_0$ for long strings.}
\begin{align}
	&z_0^ 2 =\frac{\mu_3}{k_1^2} +\frac{k_1^2 -\mu_3}{k_1^2} \, \hbox{sn}^2
\big( \tau \sqrt{(1-q^2) (\mu_3 -\mu_1)} , \nu \big) \ , \\
	&z_1^ 2 =\frac{\mu_3 - k_1^2}{k_1^2} \, \hbox{cn}^2
\big( \tau \sqrt{(1-q^2) (\mu_3 -\mu_1)} , \nu \big)=\cosh ^2 \rho_0 \, \hbox{cn}^2
\big( \tau \sqrt{(1-q^2) (\mu_3 -\mu_1)} , \nu \big) \ , \notag
\end{align}
where $\nu=\frac{\mu_3 -k_1^2}{\mu_3 -\mu_1}$. It is also interesting to analyse the behaviour of the time coordinate. To do that we use
\begin{equation}
	\dot{\beta}_0=\dot{t}=\frac{u_0 +q z_1^2 k_1}{g^{00} z_0^2}=\frac{u_0-q k_1}{-z_0^2} +q k_1 \ ,
\end{equation}
which we can integrate to
\begin{align}
	t &=q k_1 \tau-\int_0^\tau{d\tau ' \frac{u_0-q k_1}{z_0^2(\tau ')}}=-q k_1 \tau+T(\tau ) \\
	&=q k_1 \tau -(u_0-q k_1) \frac{\Pi \left(\tanh ^2 ( \rho _0 ) , \hbox{Am}[ \tau \sqrt{(1-q^2) (\mu_3 -\mu_1)} ,\nu] , \nu \right) }{\cosh ^2 (\rho _0) \sqrt{1-\nu \, \hbox{sn}^2 (\tau \sqrt{(1-q^2) (\mu_3 -\mu_1)} , \nu)}}\times \notag \\
	&\times \hbox{dn}(\tau \sqrt{(1-q^2) (\mu_3 -\mu_1)} ,\nu) \ .
\end{align}
For later comparisons it is more interesting to write this coordinate as
\begin{equation}
	\hbox{sc}(t, \nu)=\frac{\hbox{sc}(T (\tau ),\nu) \hbox{dn}(q k_1 \tau ,\nu)+\hbox{sc}(q k_1 \tau ,\nu) \hbox{dn}(T(\tau ),\nu)}{1-\hbox{sc}(T(\tau ),\nu) \hbox{sc}(q k_1 \tau ,\nu) \hbox{dn}(T(\tau ),\nu) \hbox{dn}(q k_1 \tau ,\nu)} \ ,
\end{equation}
where we have used the property
\begin{displaymath}
	\hbox{sc}(x+y)=\frac{\hbox{sc}(x) \hbox{dn}(y)+\hbox{sc}(y) \hbox{dn}(x)}{1-\hbox{sc}(x) \hbox{sc}(y) \hbox{dn}(x) \hbox{dn}(y)} \ .
\end{displaymath}
These expressions are not very illuminating at first sight. However taking the limit $q=1$, where $u_0+k_1= \pm \cosh \rho_0$, we have
\begin{equation}
	\tan (t)=\frac{\tan (k_1 \tau) \mp \tan (n \tau) \cosh (\rho_0)}{1 \pm \tan (k_1 \tau) \, \tan (n \tau) \cosh (\rho_0)} \ ,
\end{equation}
which is equivalent to equation (34) of \cite{Moo}\footnote{Note that there is a typo in eq. (34) in that article. This can easily be seen as it is inconsistent with eq. (44) in the same article. It can be seen that the analytically continued version of this equation is indeed equivalent to eq. (44).}.

Now we have all the elements to calculate the dispersion relation. To do that we set the argument of the elliptic sine to $n \tau$ and extract from there the energy,
\begin{displaymath}
	n=\sqrt{(1-q^2) (\mu_3 -\mu_1)} \Longrightarrow n^4 =((1-q^2) f_1^2 -(u_0+q k_1)^2)^2 +4(1-q^2)(u_0+q k_1)^2 k_1^2 \ .
\end{displaymath}
If we now substitute the value of $f_1$ and the value of $\tilde{F}_1$ found in the Virasoro constraint and solve the quadratic equation we get
\begin{align*}
	& 4 k_1^2 (u_0+q k_1)^2 -4 q k_q (k_1^2 -\omega^2) (u_0+q k_1) +(k_1^2 -\omega ^2)^2 -n^4=0 \ , \\
	&u_0+qk_1=\frac{k_1^2 -\omega^2}{2 k_1} \left[ q \pm \sqrt{q^2-1 +\frac{n^4}{(k_1^2 -\omega^2)^2}} \right]=-\frac{E}{\sqrt{\lambda}}+q k_1 \ , \\
	&E=q k_1 \sqrt{\lambda} +\sqrt{\lambda} \frac{\omega^2 -k_1^2}{2 k_1} \left[ q \mp \sqrt{q^2-1 +\frac{n^4}{(k_1^2 -\omega ^2)^2}} \right] \ .
\end{align*}
Where the upper sign correspond to our solution and the lower sign correspond to the analytic continuation of our solution found by performing the change $n \rightarrow in$, which is also a solution.
Note that this dispersion relation in the limit $q=1$ reduces to
\begin{displaymath}
	E=\frac{k_1 \sqrt{\lambda}}{2}  +\frac{(\omega^2 \mp n^2) \sqrt{\lambda}}{2 k_1} \ ,
\end{displaymath}
which corresponds to the dispersion relations in \cite{Moo}\footnote{Note that eq. (33) has a factor of two dividing $-k\alpha^2$ missing, as can be seen by multiplying eq. (32) by two.}. The upper sign corresponds to ``short strings'' and the lower sing to ``long strings'' in that reference.

One interesting result is that the energy of our string is real only if $n$ is above a particular value, that is, if $n^4 \geq n_{min}^4 = (1-q^2) (k_1^2 -H_{S^3})^2$, which also implies that the maximum radius of the string has to have a minimum $\rho_0 \geq \rho_{0,min}$. Note that these restrictions disappear when $q=1$.

\subsection{The pulsating $S^3 \times \mathbb{R}$ solution}

In this section we are going to study the solution when we set $z_1=\beta_1=k_1=0$, $z_0=1$ and $\beta_0=u_0 \tau$ and the degrees of freedom of the sphere remains unconstrained. For this ansatz the constrains from the Virasoro conditions can be written as
\begin{align}
	&z_1^2 k_1 \dot{\beta}_1 +\sum_i{r_i^2 \dot{\alpha}_i m_i}=z_0^2 k_0 \dot{\beta}_0 \Longrightarrow \sum{m_i v_i}=0 \ , \\
	&H_{S^3}+H_{AdS_3}=\frac{\sqrt{\lambda}}{4 \pi} \left[ \sum_i{ \dot{r}_i^2 +r_i^2 \dot{\alpha}_i^2 +r_i^2 m_i^2)} +u_0^2\right]=0 \Longrightarrow \\
	&\Longrightarrow (m_2^2 -m_1^2) \tilde{I}_1=m_2^2 (1-q^2 ) +v_1^2 +(v_2-q m_1)^2 -u_0^2 \ .
\end{align}
The solution to the differential equation given by the Uhlenbeck constant reads now
\be
r_1^2(\sigma) = \frac{\zeta_3 - \omega_1^2}{\omega_2^2 - \omega_1^2} + \frac{\zeta_2 - \zeta_3}{\omega_2^2 - \omega_1^2} \, \hbox{sn}^2
\left( \tau \sqrt{(1-q^2) (\zeta_3 -\zeta_1)} , \frac{\zeta_3 -\zeta_2}{\zeta_3 -\zeta_1} \right) \ .
\label{r1ellipticpulsating}
\ee
As we did in the previous section, we can set the argument of the elliptic sine to $n \tau$ and use the Virasoro constraints to find the dispersion relation. However, this is very difficult to do in the general case but we can look at two particular configurations: $v_2=0 \leftrightarrow J_2=0$ and $v_1=q m_2 \leftrightarrow J_1=\sqrt{\lambda} q m_2$. For the first configuration the dispersion relation reads
\begin{equation}
	-\frac{E^2}{\lambda}=-u_0^2 =m_2^2 -2 m_2 v_1 q \pm \sqrt{n^4 -4 m_2^2 (1-q^2) (v_1 -q m_2)^2} \ ,
\end{equation}
where we have set $m_1=0$ as a consequence of the first Virasoro constraint. While for the second one we get
\begin{equation}
	-\frac{E^2}{\lambda}=-u_0^2 =m_2^2 +q^2 m_1^2 +(m_2^2 -m_1^2 )(1-q^2) \pm \sqrt{n^4 -4 q^2 m_1^2 (m_1^2 - m_2^2) (1-q^2) } \ ,
\end{equation}
where we have set $v_2=-q m_1$ again from the Virasoro constraint.

\chapter{$\eta$-deformed Neumann-Rosochatius system} \chaptermark{$\eta$-deformed N-R system}

	\begin{chapquote}{M. Staudacher, replying to A. A. Migdal at the Itzykson Meeting 2007, \cite{Torrielli_Yangians}}
 		-What makes you think that the theory will still be integrable?\\
		-Unlimited optimism.
	\end{chapquote}

In this chapter we are going to present other possible way of deforming the $AdS_3 \times S^3$ background that does not break the integrability of the string Lagrangian. This deformation can be classified among the \emph{Yang-Baxter sigma models}. These kind of models were first proposed by Klimcik in \cite{Klimcik_2002}, and later developed in \cite{Klimcik_2009,Squellari_2014}, as a way to construct integrable deformations of the PCM using classical R-matrices that solve the modified classical Yang-Baxter equation. This method has been extended to bosonic coset models \cite{Delduc_2013} and, in particular, to the $AdS_5 \times S^5$ string action \cite{etaDMV,DelducActionandSymmetries}. In recent years there has been a great interest in these kind of deformations, also studied in\cite{KawaguchiJordanianDef,TowardsCYBE,MatsumotoNoncommutative,TongerenYBAdS5S5,%
TongerenNoncommutative,AlmostAbelian,AraujoYBSM,AraujoConfTwists}

In the first section we will review the construction and integrability of one particular deformation of the $AdS_5 \times S^5$ string action called $\eta$-deformation, obtained by using a Drinfeld-Jimbo R-matrix. In that presentation we will mostly follow \cite{etaArutyunov,ArutyunovMedinaRincon}. In the second section we will truncate the $( AdS_5 \times S^5 )_\eta$ Lagrangian to a $( AdS_3 \times S^3 )_\eta$ Lagrangian. With that simplification we will be able to construct an easier-to-handle deformed Uhlenbeck constant which will allow us to construct the solutions to the equations of motion (additional solutions corresponding to other string configurations in the full $( AdS_5 \times S^5 )_\eta$ background have been studied before using diverse approaches in references \cite{Stijn2014a,Stijn2014b,Panigrahioscillating,Kameyama2014,PanigrahiPulsatingads3kappa,%
Panigrahispikykappa,Panigrahicircularkappa,Roychowdhury2016,Roychowdhury2017,Banerjee2017}). Although we will be able to find the dispersion relation as a series in inverse powers of the total angular momentum, as in previous cases dispersion relation for the elliptic string will be lengthy, complicated and not very illuminating. Instead we wil dedicate the last part of the section to some particular limits where solutions are easier to construct. The results obtained for these limits from the point of view of the deformed Uhlenbeck constant are compared in Appendix \ref{appendixkappa} with results obtained directly from the Lagrangian for the same limits. The results of this last section were published in \cite{EtaSpinningstrings}.

\section{Neumann system in $\eta$-deformed $AdS_5 \times S^5$}\sectionmark{Neumann system in $\eta$-deformed $A\MakeLowercase{d}S_5 \times S^5$}

There are two known classes of integrable deformations of the $AdS_5 \times S^5$. On one hand we have deformations that can be conveniently described in terms of the original theory, where the deformation parameter only appears as integrable quasi-periodic boundary conditions for the world-sheet fields. Examples of these deformations are orbifolding (like the one studied in \cite{Kachru_1998}) and TsT deformations, obtained by the successive application of a T-duality, a shift and another T-duality (examples of this deformation are \cite{Lunin_2005} or \cite{SfetsosNATduals}). On the other hand we have deformations of the underlying symmetry algebra in the sense of quantum groups \cite{Beisert_2008,Beisert_2012,HollowoodIntDeform}. That is the case of the $\eta$-deformation we will present here, where we deform the underlying $\mf{psu}(2|2)$ to $\mf{psu}(2|2)_q$ with $q=\exp \left[ -\frac{2\eta}{g(1+\eta^2)} \right]$. 

The action for the superstring on the deformed $AdS_5 \times S^5$ depending on the real deformation parameter $\eta$ is given by\footnote{We have chosen to follow the normalization of \cite{etaArutyunov} instead of the one used in \cite{etaDMV}.}
\begin{equation}
	S=\frac{\rho (1+\eta^2)}{4} \int{d^2 \sigma (\eta^{\alpha \beta} -\epsilon^{\alpha \beta}) \text{ STr} \left[ \tilde{d} (j_\alpha ) \frac{1}{1-\eta R_{\mf{g}} \circ d} (j_\beta) \right]} \ ,
\end{equation}
where we have chosen the conformal gauge, $j$ is the left invariant current, $\rho$ is a coupling constant, $\epsilon^{\tau \sigma}=+1$ and operators $d$ and $\tilde{d}$ are defined as linear combinations of projections over the different components of the $\mathbb{Z}_4$ decomposition of the algebra
\begin{align*}
	d &=P_1 +\frac{2}{1-\eta^2} P_2 -P_3 \ , \\
	\tilde{d} &=-P_1 +\frac{2}{1-\eta^2} P_2 +P_3 \ .
\end{align*}
The action of the operator $R_{\mf{g}}$ on the algebra is given by
\begin{equation}
	R_{\mf{g}} \circ (M)=\Ad^{-1}_{\mf{g}} \circ R_{\mf{g}} \circ \Ad_{\mf{g}} \circ (M)=g^{-1} R(g M g^{-1}) g
\end{equation}
where we used the same $g$ that defines the current and $R(M)=\STr_2 \{r (1\otimes M) \}$ is a linear operator obtained by aplying a supertrace (over the second space) to a classical R-matrix. If this R-matrix satisfy the modified classical Yang-Baxter equation \cite{Bordemann_1990}, this operator also satisfy them\footnote{The MCYBE written in \cite{Bordemann_1990} and the one written in \cite{Klimcik_2009} differ by a sign in the right-hand side. We have chosen the sign convention of the later.}
\begin{equation}
	[R(X),R(Y)]-R([R(X),Y]+[X,R(Y)])=[X,Y] \ .
\end{equation}
There are several choices for this operator, each of which generates a different deformation. For a classification of these operators and models we refer to \cite{Matsumoto_2014}. We will be interested in the deformation generated by
\begin{equation}
	R(M)_{ij}=-i\hat{\epsilon}_{ij} M_{ij} \ , \quad \text{ where } \quad \hat{\epsilon}_{ij}=\left\{\begin{array}{l}
	1 \text{ if } i<j \\
	0 \text{ if } i=j \\
	-1 \text{ if } i>j \\
\end{array}	  \right. \ , 
\end{equation}
which is a particular case of a \emph{Drinfeld-Jimbo R-matrix}.

From now on we will focus on the bosonic sector of the theory. The Lagrangian with the fermionic degrees of freedom switched off simplifies to
\begin{equation}
	L=-\frac{\rho\sqrt{1+\varkappa^2}}{2} (\eta^{\alpha \beta} - \epsilon^{\alpha \beta} ) \text{ STr} \left[ j^{(2)}_\alpha  \frac{1}{1-\varkappa R_{\mf{g}} \circ P_2} j_\beta \right] \ ,
\end{equation}
where
\begin{equation}
	\varkappa=\frac{2\eta}{1-\eta^2} \ ,
\end{equation}
it is a convenient rewriting of the deformation parameter. Let us choose the ansatz that gave us the Neumann Lagrangian in the previous chapter,
\begin{equation}
X_1 + i X_2 = r_1 (\sigma) \, e^{i \omega_1 \tau} \ , \quad X_3 + i X_4 = r_2 (\sigma) \, e^{i \omega_2 \tau } \ , \quad X_5 + i X_6 = r_3 (\sigma) \, e^{i \omega_3 \tau } \ ,
\end{equation}
after some algebra, which can be checked in \cite{etaDMV} and \cite{ArutyunovMedinaRincon}, the Lagrangian acquires the form
\begin{align}
	L &=-\frac{\rho \sqrt{1+\varkappa^2}}{2} \left[ \frac{(r'_1 r_2 -r_1 r'_2)^2}{(r_1^2 +r_2^2) (1+\varkappa^2 r_2^2 (r_1^2 +r_2^2))} +\frac{2\varkappa \omega_1 r_1 r_2 (r_1 r'_2 -r_2 r'_1)}{1+\varkappa^2 r_2^2 (r_1^2 +r_2^2)} \right. \notag \\
	&+\frac{r_3^{\prime 2}}{(r_1^2 +r_2^2) (1+\varkappa^2 (r_1^2 +r_2^2))} +\frac{2\varkappa \omega_3 r_3 r'_3}{1+\varkappa^2 (r_1^2 +r_2^2)} -\frac{\omega_1^2 r_1^2}{1+\varkappa^2 r_2^2 (r_1^2 +r_2^2)} -\omega_2^2 r_2^2 -\notag \\
	&\left. -\frac{\omega_3^2 r_3^2}{1+\varkappa^2 (r_1^2 +r_2^2)} \right] - 
\frac{\Lambda}{2} (r_1^2 +r_2^2 +r_3^2 -1) \ .
\end{align}

The Lax connection for the bosonic Lagrangian is constructed in the following way \cite{etaDMV}
\begin{align}
	L_\alpha &= \frac{\eta^{\alpha \beta} +\epsilon^{\alpha \beta}}{2} \left[-\left( \frac{1}{1+\varkappa R_{\mf{g}} \circ P_2} j_\alpha \right)^{(0)} -\frac{\sqrt{1+\varkappa^2}}{\lambda} \left( \frac{1}{1+\varkappa R_{\mf{g}} \circ P_2} j_\alpha \right)^{(2)} \right] \notag \\
	&+\frac{\eta^{\alpha \beta} -\epsilon^{\alpha \beta}}{2} \left[-\left( \frac{1}{1-\varkappa R_{\mf{g}} \circ P_2} j_\alpha \right)^{(0)} -\lambda \sqrt{1+\varkappa^2} \left( \frac{1}{1-\varkappa R_{\mf{g}} \circ P_2} j_\alpha \right)^{(2)} \right] \ .
\end{align}
Using the Lax formalism we have explained in section~\ref{PCMintegrability}, we can generate integrals of motion. However these integrals of motion are not the direct generalization of the Uhlenbeck but a linear combination of them. In \cite{ArutyunovMedinaRincon} these generalized Uhlenbeck constants were obtained as a deformation of the original ones
\begin{displaymath}
	{\mathscr I}_j = I^N_{j} +\mc{O} (\varkappa) \ .
\end{displaymath}
A generalization of these Uhlenbeck integral to a $\eta$-deformed Neumann-Rosochatius was later presented in \cite{AMRH}.

\section[$\eta$-deformed Neumann-Rosochatius system. Spinning strings in $\eta$-deformed $\mathbb{R}\times S^3$]{$\eta$-deformed Neumann-Rosochatius system. Spinning strings in $\eta$-deformed $\mathbb{R}\times S^3$\sectionmark{Spinning strings in $(\mathbb{R}\times S^3)_\eta$}} \sectionmark{Spinning strings in $(\mathbb{R}\times S^3)_\eta$}

We will now move to the case of a spinning string in the $\eta$-deformation of $AdS_{5} \times S^{5}$ with the more general ansatz that gave us the Neumann-Rosochatius Lagrangian in the previous chapter
\begin{equation}
	X_{2j-1}+i X_{2j}=r_j (\sigma ) e^{i\omega_j \tau +i\alpha_j (\sigma)} \ ,
\end{equation}
which gives us the following Lagrangian for the sphere component, the $\eta$-deformed Neumann-Rosochatius system\footnote{
Note that we can use the constraint $r_1^2 +r_2^2 +r_3^2 =1$ to bring the first term in (\ref{N=3etalagrangian}) to the form
\[
\frac{(r_1 r'_2 -r'_1 r_2)^2}{(r_1^2 +r_2^2)[1+\varkappa ^2 (r_1^2 +r_2^2)r_2^2]}= \frac{r_1^{\prime 2} +r_2^{\prime 2} 
+r_3^{\prime 2}}{1+\varkappa ^2 (r_1^2 +r_2^2)r_2^2}- \frac{r_3^{\prime 2}}{(r_1^2 +r_2^2) [1+\varkappa^2 (r_1^2 +r_2^2) r_2^2]} \ .
\]
Furthermore the term before the Lagrange multiplier is just a total derivative,
\[
\frac{2\varkappa \omega_3 r_3 r'_3}{1+\varkappa^2 (r_1^2+r_2^2)}= -\frac{\omega_3}{\varkappa} \left[ \ln (1+\varkappa^2 (r_1^2+r_2^2)) \right]' \ .
\]
}
\ba
L & = & \frac{\sqrt{\lambda}}{2\pi} \left[ \frac{(r_1 r'_2 -r'_1 r_2)^2}{(r_1^2 +r_2^2)[1+ \varkappa ^2 (r_1^2 +r_2^2)r_2^2]} 
+ \frac{r_3^{\prime 2}}{(r_1^2 +r_2^2) [1+\varkappa^2 (r_1^2 +r_2^2)]}  \right. \notag \\
& + & \frac{r_1^2 (\alpha_1^{\prime 2} -\omega_1^2)}{1+\varkappa ^2 (r_1^2 +r_2^2)r_2^2} 
+ r_2^2 (\alpha_2^{\prime 2} -\omega_2^2)+\frac{r_3^2 (\alpha_3^{\prime 2} -\omega_3^2)}{1+\varkappa^2 (r_1^2 +r_2^2)} 
+  \frac{ 2\varkappa \omega_1 r_1 r_2 (r_1 r'_2 -r_2 r'_1)}{1+\varkappa ^2 (r_1^2 +r_2^2)r_2^2} \notag \\
& + & \left. \frac{2\varkappa \omega_3 r_3 r'_3}{1+\varkappa^2 (r_1^2+r_2^2)} - 
\frac{\Lambda}{2} (r_1^2 +r_2^2 +r_3^2 -1) \right] \ , \label{N=3etalagrangian}
\ea
where, for convenience, we have defined $\lambda=\rho^2 \pi^2 (1+\varkappa^2)$. It is immediate to write down the complete equations of motion for the radial and angular coordinates coming from this Lagrangian. However in this chapter we will only be interested in the case of a string spinning on an $\eta$-deformed three-sphere. Therefore, rather than presenting the general set of equations 
we will focus on how we should perform a consistent reduction to capture the dynamics on a deformed three-sphere. 
We can clarify this by inspecting the equation of motion for $r_3$, which is given by
\begin{equation}
\left[ \frac{r'_3}{(r_1^2+r_2^2) (1+\varkappa^2 (r_1^2+r_2^2))} \right] '
=\Lambda r_3 +\frac{r_3 (\alpha_3^{\prime 2} -\omega_3^2)}{1+\varkappa^2 (r_1^2 +r_2^2)} \ ,
\end{equation}
We see that $r_3=0$ is a solution independently of the behaviour of the other two coordinates. 
This means that setting $r_{3}=0$ is a consistent truncation from the $\eta$-deformed five-sphere to an $\eta$-deformed 
three-sphere\footnote{
However this is not the only reduction that we can perform to obtain a consistent truncation from the $S^{5}_{\eta}$ to $S^{3}_{\eta}$. 
For instance, from the equation of motion for $r_{1}$,
\begin{align*}
\frac{r''_1}{r}&=\varkappa^2 \frac{2(r_1^2 + r_2^2) r'_1 r_2 r'_2 +r_1 r_1^{\prime 2} r_2^2 +2 r'_1 r_2^3 r'_2 -r_1 r_2^2 xr^{\prime 2}_2}{r^2} \notag \\
& - 4 \varkappa \omega_1 \frac{r_1 r_2 r'_2}{r^2}+\Lambda r_1+\frac{r_1 (\alpha_{1}^{\prime 2} - \omega_1^2)}{r} \left( 1-\varkappa^2 \frac{r_1^2 r_2^2}{r} \right) \ ,
\end{align*}
with $r=1+\varkappa^2 r_2^2 (r_1^2+r_2^2)$, we conclude that the choice $r_1=0$ provides indeed another possible truncation. 
When we set $r_{1}=0$ the Lagrangian becomes
\[
L=\frac{1}{2} \left[ \frac{r_3^{\prime 2}}{r_2^2 (1+\varkappa^2 r_2^2)} +r_2^ 2 (\alpha_2^{\prime 2} -\omega_2^2) 
+ \frac{r_3^2 (\alpha_3^{\prime 2} -\omega_3^2)}{1+\varkappa^2 r_2^2} \right] +\frac{\Lambda}{2} (r_2^2+r_3^2-1) \ ,
\]
which can be easily seen to be equivalent to the one for the $r_{3}=0$ truncation.
}. 
The Lagrangian simplifies to
\be
L = \frac{\sqrt{\lambda}}{2\pi} \left[ \frac{r_1^{\prime 2} + r_2^{\prime 2} + r_1^ 2 (\alpha_1^{\prime 2} -\omega_1^2)}{1+\varkappa^2 r_2^2} 
+ r_2^ 2 (\alpha_2^{\prime 2} - \omega_2^2) 
- \frac{\Lambda}{2} (r_1^2 + r_2^2-1) \right] \ .
\ee
The equations of motion for the radial coordinates are given by
\begin{align}
& \frac{r''_1}{1 + \varkappa^2 r_2^2} + 2 \varkappa^2 \frac{r_1 r_1^{\prime 2}}{(1+\varkappa^2 r_2^2)^2} = \frac{r_1( \alpha_1^{\prime 2} -\omega_1^2)}{1 + \varkappa^2 r_2^2} 
+ \Lambda r_1 \label{x1S3equation} \ , \\
& \frac{r''_2}{1+\varkappa^2 r_2^2} - 2 \varkappa^2 \frac{r_2 r_2^{\prime 2}}{(1 + \varkappa^2 r_2^2)^2} = r_2( \alpha_2^{\prime 2} - \omega_2^2 )
- \varkappa^2 r_2 \frac{r_1^{\prime 2} + r_2^{\prime 2} + r_1^2 (\alpha_1^{\prime 2} - \omega_1^2)}{(1+\varkappa^2 r_2^2)^2} + \Lambda r_1 \ , \label{x2S3equation}
\end{align}
and for the angular functions we find
\be
\alpha '_1 = \frac{v_1}{r_1^2} (1+\varkappa ^2 r_2^2 (r_1^2 +r_2^2)) \ , \quad 
\alpha '_2 = \frac{v_2}{r_2^2} \ . 
\ee
The Virasoro constraints become
\begin{align}
& \frac{r_1^{\prime 2} + r_2^{\prime 2} + r_1^ 2 (\alpha_1^{\prime 2} + \omega_1^2)}{1+ \varkappa^2 r_2^2} + r_2^ 2 (\alpha_2^{\prime 2} +\omega_2^2) = w_0^{2} \ , \label{Virasoro1}\\
& \frac{r_1^2 \alpha'_1 \omega_1}{1 + \varkappa^2 r_2^2} + r_2^2 \alpha '_2 \omega_2 = 0 \label{Virasoro2} \ ,
\end{align}
and the energy and the angular momenta are given now by
\be
E = \sqrt{\lambda} w_{0} \ , \quad J_1 = \sqrt{\lambda} \int{\frac{d\sigma}{2\pi} \frac{r_1^2 \omega_1}{1+\varkappa^2 r_2^2}} \ , \quad J_2 = \sqrt{\lambda} \int{\frac{d\sigma}{2\pi} r_2^2 \omega_2} \ .
\ee

We can exhibit that integrability remains a symmetry of the system after the $\eta$-deformation by constructing a deformation $\tilde{I}_{i}$ of the Uhlenbeck constants which makes 
them constants of motion again. To find this deformation we are going to assume that
\be
\tilde{I}_{1}=\frac{1}{\omega_1^2 -\omega_2^2} \left[ f(r_1,r_2) [r_1^{\prime 2} + r_2^{\prime 2}] + \frac{v_1^2 r_2^2}{r_1^2} +\frac{v_2^2 r_1^2}{r_2^2} +h(r_1,r_2) \right] \ ,
\ee
and impose that $\tilde{I}'_1=0$. By doing this we find that
\be
- 2 \varkappa^2 f \left( \frac{r_1 r_1^{\prime 3}}{1+\varkappa^2 r_2^2} +\frac{r_1 r'_1 r_2^{\prime 2}}{1+\varkappa^2 r_2^2} \right) +f' r_1^{\prime 2} +f' r_2^{\prime 2}=0 \ ,
\ee
where we have made use of the equations of motion (\ref{x1S3equation}) and (\ref{x2S3equation}). We can easily integrate this relation to get
\be
f(r_2)=\frac{A}{1+\varkappa^2 r_2^2} \ ,
\ee 
where $A$ is an integration constant that we will set to 1. We can proceed in the same way to obtain the function $h$. We finally conclude that
\footnote{
The Uhlenbeck constants for the $(AdS_5 \times S^5)_\eta$ Neumann-Rosochatius system were constructed using the Lax representation in \cite{AMRH}. Some immediate algebra 
shows that those more general constants reduce to the one we present in here along the $r_{3}=0$ truncation.}
\begin{equation}
\tilde{I}_{1}=\frac{1}{\omega_1^2 - \omega_2^2} \left[ \frac{r_1^{\prime 2} + r_2^{\prime 2} + r_1^2 \omega_1^2}{1+\varkappa^2 r_2^2}
- r_1^2 \omega_2^2 + (1 + \varkappa^2) \frac{v_1^2 r_2^2}{r_1^2} + \frac{v_2^2 r_1^2}{r_2^2} \right] \ .
\end{equation}

We will now focus on the construction of general solutions of the $\eta$-deformed Neumann-Rosochatius system. In order to do so we will introduce 
an ellipsoidal coordinate defined as the root of the equation 
\begin{equation}
\frac{r_1^2}{\zeta - \omega_1^2} + \frac{r_2^2}{\zeta - \omega_2^2} =0 \ . 
\end{equation}
If we assume that $\omega_1 <  \omega_2$, then the ellipsoidal coordinate will vary from $\omega_1^2$ to $\omega_2^2$. When we replace the radial coordinates by the ellipsoidal one 
in the equations of motion we are left with a second-order differential equation for $\zeta$. Following the strategy used in the previous chapter, we can more conveniently reduce the problem to the study of a first-order equation 
by writing the Uhlenbeck constant in terms of the ellipsoidal coordinate. We find that
\be
\zeta^{\prime 2} = - 4 P_4 (\zeta) \ ,
\label{zetaequation}
\ee
where $P_4(\zeta)$ is the fourth-order polynomial
\begin{align}
P_4 (\zeta) & = - \frac{\varkappa^2 \omega_2^2}{(\omega_1^2-\omega_2^2)^2} (\zeta -\omega_1^2)^2 (\zeta-\omega_2^2)^2 
+ \big( \omega_1^2-(1+\varkappa^2) \omega_2^2+\varkappa^2 \zeta \big) \Big[ \tilde{I}_{1} (\zeta -\omega_1^2) (\zeta-\omega_2^2) \nonumber \\
& + \frac{(1+\varkappa^2) v_1^2}{\omega_1^2-\omega_2^2} (\zeta-\omega_2^2)^2 
+ \frac{v_2^2}{\omega_1^2-\omega_2^2} (\zeta-\omega_1^2)^2 \Big]+(\zeta -\omega_1^2)^2 (\zeta -\omega_2^2)= \nonumber \\
&= - \frac {\varkappa^2 \omega_2^2}{(\omega_1^2-\omega_2^2)^2} \prod_{i=1}^4{(\zeta-\zeta_i )} \ . \label{P4definition}
\end{align}
We can solve this equation if  we change variables to
\begin{equation}
\eta^2=\frac{\zeta -\zeta_4}{\zeta_3 -\zeta_4} \ ,
\end{equation}
which transforms equation (\ref{zetaequation}) into
\begin{equation}
\eta^{\prime 2}=\frac{\varkappa^ 2 \omega_2^2 \zeta_{34}^2}{(\omega_1^ 2 - \omega_2^2)^2} (\eta^2 -1) (\eta^2 -\eta_1^2) (\eta^2 -\eta_2^2) \ ,
\end{equation}
where we have defined $\zeta_{ij}=\zeta_i-\zeta_j$ and $\eta_i^2=\zeta_{i4}/\zeta_{34}$. The solution to this equation is
\be
\eta (\sigma) = \frac{-i \text{ sn}\left[ \eta_1 \sqrt{(1-\eta_2^2)} \left(\pm \frac{i \varkappa \omega_2 \zeta_{34} (\sigma -\sigma_0)}{\omega_1^2 -\omega_2^2} \right), 
\frac{(1-\eta_1^2) \eta_2^2}{(1-\eta_2^2) \eta_1^2} \right]}{\sqrt{1-\frac{1}{\eta_2^2}-\text{ sn}^2 \left[ \eta_1 \sqrt{(1-\eta_2^2)} 
\left(\pm \frac{i \varkappa \omega_2 \zeta_{34} (\sigma -\sigma_0)}{\omega_1^2 -\omega_2^2} \right), \frac{(1-\eta_1^2) \eta_2^2}{(1-\eta_2^2) \eta_1^2} \right]}} \ ,
\ee
where $\sigma_0$ is an integration constant that we can set to zero by performing a rotation. Therefore we conclude that
\footnote{We should note that this solution is well defined not only for real values of the parameter $\varkappa$, 
but also for pure imaginary values of this parameter (although an analytical continuation of the $r_i$ coordinates may be needed for that). If we define $\varkappa=i\hat{\varkappa}$ we have
\[
r_1^2 (\sigma) =\frac{\omega_1^2 -\zeta_4}{\omega_1^2 -\omega_2^2} + \frac{\zeta_{34}}{\omega_1^2 -\omega_2^2} \frac{ \zeta_{24} 
\text{ sc}^2\left[ \pm \frac{\hat{\varkappa} \omega_2 \sqrt{\zeta_{14} \zeta_{23}} \sigma}{\omega_1^2 -\omega_2^2} , 
-\frac{\zeta_{12} \zeta_{34}}{\zeta_{14} \zeta_{23}} \right]}{  \zeta_{23} - \zeta_{24}  
\text{ sc}^2 \left[ \pm \frac{\hat{\varkappa} \omega_2 \sqrt{\zeta_{14} \zeta_{23}} \sigma}{\omega_1^2 -\omega_2^2} , 
-\frac{\zeta_{12} \zeta_{34}}{\zeta_{14} \zeta_{23}} \right]} \ . 
\]
}
\be
r_1^2 (\sigma) =\frac{\omega_1^2 -\zeta_4}{\omega_1^2 -\omega_2^2} - \frac{\zeta_{34}}{\omega_1^2 -\omega_2^2} \frac{ \zeta_{24} 
\text{ sn}^2\left[ \pm \frac{\varkappa \omega_2 \sqrt{\zeta_{14} \zeta_{23}} \sigma}{\omega_1^2 -\omega_2^2} , 
\nu \right]}{  \zeta_{23} + \zeta_{24}  
\text{ sn}^2 \left[ \pm \frac{\varkappa \omega_2 \sqrt{\zeta_{14} \zeta_{23}} \sigma}{\omega_1^2 -\omega_2^2} , 
\nu \right]} \ . 
\label{ellipticr1} 
\ee
where $\nu=\frac{\zeta_{13} \zeta_{24}}{\zeta_{14} \zeta_{23}}$. Now we could use this expression to write the energy as a function of the winding numbers and the angular momenta. However, the first step in this direction, 
which is finding the winding numbers and the momenta in terms of the integration constants $v_i$ and the angular frequencies $\omega_i$, 
already leads to complicated integrals. Instead of following this path, which leads to cumbersome and non-illuminating expressions, in what follows we will analyse 
the problem in several interesting regimes of $\varkappa$.

But before we move to the study of the fate of solutions (\ref{ellipticr1}) for some limiting values of the deformation parameter, we will consider 
the case where the radii are taken to be constant. When we set to zero the derivatives in the equations of motion 
and solve for the Lagrange multiplier we find that
\be
\frac{\alpha_1^{\prime 2}-\omega_1^2}{1+\varkappa^2 r_2^2} = \alpha_2^{\prime 2} - \omega_2^2 - \varkappa^2 \frac{r_1^ 2 (\alpha_1^{\prime 2} - \omega_1^2)}{(1+\varkappa^2 r_2^2)^2} \ .
\ee
We can rewrite this expression as\footnote{It is immediate to see that in the limit $\varkappa \rightarrow i$ the solution reduces to $r_1=0$, $r_2=1$, together with either zero total angular momentum 
or zero winding $m_2$ because of the Virasoro constraint (\ref{Virasoro2}).}
\begin{equation}
1 + \varkappa^2 r_2^2 =\pm \sqrt{(1+\varkappa^2) \frac{m_1^2 -\omega_1^2}{m_2^2 -\omega_2^2}} \ ,
\end{equation}
where we have used the constraint $r_1^2 + r_2^2=1$ and the fact that $m_i=\alpha'_i$ because the winding velocities are constant when the radii are constant. However solving 
exactly this equation together with the Virasoro constraint leads to an algebraic equation of sixth degree. 
Instead of trying to solve the problem directly, we can write the solution as a power series expansion in inverse powers of the total angular momentum. We get \footnote{There is an additional 
possible expansion, depending on the choice of signs of the winding numbers.}
\begin{align}
r_1^2 & = \frac{km_2}{k m_2-m_1} + \frac {\lambda}{2 J^{2}} 
\frac{k m_1 m_2 (m_1 +m_2) (m_1 -m_2)^3 (m_1^2 - 2 k m_1 m_2 +m_2^2)}{(k m_1 -m_2)^2 (m_1 -k m_2)^4} + \dots \ , \\
r_2^2 &= \! \frac{m_1}{m_1-km_2} - \frac {\lambda}{2 J^{2}} 
\frac{k m_1 m_2 (m_1+m_2) (m_1 -m_2)^3 (m_1^2 -2 k m_1 m_2 +m_2^2)}{(k m_1 -m_2)^2 (m_1 -k m_2)^4} + \dots \ , 
\end{align}
for the radial coordinates, and 
\begin{align}
\omega_1 & = \frac{J}{\sqrt{\lambda}} \frac{k m_1 -m_2}{m_1 -m_2} 
+ \frac {\sqrt{\lambda}}{2 J} \frac{k m_1 (m_1 +m_2) (m_1 -m_2)^2 (m_1^2 -2 k m_1 m_2 +m_2^2)}{(km_1 -m_2)^2 (m_1 -km_2)^2} + \dots \ , \\
\omega_2 & = \frac{J}{\sqrt{\lambda}} \frac{m_1 -k m_2}{m_1 -m_2} 
+ \frac {\sqrt{\lambda}}{2 J} \frac{k m_2 (m_1 +m_2) (m_1 -m_2)^2 (m_1^2 -2 k m_1 m_2 +m_2^2)}{(km_1 -m_2)^2 (m_1 -km_2)^2} + \dots \ ,
\end{align}
for the angular frequencies, where we have introduced $k=\sqrt{1+\varkappa^2}$. 
Using now equation~(\ref{Virasoro1}) it is immediate to write the dispersion relation,
\begin{equation}
E^2 = J^2 \frac{(m_1^2 -2 k m_1 m_2 +m_2^2)}{(m_1-m_2)^2} + \lambda \frac{m_1 m_2 (m_1^2 -2km_1 m_2 +m_2^2)}{(km_1 -m_2) (km_2 -m_1)} + \dots \ .
\end{equation}

\section{Limiting cases of the $\eta$-deformed N-R system}
\label{limitingkappa}

In this section we are going to see how the solution we have found in the previous section behave in some particular limits of the deformation parameter from the Uhlenbeck constant point of view. Upon inspection of the polynomial (\ref{P4definition}) it is clear that the limits $\varkappa=\infty$ and $\varkappa =i$ simplify the evaluation of the roots. The backgrounds obtained from taking these limit has been studied in \cite{deformationsupercosets,Hoare2015}. In the case $\varkappa=\infty$ 
the deformed ten-dimensional metric is T-dual to de Sitter space times the hyperboloid, $dS_5 \times H^5$, which can also be understood as a flipped double Wick rotation of $AdS_5 \times S^5$. On the other hand, in the limit $\varkappa=i$\footnote{From an algebraic point of view, the $\varkappa=i$ limit behaves in the same way as the limit of pure NS-NS flux in the analysis of the deformation by flux of the Neumann-Rosochatius system presented in the previous chapter.} the deformed ten-dimensional metric turns into a pp-wave type background. We will also analyse these limits from the point of view of the Lagrangian formalism in Appendix \ref{appendixkappa}, where some relations between these limits are more transparent.

We will first focus on the $\varkappa\rightarrow \infty$ limit and the case where $v_2=\omega_1=0$. We can see that the four roots, as series on $\varkappa$, behave in the following way
\begin{align*}
	\zeta_1 &=-v_1^2 \varkappa^2 -\left[ v_1^2 +\omega_2^2 (1-\tilde{I}_1) \right]+\frac{\omega_2^ 2 (1-\tilde{I}_1)}{v_1^2 \varkappa^2}+\dots \ , & \zeta_4&=\omega_2^2 \ , \\
	\zeta_2 &=\omega_2^2 +\frac{\omega_2^4 (\tilde{I}_1-1)}{v_1^2 \varkappa^2}+\frac{\omega_2^4 (\tilde{I}_1-1) (\tilde{I}_1 \omega_2^2-2\omega_2^2 -v_1^2)}{v_1^4 \varkappa^4}+\dots \ , & \zeta_3 &=\frac{\omega_2^2 (1+\varkappa^2)}{\varkappa^2} \ ,
\end{align*}
Actually the series for the two first solutions come from the solution to a quadratic equation
\be
\zeta_{1,2}=\frac{\tilde{I}_1 \omega_2^2-v_1^2 (1+\varkappa^2) \pm \sqrt{[\tilde{I}_1 \omega_2^2-v_1^2 (1+\varkappa^2)]^2+4 v_1^2 \omega_2^2(1+\varkappa^2)}}{2} \ .
\ee
Therefore in the $\varkappa\rightarrow\infty$ limit one of the roots goes to $-\infty$ and we reduce the degree of our polynomial to $3$. 
In order to take the limit at the level of the solution, we need to make sure that $\zeta_1\rightarrow -\infty$. This requires writing (\ref{ellipticr1}) in the form
\be
r_1^2 (\sigma) =\frac{\omega_1^2 -\zeta_4}{\omega_1^2 -\omega_2^2} - \frac{\zeta_{14}}{\omega_1^2 -\omega_2^2} \frac{ \zeta_{24} 
\text{ sn}^2 \left[ \pm \frac{\varkappa \omega_2 \sqrt{\zeta_{34} \zeta_{21}} \sigma}{\omega_1^2 -\omega_2^2} , 
\frac{\nu}{\nu -1} \right]}{  \zeta_{13} + \zeta_{34}  
\text{ sn}^2 \left[ \pm \frac{\varkappa \omega_2 \sqrt{\zeta_{34} \zeta_{21}} \sigma}{\omega_1^2 -\omega_2^2} , 
\frac{\nu}{\nu -1} \right]} \ ,
\label{ellipticr1prime}
\ee
which, after taking $\omega_1=0$ and $\zeta_1 \rightarrow -\infty$, becomes
\be
r_1^2 (\sigma)=\frac{\zeta_4}{\omega_2^2} + \frac{\zeta_{34}}{\omega_2^2} \text{ sn}^2 \left[ \pm \frac{\varkappa \omega_2 \sqrt{\zeta_{24} \zeta_{21}} \sigma}{\omega_2^2} , 
\frac{\zeta_{34}}{\zeta_{24}} \right] + \dots \ ,
\ee
and substituting explicitly the remaining roots we arrive to
\be
r_1^2 (\sigma)=1+\frac{1}{\varkappa^2} \text{ sn}^2 \left[ \pm \varkappa \sqrt{\omega_2^2 (\tilde{I}_1-1)} \sigma , 
\frac{v_1^2}{\omega_2^2(\tilde{I}_1-1)} \right]+\dots \ .
\label{gensolinfty}
\ee

The second interesting $\varkappa\rightarrow \infty$ limit is the $\omega_2=v_1=v_2=0$ case. In this limit the roots become
\be	
\zeta_1 = - \infty \ , \quad \zeta_2 = 0 \ , \quad \zeta_3 =\omega_1^2 \frac{1-\tilde{I}_1}{1+\varkappa^2 \tilde{I}_1} \ ,  \quad \zeta_4 = \omega_1^2 \ .
\ee
and thus the degree of the polynomial is reduced again from four to three. The general solution, after a reordering of the roots in a similar way as in the previous case, is given by\footnote{A similar result was obtained for the pulsating string ansatz in \cite{Panigrahicircularkappa}.}
\begin{align}
r_2^2 (t) &
=\frac{\zeta_2}{\omega_1^2} +\frac{\zeta_4 -\zeta_2}{\omega_1^2} \text{ sn}^2 \left( \sqrt{(\zeta_2-\zeta_3) (1+\varkappa^2 \tilde{I}_1 )} \sigma , \frac{\zeta_2 - \zeta_4}{\zeta_2 - \zeta_3} \right) \notag \\
&=\text{ sn}^2 \left( \sqrt{-\omega_1^2 (1- \tilde{I}_1)} \sigma , \frac{1+\varkappa^2 \tilde{I}_1}{1-\tilde{I}_1}\right) \notag \\
&=\frac{1-\tilde{I}_1}{1+\varkappa^2 \tilde{I}_1} \text{ sn}^2 \left( \sqrt{-\omega_1^2 (1+\varkappa^2 \tilde{I}_1)} \sigma , \frac{1-\tilde{I}_1}{1+\varkappa^2 \tilde{I}_1}\right)\ . \label{omega1notzero}
\end{align}
where, in the last line, we have made use of the relation $\sqrt{m}$ sn$(u, m) = $sn$(\sqrt{m} u, \frac{1}{m} )$.  We must note
that this solution contains four different regimes. We can see that our solution has four different regimes: $\tilde{I}_1 \geq 1$ 
where we have to analytically continue the $r_2$ coordinate to $ir_2$. The same happens with the $\tilde{I}_1 < \frac{-1}{\varkappa^2}$, while the region $0<\tilde{I}_1 <1$ requires the continuation of the $r_1$ coordinate instead. Finally in the region $\frac{-1}{\varkappa^2} < \tilde{I_1} <0$ the solutions is circular, which completely disappears in the $\varkappa \rightarrow \infty$ limit, which is agreement with the transformation of the sphere into the hyperbolic plane for $\varkappa\rightarrow \infty$ showed in \cite{deformationsupercosets}.

To conclude our analysis we are going to study the $\varkappa \rightarrow i$ limit. In this limit the contribution from $v_1$ is negligible and $\omega_1$ becomes a shift in the Uhlenbeck constant. Therefore $\omega_2$ and $v_2$ are the only important free parameters. First we are going to consider the case $v_2=0$, where the roots behave like
\begin{align}
	\zeta_1 &=0+\mathcal{O} (\varkappa -i) \ , & \zeta_2 &=0+\mathcal{O} (\varkappa -i) \ , \\
	\zeta_3 &= \tilde{I}_1 \omega_2^2 \ , & \zeta_4 &=\omega_2^2 \ .
\end{align}
Substituting and performing some manipulations we arrive to the solution
\begin{equation}
	r_1^2= \frac{\tilde{I}_1}{1-(\tilde{I}_1 -1) \cosh^2 \left( \sqrt{\omega_2^2 \tilde{I}_1} \sigma \right)} \label{omega2notzero}\ .
\end{equation}

The other limit, $\omega_2=0$, is characterized by the roots
\begin{align}
	\zeta_1 & \approx \zeta_2= \omega_1^2 +\mathcal{O} (\varkappa-i) \ , \\
	\zeta_{3} &= \frac{v_2^2 \omega_1^2}{v_2^2 +\omega_1^2 (\tilde{I}_1 -1)} \ , & \zeta_4=-\infty \ ,
\end{align}
where we have kept $\omega_1 \neq 0$ to simplify our computations, as the Uhlenbeck constant has to be modified when both $\omega_i$ vanish. Substituting and performing some manipulations we arrive to the solution
\begin{equation}
	r_1^2= \frac{\omega_1^2 -\zeta}{\omega_1^2}= \frac{\omega_1^2(1-\tilde{I}_1 )}{v_2^2 +\omega_1^2 (\tilde{I}_1 -1)} \sech^2 \left[ \left(1-\tilde{I}_1 \right) \sigma \right] \label{omega2zero}\ .
\end{equation}
Now we can eliminate the $\omega_1$ factor by the redefinition $v_2=\tilde{v}_2 \omega_1$. This happens, as we said, because the term encoding the dependence with $\omega_1^2$ in the Uhlenbeck becomes a constant in the $\varkappa=i$ limit, making it a dummy variable.

Although the solutions for $v_2=0$ and $\omega_2=0$ seem completely different, they are deeply related. This relation is not explicit from the Uhlenbeck constant, but it is evident when we write the Lagrangian systems associated to both limits. We will explore this direction in Appendix \ref{appendixkappa}.

\part{Integrability on the Field Theory side. Spin chains}

\chapter[Introduction: The two Bethe Ansätze\protect\footnote{Without wanting to create a debate about the plural form of ansatz.}]{Introduction: The two Bethe Ansätze}\chaptermark{Introduction: The two Bethe Ansätze}
\label{Betheansatze}
	\begin{chapquote}{Abstruse Goose 342, \textit{Moment of Clarity(?) - part 2}}
 		Quantum Mechanical Spin. What is it? And don't give me any of that bullshit about Pauli Matrices, Fermi-Dirac and Bose-Einstein statistics, etc. Any Monkey can do the Math but what does the math mean? I mean, WHAT IS SPIN REALLY?!!
	\end{chapquote}

In this chapter we are going to review two techniques to perform computations in quantum theories in general, and spin chains and theories in a lattice in particular: the Coordinate Bethe Ansatz (CBA) and the Algebraic Bethe Ansatz (ABA). Although in the recent years there has been a boom of new and powerful methods, like the Analytic Bethe Ansatz \cite{Reshetikhin_1983}, Off-diagonal Bethe Ansatz \cite{Wang2015}, Separation of Variables (SoV) \cite{Sklyanin1995}... The two Bethe Ansätze we are going to focus on are the cornerstones of quantum integrability and still very powerful by themselves.

This chapter is divided in five sections. In the first one we will introduce the CBA. In particular we are going to present it by solving the Heisenberg Hamiltonian. For this section we will mostly follow \cite{9780521460651}. In the second section we will present the second of the ansätze, the ABA. We will construct it from first principles as a lattice version of the Lax formalism presented in section~\ref{PCMintegrability}. After that we will specify the particular case of the Heisenberg Hamiltonian and explicitly solve it. The last part of this section is devoted to understanding the relation between operators in both ansätze and how to compute scalar products in the ABA. For this section we will mostly follow references  \cite{FADDEEValgebraic,Howthealgebraicbetheansatz,NEPOMECHIE_1999}. The third section deals with the scalar products in both the CBA and the ABA. We will present the computation of the same correlation functions in both formalism and compare the different normalizations. For this section we will mostly follow \cite{formfactorsnieto}. After that we will present the so-called BDS spin chain, a spin chain hamiltonian that reproduces $\mc{N}=4$ SYM at all loops up to the lenght of the chain, as it does not take into account wrapping effects. In the last section we will present the details of the bootstrap program and the form factor axioms. For this section we will follow references \cite{9780521460651,Babujian_2006}.

\section{Coordinate Bethe Ansatz}

Let us consider the spin $\frac{1}{2}$ Heisenberg Hamiltonian on a chain of length $L$. Our Hilbert space will be $L$ copies of the Hilbert space of a spin $\frac{1}{2}$ particle, that is, $\left(\mathbb{C}^2 \right)^L$ and the Hamiltonian will be given by
\begin{align}
	\mathrm{H}&=J\sum_{l=1}^L{(\sigma_l^x \sigma_{l+1}^x +\sigma_l^y \sigma_{l+1}^y +\sigma_l^z \sigma_{l+1}^z  )} \\
	&=J\sum_{l=1}^L{[2(\sigma_l^+ \sigma_{l+1}^- +\sigma_l^- \sigma_{l+1}^+ )+\sigma_l^z \sigma_{l+1}^z ]} \ , \label{HesienbergH}
\end{align}
where $J<0$ is a coupling constant\footnote{This is called the ferromagnetic Heisenberg Hamiltonian because, as we will show, its ground state is a saturated state with all spins aligned. When $J>0$ the behaviour changes radically. This other regime is called antiferromagnetic or Néel phase. The Hilbert space can be constructed in a similar way as the one we are going to present but starting from the state given by a chain with alternated spins called ``Néel state'' \cite{Karbach_1998}, although it is not the ground state (this happens because because it minimizes the expectation of $\langle s^z_n s^z_{n+1} \rangle$ but it does not minimize it for the other two components).}  and the $\sigma^i$ are the Pauli matrices. The subindices indicate in which chain site the Pauli matrices act non-trivially, that is,
\begin{equation}
	\sigma^i_l=\overbrace{\mathbb{I} \otimes \dots \otimes \mathbb{I}}^{(l-1) \text{ times}} \otimes \sigma^i \otimes \mathbb{I} \otimes \dots \otimes \mathbb{I} \ .
\end{equation}
We also want to impose periodic boundary conditions, which means $\sigma^i_{L+1}=\sigma^i_1$.

A generalization of this Hamiltonian is the so-called XYZ spin chain Hamiltonian, where each of the three terms of the Hamiltonian have different couplings. We recover the Heisenberg Hamiltonian by putting $J_x=J_y=J_z$ and hence it is usually called the XXX spin chain. Another case of interest is the XXZ spin chain where $J_x=J_y=J\neq J_z=J\Delta$.

This particular Hamiltonian is of interest for us because, by using the isomorphism between anomalous dimensions of single-trace operators and spin chains, it corresponds to the restriction to the scalar $SU(2)\subset SO(6)$ sector of the matrix of anomalous dimensions/spin chain Hamiltonian presented in equation (\ref{MZHamiltonian}). The isomorphism works in the following way: each operator is build out from a single trace of products of two complex scalars and is mapped to a spin chain state. One of these complex scalars is interpreted as the spin up (usually identified with the complex scalar $Z$), while the other as the spin down (usually identified with the complex scalar $X$). This restriction to the $SU(2)$ subgroup is consistent because it is a \emph{closed sector} at one-loop\footnote{A closed sector is a subgroup of the full $PSU(2,2|4)$ spin chain such that the operator mixing can only occur between operators inside said subgroup. We will talk a little more in depth about it in section \ref{BDSspinchain}.}.

There are two obvious symmetries on this Hamiltonian. The first one is the shift operator $U=e^{-ip}$, which shifts the states on the chain one lattice position to the right. It is immediate to prove that $[\mathrm{H},U]=0$ from the form of the Hamiltonian and its periodicity. Note that periodicity also imposes $U^L=e^{-ipL}=\mathbb{I}$. The second symmetry is the total spin in one direction, which we are going to choose as the $z$ direction $[\mathrm{H},s^z]=0$, where the total spin $s^z$ is computed as the sum of the individual spins of each lattice site, $s^z=\sum_{l=1}^L{\frac{\sigma_l^z}{2}}$. The existence of these two symmetries implies that we can classify the eigenstates of the Hamiltonian by the total spin in the $z$ direction and the total momentum.

Also, the second writing of the Hamiltonian (\ref{HesienbergH}) hints us two of its eigenstates, the one with all spins up in the $z$ direction and the one with all spins down. These two states are annihilated by the $\sigma^+_j$ and $\sigma^-_j$ operators respectively for every value of the lattice coordinate $j$. This imply that, when we apply the Hamiltonian operator to these states, the two first terms of every addend in the sum vanish, leaving only the contribution of $\sum_j\sigma^z_j \sigma^z_{j+1}$, of which they are already an eigenstate. We are going to take the first one as our reference state for our construction of the Hilbert space as it is a ground state. We can see that
\begin{equation}
	\mathrm{H} \ket{0}=\mathrm{H} \ket{\uparrow \uparrow \dots \uparrow}=\frac{JL}{4}\ket{\uparrow \uparrow \dots \uparrow} \ ,
\end{equation}
while if one of the spins if flipped two of the contributions pick the opposite sign and the energy grows. The  following construction can be done in a symmetrical way if we choose $\ket{\downarrow \downarrow \dots \downarrow}$ as our reference state.

If we have only one spin flipped down we can write an ansatz for the eigenvector of the form
\begin{equation}
	\ket{P}_{(1)}=\sum_{x=1}^L \psi_P (x) \ket{x} \ ,
\end{equation}
where the subindex indicates the number of spins down and $\ket{x}=\sigma_x^- \ket{0}$ is the state with all spins up except for the one at lattice site $x$. Translational invariance tells us that we should choose
\begin{equation}
	\psi (x)=A e^{iPx} \ ,
\end{equation}
where $A=1$ is a normalization constant\footnote{There are two usual normalizations of these states, either $A=1$ or $A=\frac{1}{\sqrt{L}}$. We are going to choose the first one for simplicity even though the states are not properly normalized in that way. A third less usual normalization, but useful in the AdS/CFT context, is $A=\frac{1}{\sqrt{L} N_c^{(L+M)/2}}$, where $N_c$ is the range of the gauge group and $M$ the number of flipped spins. This last one is used, for example, in \cite{BMN}.} and $P=\frac{2\pi k}{L}$ with $k\in \mathbb{Z}_{L}$ because of periodicity. Therefore there are $L$ states in the Hilbert space of one spin down, as expected. These states are usually called \emph{one magnon states}, as they are constructed by one excitation that takes the shape of a ``spin wave''. The energy of these states is
\begin{equation}
	\mathrm{H} \ket{P}_{(1)}= \frac{1}{4}[JL-4J(1-\cos \, P)] \ket{P}_{(1)}= (E_0 +E (P)) \ket{P}_{(1)} \ ,
\end{equation}
which we have separated it into the energy of the ground state $E_0$ and the energy of the excitation $E(P)$.

In principle we can repeat part of the strategy for the case of two spins flipped down
\begin{equation}
	\ket{P}_{(2)}=\sum_{\substack{x,y=1\\x<y}}^L \psi_P (x,y) \ket{x,y} \ ,
\end{equation}
where $\ket{x,y}=\sigma_x^- \sigma_y^- \ket{0}$
. Translational invariance implies in this case that $\psi_P (x,y)=\psi_P (y,x+L)$, but it doesn't provide enough information to find the eigenvector. However, 
Bethe \cite{Bethe_1931} proposed the product of plane-waves as ansatz 
\begin{equation}
	\psi_P (x,y)=A_{12} e^{i(p_1 x + p_2 y)}+A_{21} e^{i (p_1 y +p_2 x)} \ ,
\end{equation}
which solves the condition of being an eigenstate if $x+1 \neq y$. Translational invariance transforms into the condition $P=p_1+p_2$ and the periodicity condition now reads
\begin{equation}
	\frac{A_{21}}{A_{12}}=e^{ip_2 L}=\frac{1}{e^{ip_1 L}}=S_{12}=\frac{1}{S_{21}} \Longrightarrow \left\{\begin{array}{l}
	e^{ip_1 L} S_{12}=1 \\
	e^{ip_2 L} S_{21}=1
\end{array}	  \right. \ . 
\end{equation}
The physical interpretation of this equation is that the total phase shift undergone by a spin wave after traveling around the closed chain, which should be trivial, is given by a kinematical factor (like in the case of one magnon) and a phase shift produced by the interchange of the two spin waves, which we are going to call \emph{$S$-matrix} (although in this cases it is only a number, not a matrix). But we still have to fix the value of this $S$-matrix. To do so we have to impose the eigenstate condition for $x+1=y$, which gives us
\begin{equation}
	S_{12}=\frac{1-2e^{ip_2}+e^{i(p_1+p_2)}}{1-2e^{ip_1}+e^{i(p_1+p_2)}} \ .
\end{equation}
Joining this result with the periodicity condition we obtain the equation
\begin{equation}
	e^{ip_2 L}=\frac{1-2e^{ip_2}+e^{i(p_1+p_2)}}{1-2e^{ip_1}+e^{i(p_1+p_2)}} \ ,
\end{equation}
which called in the literature the \emph{Bethe Ansatz Equation}. It is important to notice here that there are two kinds of solutions to these equations, something already noticed by Bethe in his original article, namely, one set of solutions with real momenta and other set of solutions with complex conjugated momenta, which can be understood as a bound state of the magnons. We still have to find a normalization for the coefficients. We are going to choose $A_{12}=1$ for convenience, as explained in a previous footnote. The energy of these states, after some algebra, is
\begin{align}
	\mathrm{H} \ket{P}_{(2)} &=\frac{1}{4} [JL-4J(1-\cos \, p_1)-4J(1-\cos \, p_2)] \ket{P}_{(2)} \notag \\
	&=(E_0 +E(p_1) +E(p_2)) \ket{P}_{(2)} \ ,
\end{align}
that is, the total energy of the two particle excitation is just like the two particles were completely independent. This is already a sign of the underlying integrability of the model, but we still have to present the tools to prove it.

Moving now to the case of three magnons, we can again write the eigenfunction as a sum over the possible positions of the spins down weighted with a wave function
\begin{equation}
	\ket{P}_{(3)}=\sum_{\substack{x,y,z=1\\x<y<z}}^L \psi_P (x,y,z) \ket{x,y,z} \ ,
\end{equation}
for which Bethe proposed again a plane-wave ansatz,
\begin{equation}
	\psi_P (x,y,z)=\sum_{\sigma \in S_3}{A_{\sigma} e^{i(p_{\sigma (1)} x + p_{\sigma (2)} y +p_{\sigma (3)} z)}} \ ,
\end{equation}
where translational invariance imposes $P=p_1+p_2+p_3$ and the periodicity condition imposes
\begin{align}
	e^{ip_1 L}&=\frac{A_{123}}{A_{132}}= \frac{A_{132}}{A_{321}} \ , \notag \\
	e^{ip_2 L}&=\frac{A_{231}}{A_{312}}= \frac{A_{213}}{A_{132}} \ , \\
	e^{ip_3 L}&=\frac{A_{312}}{A_{123}}= \frac{A_{321}}{A_{213}} \ , \notag 
\end{align}
which determines all the coefficients up to one of them, which depends on the normalization we choose. These equations hide the most important condition of quantum integrability, the factorized scattering and the Yang-Baxter equation. We will talk more about it in the next section. For the moment we are going to define the S-matrices as the quotient of two prefactors with one of the indices fixed, that is
\begin{align}
	S_{12} &=\frac{A_{213}}{A_{123}}=\frac{A_{321}}{A_{312}} \ , & S_{13} &=\frac{A_{312}}{A_{132}}=\frac{A_{231}}{A_{213}} \ , & S_{23} &=\frac{A_{321}}{A_{231}}=\frac{A_{132}}{A_{123}} \ .
\end{align}
Hence the Bethe ansatz equations can be written now
\begin{equation}
	e^{ip_l L}= \prod_{\substack{k=1 \\k\neq l}}^{3} S_{kl} \ ,
\end{equation}
where the form of the S-matrices is fixed again by imposing the eigenstate condition when two magnons are not separated and the third is well separated
\begin{equation}
	S_{ij}=\frac{1-2e^{ip_j}+e^{i(p_i+p_j)}}{1-2e^{ip_i}+e^{i(p_i+p_j)}} \ ,
\end{equation}
which is the same as in the case of two magnons. The energy can be obtained from the eigenstate condition with all magnons well separated. We get
\begin{equation}
	E=E_0 +\sum_{l=1}^3 E(p_l) \ ,
\end{equation}
so again it is the sum of the contribution of each individual magnon.

From the case of three magnons it is easy to see how to generalize the formalism to an arbitrary number of magnons $M$. The most important formulas are
\begin{align}
	&\psi_P (x_1, x_2 , \dots ,x_M)=\sum_{\sigma \in S_M}{A_{\sigma} e^{i \sum_{j=1}^M p_{\sigma (j)} x_j }} \ , \\
	&e^{ip_l L}= \prod_{\substack{k=1 \\k\neq l}}^{M} S_{kl} \, \text{ where } \, S_{ij}=\frac{1-2e^{ip_j}+e^{i(p_i+p_j)}}{1-2e^{ip_i}+e^{i(p_i+p_j)}} \ , \label{CBAE} \\
	&E=E_0 +\sum_{l=1}^M E(p_l) \ .
\end{align}
As we can see they are a direct generalization of the formulas for three magnons\footnote{This method can be directly applied in other models like XXZ spin chain or the Lieb-Liniger model (bosonic particles in a 1-dimensional box with delta interactions) \cite{0521373204}. However it needs some modification to solve the XYZ spin chain.}.

This construction can easily be generalized to groups with higher rank. In particular a general construction for the $SU(N)$ spin chain Hamiltonian can be found in chapter 1.1 of \cite{Volin_2011}.


We want to end this section by recalling that we will be interested in the case of spin chains representing $\mc{N}=4$ supersymmetric Yang-Mills operators. These kinds of spin chains should correspond to single-trance operators, so a more restrictive periodicity condition has to be imposed as traces are cyclic. In particular cyclicity of the trace implies that the shift operator we presented above should be trivial, which imposes a zero total momentum condition over the spin chain,
\begin{equation}
	U^{-1}=e^{iP}=\prod_{i=1}^M{e^{ip_j}}=e^{i \sum_{j=1}^M{p_j}}=1 \Longleftrightarrow P=\sum_{j=1}^M{p_j}=0 \ . \label{trace}
\end{equation}

\section{Algebraic Bethe Ansatz}

\subsection{Lax formalism in the lattice and the ABA}

We are going to start this section by writing lattice versions of the Lax formalism whose classical versions we have already presented in section~\ref{PCMintegrability}. First of all, we are going to present a lattice equivalent of the infinitesimal parallel transport, that is, the space component of the linearized equations for the classical wave function~(\ref{classicalwaveequation}),
\begin{equation}
	\phi_{n+1} =L_{n,a} (\lambda) \phi_n \ ,
\end{equation}
where $\phi_n\in \mathcal{H}_n \otimes V$ is a vector from the tensor product of the Hilbert space of site $n$ (usually called ``physical space'') and an auxiliary space $V$, and $L_{n,a} (\lambda) $ is the Lax operator acting on  $\mathcal{H}_n \otimes V$ with spectral parameter $\lambda$. Hence the definition of the monodromy matrix is also the lattice version of eq.~(\ref{monodromymatrix})\footnote{There are two ways of ordering the Lax operators inside the monodromy matrix. We are going to use the most common in the literature, which can be found in \cite{9780521460651}, \cite{FADDEEValgebraic} and \cite{0521373204} among other. The opposite ordering, $T_a(\lambda) = L_{1,a} (\lambda) L_{2,a} (\lambda) \dots L_{L,a} (\lambda)$, can be found, for example, in \cite{EscobedoTailoring}.},
\begin{equation}
	T_a(\lambda) = L_{L,a} (\lambda) L_{L-1,a} (\lambda) \dots L_{1,a} (\lambda) \ .
\end{equation}
Actually, this is not the most general monodromy matrix we can write as we can add a set of extra degrees of freedom in two ways: first, by adding a set of fixed parameters to each of the arguments of the Lax operator which are usually called \emph{inhomogeneities}, introduced in the following way 
\begin{equation}
	T_a(\lambda, \{ \xi \} ) = L_{L,a} (\lambda -\xi_L) L_{L-1,a} (\lambda -\xi_{L-1}) \dots L_{1,a} (\lambda -\xi_1) \ ; \label{abamonodromymatrix}
\end{equation}
second, by adding a twist matrix $K\in \mf{sl} (2)$,
\begin{equation}
	T_{a,K} (\lambda ) =K L_{L,a} (\lambda) L_{L-1,a} (\lambda) \dots L_{1,a} (\lambda) \ ,
\end{equation}
although in principle we could have written a more general expression with left and right twists,
\begin{equation}
T_{a,K_1|K_2} (\lambda ) =K_1 L_{L,a} (\lambda) L_{L-1,a} (\lambda) \dots L_{1,a} (\lambda) K_2 \ .
\end{equation}
The left and right twists can be related so we have to care only about having a twist in one of the ends \cite{TwistsABA}. Most of the time we will set these inhomogeneities to be the same and the twists to identity. The transfer matrix is defined as the trace of the monodromy matrix $\mathcal{T}(\lambda )=\Tr_a$~$(T_a (\lambda))$.  To prove that it actually generates a tower of conserved operators we have to introduce the R-matrix.

\begin{figure}[p]
\begin{center}
	\includegraphics[height=0.3\textheight,keepaspectratio]{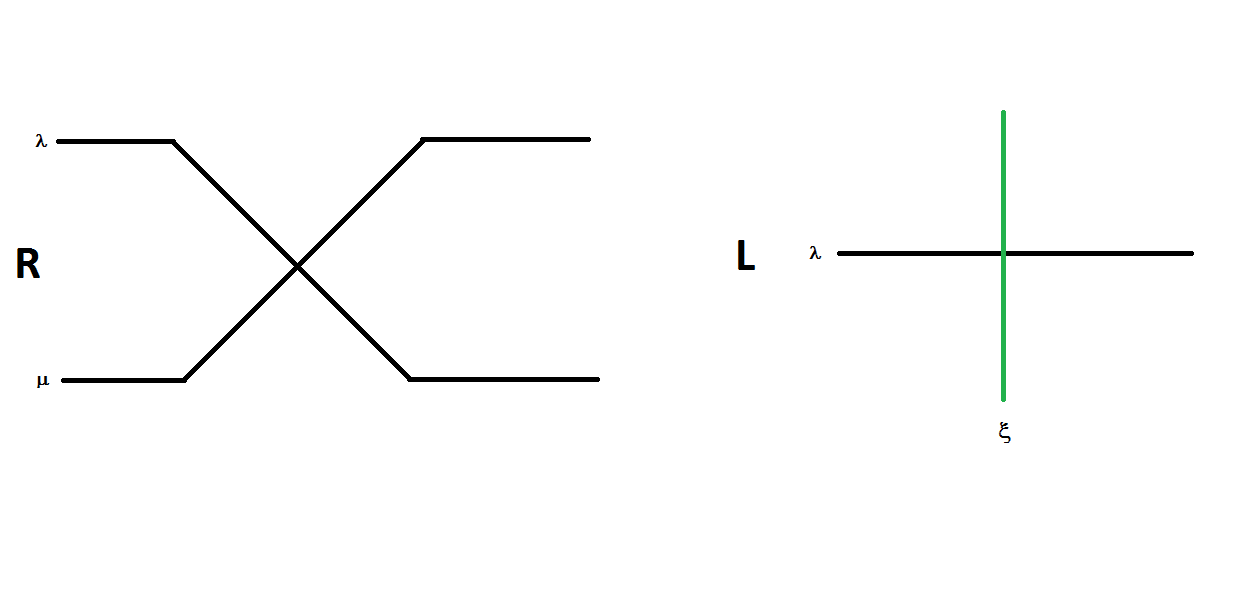}
\caption{Pictorial representation of the R matrix and the L matrix. We have represented the auxiliary spaces in black and the physical spaces in green.} \label{graphicdefinition}
\end{center}
\end{figure}

\begin{figure}[p]
\begin{center}
	\includegraphics[width=\textwidth,keepaspectratio]{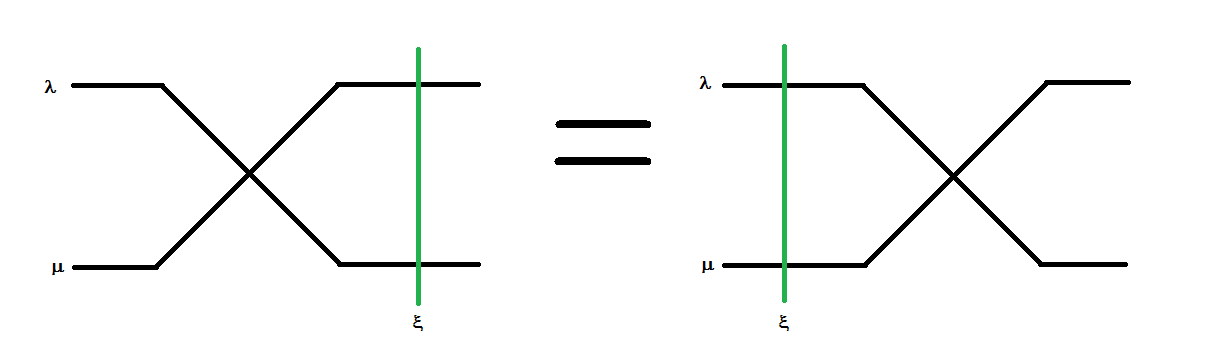}
\caption{Pictorial representation of the RLL equation.} \label{RLLeqfigure}
\end{center}
\end{figure}

If we have now two Lax operators that act in the same Hilbert space but in different auxiliary space, the commutation rule is given by
\begin{equation}
	R_{a_1 a_2} (\lambda - \mu ) L_{n,a_1} (\lambda ) L_{n,a_2} (\mu ) = L_{n,a_2} (\mu ) L_{n,a_1} (\lambda ) R_{a_1 a_2} (\lambda - \mu ) \ , \label{RLLequation}
\end{equation}
which is the quantum version of the definition of the R-matrix (\ref{rmatrixdefinition}). Indeed, there exists a quantum deformation parameter $\hbar$ such that the the first non-trivial term of the expansion of the R-matrix around $\hbar=0$ can be identified as the classical R-matrix\footnote{R-matrices that fulfil this property are called quasi-classical \cite{JIMBO1989}. Examples of not quasi-classical solutions can be found in \cite{Au_Yang_1987,Baxter_1988,Fateev_1982,Kashiwara_1986}.}, $R=\mathbb{I}+\hbar r+\dots$, so we can recover the classical R-matrix from the quantum one \cite{Kupershmidt_1999}. There is a nice and very useful graphical representation of this formula and the following ones. If we represent the real space by vertical lines and the auxiliary space by horizontal lines, the pictorial representation of the $L$ and $R$ matrices take the form displayed in figure~\ref{graphicdefinition}. Figure~\ref{RLLeqfigure} represents the RLL equation using this identification. This pictorial representation also highlights the application of the Bethe Ansatz to solve problems involving Temperley-Lieb algebras (see, for example, \cite{Abramsky_2007} for an introduction to TL algebras and \cite{Kulish_2003} for its relation with XXZ spin chains), either using the coordinate version \cite{GHIOTTO_2000} or the algebraic version \cite{Nepomechie_TLa}.

\begin{figure}[p]
\begin{center}
	\includegraphics[width=\textwidth,keepaspectratio]{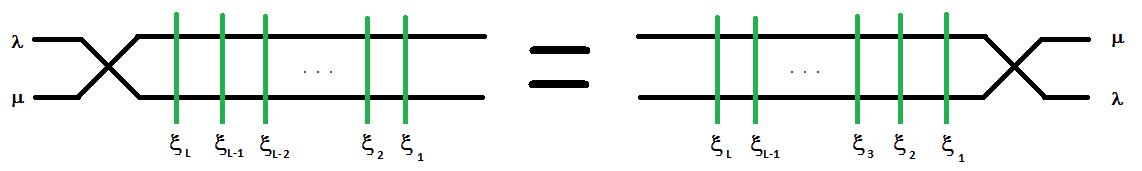}
\caption{Pictorial representation of the RTT equation (or, as refered in \cite{FADDEEValgebraic} ``Train argument'').} \label{RTTeqfigure}
\end{center}
\end{figure}

Successive applications of the commutation relation (\ref{RLLequation}) can be used to construct the commutation relation for the monodromy matrix 
\begin{equation}
	R_{a_1 a_2} (\lambda - \mu ) T_{a_1} (\lambda ) T_{a_2} (\mu ) = T_{a_2} (\mu ) T_{a_1} (\lambda ) R_{a_1 a_2} (\lambda - \mu ) \ , \label{RTT}
\end{equation}
usually called \emph{RTT relation} (sometimes also called FRT exchange relation\cite{Arnaudon2005}). This relation is the quantum version of eq.~(\ref{sklyaninrelation}) and from it we can prove the commutativity of the transfer matrices 
\begin{equation}
	[\mc{T}(\lambda ) ,\mc{T}(\mu )]=0 \ .
\end{equation}
Hence we can start constructing conserved quantities as, for example, logarithmic derivatives of the transfer matrix at some particular value of the spectral parameter. Using its pictorial representation, figure \ref{RTTeqfigure}, the RTT relation is very easy to prove from the RLL relation.

\begin{figure}[t]
\begin{center}
	\includegraphics[width=\textwidth,keepaspectratio]{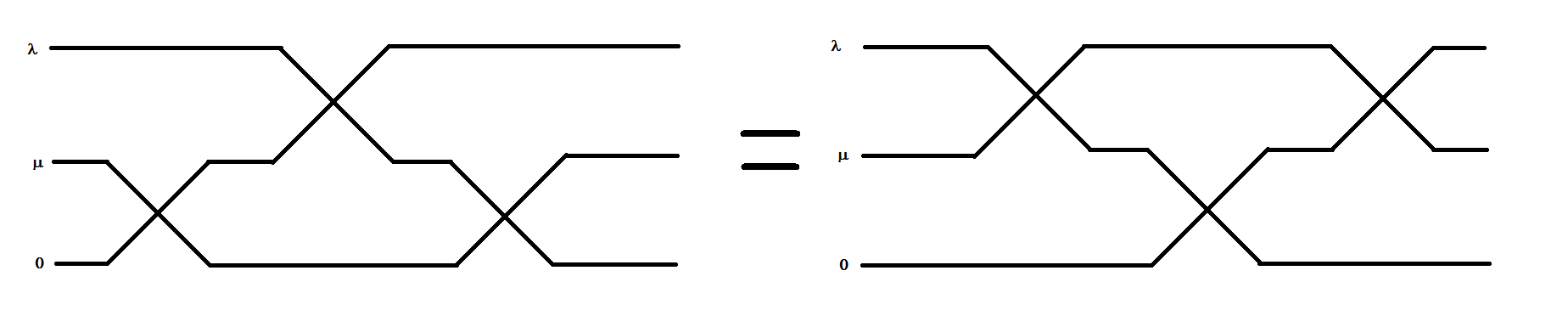}
\caption{Pictorial representation of the Yang-Baxter or RRR equation.} \label{RRReqfigure}
\end{center}
\end{figure}

Coming back to the R-matrix, we want to highlight that it fulfils the Yang-Baxter equation
\begin{equation}
	R_{a_1 a_2} (u-v) R_{a_1 a_3} (u) R_{a_2 a_3} (v) =R_{a_2 a_3} (v) R_{a_1 a_3} (u) R_{a_1 a_2} (u-v) \ , \label{Yang-Baxter}
\end{equation}
which, after performing the expansion in the quantum deformation parameter $\hbar$ we commented before, gives us the classical Yang-Baxter equation~(\ref{classicalyangbaxter})
\begin{equation}
	\left[ r_{12} (u) , r_{13} (u+v) \right] +\left[ r_{12} (u) , r_{23} (v) \right] +\left[ r_{13} (u+v) , r_{23} (v) \right] =0 \ .
\end{equation}
The Yang-Baxter equation is represented in figure~\ref{RRReqfigure}.

\subsection{Solving the Heisenberg spin chain with ABA}
\label{solvingABA}

Now let us present the particular R-matrix, Lax operator and auxiliary space which we are going to use to describe the XXX Heisenberg spin chain\footnote{In this choice the quantum deformation parameter $\hbar$ is hidden in a re-scaling of the $\lambda$ and $\xi$.}
\begin{align}
	L_{n,a} (\lambda-\xi) &=\left( \lambda -\xi_n \right) \mathbb{I}_{n,a} + i \mathcal{P}_{n,a} \ ,\\
	R_{a,b} &=\lambda \mathbb{I}_{a,b} + i \mathcal{P}_{a,b} \ ,
\end{align}
where $\mathcal{P}$ is the permutation operator $\mathcal{P} (a\otimes b) \mathcal{P}=b\otimes a$, $\lambda\in \mathbb{C}$ is the spectral parameter and we have chosen $V\cong \mathcal{H}_n \cong \mathbb{C}^2$. This is a particular case of the 6-vertex model, originally introduced by Pauling \cite{Pauling1935} to account for the residual entropy of water ice. The 6-vertex model allow us to solve also the XXZ spin chain, however the XYZ spin chain requires the more general 8-vertex model. For our particular case we are going to set all inhomogeneities to $\xi=\frac{i}{2}$ as this particular value will allow us to write
\begin{equation}
	L_{n,a} (\lambda-\frac{i}{2}) =\lambda (\mathbb{I}_n \otimes \mathbb{I}_a ) +i \sum_j{(s^j_n \otimes \sigma^j_a)} \ ,
\end{equation}
where $s^j=\frac{1}{2} \sigma^j$ is the $\frac{1}{2}$ representation of the spin operators. At this point it is important to make a comment about the notation, because some authors use as R-matrix what we call $\check{R}=\mathcal{P} R$ or has the spectral parameter shifted\footnote{An easy way to distinguish which of the notations an author used is to compute the value of $R$ or $\frac{R}{\lambda}$ in the limit $\lambda \rightarrow \infty$, depending on which is finite. An example of the second case can be found in \cite{Howthealgebraicbetheansatz,Zabrodin_2015}, where $L_{n,a} (\lambda )=R_{a,b}=\left( \lambda- \frac{i}{2} \right) I+\mc{P}$ and inhomogeneities set to zero.}.

By direct observation it is evident that the point $\lambda=\frac{i}{2}$ is a very particular point, because it simplifies the Lax operator but it does not make it the identity. We will use that in our advantage. If we compute the transfer matrix at that point we get,
\begin{equation}
	\mathcal{T} \left( \frac{i}{2} \right)= i^L \text{Tr}_a \left\{ \mathcal{P}_{L,a} \mathcal{P}_{L-1,a} \dots \mathcal{P}_{1,a} \right\} \ .
\end{equation}
Using now that $\mathcal{P}_{n,a} \mathcal{P}_{m,a}=\mathcal{P}_{m,n} \mathcal{P}_{n,a}=\mathcal{P}_{n,m} \mathcal{P}_{n,a}$ and $\Tr_a \{\mathcal{P}_{i,a}\}=\mathbb{I}$, we conclude that
\begin{equation}
	\mathcal{T} \left( \frac{i}{2} \right)= i^L \mathcal{P}_{1,2} \mathcal{P}_{2,3} \dots \mathcal{P}_{L-1,L}\mathcal{P}_{L,1}=i^L U=i^L e^{-iP} \ ,
\end{equation}
where $U$ and $P$ are the shift operator and momentum operator introduced in the previous section. Therefore $\mathcal{T} \left( \frac{i}{2} \right)$ is a conserved charge, as we expected. To compute the second conserved charge one derivative of $\mc{T}$ is needed. Using that $\frac{dL_{n,a}}{d\lambda}=\mathbb{I}_{n,a}$, we get
\begin{align}
	&\left. \frac{d \mathcal{T}}{d\lambda} \right|_{\lambda=\frac{i}{2}} =i^{L-1} \sum_{j=1}^L \text{Tr}_a \left\{ \mathcal{P}_{L,a} \dots \mathcal{P}_{j+1,a} \mathcal{P}_{j-1,a} \dots \mathcal{P}_{1,a} \right\} \ , \\
	& \left. \frac{d \mathcal{T}}{d\lambda} \mathcal{T}^{-1} (\lambda ) \right|_{\lambda=\frac{i}{2}} \propto \sum_{j=1}^L \mathcal{P}_{j,j+1}\propto \sum_{j=1}^L \vec{s}_j \cdot \vec{s}_{j+1} \propto \mathrm{H} \ ,
\end{align}
which is the Hamiltonian of the XXX spin chain. Furthermore, we can also check that the parallel transport (suitably rescaled) implies the classical spin Hamiltonian
\begin{equation}
	\phi_{n+1}=\frac{L_n \phi_n}{\lambda} \Longrightarrow \phi ' (x) = \frac{ i s(x) \phi(x)}{\lambda} \ .
\end{equation}

Now that the equivalence between working with this Lax operator and with the XXX spin chain is proven, we have to find a way to construct the states and compute its energy. To do that, a look at the monodromy matrix as a matrix on the auxiliary space is needed
\begin{equation}
	T_a (\lambda )=\left( \begin{array}{cc}
	A(\lambda ) & B(\lambda ) \\
	C(\lambda ) & D(\lambda )
	\end{array}\right) \ ,
\end{equation}
where all four operators act on the tensor product of Hilbert spaces of all sites of the chain. In terms of these operators, the transfer matrix reads $\mathcal{T} (\lambda )=A(\lambda )+D(\lambda )$. If we compute the action of these four operators over the reference state we have defined in the previous section, $\ket{0}=\ket{\uparrow \uparrow \dots \uparrow}$, we get
\begin{align}
	A(\lambda ) \ket{0} &= a(\lambda ) \ket{0}= \prod_{n=1}^\infty{(\lambda+i-\xi_n)} \ket{0}= \left( \lambda +\frac{i}{2} \right)^L \ket{0} \ , & B(\lambda ) \ket{0} &\neq 0 \ , \\
	D(\lambda ) \ket{0} &= d(\lambda ) \ket{0}= \prod_{n=1}^\infty{(\lambda-\xi_n)} \ket{0}= \left( \lambda -\frac{i}{2} \right)^L \ket{0} \ , & C(\lambda ) \ket{0} &= 0  \ .
\end{align}
This suggests us that the $C$ operator might be used as an annihilation operator and the $B$ operator might be used as a creation operator.

Before seeing if this assumption is correct, we have to write the commutation relations between the operators of the monodromy matrix. Those are obtained from the RTT relation (\ref{RTT}). The six interesting for us at this moment are
\begin{align}
	&[B(\lambda ) , B(\mu )]=0 \ , \\
	&A(\lambda ) B(\mu )=f(\mu , \lambda) B(\mu ) A(\lambda ) +g(\lambda , \mu ) B(\lambda ) A(\mu ) \ ,\label{commAB}\\
	&D(\lambda ) B(\mu )=f(\lambda, \mu) B(\mu ) D(\lambda ) + g(\mu , \lambda ) B(\lambda ) D(\mu ) \ , \\
	&C(\mu ) A(\lambda )=f(\mu , \lambda) A(\lambda ) C(\mu ) +g(\lambda , \mu ) A(\mu ) C(\lambda ) \ ,\label{commCA}\\
	&C(\mu ) D(\lambda )=f(\lambda, \mu) D(\lambda ) C(\mu ) + g(\mu , \lambda ) D(\mu ) C(\lambda ) \ , \label{commCD}\\
	&[C(\lambda ) , B(\mu) ]=g(\lambda , \mu ) \left[ A(\lambda ) D(\mu ) - A(\mu ) D(\lambda ) \right] \ , \label{commCB}
\end{align}
where, for convenience, we have introduce the functions
\begin{align}
	f(\lambda , \mu) &=\frac{\lambda - \mu +i}{\lambda - \mu} \ , & g(\lambda , \mu )&=\frac{i}{\lambda - \mu} \ , \\
	f(\mu , \lambda) &=\frac{\lambda - \mu -i}{\lambda - \mu} \ , & g(\mu , \lambda )&=\frac{-i}{\lambda - \mu} \ .
\end{align}
The complete list of commutation relations can be found, for example, in section 2.2 of \cite{EscobedoTailoring} for the XXX spin chain or in chapter VII of \cite{0521373204} for a general 6-vertex R-matrix. The first relation tells us that the ordering of the operators defining the state is irrelevant. The second and the third one tell us how the transfer matrix commutes with the B operator, giving us two kind of terms: a ``wanted'' term where the operators conserve their arguments, and an ``unwanted'' term where they exchange them. For our states to be eigenstates of the transfer matrix we have to impose the cancellation of these unwanted terms. The fourth and the fifth one are the same but with the C operator instead of the B operator. We will comment about the sixth one in the next section as it will be used for computing scalar products.

Applying now the transfer matrix $A(\mu ) + D(\mu )$ to a general state
\begin{equation}
\ket{\lambda_1 \dots \lambda_M}=\prod_{i=1}^M B(\lambda_i ) \ket{0} \ ,
\end{equation}
and taking into account these commutation relations, the wanted term and the unwanted term involving the transfer matrix with argument $\lambda_1$ (that is, the term obtained by commutating in an unwanted way $A+D$ with the first $B$ operator but in a wanted way with the rest of the operators) read
\begin{align}
	&\mathcal{T} (\mu ) \ket{\lambda_1 \dots \lambda_M}= \notag \\
	&=\left[ \left( \mu +\frac{i}{2} \right)^L \prod_{i=1}^M \frac{\lambda_i -\mu +i}{\lambda_i -\mu}  + \left( \mu -\frac{i}{2} \right)^L \prod_{i=1}^M \frac{\lambda_i -\mu -i}{\lambda_i -\mu}  \right]\ket{\lambda_1 \dots \lambda_M} +\notag \\
	&+ \left[ \left( \lambda_1 +\frac{i}{2} \right)^L \prod_{i=1}^M \frac{\lambda_i -\lambda_1 +i}{\lambda_i -\lambda_1}  +\left( \lambda_1 -\frac{i}{2} \right)^L \prod_{i=1}^M \frac{\lambda_i -\lambda_1 -i}{\lambda_i -\lambda_1}  \right] \ket{ \mu \lambda_2 \dots \lambda_M}+\dots \label{eigenvaluemonodromy}
\end{align}
It is immediate to see that no other term involving the state $\ket{ \mu \lambda_2 \dots \lambda_M}$ can appear. To compute the rest of the unwanted terms we can use the reordering symmetry of the $B$ operators that allows us to put any factor we want the first. Hence all factors in front of the unwanted states have the same structure\footnote{This structure is not easy to see if we compute the coefficients by brute force. Terms involving the state $\ket{\lambda_1 \mu \lambda_3 \dots \lambda_M}$ can appear from wanted-unwanted-wanted-...-wanted or from unwanted-unwanted-wanted-...-wanted sequences and we have to sum both contributions to get the same answer. So for the state with $\lambda_i$ substituted by $\mu$ we have to sum $2^{i-1}$ terms, making it more difficult to get the general formula via this procedure.}. We can rewrite the vanishing of the unwanted terms as
\begin{equation}
	\left( \frac{\lambda_j +\frac{i}{2}}{\lambda_j -\frac{i}{2}} \right)^L=\prod_{\substack{i=1\\ i\neq j}}^M \frac{\lambda_i -\lambda_j +i}{\lambda_i -\lambda_j -i} \ . \label{ABAE}
\end{equation}
Note that the equations we are getting here are exactly the condition for the cancellation of the poles of the eigenvalues of the monodromy matrix, which is the first line of (\ref{eigenvaluemonodromy}).

The question that arises now is the meaning of the arguments $\lambda_i$, which from now on we are going to call \emph{rapidities}. To see their meaning we are going to compute the eigenvalue of the shift operator $U=e^{-iP}=i^{-L} \mathcal{T} \left( \frac{i}{2} \right)$,
\begin{equation}
	U \ket{\lambda_1 \dots \lambda_M}= \prod_{i=1}^M \frac{\lambda_i +\frac{i}{2}}{\lambda_i -\frac{i}{2}} \ket{\lambda_1 \dots \lambda_M} \ .
\end{equation}
As this has to hold for any number of operators,  rapidity and momentum can be directly related,
\begin{equation}
	p_j= \frac{1}{i} \ln \left( \frac{\lambda_j -i/2}{\lambda_j +i/2} \right) \ , \label{momentumABA}
\end{equation}
which gives $p\in [0,2\pi ]$ for real rapidity. If we substitute this relation in the eigenvector condition, eq.~(\ref{ABAE}), we obtain that it is the same as the Bethe Ansatz Equations (\ref{CBAE}) we have obtained in the previous section. We can also compute the spectrum of the Hamiltonian using the eigenvalue of the transfer matrix, and in the same way it is also additive
\begin{equation}
	\text{H}\ket{\lambda_1 \dots \lambda_M}= \sum_{i=1}^M \frac{-2J}{\lambda_i^2 +1/4} \ket{\lambda_1 \dots \lambda_M} \ .
\end{equation}
If we substitute now the rapidity with the momentum, we get the same answer as for the coordinate Bethe ansatz, $E (p)=4J (\cos \, p -1)$.

However we still have to prove that the $B$ operator is an operator that flips the spin of the reference state. The easiest way to do that is to find the spin operator in the monodromy matrix. In particular it doesn't appear in the expansion around $\lambda=\frac{i}{2}$ but in the expansion around $\lambda \rightarrow \infty$,
\begin{equation}
	T_a(\lambda )= \lambda^L \left( \mathbb{I} + \frac{\vec{s}\cdot \vec{\sigma}_a}{\lambda} +\dots \right) \ ,
\end{equation}
where $\vec{s}=\sum_{j=1}^L \vec{s}_j=\sum_{j=1}^L \frac{\vec{\sigma}_j}{2}$ and the following terms are related with the non-zero levels of the Yangian symmetry \cite{BERNARD_1993}\footnote{It was shown independently by Drinfeld \cite{DRINFELD_1990} and Jimbo \cite{Jimbo_1985} that the Yang-Baxter equation (\ref{Yang-Baxter}) and the RTT relations (\ref{RTT}) are related to the Hopf algebra structures and to the deformations of universal enveloping Lie algebras. The R-matrix and the Lax matrix appear to be representation of some universal object called \emph{universal R-matrix}.}. Taking RTT relations (\ref{RTT}) in the limit $\lambda \rightarrow \infty$ we get
\begin{equation}
	\left[ \frac{1}{2} \vec{\sigma}_a+\vec{s}, T_a (\mu) \right]=0 \Longrightarrow \left\{\begin{array}{l}
	s^z B(\mu )=B(\mu) (s^z -1) \\
	\null [s^+ , B(\mu )]=A(\mu ) -D(\mu )
\end{array}	 \right. \ , 
\end{equation}
which implies that the $B$ operators turns down one spin.

There is also a pictorial way to interpret the state, but we are not going to talk about it here as it is more complex than the ones we have presented before. Instead we refer to figure 7 of \cite{Foda_2013}.

In the same way we have constructed the states from products of $B$ operators, the reference state and the Bethe ansatz equations, the dual states can be constructed with the dual reference state $\bra{\uparrow \uparrow \dots \uparrow}$ and products of $C$ operators. It can be checked that the eigenvector condition gives the same BAE.

In the same way we could have started with the vacuum with all spins pointing down $\ket{0'}=\ket{\downarrow \downarrow \dots \downarrow}$ and applied products of $C(u_i)$ operators. In a spin chain of finite length there is an isomorphism between both constructions and it can be shown the equivalente $\prod_{i=1}^N B(u_i) \ket{0} \equiv \prod_{i=1}^{L-N} C(v_i) \ket{0'}$. The precise relation between both sets of rapidities can be found, for example, in \cite{BGood}.

\subsection{The inverse scattering problem and scalar products} \label{inversescattering}

We have seen in the previous section that we can compute quantities like the momenta and the spectra both in the ABA and the CBA and that we get the same answers. But there is still one last thing to do: compute the relations between the entries of the monodromy matrix and the local spin operators appearing in the CBA. This is the last step of the construction of states and correlation functions we have presented, sometimes called \emph{quantum inverse scattering method} (QISM) for historical reasons related with classical integrability
.

The relation we are looking for was found in \cite{KMT} using the method of factorizing F-matrices 
for the case of the general XXZ inhomogeneous spin chain and later in \cite{GK} for the XYZ homogeneous spin chain using the properties of the R-matrix and the monodromy matrix. The solution for the inhomogeneous spin chain is 
\begin{align}
	\sigma^+_k &=\prod_{i=1}^{k-1}{ (A+D) (\xi_i)} \, C(\xi_k) \, \prod_{i=k+1}^L{ (A+D) (\xi_i)} \ , \label{sigmaplus} \\
	\sigma^-_k &=\prod_{i=1}^{k-1}{ (A+D) (\xi_i)} \, B(\xi_k) \, \prod_{i=k+1}^L{ (A+D) (\xi_i)} \ , \label{sigmaminus} \\
	\sigma^z_k &=\prod_{i=1}^{k-1}{ (A+D) (\xi_i)} \, (A-D)(\xi_k) \, \prod_{i=k+1}^L{ (A+D) (\xi_i)} \ , \label{sigmaz}
\end{align}
where $k$ is a given site of the spin chain. Instead of the Pauli matrices we can write the solution in a compact way in terms of the elementary operators $E^{ij}_k$, which act on site $k$ as the $2\times 2$ matrices $(E^{ij})_{ab}=\delta^i_a \delta^j_b$ \cite{Maillet_2000},
\begin{equation}
	E^{ij}_k =\prod_{i=1}^{k-1}{ (A+D) (\xi_i)} \, \left[ T_a (\xi_k) \right]_{ij} \, \prod_{i=k+1}^L{ (A+D) (\xi_i)} \ . \label{inversegeneral}
\end{equation}
As we will show in the next section and mainly in the next chapter, these expressions will allow us to calculate expectation values of local operators by means of the Yang-Baxter algebra. A more detailled explanation of the quantum inverse scattering problem can be found in \cite{Maillet_2000}, with an emphasis of its generalization to fundamental and fused lattice models.

Apart form the explicit form of the operators, we need a method to compute the scalar products of states. For the case of the CBA it is easy to do that because we already have an explicit form of the wave function. Therefore we are going to focus in the computation for the ABA. There is a large amount of literature devoted to this kind of computation (see, for example, \cite{Korepin_1982} and references therein).

The scalar products we are going to be interested in are defined as
\begin{equation}
	S_N (\{ \mu_j \} , \{\lambda_k \}) = \langle 0 | \prod_{j=1}^N C(\mu _j) \prod_{k=1}^N B(\l _k) | 0 \rangle \ ,
\end{equation}
where the set of rapidities $\{ \l _k\}$ is a solution to the Bethe equations (we are going to say that it is an ``on-shell Bethe vector'') and $\{ \mu _j \}$ is an arbitrary set of parameters (``off-shell Bethe vector''). These scalar products can be constructed as a ratio of two determinants\footnote{N. A. Slavnov proved this formula by induction in \cite{Slavnov_1989}. There are other ways to prove it, like the use of the \emph{F-basis} \cite{KMT} or by direct application of eq.~(\ref{commCB})
.}, 
\begin{equation}
S_N (\{ \mu_j \} , \{\lambda_k \})=\frac{\det T}{\det V} \ , 
\label{scalarproduct}
\end{equation}
where $T$ and $V$ are $M\times M$ matrices given by
\begin{align}
T_{ab} &=\dpartial{\tau (\mu_b,\{ \l \})}{\lambda_a} \ , & &\tau (\mu,\{ \l \}) 
=a(\mu) \prod_{k=1}^M{\frac{\l _k-\mu+i}{\l_k -\mu}}+d(\mu) \prod_{k=1}^M{\frac{\l _k-\mu-i}{\l_k -\mu}} \ , \notag\\
V_{ab} &=\frac{1}{\mu_b-\l _a} \ , & &\det V =\frac{\prod_{a<b}{(\l _a -\l _b)} \prod_{j<k}{(\mu_k-\mu_j)}}
{\prod_{a,k=1}^M{(\mu_k -\lambda_a)}} \ . 
\label{howtoscalarproduct}
\end{align}
An equivalent expression holds if we put the set of $C$ operators on-shell and the set of $B$ operators off-shell.

If we take the limit $\mu_a \rightarrow \lambda_a$ in these expressions we recover the Gaudin formula for the square of the norm 
of a Bethe state \cite{Korepin_1982,Gaudin_1981},
\begin{gather}
S_N (\{\lambda_k \}, \{\lambda_k \})= 
i^N \prod_{j \neq  k}{\frac{\l _j -\l _k +i}{\l _j -\l _k}} \det \Phi '(\{\lambda_k \}) \ , \nonumber \\
\Phi '_{ab} (\{\lambda_k \}) =-\dpartial{}{\l _b} \ln \left( \frac{a (\l _a)}{d(\l _a)} \prod_{c\neq a}{\frac{\l _a-\l _c+i}{\l _a -\l _c -i}} \right) \ .
\label{gaudin}
\end{gather}
This way of calculating scalar products is valid for the case of a finite spin chain. 
The generalization of these expressions to the thermodynamical limit of very long chains can be found, for example, in reference \cite{Kitanine_2009}.

We must note that there is a more general expression for computing the scalar product of two off-shell states \cite{Korepin_1982}. However, we are not going to explicitly write it here nor use it as it cannot be written as a determinant, only as a sum over partitions, making it more difficult to handle.

To end this section we are going to propose a way of simplifying a little bit the computation of the scalar products. To do that we need the following property of the R and L matrices: the Yang-Baxter and RLL equations define these matrices up to an scalar factor, therefore we can change our normalization choice without changing the underlying physics. One good choice is to make, which we are going to use for the rest of this thesis
\begin{equation}
	L_{n,a} (\lambda -\xi_n) =\frac{\lambda -\xi_n}{\lambda +\frac{i}{2}} \mathbb{I}_{n,a} + \frac{i}{\lambda +\frac{i}{2}} \mathcal{P}_{n,a} \label{newnormalization}\ .
\end{equation}
This choice makes the eigenvalues of the $A$ and $D$ operators on the reference state $a(\lambda )=1$ and $d(\lambda )=\left( \frac{\lambda -\frac{i}{2}}{\lambda +\frac{i}{2}} \right)^L$. Hence we only have to care about one of the factors and the periodicity condition simplifies to $\prod_{\mu\in \{\lambda \}} d(\mu )=1$.

\section{Normalization issues}
	\label{normalizationisues}

States in the algebraic and the coordinate Bethe ansatz are normalized differently. 
As a consequence, any correlation function evaluated using the ABA will differ from the corresponding CBA computation by some global factor. We will see that this global factor is going to arise from the states not being properly normalized and an extra global phase.

The simplest correlation function that exhibits this issue is  $\granesperado{\lambda}{\sigma^+_k \sigma^-_l}{\lambda}$. 
In the CBA this correlation function is given by $e^{ip(l-k)}$. In order to approach the calculation of this correlator in the ABA 
we just need to write the spin operators in terms of elements of the monodromy matrix, 
\begin{align}
\granesperado{\lambda}{\sigma^+_k \sigma^-_l}{\lambda} &= \granesperado{0}{C(\lambda) \, 
(A+D)^{k-1} \, B(\xi) \, (A+D)^{L-k+l-1} \, C(\xi) \, (A+D)^{L-l} \, B(\lambda)}{0} \nonumber \\
& =e^{-ip(L-l+k-1)} \granesperado{0}{C(\lambda) \,  B(\xi) \, (A+D)^{L-k+l-1} \, C(\xi) \, B(\lambda)}{0} \ .
\end{align}
From the commutation relation~(\ref{commCB}) we find that
\begin{equation}
\bra{0} C(\lambda) \, B(\xi)=i\frac{d(\lambda)}{\lambda-\xi} \bra{0} \ , 
\end{equation}
with an identical result for $C(\xi) \, B(\lambda) \ket{0}$. Recalling that the Bethe ansatz equation for a single-magnon state, which the normalization presented in eq.~(\ref{newnormalization}), reads $d(\lambda)=1$ we conclude that 
\begin{equation}
\granesperado{\lambda}{\sigma^+_k \sigma^-_l}{\lambda} = \frac{i^2 e^{ip(l-k+1)}}{(\l-\xi)^2} \ .
\end{equation}
We can try to solve the disagreement with the CBA dividing this result by the norm of the state. 
This can be easily computed using the Gaudin formula~(\ref{gaudin}), 
\begin{equation}
\langle \lambda | \lambda \rangle =i\dpartial{d}{\lambda}=\frac{i^2 L}{\lambda^2-\xi^2} \ . 
\end{equation}
Therefore
\begin{align}
	& \frac {\granesperado{\lambda}{\sigma^+_k \sigma^-_l}{\lambda}}{\langle \lambda | \lambda \rangle}
	= \frac {e^{ip(l-k)}}{L} \left( \frac{\lambda+\xi}{\lambda-\xi} e^{ip} \right) = \frac {e^{ip(l-k)}}{L} \ , 
\end{align}
which is the result in the CBA provided we divide by the norm of the state in there. 
Thus we conclude that indeed both ansätze were not properly normalized.

However we can see that the procedure of dividing the correlation function by the norm of the states is not enough to cure the disagreement for general cases. We can easily exhibit that dividing by the norm is not enough if for instance we calculate the form factor $\granesperado{0}{\sigma^+_k}{\lambda}$ and 
divide by $\sqrt{\langle \lambda | \lambda \rangle}$,
\begin{equation}
\frac{\granesperado{0}{\sigma^+_k}{\lambda}}{\sqrt{\langle \lambda | \lambda \rangle}}=
\frac{e^{ipk}}{\sqrt{L}} \ \sqrt{\frac{\lambda+\xi}{\lambda-\xi}}=\frac{e^{ip(k-\frac{1}{2})}}{\sqrt{L}} \ .
\end{equation}
The reason for the additional $1/2$ factor is that besides the different normalization there is also an additional phase which 
depends on the rapidity (See reference \cite{EscobedoTailoring} for a discussion on this point).

In order to fix the normalization of states in the ABA with respect to the normalization of states in the CBA 
we will go back to the definition of the transfer matrix, equation (\ref{abamonodromymatrix}),
and apply it to the ground state,
\begin{align}
	& L_{n,a} \ket{\uparrow}_n =\left( \begin{matrix}
	1 & \frac{i}{\lambda-\xi+i} s^-_n \\
	0 & \frac{\lambda-\xi+i}{\lambda-\xi} \end{matrix} \right) \ket{\uparrow}_{n+1} \ .
\end{align}
If we focus on the operator $B(\lambda)$, we can write\footnote{The operator $B(\lambda)$, when not applied over the vacuum, has extra terms which are schematically of the form $(s^-)^{n+1} (s^+)^n$ with $1\leq n \leq L-1$, a combination of rising and lowering operators applied all at different sites in the lattice with always one more lowering operator than raising operators.}
\begin{align}
B(\lambda) & =\frac{i}{\lambda+\xi} \left[ s^-_1+ s^-_2 \left( \frac{\lambda-\xi}{\lambda+\xi} \right) 
+ s^-_3 \left( \frac{\lambda-\xi}{\lambda+\xi} \right)^2 + \dots \right] \ket{0} \nonumber \\
& = \frac{i}{\lambda+\xi} \sum_{n=1}^L{ s^-_n \left( \frac{\lambda-\xi}{\lambda+\xi} \right)^{n-1} }
\ket{0} = \frac{i}{\lambda-\xi} \sum_{n=1}^L{ s^-_n e^{ipn}} \ket{0} \ .
\end{align}
Therefore states with a single magnon in the ABA, $\ket{\lambda}^{\text{a}}$, relate to states in the CBA through
\begin{equation}
B(\lambda)\ket{0}=\ket{\lambda}^{\text{a}}=\frac{i}{\lambda-\xi} \ket{\lambda}^{\text{c}} \ . 
\end{equation}
When we repeat this with the state $^{\text{a}} \bra{\lambda}$ we conclude that 
\begin{equation}
^{\text{a}} \bra{\lambda}=i\frac{d(\lambda)}{\lambda+\xi} \  ^{\text{c}} \! \bra{\lambda} \ , 
\end{equation}
because for bra states
\begin{equation}
_n \! \bra{\uparrow} L_{n,a} =  \, _{n+1} \! \bra{\uparrow} \left( \begin{matrix}
\frac{\lambda-\xi+i}{\lambda-\xi} & 0 \\
\frac{i}{\lambda-\xi+i} s^+_n & 1 \end{matrix} \right)	\ , 
\end{equation}
and therefore\footnote{Extra terms of the $C(\lambda)$ operator will have the same structure than the one for the $B(\lambda)$ operator but with one more rising operator than lowering operators, $(s^-)^{n} (s^+)^{n+1}$.}
\begin{align}
\bra{0}C(\lambda) & = \bra{0} \frac{i}{\lambda+\xi} \left[ s^+_1 \left( \frac{\lambda-\xi}{\lambda+\xi} \right)^{L-1} + 
s^+_2 \left( \frac{\lambda-\xi}{\lambda+\xi} \right)^{L-2} +s^+_3 \left( \frac{\lambda-\xi}{\lambda+\xi} \right)^{L-3} + \cdots \right] \nonumber \\
& = \bra{0} \frac{i\, d(\lambda)}{\lambda+\xi} \sum_{n=1}^L{ s^+_n \left( \frac{\lambda-\xi}{\lambda+\xi} \right)^{-n} }
= \bra{0} \frac{i\, d(\lambda)}{\lambda+\xi} \sum_{n=1}^L{ s^+_n e^{-ipn}} \ .
\end{align}
An identical discussion holds in the case of states with more than one magnon, so in general we conclude that
\begin{gather}
	\ket{\l _1 , \l _2 , \dots , \l _N}^{\text{a}}= \prod_{j=1}^N{ \frac{i}{(\l _j-\xi )}}
	\prod_{i<j}{\frac{\l _j-\l _i+i}{\l _j-\l _i}} \, \ket{\l _1 , \l _2 , \dots , \l _N}^{\text{c}} \ , \\	
	\bra{\l _1 , \l _2 , \dots , \l _N}^{\text{a}}= \prod_{j=1}^N{ i \frac{d(\l_j )}{(\l _j+\xi )}}
	\prod_{i<j}{\frac{\l _j-\l _i-i}{\l _j-\l _i}} \, \bra{\l _1 , \l _2 , \dots , \l _N}^{\text{c}}\ .
\end{gather}
The first factor can be removed by an appropriate normalization of the states, and thus there will only remain a shift 
in the position of the coordinates by $-\frac{1}{2}$. The second factor is related to the fact that CBA states 
are not symmetric if we interchange two magnons. In fact they pick up a phase which is equal to the $S$-matrix. 
On the other hand ABA states are symmetric under exchange of two magnons. Therefore if we want to obtain 
the same result from the CBA and the ABA we will have to normalize carefully the states. This can 
be done if we choose the phase in such a way that the correlation functions have the structure $\sqrt{\prod_{\mu_i<\mu_j}{S_{ij}}}\cdot \{\text{term symmetric in the rapidities}\}$, for reasons we will explain later\footnote{Actually there is more freedom in this choice. If we write the S-matrix as $S(u,v)=\frac{h(u,v)}{h(v,u)}$, we can use instead the product $\prod_{i<j}{h(\mu_i , \mu_j)}$. Although the easiest choice is $h(u,v)=f(u,v)$, this function can be redefined by a multiplication by a function symmetric in $u$ and $v$. In particular $\hat{h}(u,v)=\frac{h(u,v)}{\sqrt{h(u,v) h(v,u)}}=\sqrt{S(u,v)}$ is another possible choice. We want to thank I. Kostov for pointing us this fact.}. Despite being a very ad hoc solution, we are going to keep this idea in mind.

An alternative argument can be obtained if instead of using {\em B-states} to define the excitations 
we use {\em Z-states}, where\footnote{Here we are going to follow the definition given in \cite{FADDEEValgebraic}. However a different definition was used in \cite{Kulish1984}.}
\begin{equation}
	Z(\lambda)=B(\lambda) A^{-1} (\lambda) \ .
\end{equation}
In fact it is natural to use these states because they generate a Zamolodchikov-Faddeev algebra \cite{Howthealgebraicbetheansatz}, that is, the operators commute up to an $S$-matrix,
\begin{equation}
Z(\l) Z(\m)=Z(\m) Z(\l) S_{\m \l}=Z(\m) Z(\l) \frac{\m -\l -i}{\m -\l +i} \ . \label{ZFalgebraSU2}
\end{equation}
Later we will define the Zamolodchikov-Faddeev algebra in more detail, but for the moment we are going to use this definition as the $SU(2)$ sector does not have any index structure. Note that in this way states in the ABA will have the same behavior under the exchange of two magnon states in the CBA.

In order to be able to work with $Z$-states we will have first to calculate the commutation 
relation between the operator $A^{-1}$ and the $B$ operator. To find this commutator 
we will start by taking the commutation relations between $A$ and $B$,
\begin{align*}
A(\l) B(\m) &= \left( 1- \frac{i}{\l-\m} \right) B(\m) A (\l) + \frac{i}{\l-\m} B(\l) A (\m) \ ,\\
B(\m) A(\l) &= \left( 1+ \frac{i}{\l-\m} \right) A(\l) B (\m) - \frac{i}{\l-\m} A(\m) B (\l) \ .
\end{align*}
Now if we left and right-multiply both expressions by $A^{-1} (\l)$, and commute a factor $A(\m) B(\l)$ arising 
in the second equation, we obtain
\begin{align*}
B(\m) A^{-1} (\l) &= \frac{\l-\m-i}{\l-\m} A^ {-1} (\l) B(\m) 
+ \frac{i}{\l-\m} A^ {-1} (\l) B(\l) A (\m) A^ {-1} (\l) \ , \\
A^{-1} (\l) B(\m) &=\frac{\l-\m}{\l-\m -i} B(\m) A^{-1} (\l) - \frac{i}{\l-\m-i} 
A^{-1} (\l) B(\l) A (\m) A^{-1} (\l) \ .
\end{align*}
We also need the action of $A^{-1}$ over the vacuum state, which can be easily proven to be trivial.
We thus conclude that there is a relationship between the $Z$-states and the $B$-states,
\begin{equation}
A^{-1} (\lambda) \prod_i{B(\mu_i)}\ket{0} = \prod_i{\frac{\lambda-\m}{\lambda-\m-i}} 
\prod_i{B(\mu_i)} \ket{0} \ ,
\end{equation}
where we have used that if we have two magnons with the same rapidity the state must vanish. Therefore
\begin{equation}
\mathcal{R} \left[ \prod_i{Z(\mu_i) \ket{0}} \right] = \prod_{i<j}{\frac{\mu_j-\mu_i}{\mu_j-\mu_i+i}} 
\prod_i{B(\mu_i) \ket{0}} \ ,
\end{equation}
where $\mathcal{R}$ denotes just an ordering operator in the rapidities. Hence using the Zamolodchikov-Faddeev states 
instead of the usual magnon states introduces a phase shift. In fact this phase is the factor we wanted to introduce ad hoc.

However there could still be a problem if the norm of our states behaves in the same way. 
We can exclude this possibility if we introduce the operators $F(\lambda)=d(\lambda) \, D^{-1} (\lambda) \, C(\lambda)$. 
To prove that this is the adequate operator we need in order to define the correct left-state, we first have to calculate the commutation 
relations of $D$ with $C$. Using the same procedure as before we find that
\begin{align}
	D^{-1} (\l) C(\m) &= \frac{\m-\l-i}{\m-\l} C(\m) D^{-1} (\l) - \frac{i}{\l-\m} 
	D^{-1} (\l) D (\m) C(\l) D^{-1} (\l) \ , \nonumber \\
	C(\m) D^{-1} (\l) &=\frac{\m -\l}{\m -\l -i} D^{-1} (\l) C(\m) - \frac{i}{\m -\l -i} 
	D^{-1} (\l) D (\m) C(\l) D^{-1} (\l) \ . \nonumber 
\end{align}
With these equations at hand we can easily prove that $F$ generates a Zamolodchikov-Faddeev algebra, 
$F(\l) F(\m) = F(\m) F(\l) S_{\m \l}$, and also that 
\begin{equation}
\bra{0} F(\m) F(\l)=\frac{\m -\l}{\m -\l -i} 
\bra{0} C(\l) C(\m) \ , 
\end{equation}
so that
\begin{equation}
	\granesperado{0}{F(\m) F(\l) Z(\l) Z(\m)}{0}=(\m -\l)^2 \, 
	\frac{\granesperado{0}{C(\m) C(\l) B(\l) B(\m)}{0}}{(\m -\l -i)(\m -\l +i)} \ ,
\end{equation}
which is indeed symmetric under exchange of $\l$ and $\m$ as we wanted\footnote{Although it is very similar to the Zamolodchikov-Faddeev, in reality it is more similar to the algebra proposed by \cite{Pakuliak_complete} based on the Yangian double of $SU(2)$. The difference between both algebras is the commutation relation of the annihilation operators with the creation operators.}.

\section{BDS spin chain}
\label{BDSspinchain}

In the introduction \ref{MinahanZaremboargument} we presented the equivalence between the dilatation operator at one-loop and the Heisenberg spin Hamiltonian. We can make the exercise of computing higher-loop order planar diagrams and try to map the new dilatation operator to another spin chain. The two-loop calculation of the dilatation operator for the $SU(2)$ sector was performed in \cite{Gross_2003}  by computing only the flavour-permutation diagrams and determining the term proportional to the identity by imposing the vanishing of the contribution for the ground state. This contribution can be written in terms of the permutation operator we defined in subsection~\ref{solvingABA}
\begin{equation}
	D_4 \propto -4\mb{I}+6\mc{P} - \sum_{k=1}^L {\left( \mc{P}_{k,k+1} \mc{P}_{k+1,k+2} + \mc{P}_{k+1,k+2} \mc{P}_{k,k+1} \right)} \ ,
\end{equation}
so already at two-loops we have interactions beyond nearest-neighbours. One of the first things that can be shown is that at a given loop order $K$ the relevant Feynman diagrams can only generate permutation structures involving $n\leq K$ nearest-neighbour, as each nearest-neighbour permutation is associated with a least one-loop in the underlying Feynman graph.

As we commented before, we were able to perform the analysis of the $SU(2)$ sector at one-loop because it is what is called \emph{closed sector}, that is, there is no mixing with other sectors. The two-loop computation proved that it is also closed at this level. In fact the $SU(2)$ sector is closed at all loops\footnote{This is a consequence of the commutation of the dilatation operator with the Lorentz and R-symmetry generators, so mixing can only appear between operators with the same R-charges, Lorentz charges and bare dimensions. Examples of other closed sectors are the $SU(2|3)$ sector formed by three types of scalars and two types of fermions, and the $SU(1,1)=SL(2)$ sector formed by one type of scalar field and covariant derivatives with one type of polarization. The $SO(6)$ sector we presented in the introduction is only closed at one-loop.}.

The first attempt to construct a long-range spin chain to model the all-loop dilatation operator was \cite{Serban_2004}, where it was found that, up to three-loop order, the dilatation operator in the $SU(2)$ sector may be constructed from the conserved charges of the Inozemtsev model \cite{Inozemtsev_1990,Inozemtsev_2002}. This proposal lead to an extension to all-loop order made by Beisert, Dippel and Staudacher \cite{BDS}, who proposed to extend the three-loop $SU(2)$ Bethe ansatz in a way that matches the prediction of the string sigma model and with the BMN dispersion relation (\ref{BMNdispersion}). This would mean modifying the $SU(2)$ Bethe Ansatz equations in the following way
\begin{equation}
	e^{ip_k L}=\left( \frac{x^+ (\lambda_k)}{x^- (\lambda_k)} \right)^L=\prod_{\substack{l=1 \\ l\neq k}}^M{\frac{\lambda_k -\lambda_l +i}{\lambda_k -\lambda_l -i}} \ , \label{BDSequation}
\end{equation}
where
\begin{align}
	\lambda & =\frac{1}{2} \cot \frac{p}{2} \sqrt{1+16g^2\sin^ 2 \frac{p}{2}} \ , & \lambda &=x+\frac{g^2}{x} \ , & x^{\pm}(\lambda) &= x(\lambda\pm i/2) \ .
\end{align}
All the conserved quantities associated with this spin chain can be written in the form
\begin{equation}
	q_r=g^2 \sum_{i=1}^M \left( \frac{i}{(x_i^+)^{r-1}}-\frac{i}{(x_i^-)^{r-1}} \right) \ ,
\end{equation}
being the only exception the momentum, defined above. In particular, the energy is given by
\begin{equation}
	E=g^2 \sum_{i=1}^M \left( \frac{i}{(x_i^+)}-\frac{i}{(x_i^-)} \right)=\sum_{i=1}^M \left( \sqrt{1+16g^2 \sin^ 2 \frac{p_k}{2}} -1 \right) \ .
\end{equation}
Those equations were obtained by redefining the coupling constant and the charges of the Inozemtsev spin chain, therefore this has to be a long-range spin chain by definition (that is, the Hamiltonian should involve interactions beyond first nearest neighbours). Interestingly this homogeneous long-range spin chain can be mapped to a short-range inhomogeneous spin chain up to wrapping corrections. The basic idea is that the left hand side of equation~(\ref{BDSequation}) can be expanded as a polynomial and the first $L$-th terms can be absorbed into the degrees of freedom of the inhomogeneities, therefore being equivalent up to wrapping order. Hence we can rewrite the Bethe equations in the following way
\begin{equation}
	\prod_{n=1}^L{\frac{\lambda+i-\xi_n}{\lambda-\xi_n}}=\prod_{\substack{l=1\\ l\neq k}}^M{\frac{\lambda_k -\lambda_l +i}{\lambda_k -\lambda_l -i}} \ ,
\end{equation}
where the inhomogeneities have to be chosen in the following way
\begin{equation}
	\xi_n =\frac{i}{2}+\sqrt{2} g \cos \left( \frac{(2n-1)\pi}{2L} \right) \ .
\end{equation}
Because an inhomogeneous spin chain is easier to handle than a long-range one, we are going to use this second realization in our following computations.

Shortly after this, Beisert and Staudacher \cite{Beisert_2005} conjectured the $SU(1|1)$ symmetric $S$-matrix for the $SU(2|1)$ spin chain\footnote{This is because the excitations of a $SU(N)$ spin chain over the vacuum have a residual $SU(N-1)$ symmetry from fixing the vacuum state.} sector at all loops, and, from the consistency relations which the nested Bethe ansatz has to satisfy, they conjectured the full $PSU(2,2|4)$ Bethe ansatz at all loops. However there is still an element missing because BMN scaling breaks down at the four-loop order. This is because the $S$-matrix is fixed by symmetry and Yang-Baxter up to an scalar factor. The element missing is the so called \emph{dressing phase} \cite{Beisert_2006,BESphase,Reviewdressing}. However we are not going to include it in our computations.

\section{The Bootstrap program. Form factors}
\label{bootstrap}

The bootstrap program is a non-perturbative method to construct a quantum field theory in $1+1$ dimensions, not from a Lagrangian, but from the symmetries of the theory and a set of properties we assume the theory fulfils. It can be summed up in three steps.
\begin{enumerate}
	\item Computing the \emph{$S$-matrix}: for that we have to assume unitarity, crossing, an underlying symmetry (model dependent), fulfilling of the Yang-Baxter equation (if the theory is integrable) and \emph{maximal analiticity} (sometimes called \emph{Landau property}). The last one means that the two-particle $S$-matrix is an meromorphic function in the physical plane (usually called $s$-plane, where $s=(p_1+p_2)^2$ is the only independent Mandelstam variable in $1+1$ dim., which we are going to analytically continue to the complex plane) that only has poles and cuts of physical origin. In particular there has to be two cuts $(-\infty , 0)$ and $(4m^2 , \infty )$ with $m$ the mass of the lightest particle in the theory, so the $S$-matrix has to be analytic in the segment $[0,4m^2]$ except for single poles associated to bound states.
	
	\item Computing the \emph{generalized form factors}: form factors are matrix elements of local operator evaluated between two asymptotic states, one incoming and one outgoing. We will talk more about them below.
	
	\item Computing the \emph{Wightman functions}: they can be computed from form factors by inserting a complete set of intermediate states. In particular, the two-point function for a Hermitian operator $\O (x)$ is given by
	\begin{equation}
		\granesperado{0}{\O (x) \O (0)}{0}=\sum_{n=0}^\infty{\frac{1}{n!} \int{\frac{d\theta_1}{4\pi} \dots \frac{d\theta_n}{4\pi} |\granesperado{0}{\O (0)}{\theta_1 , \dots , \theta_n}|^2 e^{-ix\sum{p_i}}}} \ .
	\end{equation}
\end{enumerate}
And, by the Wightman Reconstruction Theorem, there have to exists a (Wightman) QFT for which these functions are Wightman distributions \cite{9781400884230}. This program has already been accomplished for some models like the Sine-Gordon model \cite{Babujian_1999,Babujian_2002}, the Thirring model \cite{Nakayashiki_2002,Takeyama_2003} and the SU(N) Lieb-Lininger model \cite{Babujian_2006}. Concerning the $\mathcal{N}=4$ SYM, the first step is already completed. Nowadays we are in the second step of the program, as we are trying to understand the structure of the form factors of the theory \cite{KM1,KM2,Bork_2014,HHLJanik}.

For local operators, the generalized form factors are defined as
\begin{equation}
	F^{\O}_{\alpha_1 \dots \alpha_n} (\theta_1 , \dots , \theta_n)=\granesperado{0}{\O (0)}{\theta_1, \dots , \theta_n}^{\text{in}}_{\alpha_1 \dots \alpha_n} \ ,
\end{equation}
where $\alpha_i$ are possible quantum numbers, and $\theta_i$ are the rapidity variables, defined as $p_0=m \sinh \theta$ and $p_1=m \cosh \theta$. Therefore if we want a well defined ``in'' state we will have to order the rapidities as $\theta_1 >\theta_2 >\dots >\theta_n$. In the same way we can define form factors using out states,
\begin{equation}
	\tilde{F}_{\O}^{\alpha_1 \dots \alpha_n} (\theta_1 , \dots , \theta_n)=\null^{\text{out}}_{\alpha_1 \dots \alpha_n}\granesperado{\theta_1, \dots , \theta_n}{\O (0)}{0} \ ,
\end{equation}
where the rapidities have to be ordered in the opposite way, that is, $\theta_1 <\theta_2 <\dots <\theta_n$. These two choices are consistent as the fastest particle starts at the left, interacts with every other particle, and ends at the right. We will define our form factors using the in-states because we will see we can write $\tilde{F}$ as a function of $F$. Maximal analiticity (taking into account that the segment $[0,4m^2]$  in $s$ has the segment $[0,2\pi i]$ as image in $\theta$) and LSZ assumptions impose the following constraints to the form factors:
\begin{enumerate}
	\item Watson's equations \cite{WatsonEquations}: they impose the transformation under the permutation of two variables (giving us how to analytically continue to other orderings of the rapidities)
	\begin{equation}
		F^{\O}_{\dots \alpha_i \alpha_{i+1} \dots} (\dots , \theta_i , \theta_{i+1} , \dots)= \sum_{\beta_i , \beta_{i+1}} S_{\alpha_i , \alpha_{i+1}}^{\beta_i \beta_{i+1}} (\theta_{i}, \theta_{i+1} ) F^{\O}_{\dots \beta_{i+1} \beta_i \dots} (\dots , \theta_{i+1}, \theta_i  , \dots)  \ .\label{watsonequation}
	\end{equation}
	
	\item Crossing relation; also known as \emph{deformed cyclicity}. Crossing defines how to transform an outgoing particle into an ingoing particle
	\begin{equation}
		F^{\O}_{\alpha_1 \dots \alpha_n} (\theta_1 , \theta_2  , \dots , \theta_n +2\pi i)=F^{\O}_{\alpha_n \alpha_1 \dots \alpha_{n-1}} (\theta_n , \theta_1 , \dots , \theta_{n-1} ) \ .
	\end{equation}
	This equation is a consequence of the mapping between the Mandelstam variable $s$ and the rapidity $\theta$, as the second one covers two times the first one, so $\theta$ maps to $s+0i$ while $\theta+2\pi i$ maps to $s-i0$.
	
	It is interesting to use both axioms at the same time to write a tensor-valued Riemann-Hilbert problem,
	\begin{align}
		&F^{\O}_{\alpha_1 \dots \alpha_n} (\theta_1 , \theta_2  , \dots , \theta_n +2\pi i) = F^{\O}_{\beta_1 \dots \beta_n} (\theta_1 , \theta_2  , \dots , \theta_n) \notag \\
		&\times S^{\beta_n \beta_{n-1}}_{\alpha_{n-1} \tau_1} (\theta_{n-1}, \theta_n ) S^{\tau_1 \beta_{n-2}}_{\alpha_{n-2} \tau_2} (\theta_{n-2}, \theta_n ) \dots S^{\tau_{n-2} \beta_{1}}_{\alpha_{1} \alpha_n} (\theta_{1}, \theta_n ) \ .
	\end{align}
	
	\item Decoupling condition or particle-antiparticle poles. This kind of poles appear when we have a particle-antiparticle pair with opposite momenta, so $\alpha_2=\bar{\alpha}_1$ and $\theta_1=\theta_2 +i\pi$. These poles have residue
	\begin{multline}
		\Res_{\theta_1-\theta_2=i\pi} F^{\O}_{\alpha_1 \bar{\alpha}_ 1\dots \alpha_n} (\theta_1 , \theta_2 , \dots , \theta_n)= \\ =2i\hat{C}_{1} (\delta_{\alpha_3}^{\beta_3} \dots \delta_{\alpha_n}^{\beta_n} -\sigma_{\bar{1}}^{\O} S_{\alpha_n \alpha_1}^{\beta_n \gamma_1} S_{\alpha_{n-1} \gamma_1}^{\beta_{n-1} \gamma_2} \dots S_{\alpha_3 \gamma_{n-2}}^{\beta_3 \bar{\alpha}_1} ) F^{\O}_{\beta_3 \dots \beta_n} (\theta_3 , \dots , \theta_n) \ , \label{Decouplingcondition}
	\end{multline}
where $\hat{C}$ is the charge conjugation transformation and $\sigma_{\bar{1}}^{\O}$ takes into account the statistics of the operator with respect to excitation $\bar{1}$.
	
	\item Bound state poles. For models with bound states there are additional poles on the $S$-matrix with residue $\Res_{\theta_1 -\theta_2=i\theta^3_{12}} S^{\alpha_1 ' \alpha_2 '}_{\alpha_1 \alpha_2}=\Gamma_{\alpha_1 \alpha_2}^{\alpha_3} \Gamma^{\alpha_1 ' \alpha_2 '}_{\alpha_3}$. It is sufficient to indicate these poles in the strip $0<$ Im $\theta<\pi$ because the first two axioms can be used to obtain the rest of them. Then the form factors should have simple poles at this bound states with residue
	\begin{equation}
		\Res_{\theta_i -\theta_{i+1}=i\theta^{\text{b}}_{i,i+1}} F^{\O}_{\dots \alpha_i \alpha_{i+1} \dots } (\dots ,\theta_i , \theta_{i+1}  , \dots )=\Gamma^{\alpha_{\text{b}}}_{\alpha_i \alpha_{i+1}} F^{\O}_{\dots \alpha_{\text{b}} \dots } (\dots ,\theta^{\text{b}}_{i,i+1} , \dots ) \ .
	\end{equation}

	Note that form factors containing bound states are secondary objects as they can be obtained as residues of form factors of the main particles of our theory. Therefore we do not need to compute them.

\end{enumerate}

Finally, not as an axiom but as a consequence of its analytical properties, form factors of general kind (that is, where the operator is evaluated between non-trivial in and out states) can be written as an usual form factor as
\begin{multline*}
	F^{\alpha '_m , \dots , \alpha '_1}_{\alpha_1 , \dots , \alpha_n} (\O | \theta_m',  \dots , \theta_1' | \theta_1 , \dots , \theta_n) \\
	\shoveright{=\prod_{j=1}^m C^{\alpha_j ' \alpha_j ''} F^{\O}_{\alpha ''_m , \dots , \alpha ''_1, \alpha_1 , \dots , \alpha_n} (\theta_m' -i\pi ,  \dots , \theta_1'-i\pi , \theta_1 , \dots , \theta_n)} \ \hphantom{.} \\
	=\prod_{j=1}^m C^{\alpha_j ' \alpha_j ''} F^{\O}_{\alpha_1 , \dots , \alpha_n , \alpha ''_m , \dots , \alpha ''_1} (\theta_1 , \dots , \theta_n , \theta_m' +i\pi ,  \dots , \theta_1'+i\pi)  \ .
\end{multline*}
This relation provides another way of writing the crossing condition where the crossing is more explicit,
	\begin{multline}
		\null^{\text{out}}_{\bar{\alpha_1}}\granesperado{p_1}{\O (0)}{p_2 , \dots , p_n}^{\text{in}}_{\alpha_2 \dots \alpha_n}= \\=\hat{C}_{1} \sigma_1^{\O} F^{\O}_{\alpha_1 \dots \alpha_n} (\theta_1 +i\pi, \theta_2  , \dots , \theta_n)=F^{\O}_{\alpha_2 \dots \alpha_n \alpha_1} (\theta_2 , \dots , \theta_n , \theta_1-i\pi) \hat{C}_{1} \ . \label{formfactorcrossing}
	\end{multline}

A convenient way to construct states that obey these four axioms is to use creation and annihilation operators that satisfy the Zamolodchikov-Faddeev algebra\footnote{We have already presented the ZF algebra for the $SU(2)$ spin chain in eq.~(\ref{ZFalgebraSU2}). However in this case we need the ZF algebra for a bigger group so we have to care about the index structure of the operators and of the $S$-matrix. A particular index structure is not widely accepted, so we find several different choices in the literature \cite{Caudrelier_2004,Kulish_1991,Doyoncourse}. Throughout this thesis we are going to follow the same definition as \cite{Smirnov}, as it will prove useful later.}, defined in the following way
\begin{align}
	A_i^\dagger (u) A_j^\dagger (v) &=S_{ij}^{kl} (u,v) A_l^\dagger (v) A_k^\dagger (u) \ , \notag \\
	A^i (u) A^j (v) &=S^{ij}_{kl} (u,v) A^l (v) A^k (u) \ , \notag \\
	A^i (u) A_j^\dagger (v) &=S^{li}_{jk} (v,u) A_l^\dagger (v) A^k (u) +\delta_j^i \delta (u-v) \ .
\end{align}
This set of operators will automatically assure the fulfilling of the first, third and fourth axioms because they are properties implemented directly by the presence of the $S$-matrix. The second axiom comes from the writing of out-states as a function of in-states, so it is also fulfilled.

We want to end this section by pointing out one of the most important properties of having form factors that satisfy this set of axioms: the \emph{local commutative theorem}. This theorem \cite{Smirnov} assures us that two local operators will commute on a space-like interval if their form factors satisfy these axioms.

\chapter{Two-points functions and ABA}

\begin{chapquote}{Carl Von Clausewitz, \textit{On War} \cite{9780691018546}, Book II, Chapter II}
	Taking this point of view, there is a possibility afforded of a satisfactory, that is, of a useful theory [...], never coming into opposition with the reality, and it will only depend on national treatment to bring it so far into harmony with action, that between theory and practice there shall no longer be that absurd difference which an unreasonable theory, in defiance of common sense, has often produced, but which, just as often, narrow-mindedness and ignorance have used as a pretext for giving way to their natural incapacity.
\end{chapquote}

In this chapter we are going to apply the inverse scattering techniques presented in the previous chapter to compute correlation functions in general and form factors in particular. This will provide an understanding of generic correlation functions which could be employed to shed some light on the spectrum of correlation functions in the AdS/CFT correspondence.

The ABA and the solution to the inverse scattering problem were first used in \cite{RV} to evaluate three-point functions of scalar operators in $\mc{N}=4$ SYM as inner products of Bethe states. This lead to the later expressions of structure constants in terms of some elegant determinant expressions and integrals \cite{EscobedoTailoring,Foda_2012,KostovClassical3point,Kostov_2012,FodaWheeler2012,%
SerbanNoteEigenvectors,FodaPartialDomainWall,KostovDomainWall,FodaScalarProduct,Bissi_2012,%
SerbanEigenvectorsII,Foda_2013,Kazama_2013,Wheeler_2013,Bettelheim_2014}. We will review some of these works in the following chapter.

We will consider here the case of correlation functions with spin operators located at non-adjacent sites in the $SU(2)$ sector of $\mc{N}=4$ SYM. In the first section we will present how computations of spin operators are related to scalar products of one on-shell Bethe state with one off-shell Bethe state. As an example we will compute the correlation function $\granesperado{\lambda}{\sigma^+_k }{\mu_1 \mu_2}$. In the second section we will move to the slightly more difficult computation of correlation functions involving two spin operators. This will require some care because some apparent singularities have to be removed for the homogeneous spin chain case, but the final answer has to be finite. Our method to obtain a finite answer consists on rewriting the problem in a recursive way. This approach will be the central part of this chapter. The following section will be dedicated to the application of this method to the case of three spin operators. In the fourth section we will repeat the same computation but in the BDS spin chain presented in section \ref{BDSspinchain} by using its inhomogeneous short-range spin representation. The results presented in this chapter are contained in \cite{formfactorsnieto}.

\section{Correlation functions involving one operator}
\label{lambdasigmamumusection}

We are going to begin the evaluation of correlation functions of spin operators by explicitly computing the case of the three-magnon form factor 
$\granesperado{\lambda}{\sigma^+_k }{\mu_1 \mu_2}$. At the end of the section we will explain the extension to form factors with $n-1$ outgoing magnons and $n$ ingoing magnons, as it is a generalization of the computation below. 
Using relation (\ref{sigmaplus}) we can bring the problem to a computation in the ABA,
\ba
\granesperado{\lambda}{\sigma^+_k }{\mu_1 \mu_2}^{\text{a}} \!\! & = & \!\! 
\granesperado{0}{C(\lambda) \, (A+D)^ {k-1} (\xi) \, C(\xi) \, (A+D)^ {L-k} (\xi) \, B(\mu_1) \, B(\mu_2)}{0} \nonumber \\
& = & \!\! e^{-i [(p_1+p_2)\cdot (L-k)+p_{\lambda}(k-1)]} \granesperado{0}{C(\lambda) \,  C(\xi) \, B(\mu_1) \, B(\mu_2)}{0} \ .
\ea
Note that although $\l$ satisfies the Bethe equations for a single-magnon state, the pair $\{ \l, \xi\}$ does not define a Bethe state. 
Therefore to find this form factor we need to calculate the scalar product of an arbitrary vector with a Bethe state. 
This can be done following the recipe we stated in section \ref{inversescattering}. The first step is to write (recall that $\xi=i/2$ for the Heisenberg chain)
\begin{align}
	\tau (\xi,\{\mu_1 , \mu_2\} ) &=\frac{\mu_1-\xi+i}{\mu_1-\xi} \, \frac{\mu_2-\xi+i}{\mu_2-\xi} 
	= \frac{\mu_1+\xi}{\mu_1-\xi} \, \frac{\mu_2+\xi}{\mu_2-\xi} \ , \nonumber \\
	\tau (\lambda,\{\mu_1 , \mu_2\} ) &=\frac{\mu_1-\lambda+2\xi}{\mu_1-\lambda} \, \frac{\mu_2-\lambda+2\xi}{\mu_2-\lambda}+d(\lambda) \, 
	\frac{\mu_1-\lambda-2\xi}{\mu_1-\lambda} \, \frac{\mu_2-\lambda-2\xi}{\mu_2-\lambda} \ ,
\end{align}
so that the $T$ and $V$ matrices are given by
\begin{align}
	T_{11} &	= \frac{-2\xi}{(\mu_1-\xi)^2} \frac{\mu_2+\xi}{\mu_2-\xi} \ , \quad 
	T_{21} = \frac{\mu_1+\xi}{\mu_1-\xi} \frac{-2\xi}{(\mu_2-\xi)^2} \ , \nonumber \\
	T_{12} &=\frac{-2\xi}{(\mu_1-\lambda)^2} \, \frac{\mu_2-\lambda+2\xi}{\mu_2-\lambda}+ \frac{2\xi}{(\mu_1-\lambda)^2} \, 
	\frac{\mu_2-\lambda-2\xi}{\mu_2-\lambda} \ , \nonumber \\
	T_{22} &= \frac{\mu_1-\lambda+2\xi}{\mu_1-\lambda} \, \frac{-2\xi}{(\mu_2-\lambda)^2}+\frac{\mu_1-\lambda-2\xi}{\mu_1-\lambda} \, 
	\frac{2\xi}{(\mu_2-\lambda)^2} \ , \nonumber \\
	\frac{1}{\det V} &=\frac{(\mu_1 -\xi) (\mu_1 -\lambda)(\mu_2 -\xi)(\mu_2 -\lambda)}{(\lambda-\xi) (\mu_1 -\mu_2)} \ ,
\end{align}
where we have used that $d(\xi)=0$ and that for a single-magnon the Bethe ansatz equations imply $d(\lambda)=1$. 
After some immediate algebra the form factor becomes
\begin{equation}
	\granesperado{\lambda}{\sigma^+_k }{\mu_1 \mu_2}^{\text{a}}=\frac{16\xi^3 \, e^{i(p_1+p_2-p_\lambda )k}}{(\lambda+\xi) (\mu_1 -\mu_2)} \, 
	\left[ \frac{\mu_2+\xi}{(\mu_1-\xi)(\mu_2-\lambda)} -\frac{\mu_1+\xi}{(\mu_2-\xi)(\mu_1-\lambda)} \right] \ .
\end{equation}
Now if we want to read this result in the normalization of the CBA we need to recall the discussion in section~\ref{normalizationisues}. In the case at hand
\be
\granesperado{\lambda}{\sigma^+_k}{\mu_1 \mu_2}^{\text{a}}=\frac{i\, d(\lambda)}{\lambda+\xi} \, 
\frac{\mu_2-\mu_1+i}{\mu_1-\mu_2} \, \frac{1}{(\mu_1-\xi)(\mu_2-\xi)} \granesperado{\lambda}{\sigma^+_k}{\mu_1 \mu_2}^{\text{c}} \ .
\ee
Therefore
\begin{equation}
	\granesperado{\lambda}{\sigma^+_k}{\mu_1 \mu_2}^{\text{c}}=e^{i(p_1+p_2-p_\lambda )k}\frac{-2}{\mu_2-\mu_1+i} \, 
	\left[ \frac{\mu_2^2-\xi^2}{(\mu_2-\lambda)} -\frac{\mu_1^2-\xi^2}{(\mu_1-\lambda)} \right] \ .
\end{equation}
Now we have to divide by the norm of the states in both cases,
which can be easily calculated using the Gaudin formula~(\ref{gaudin}). 
In the ABA,
\be
\scal{\mu_1 ,\mu_2}{\mu_1 , \mu_2}^{\text{a}} = \frac{16\xi ^4 L^2 \, \big[ (\mu_2 -\mu_1)^2 -4\xi^2 \big]}{(\mu_2-\mu_1)^2 
    \left( \mu_1^2 -\xi^2 \right) \left( \mu_2^2 -\xi^2 \right)} \left( 1 -\frac{2}{L} \cdot 
    \frac{\left( \mu_1^2 +\mu_2^2 -2\xi^2 \right)}{\left[ (\mu_2 -\mu_1)^2 -4\xi^2 \right]} \right) \ . 
\label{algebraicnorm}
\ee    
Recalling again section~\ref{normalizationisues}, states in the CBA and the ABA  are related through 
\be
\scal{\mu_1 ,\mu_2}{\mu_1 , \mu_2}^{\text{a}} = \left( \frac{\mu_2-\mu_1+i}{\mu_1-\mu_2} \right) \left( \frac{\mu_2-\mu_1-i}{\mu_1-\mu_2} \right) 
\frac{\scal{\mu_1 ,\mu_2}{\mu_1 , \mu_2}^{\text{c}} }{{\left( \mu_1^2 -\xi^2 \right) \left( \mu_2^2 -\xi^2 \right)}} \ ,
\ee
and thus we conclude that
\be	
\scal{\mu_1 ,\mu_2}{\mu_1 , \mu_2}^{\text{c}} =  16\xi ^4 L^2 \left( 1 -\frac{2}{L} \cdot 
\frac{\left( \mu_1^2 +\mu_2^2 -2\xi^2 \right)}{\left[ (\mu_2 -\mu_1)^2 -4\xi^2 \right]} \right) \ . 
\ee
Therefore at leading order the norm contributes with a factor $\sqrt{L}$ for each magnon and it does not contain any momentum dependence. 
The properly normalized form factor will be
\begin{align}
	\frac{\granesperado{\lambda}{\sigma^+_k}{\mu_1 \mu_2}^{\text{c}}}{\sqrt{\scal{\l}{\l}^{\text{c}} \scal{\mu_1 ,\mu_2}{\mu_1 , \mu_2}^{\text{c}}}} 
	&=\frac{e^{i(p_1+p_2-p_\lambda )k}}{\sqrt{L^3}} \frac{2}{\mu_2-\mu_1+i} \, 	\left[ \frac{\mu_2^2-\xi^2}{(\mu_2-\lambda)} -\frac{\mu_1^2-\xi^2}{(\mu_1-\lambda)} \right] \notag \\
	&\times \left( 1 -\frac{2}{L} \cdot \frac{\left( \mu_1^2 +\mu_2^2 -2\xi^2 \right)}{\left[ (\mu_2 -\mu_1)^2 -4\xi^2 \right]} \right)^{-1/2} \ .
\end{align}
At this point there are two important points we should stress. The first one is that the form factor in the CBA agrees with the computation in 
the ABA when using Zamolodchikov-Faddeev states if we also perform the change $k\rightarrow k-\frac{1}{2}$ and we include a global minus sign. 
The second one is that our expression for $\granesperado{\lambda}{\sigma^+_k }{\mu_1 \mu_2}$ (regardless of whether it is the algebraic or the coordinate one), 
conveniently normalized, is \textsl{valid to all orders in $L$} provided that we use an expression for the rapidities valid to all orders in $L$. 
We can thus write the rapidities in terms of the momenta, 
$\mu=- \frac{1}{2} \cot \left( \frac{p}{2} \right)$ and expand in the length of the chain. In the single-magnon state the momentum is quantized as 
\be
p_{\l} =\frac{2\p n_{\l}}{L} \ .
\ee
In the two-magnon state the solution to the Bethe equations can be expanded as
\be
p_{1} = \frac{2\pi n_1}{L} + \frac{4\pi}{L^2} \frac{n_1 n_2}{n_1-n_2} + \order{L^{-3}} \ , \quad 
p_{2} = \frac{2\pi n_2}{L} - \frac{4\pi}{L^2} \frac{n_1 n_2}{n_2-n_1}+\order{L^{-3}}  \ .
\ee
We conclude that for the case of $k=1$
\begin{align}
    & \granesperado{\lambda}{\sigma^+_{k=1}}{\mu_1 \mu_2}^{\text{c}} = \frac{1}{\sqrt{L^3}} \frac{2 n_\l (n_1+n_2-n_\l)}{(n_\l -n_1) (n_\l -n_2)}
    \left\{ 1 + \frac {1}{L \ (n_1-n_2)^2}  \Big[ (n_1^2 +n_2^2) \right. \notag \\
    & \left. + \: \frac{4 n_1^2 n_2^2}{(n_\l-n_1)(n_\l-n_2)} + 2 i \pi (n_1-n_2) ( n_1^2 -n_2^2 +n_1 n_2 - n_\l (n_1-n_2) ) \Big] + \dots \right\} \ .
\end{align}
The leading order term in this expression is the three-particle form factor obtained in \cite{KM1} using the CBA
with one particle of momentum $p_{\l}$ and two external particles of momenta $p_1$ and $p_2$. 
In order to obtain the subleading term we need to take into account the ${\cal O}(L^{-3})$ contributions to $p_1$ and $p_2$. 

We can get a more compact result, valid to all orders in $L$, if we take into account the trace condition (\ref{trace}). 
Then in the two-magnon state we have $\mu_1=-\mu_2$, and the Bethe equations can be solved analytically, 
\footnote{We impose the trace condition on the two-magnon state rather than on both states, 
because in this later case the correlation function becomes zero. From the CFT point of view this happens because the one-excitation state is not a new primary operator but a descendent of the vacuum.}
\be
\mu_1 = -\mu_2=-\frac{1}{2} \cot \left( \frac{n \, \pi}{L-1} \right) \ , \quad n\in \mathbb{Z} \ . \label{rapidityroiban}
\ee
Substituting we obtain
\begin{align}
& \frac{\granesperado{\lambda}{\sigma^+_k}{\mu, -\mu}^{\text{c}}}{\scal{\l}{\l}^{\text{c}} \scal{\mu ,-\mu}{\mu , -\mu}^{\text{c}}} 
= \frac{e^ {-ip_\l k}}{L \sqrt{(L-1)}} \frac{2\m (\m + \xi )}{\m ^2 -\l ^2} \nonumber \\
& = e^ {-2\pi i n_\l k/L}\frac{\cot \left( \frac{n \, \pi}{L-1} \right) }{L \sqrt{(L-1)}} 
\frac{ 2\left[ \cot \left( \frac{n \, \pi}{L-1} \right) -i\right]}{\cot ^2 \left( \frac{n \, \pi}{L-1} \right) -\cot ^2 \left( \frac{n_\l \, \pi}{L} \right) } \ ,
\end{align}
where $n$ and $n_\l$ are integer numbers.

The same computation can be carried out for the operator $\sigma^+_k$ evaluated on on-shell Bethe states with higher number of magnons. First of all, the left and right $(A+D)^n$ factors act in a known way over the states, as they are Bethe states. Hence the only difficult step is to compute the scalar product between an on-shell state, $\{\mu \}$, and an off-shell state, $\{\{\lambda\},\xi\}$, using equation~(\ref{scalarproduct}). A detailed computation of these form factors can be found in section 3.1.6 of \cite{MosselThesis}.

Of course, the computation of correlation functions of the operator $\sigma^-_k$ can be obtained as the conjugate of the computation we have presented in this section. The way of proceeding for the $\sigma^z_k$ operator is a little bit different, but in the following section we are going to talk about the operators $\sigma_k^+ \sigma_k^-$ and $\sigma_k^- \sigma_k^+$ that can be evaluated in a similar way to this one.


\section{Correlation functions involving two operators}
\label{evaluationcorrelationsection}

In the previous section we have described how the ABA can be employed to calculate correlation functions for one spin operator. However there seems to be problems with this method when we want to perform computations involving two or more operators. 
This is because most correlation functions have the general form
\begin{displaymath}
	\granesperado{0}{\dots C(\xi) (A+D)^n (\xi) \dots}{0} \ .
\end{displaymath}
Therefore, according to the algebra~(\ref{commCB}), whenever we try to commute the $(A+D)$ operators with the $C$ operator 
a divergence should appear. In this section we are going to show that actually there are no divergences at all. We will describe how to deal with these apparent divergences. We will first show how to proceed 
in the most simple case, that is, when we only have the operator $C$ at the left of the $(A+D)^n$ factor. Later on we will extend 
the computation to more general correlation functions involving additional factors.

\subsection{Evaluation of $\langle \{\mu\}|\sigma^+_k \sigma^-_k |\{\lambda\}\rangle$ (in an easy way)}

We are going to start by evaluating the correlation function $\langle \{\mu\}|\sigma^+_k \sigma^-_k |\{\lambda\}\rangle$. By using equation~(\ref{inversegeneral}) we can write
\begin{equation}
	\sigma^+_k \sigma^-_k= \left(\begin{array}{cc}
	1 & 0 \\
	0 & 0
\end{array} \right)_k= (A+D)^{k-1} (\xi ) A(\xi ) (A+D)^{L-k} (\xi ) \ .
\end{equation}
As in the previous section, the action of the $(A+D)^n$ factors over the Bethe states are known to give exponential of the momenta,
\begin{equation}
	\granesperado{ \{\mu\}}{\sigma^+_k \sigma^-_k }{\{\lambda\}}=\granesperado{ \{\mu\}}{A(\xi ) }{\{\lambda\}}e^{i\sum_{i}{p_{\mu_i} (k-1)}-i\sum_{i}{p_{\lambda_i} (L-k)}} \ .
\end{equation}
Now we can apply the commutation relation~(\ref{commAB}) to move the $A(\xi)$ operator all the way to the right (similarly, we can do it by moving it to the left). Using the reordering symmetry of the $B$ operators as we did previously to compute equation~(\ref{eigenvaluemonodromy}) we can write
\begin{equation}
	A(\xi) \ket{\{\lambda \}}=\prod_{i=1}^M{\frac{\xi-\lambda_i-i}{\xi-\lambda_i}} \ket{\{\lambda\}} +\sum_{i=1}^M{ \frac{i}{\xi-\lambda_i} \prod_{j\neq i}{\frac{\lambda_i-\lambda_j-i}{\lambda_i-\lambda_j}} \ket{\{\hat{\lambda}_i \}, \xi}} \ ,
\end{equation}
where $\{\hat{\lambda}_i \}$ means that the rapidity $\lambda_i$ is missing from the set $\{\lambda\}$. If we now apply this expression to the bra state $\bra{\{\mu\}}$, the first term of the sum will involve the scalar product between both on-shell states, which means that it will only contribute when $\{\lambda\}=\{\mu\}$. The rest of the terms will involve the computation of off-shell-on-shell scalar products.  The final answer is then
\begin{equation}
	\granesperado{ \{\mu\}}{\sigma^+_k \sigma^-_k }{\{\lambda\}}=\scal{\{\mu\}}{\{\lambda\}} \prod_{i=1}^M{\frac{\xi-\lambda_i-i}{\xi-\lambda_i}} +\sum_{i=1}^M{ \frac{i\Bigscal{\{\mu\}}{\{\hat{\lambda}_i \}, \xi}}{\xi-\lambda_i} \prod_{j\neq i}{\frac{\lambda_i-\lambda_j-i}{\lambda_i-\lambda_j}}} \ .
\end{equation}
Although at first sight it seems that we can use equation~(\ref{scalarproduct}) to compute it, there are some subtleties. To get a better understanding of this formula and some of the difficulties related with its explicit computation, we are going to see some particular cases of a low number of magnons. The most trivial case is the empty case $\{\lambda\}=\{\mu\}=\emptyset$, which reads
\begin{equation}
	\granesperado{0}{\sigma^+_k \sigma^-_k }{0}=\granesperado{0}{A(\xi ) }{0}=a(\xi )=1 \ .
\end{equation}
Which trivially agrees with the result we get using CBA. The first non-trivial case is the one with a single magnon, which reads
\begin{align}
	&\granesperado{ \mu}{\sigma^+_k \sigma^-_k }{\lambda}=\granesperado{ \mu}{A(\xi ) }{\lambda}e^{ip_{\mu} (k-1)-ip_{\lambda} (L-k)} = \notag \\
	&=e^{-ip_{\mu} (k-1)-ip_{\lambda} (L-k)} \left[ \frac{\lambda +\xi}{\lambda-\xi} \frac{i}{\mu-\lambda} [d(\lambda)-d (\mu) ] +\frac{i}{\xi-\lambda} \frac{i}{\mu-\xi} [d(\xi)-d(\mu)] \right] \ .
\end{align}
If we impose now the Bethe equations $d(\lambda )=e^{ip_{\lambda}L}=d(\mu )=e^{ip_{\mu}L}=1$, we find two well differentiated cases: either $\lambda \neq \mu$ and everything but the last term cancels, or $\lambda =\mu$ and the first term becomes the norm of the state. So the final answer can be written as
\begin{equation}
	\granesperado{ \mu}{\sigma^+_k \sigma^-_k }{\lambda}=\left\{ \begin{array}{cc}
	\frac{-e^{i(p_\lambda - p_\mu) k} }{(\lambda + \xi ) (\mu-\xi )} &\text{ if } \lambda \neq \mu\\
	\frac{L}{(\lambda + \xi ) (\mu-\xi )} \left( 1-\frac{1}{L} \right) &\text{ if } \lambda = \mu
\end{array}	  \right. \ .
\end{equation}
The second one is easily comparable with the CBA, because the result can be expressed as $\granesperado{ \lambda}{\sigma^+_k \sigma^-_k }{\lambda}=\scal{\lambda}{\lambda} \left( 1-\frac{1}{L} \right)$, which agrees with the CBA. In the same way, after properly normalizing, we obtain that the case $\lambda \neq \mu$ is equal to $e^{i(p_\lambda - p_\mu) (k-\frac{1}{2})}$, which is the same expression that can be obtained using the CBA up to the already expected shift of half a lattice spacing.

Although the one-magnon case has no problems, the case $\granesperado{ \lambda_1 \lambda_2}{\sigma^+_k \sigma^-_k }{\lambda_1 \lambda_2}$ already contains some subtleties that will be present for all cases with higher number of magnons. In particular this computation will involve the calculation of the correlation functions $\scal{ \lambda_1 \xi}{\lambda_1 \lambda_2}$ and $\scal{ \xi \lambda_2}{\lambda_1 \lambda_2}$ which apparently diverge when computed using the Slavnov determinant (\ref{scalarproduct}). This apparent divergence appears because this scalar product have already been simplified using the Bethe equations. Therefore to find the correct answer we have to compute first $\scal{ \mu_1 \mu_2}{\lambda_1 \lambda_2}$ as off-shell rapidities and, without imposing the Bethe equations, take the limits $\mu_i \longrightarrow \{ \lambda_1 , \xi \}$ and $\mu_i \longrightarrow \{ \xi , \lambda_2 \}$ respectively\footnote{We want to thank N. A. Slavnov for discussions about this subject.}.

\subsection{Evaluation of $\langle 0|\sigma^+_k \sigma^-_l |0 \rangle$ (in a non-easy way)}
\label{noneasyway}

Now we are going to repeat the computation from the previous section but instead of using the definition of $\sigma^+_k \sigma^-_k$ we are going to use the definition of $\sigma^+_k$ and $\sigma^-_k$ in a separate way. The computation is going to be more cumbersome and longer, but it has two advantages with respect to the previous one: firstly it can be generalized to the computation of correlation functions of $\sigma^+_k \sigma^-_l$, that is, where the two operators are not placed in the same site, which cannot be done with the procedure presented in the previous section; and secondly it is going to lighten us the way to compute the correlation function of $\sigma^+_k \sigma^+_l$ operator. As we only want to present the basic computations and the procedure to remove the apparent singularities, we are going to focus mainly on the easiest correlation function to compute $\langle 0|\sigma^+_k \sigma^-_l |0 \rangle$.

Again, the starting point in the ABA are the relations between local spin operators in the CBA and the elements of the monodromy matrix.
If we recall that $(A+D) (\xi_i) \ket{0}=\ket{0}$ for the Heisenberg chain, we will need to evaluate
\begin{equation}
	\langle 0 | \sigma^+_k \sigma^-_l |0\rangle =
	\langle 0| {C (\xi) (A+D)^{L+l-k-1} (\xi) B (\xi)} |0 \rangle \ .
\end{equation}
In order to evaluate this correlation function we have to commute the operators $(A+D)$ with $C$ or $B$ using equation (\ref{commCB}).
However, although it seems that when trying to commute $(A+D)^n$ we should obtain a pole of order $n$ because of the divergence 
of the commutation relations when the two rapidities are equal, the residue turns out to be zero for all $n$ and the expression is finite. 
In order to understand this cancellation some care will be needed. Let us first introduce 
some notation. We will define
\begin{align}
{\cal F}^L_{n}(\alpha,\delta) &= \granesperado{0}{C(\xi+\alpha) \mathcal{O}(\delta)}{0} \ , \nonumber \\
{\cal F}^L_{n+1}(\alpha,\delta) &= \lim_{\beta\rightarrow \alpha} 
\granesperado{0}{C(\xi+\alpha) (A+D)(\xi+\beta) \mathcal{O}(\delta)}{0}= \lim_{\beta\rightarrow \alpha} f^L_{n+1}(\a , \b , \d )\ , 
\label{F}
\end{align}
where $\mathcal{O}(\delta)$ denotes any operator. The reason for the subindex $n$ is that in all the cases that we will consider that $\mathcal{O}(\delta)$ includes a factor $(A+D)^n$. Then using (\ref{commCA}) and (\ref{commCD}) we can write
\begin{align}
	{\cal F}^L_{n+1} (\alpha,\delta)&=\big[ 1+d(\xi+\alpha) \big] {\cal F}^L_{n}(\alpha,\delta) 
	+ \lim_{\b \rightarrow \a}\frac{i}{\b-\a} \left\{ \big[d(\xi+\b)-1 \big] {\cal F}^L_{n}(\alpha,\delta) \right. \nonumber \\
	&-\left. \big[ d(\xi+\alpha)-1 \big] {\cal F}^L_{n}(\b,\delta) \right\} \ .
\label{Fn1}
\end{align}
Now if we expand in a Taylor series we find that all terms of order $1/(\b-\a)$ cancel themselves. 
Therefore we can safely take the limit $\b \rightarrow \a$ to get 
\begin{equation}
	{\cal F}^L_{n+1} (\alpha,\delta)=\left[1+d(\xi+\alpha)+i\left.\dpartial{d}{\lambda} \right|_{\xi+\alpha} \right] {\cal F}^L_n (\alpha, \delta) + 
	i \big[ 1-d(\xi+\alpha) \big] \dpartial{{\cal F}^L_n (\alpha, \delta)}{\alpha} \ .
\label{Frecurrence}
\end{equation}
We should stress that in this expression the derivative in $\a$ must be understood with respect to the argument of the $C$ operator. 
As a consequence it does not act on the rest of the operators. This will introduce some subtleties in the next step of the calculation.
The idea now is to use (\ref{Frecurrence}) as a recurrence equation to find $\langle 0|\sigma^+_k \sigma^-_l |0\rangle$.
However this is not straightforward, as it requires information on correlation functions of the form 
\be
\granesperado{0}{C(\xi+\alpha) (A+D) (\xi+\delta) \dots}{0} \ ,
\label{fn}
\ee
but returns instead information about correlators of the form 
\be
\granesperado{0}{C(\xi+\alpha) (A+D) (\xi+\alpha) (A+D) (\xi+\delta) \dots}{0} \ . 
\label{fn1}
\ee
We must note also that the argument of the first $(A+D)$ factor in (\ref{fn1}) depends on $\a$ and thus 
in order to find the correct correlation function we should take the derivative with respect to $\a$ in $f^L_{n+1} (\a ,\b ,\d)$, and then take the limit $\b \rightarrow \a$, 
instead of taking directly the derivative in ${\cal F}^L_{n+1}(\alpha,\delta)$.
Therefore using~(\ref{Fn1}),
\begin{align}
	\lim_{\b \rightarrow \a} \dpartial{ f^L_{n+1} (\alpha,\b ,\delta)}{\alpha} & = 
	\big[ 1+d(\xi+\alpha) \big] \dpartial{{\cal F}^L_{n}(\alpha,\delta)}{\alpha} + \lim_{\b \rightarrow \a}\frac{i}{\b-\a} 
	\left\{ \big[d(\xi+\b)-1 \big] \dpartial{{\cal F}^L_{n}(\alpha,\delta)}{\alpha} \right. \nonumber \\
	& - \dpartial{d(\xi+\alpha)}{\alpha} {\cal F}^L_{n}(\b,\delta) 
	+ \frac {1}{(\b-\a )^2} \Big[ \big[d(\xi+\b)-1 \big] {\cal F}^L_{n}(\alpha,\delta) \nonumber \\
	& \left. - \big[d(\xi+\alpha)-1 \big] {\cal F}^L_{n}(\b,\delta) \Big] \right\} \ .
\end{align}
The remaining piece of the calculation is similar 
to the previous one. In this case after a series expansion we find a pole of order two and a pole of order one, but they cancel themselves. 
The final result is
\begin{align}
	\lim_{\b \rightarrow \a} \dpartial{ f^L_{n+1} (\alpha,\b ,\delta)}{\alpha}&=\big[ 1+d(\xi+\alpha) \big] 
	\dpartial{{\cal F}^L_{n} (\alpha,\delta)}{\alpha}+\frac{i}{2} \dpartial[2]{d}{\alpha} {\cal F}^L_{n} (\alpha,\delta) \nonumber \\
	&+ \frac{i}{2} \big[ 1-d(\xi+\alpha) \big] \dpartial[2]{{\cal F}^L_{n} (\alpha,\delta)}{\alpha} \ .
\end{align}
So far we have proven that when we have one derivative and we commute one $(A+D)$ factor we get another 
derivative over the correlation function. In general if we have $m$ derivatives we get
\begin{align}
	\lim_{\b \rightarrow \a} \dpartial[m]{ f^L_{n+1} (\alpha,\b ,\delta)}{\alpha}& = 
	\big[ 1+d(\xi+\alpha) \big] \dpartial[m]{{\cal F}^L_{n} (\alpha,\delta)}{\alpha}+\frac{i}{m+1} 
	\dpartial[m+1]{d}{\alpha} {\cal F}^L_{n} (\alpha,\delta) \nonumber \\
	&+\frac{i}{m+1} \big[ 1-d(\xi+\alpha) \big] \dpartial[m+1]{{\cal F}^L_{n} (\alpha,\delta)}{\alpha} \ ,
\label{Fgeneral}
\end{align}
that can be easily proved if we assume that the left-hand side of the equation has no poles. 
Under this assumption, when we expand in a Taylor series we only need to track the terms without a factor $\b-\a$,
\begin{align}
	& \lim_{\b \rightarrow \a} \dpartial[m]{ f^L_{n+1} (\alpha,\b ,\delta)}{\alpha}= \big[1+d(\xi+\alpha) \big] 
	\dpartial[m]{{\cal F}^L_n (\a , \d)}{\alpha} \nonumber \\ 
	&+ \lim_{\b \rightarrow \a} \, \dpartial[m]{}{\alpha}
	\left\{ \frac{i}{\b-\a} \left[ \Big( d(\xi+\b)-1\Big) {\cal F}^L_{n}(\alpha,\delta) \right. \right.
	-\left. \left. \Big(d(\xi+\alpha)-1\Big) {\cal F}^L_{n}(\b,\delta) \right] \right\} \ .
\label{Fprove}	
\end{align}
The second term on the right hand side of this expression can be written as
\[
\lim_{\b \rightarrow \a} \sum_j{\binom{m}{j} \frac{i}{(\b -\a)^{j+1} \, (j+1)} } \cdot 
\left[ \dpartial[j+1]{d}{\alpha} \dpartial[m-j]{{\cal F}^L_{n}}{\alpha}  
- \dpartial[m-j]{(d-1)}{\alpha} \dpartial[j+1]{{\cal F}^L_{n}}{\alpha} \right] (\b -\a)^{j+1} + \dots \ ,
\]
where the dots stand for terms proportional to $(\b-\a)^k$. Now it is clear that 
the terms in $j$ are canceled by the terms in $m-j-1$. Therefore the only term surviving is the one with 
$j=m$, which does not have a partner. This is expression (\ref{Fgeneral}).

Let us summarize our results up to this point. We have obtained a complete set of recurrence equations,
\begin{align}
	{\cal F}^L_{n+1}(\alpha)&= \left[1+d(\xi+\alpha)+i\left.\dpartial{d}{\lambda} \right|_{\xi+\alpha} \right] {\cal F}^L_n (\alpha) 
	+ i\big[ 1-d(\xi+\alpha) \big] \mathcal{D} {\cal F}^L_n (\alpha) \ , \nonumber \\
	\mathcal{D}^m {\cal F}^L_{n+1} (\alpha)&=\big[ 1+d(\xi+\alpha) \big] \mathcal{D}^m {\cal F}^L_{n} (\alpha) 
	+ \frac{i}{m+1} \dpartial[m+1]{d}{\alpha} {\cal F}^L_{n} (\alpha) \nonumber \\
	&+\frac{i}{m+1} \big[ 1-d(\xi+\alpha) \big] \mathcal{D}^{m+1} {\cal F}^L_{n} (\alpha) \ , \nonumber \\
	\mathcal{D}^m {\cal F}^L_{0} (\alpha) &=\dpartial[m]{{\cal F}^L_{0} (\alpha)}{\alpha} \ , \quad \hbox{with} \quad
	{\cal F} (\alpha )=\lim_{\delta\rightarrow \alpha}{\cal F} (\alpha , \delta) \ , 
\label{recurrenceeq}
\end{align}
where $\mathcal{D}$ is just a convenient notation to refer both to the derivative and the limit, 
\be 
\mathcal{D}^m {\cal F} (\alpha )=\lim_{\substack{\delta\rightarrow \alpha \\ \b \rightarrow \a}} \dpartial[m]{ f(\alpha , \b ,\delta)}{\alpha} \ .
\ee 

Now we are ready to calculate the correlation function provided a starting condition is given. In our case, using equation~(\ref{commCB})\footnote{It is important to stress here that this ``norm'' cannot be computed using the Gaudin determinant~(\ref{gaudin}) because it assumes the fulfilling of the Bethe equation and $\ket{\xi}$ is not a Bethe state.},
\be
{\cal F}^L_0(\alpha)=\granesperado{0}{C(\xi) B(\xi+\alpha)}{0}=-\frac{i}{\alpha} \frac{\alpha^L}{(\alpha +i)^L} \ ,
\ee
which takes values ${\cal F}_0^1 (0)=1$ and ${\cal F}^{L>1}_0=0$.

In order to find the only non-vanishing correlation function, which we know from CBA arguments and the previous subsection to be $\langle 0|\sigma^+_k \sigma^-_k |0\rangle =1$, 
we have to calculate ${\cal F}_{L-1}^L (0)$. Because ${\cal F}^L_0(0)$ has a zero of order $L-1$, the only terms that can contribute are those which 
involve a number of derivatives of ${\cal F}^L_0(\alpha)$ greater than or equal to $L-1$ (other possible terms will require many more derivatives).
In appendix \ref{A} we will construct the correlation function ${\cal F}^L_n(\alpha)$ in full generality, but in this case it is easy to see that
\begin{equation}
	{\cal F}_{L-1}^L (\alpha)=\frac{i^{L-1}}{(L-1)!} \dpartial[L-1]{{\cal F}^L_0(\alpha)}{\alpha}+\dots=\frac{i^{L-1}}{(L-1)!} \cdot i \, 
	\frac{(L-1)!}{i^L}+ \order{\alpha} \ .
\end{equation}
In the limit $\alpha \rightarrow 0$ we conclude that the value of this correlation function is one, as expected from the CBA. 

As we can see, the computations using this technique are longer. However it is ``worth'' the effort as we can prove that ${\cal F}_{n}^L (\alpha)=0$ for $0\leq n <L-1$, which cannot be computed using the procedure from the previous section and agrees with the result 
$\langle 0|\sigma^+_k \sigma^-_l |0\rangle=0$ when $k\neq l$ obtained using the CBA.


\subsection{Evaluation of $\langle 0|\sigma^+_k \sigma^+_l |\mu_1 \mu_2\rangle$}
\label{0sigmasigmamumu}

We will now evaluate the correlation function $\langle 0|\sigma^+_k \sigma^+_l |\mu_1 \mu_2\rangle$. 
Using relation (\ref{sigmaplus}) we can bring again the problem to the ABA, 
\be
\langle 0|\sigma^+_k \sigma^+_l |\mu_1 \mu_2\rangle 
= \granesperado{0}{(A+D)^{k-1} (\xi) C(\xi) (A+D)^{n} (\xi) C(\xi) (A+D)^{L-l} (\xi) }{\mu_1 \mu_2} \ ,
\label{wavefunction}
\ee
where $n=L+l-k-1$. The first factor $(A+D)$ acts trivially on the vacuum. On the contrary, the last factor $(A+D)$ 
acts on the two magnon state $\ket{\mu_1 \mu_2}=B(\mu_1) \, B(\mu_2) \, \ket{0}$ and provides a factor $e^{-i(p_1+p_2)\cdot (L-l)}=e^{i(p_1+p_2)l}$, 
where in the last equality we have used the periodicity condition for the Bethe roots. 
The contribution from the remaining factors can be obtained in a similar way to the previous correlation function. 
To continue with the notation introduced in that case, 
now we will name correlation functions with $n$ inner factors of $(A+D)$ by ${\cal G}^L_n(\alpha)$,
\be
{\cal G}^L_{n}(\a) = \granesperado{0}{C(\xi+\alpha) \mathcal{O}_n(\delta) C(\xi) B(\mu_1) B(\mu_2)}{0} \ .
\ee
As we will show, the problem can again be solved as a recurrence and thus the  
starting point will be to find the initial correlator
\be
{\cal G}^L_0(\alpha)=\granesperado{0}{C(\xi+\alpha) \, C(\xi) \, B(\mu_1) \, B(\mu_2)}{0}\text{, } \quad\lim_{\alpha \rightarrow 0} {\cal G}^L_0(\alpha)=\langle 0|\sigma^+_1 \sigma^+_{L} |\mu_1 \mu_2\rangle \ ,
\ee
which is the product of a on-shell Bethe state with an off-shell Bethe state. 
As described in section~\ref{normalizationisues} we can write
\be
 \granesperado{0}{C(\xi+\alpha) \, C(\xi) \, B(\mu_1) \, B(\mu_2)}{0} = \frac{\det T}{\det V} \ .
\ee
Now the functions $\tau(\xi)$ and $\tau(\xi+\a)$ are 
\begin{align}
	\tau (\xi,\{\mu_1 , \mu_2\} ) &=\frac{\mu_1-\xi+i}{\mu_1-\xi} \, \frac{\mu_2-\xi+i}{\mu_2-\xi}=\frac{\mu_1+\xi}{\mu_1-\xi} \, 
	\frac{\mu_2+\xi}{\mu_2-\xi} \ , \nonumber \\
	\tau (\xi+\alpha,\{\mu_1 , \mu_2\} ) &=\frac{\mu_1+\xi-\alpha}{\mu_1-\xi-\alpha} \, \frac{\mu_2+\xi-\alpha}{\mu_2-\xi-\alpha} 
	+ \frac{\alpha^L}{(i+\alpha)^L} \, \frac{\mu_1-3\xi-\alpha}{\mu_1-\xi-\alpha} \, \frac{\mu_2-3\xi-\alpha}{\mu_2-\xi-\alpha} \ ,
\end{align}
and thus the matrices $T$ and $V$ become 
\begin{align}
	T_{11} &= \frac{-2\xi}{(\mu_1-\xi)^2} \frac{\mu_2+\xi}{\mu_2-\xi} \ , \quad 
	T_{21} =\dpartial{\tau (\xi,\{\mu_1 , \mu_2\} )}{\mu_2} = \frac{\mu_1+\xi}{\mu_1-\xi} \frac{-2\xi}{(\mu_2-\xi)^2} \ , \nonumber \\
	T_{12} &=\frac{-2\xi}{(\mu_1-\xi-\alpha)^2} \, \frac{\mu_2+\xi-\alpha}{\mu_2-\xi-\alpha}+\frac{\alpha^L}{(i+\alpha)^L} \, 
	\frac{2\xi}{(\mu_1-\xi-\alpha)^2} \, \frac{\mu_2-3\xi-\alpha}{\mu_2-\xi-\alpha} \ , \nonumber \\
	T_{22} &= \frac{\mu_1+\xi-\alpha}{\mu_1-\xi-\alpha} \, \frac{-2\xi}{(\mu_2-\xi-\alpha)^2}+\frac{\alpha^L}{(i+\alpha)^L} \, 
	\frac{\mu_1-3\xi-\alpha}{\mu_1-\xi-\alpha} \, \frac{2\xi}{(\mu_2-\xi-\alpha)^2} \ , \nonumber \\
	\frac{1}{\det V} &=\frac{(\mu_1 -\xi) (\mu_1 -\xi-\alpha)(\mu_2 -\xi)(\mu_2 -\xi-\alpha)}{\alpha (\mu_1 -\mu_2)} \ .
\end{align}
After some algebra we can easily organize ${\cal G}_0^L(\a)$ as an expansion in $\a$,
\ba
{\cal G}^L_0 (\alpha) & \!\!\! = \!\!\! & \left( A_0+\alpha A_1+\alpha^2 A_2 + \dots \right)+\alpha^{L-1} 
\left( B_{L-1}+\alpha B_{L}+\alpha^2 B_{L+1} + \dots \right) \nonumber \\ 
& + \!\! & \alpha^{2L-1} \left( C_{2L-1} + \a C_{2L} + \a^2 C_{2L+1} + \dots \right) \ ,
\label{FLzero}	
\ea
with $A_q$ and $B_{L+q-1}$ given by \footnote{Because of periodicity it is unnecessary to write the explicit expression for $C$.}
\ba
A_q & \!\! = \!\! & \frac{1}{\mu_1-\mu_2} \, \frac{\mu_1^+ \mu_2^+}{\mu_1^- \mu_2^-} \, 
\left[ \frac{1}{(\mu_1^-)^{q}}    \frac{\left(\mu_2-\mu_1+i \right)}{\mu_1^- \mu_2^+} 
+ \frac{1}{(\mu_2^-)^{q}} \frac{\left(\mu_2-\mu_1-i \right)}{\mu_1^+ \mu_2^-} \right] \ , \nonumber \\
B_{L+q-1} & \!\! = \!\! & \sum_{j=0}^{q}{i^j \binom{L+j-1}{j} \beta_{q-j}} \ , 
\label{AB}
\ea
where we have defined
\ba
	\beta_0 & \!\! = \!\! & B_{L-1}=\frac{1}{i^L} \, \frac{1}{\mu_1^- \mu_2^-}\, \frac{1}{\mu_1-\mu_2} \, 
	\left( \mu_2^+ \mu_1^{---} - \mu_1^+ \mu_2^{---} \right) \ , \nonumber \\
	\beta_q & \!\! = \!\! & \frac{1}{i^L} \, \frac{1}{\mu_1-\mu_2} \, \frac{1}{\mu_1^- \mu_2^-} \, 
	\left( \frac{\mu_2^+ \mu_1^{---} -\mu_2^+ \mu_2^-}{(\mu_2^-)^q} - \frac{\mu_1^+ \mu_2^{---} -\mu_1^+ \mu_1^-}{(\mu_1^-)^q} \right) \ , 
\ea
with $\mu_i^j=\mu_i+j\xi$ and $B_q=C_p=0$ for $q<L-1$ and $p<2L-1$ respectively.
The next step is to find the general form of the correlation function ${\cal G}^L_n(\a)$. Using the recurrence equations~(\ref{recurrenceeq}) 
the first terms can be easily calculated for a general value of $\a$,
\begin{align}
	{\cal G}^L_1(\a) &=\left[ 1+d+i \dpartial{d}{\lambda}  \right] {\cal G}^L_0(\a)+i \big[ 1-d \big] \dpartial{{\cal G}^L_0(\a)}{\lambda} \ , \nonumber \\
	{\cal G}^L_2(\a) &=\left[ 1+2d+2i\dpartial{d}{\lambda}+2id\dpartial{d}{\lambda}+d^2-\left( \dpartial{d}{\lambda} \right)^2 
	- \frac{1}{2} \dpartial[2]{d}{\lambda}+\frac{d}{2} \dpartial[2]{d}{\lambda} \right] {\cal G}^L_0(\a) \nonumber \\
	& + \left[ 2i-2id^2 -\dpartial{d}{\lambda}+d\dpartial{d}{\lambda} \right] \dpartial{{\cal G}^L_0(\a)}{\lambda} 
	-\frac{(1-d)^2}{2} \dpartial[2]{{\cal G}^L_0(\a)}{\lambda} \ ,
\end{align}
where $d=d(\xi +\alpha)$ and $\dpartial{d}{\lambda}=\left. \dpartial{d}{\lambda} \right|_{\xi +\alpha}$. 
If we take now the limit $\alpha\rightarrow 0$, all the $d$ and derivatives of $d$ disappear, unless 
it is a derivative of $d$ of order greater or equal to $L$. The computation of ${\cal G}_n^L(0)$ with arbitrary $n$ is a little bit more involved. 
We have collected all details in appendix \ref{A}. We find
\begin{equation}
{\cal G}^L_n(0)=\sum_{q=0}^{n}{\binom{n}{q} \left. \frac{i^{q} \mathcal{D}^{q}}{q!} {\cal G}^L_0(\a) \right|_{\a =0}}+\theta(n-L) {\cal G}^L_{n-L}(0) \ , \label{GLn}
\end{equation}
where $\theta (x)$ is the Heaviside step function. If we use now expansion (\ref{FLzero}) 
and perform the derivatives we can write
\be
	{\cal G}^L_n(0) = \sum_{q=0}^{n}{\binom{n}{q} i^{q} \left( A_q+B_q+C_q \right)}+\theta(n-L) {\cal G}^L_{n-L}(0) \ .
\label{Ffinal}
\ee
Now we are finally ready to evaluate $\langle 0|\sigma^+_k \sigma^+_l |\mu_1 \mu_2\rangle$ for different values of $n$. 

\subsubsection*{The case $n<L-1$}

We will first consider the case where $n<L-1$, which corresponds to $l<k$. 
From (\ref{Ffinal}) it is clear that when $n<L-1$ the only contribution is from the $A_q$ terms, that can be easily summed up,
\be
\sum_{q=0}^{n}{\binom{n}{q} i^{q} A_q} = \frac{1}{\mu_1-\mu_2} \frac{\mu_1^+ \mu_2^+}{\mu_1^- \mu_2^-} \! 
\left[  \left( \frac{\mu_1^+}{\mu_1^-} \right)^{n} 
\frac{\left(\mu_2-\mu_1+i \right)}{\mu_1^- \mu_2^+} 
+ \left( \frac{\mu_2^+}{\mu_2^-} \right)^{n} \frac{\left(\mu_2-\mu_1-i \right)}{\mu_1^+ \mu_2^-} \right] .
\label{Asum}
\ee
Recalling now that the rapidities parametrize the momenta, $\mu_i^+/\mu_i^-=e^{-ip_i}$, 
equation~(\ref{wavefunction}) can be written as
\be
\granesperado{0}{\sigma^+_k \sigma^+_l}{\mu_1 \mu_2} = 
\frac{1}{\mu_1-\mu_2} \, \frac{\mu_2-\mu_1+i}{\mu_1^- \mu_2^-} \left[   e^{ip_1(k-L)+ip_2 l}  +e^{ip_2(k-L)+i p_1 l} S_{21} \right] \ ,
\ee
where we have inserted the S-matrix, 
\be
S_{21}=\frac{\mu_2-\mu_1-i}{\mu_2-\mu_1+i} \ ,
\ee
and we have taken into account that $n=L+l-k-1$. Using now the Bethe equations 
$e^{-ip_1 L}=e^{ip_2 L}=S_{21}$, we find 
\begin{equation}
	\granesperado{0}{\sigma^+_k \sigma^+_l}{\mu_1 \mu_2}=\frac{1}{\mu_1-\mu_2} \, \frac{\mu_2-\mu_1+i}{\mu_1^- \mu_2^-} 
	\left[   e^{i (p_1 k+p_2 l)} S_{21}  +e^{i(p_2 k+ p_1 l)} \right] \ . \label{y<x}
\end{equation}
Note that although this result is only true as long as $l<k$, we already find that it corresponds to what we should have obtained 
from the CBA up to the factor in front of the bracket. At the end of this section we will see how the normalization proposed 
in section~\ref{normalizationisues} allows to get rid of this factor.

\subsubsection*{The case $n=L-1$}

Our next step is the calculation of ${\cal G}^L_{L-1}(0)$, which must be identically zero, because it corresponds to the case 
where both operators are located at the same site, $k=l$. 
If we take equation~(\ref{Ffinal}), we find that this correlation function can be written as
\be
{\cal G}^L_{L-1}(0)=i^{L-1} B_{L-1}+\sum_{q=0}^{L-1}{\binom{n}{q} i^{q} A_q} \ .
\ee
The second term is already known from the previous calculation. Therefore we only have to substitute the special value 
we are interested in and make use of the Bethe equations to get 
\be
\sum_{q=0}^{L-1}{\binom{L-1}{q} i^{q} A_q} = - \frac{2}{\mu_1^- \mu_2^-} \ . 
\ee
On the other hand
\be
i^{L-1} B_{L-1}=\frac{i}{\mu_1 -\mu_2} \frac{1}{\mu_1^- \mu_2^-} 
\left( \mu_1^+ \mu_2^{---} -\mu_1^{---} \mu_2^+ \right) =
\frac{2}{\mu_1^- \mu_2^-} \ .
\ee
Therefore ${\cal G}^L_{L-1}(0)=0$, as we expected from the CBA.

\subsubsection*{The case $n>L-1$}

The last correlation functions that we will evaluate will be those with $L-1<n<2L-1$. Obviously, because of periodicity of the spin chain, 
we expect that ${\cal G}^L_{n+L}(0)$ should equal ${\cal G}^L_n(0)$. In order to prove this we will first show  
that the contribution from the $B$ terms is going to be $\binom{n-L}{q+1} i^{L+q} \beta_{q}$. Next we will  
demostrate that this coefficient cancels $\sum_{q=0}^{n}{\binom{n}{q} i^{q} A_q}$, and thus we will conclude that ${\cal G}^L_{n+L}(0)={\cal G}^L_n(0)$.
Let us see how it goes.

Recalling the expression for $B_{q}$ in (\ref{AB}) and performing the sum we find
\ba
\sum_{q=L-1}^n{\binom{n}{q} i^q B_q} 
& \!\! = \!\! & \sum_{s=0}^{n-L+1}{\sum_{t=0}^{s}{\binom{n}{s+L-1} \binom{L+t-1}{t} i^{s+t+L-1} \beta_{s-t}}} \ .
\ea
In order to obtain the coefficient of a particular $\beta_q$ we have to set $s-t=q$ in the previous expression. For instance, the coefficient of $\b_{q}$ is
 \be
i^{L+q-1} \sum_{r=0}^{n-L-q+1}{\binom{n}{L+r+q-1} \binom{L+r-1}{r} (-1)^r } \ ,
\ee
where we have taken $r=s-q$ because all terms with $s<q$ do not contribute to $\beta_q$. 
We can rewrite the sum  and the binomial coefficients in a way that will allow us to use the definition of the hypergeometric function,
\begin{align}
	& \frac{n!}{(L-1)!} \sum_{r=0}^{n-L-q+1}{\frac{(L+r-1)!}{(L+r+q-1)!} \binom{n-L-q+1}{r}  (-1)^r } \nonumber \\
	& = \, _2 F_1 \left( L,q-1+L-n;L+q;1 \right) n! = \frac {(n-L)!}{(q-1)!} \ , 
\end{align}
where in the last equality we have used Kummer's first formula,
\be
_2 F_1 \left( \frac{1}{2}+m-q,-n;2m+1;1 \right)=\frac{\Gamma (2m+1) 
\Gamma \left( m+\frac{1}{2}+q+n \right)}{\Gamma \left( m+\frac{1}{2} +q \right) \Gamma (2m+1+n)} \ .
\ee
Therefore there is no contribution from $\beta_0$. But the rest of the coefficients will contribute with $\binom{n-L}{q-1} i^{L+q-1}$. 
Now if we use that 
\be
\sum_\alpha{\binom{K-L}{\alpha} \frac{i^\alpha}{\left( \mu^- \right)^\alpha}}=\left( \frac{\mu^+}{\mu^-} \right)^{K-L} \ ,
\ee
together with $\mu_1^+ \mu_2^{---}-\mu_1^+ \mu_1^-=\mu_1^+ \left( \mu_2 -\mu_1 -i \right)$, we find that 
\ba
&\sum_{q=0}^{n-L}{\binom{n-L}{q} i^{L+q} \beta_q} \nonumber \\
&= \frac{1}{\mu_1^- \mu_2^-} \, \frac{-1}{\mu_1-\mu_2} \, 
\left[ \left( \frac{\mu_1^+}{\mu_1^-} \right)^{n-L+1} \left( \mu_2 -\mu_1 -i \right) 
+ \left( \frac{\mu_2^+}{\mu_2^-} \right)^{n-L+1} \left( \mu_2 -\mu_1 +i \right) \right] \ . 
\label{sumbeta}
\ea
If we now remove the $-L$ factor by extracting a $S$ matrix, expression (\ref{sumbeta}) 
cancels exactly the contribution from the sum of the $A$'s in (\ref{Asum}). Finally we conclude that
\ba
\granesperado{0}{\sigma^+_k \sigma^+_l}{\mu_1 \mu_2} & \!\! = \!\! & \frac{e^{i(p_1+p_2)l}}{\mu_1-\mu_2} \, \frac{1}{\mu_1^- \mu_2^-} 
\left[ e^{ip_1(k-l)}  \left(\mu_2-\mu_1+i \right) +e^{ip_2(k-l)} \left(\mu_2-\mu_1-i \right) \right] \notag \\
& = \!\! & \frac{1}{\mu_1-\mu_2} \, \frac{\mu_2-\mu_1+i}{\mu_1^- \mu_2^-} \left[   e^{i(p_1k+p_2 l)}  +e^{i(p_2k+ p_1 l)} \, S_{21} \right] 
\label{y>x} \ ,
\ea
which agrees with (\ref{y<x}), but with $k$ and $l$ exchanged because now we are in the case $k<l$.

We will end this section by normalizing properly the above correlation functions. Following the discussion in section~\ref{normalizationisues},
\be
\granesperado{0}{\sigma^+_k \sigma^+_l}{\mu_1 \mu_2}^{\text{ZF}} = \frac{-1}{\mu_1^- \mu_2^-} \left[   e^{i(p_1k+p_2 l)}  
+ e^{i(p_2k+p_1 l)} S_{21} \right] \ , 
\ee
On the other hand, the norm of the states in the ABA is given by (\ref{algebraicnorm}), while
\be	
\scal{\mu_1 ,\mu_2}{\mu_1 , \mu_2}^{\text{ZF}} = \frac{16\xi ^4 L^2}{\left( \mu_1^2 -\xi^2 \right) \left( \mu_2^2 -\xi^2 \right)} 
\left( 1 -\frac{2}{L} \cdot \frac{\left( \mu_1^2 +\mu_2^2 -2\xi^2 \right)}{\left[ (\mu_2 -\mu_1)^2 -4\xi^2 \right]} \right) \ .
\ee
Therefore, we conclude that
\begin{align} 
\left( \frac{\granesperado{0}{\sigma^+_k \sigma^+_l}{\mu_1 \mu_2}}{\sqrt{\scal{\mu_1 ,\mu_2}{\mu_1 , \mu_2}}} \right)^{ZF} 
& = \frac{e^{ip_1(k-\frac{1}{2})+ip_2 (l-\frac{1}{2})}  + e^{ip_2(k-\frac{1}{2})+i p_1 (l-\frac{1}{2})} S_{21} }{L} \nonumber \\
& \times \left( 1 -\frac{2}{L} \cdot \frac{\left( \mu_1^2 +\mu_2^2 -2\xi^2 \right)}{\left[ (\mu_2 -\mu_1)^2 -4\xi^2 \right]} \right)^{-1/2} \ .
\end{align}
Now, as in the case of the form factor calculated in the previous section, we can take into account the trace condition (\ref{trace}). 
When we replace the rapidities from equation~(\ref{rapidityroiban}) in these expressions, after some immediate algebra we obtain 
\be 
L \left( \frac{\granesperado{0}{\sigma^+_k \sigma^+_l}{\mu, -\mu}}{\sqrt{\scal{\mu ,-\mu}{\mu , -\mu}}} \right)^{ZF} = 2 \sqrt{\frac{L}{L-1}} \cos \left( \frac{(2|l-k|-1)\pi n}{L-1} \right) \ ,
\label{normalizedtwomagnon}
\ee
with $|l-k|\leq L-1$. This result extends the analysis in reference \cite{RV}, where this correlation function was calculated for the cases 
$l-k=1$ and $l-k=2$  (we have written the factor $L$ on the left hand side of (\ref{normalizedtwomagnon}) to follow conventions in there).

\subsubsection*{Higher number of magnons}

The method we have presented can still be applied to evaluate correlation functions with more than 2 magnons. The next easiest case is the correlation function with 1+3 magnons, that is, $\granesperado{\lambda}{\sigma^+_k \sigma^+_l}{\m _1 \m _2 \m _3}$. However we will see that this computation will require information about the $\granesperado{0}{\sigma^+_k \sigma^+_{l}  \sigma^+_{m}}{\m _1 \m _2 \m _3}$ correlation function. Hence we are going to begin the computation here by proving that statement.

We will start by using relation (\ref{sigmaplus}) on $\granesperado{\lambda}{\sigma^+_k \sigma^+_l}{\m _1 \m _2 \m _3}$,
\begin{multline}
	\langle \lambda |\sigma^+_k \sigma^+_l |\mu_1 \mu_2 \mu_3\rangle =\\
	 \granesperado{0}{C(\l )(A+D)^{k-1} (\xi) C(\xi) (A+D)^{n} (\xi) C(\xi) (A+D)^{L-l} (\xi) }{\mu_1 \mu_2 \mu_3} \ ,
\end{multline}
where as before $n=L+l-k-1$. 
The factor $(A+D)^{k-1}$ acts on $C(\l )$ to give $e^{-ip_\l (k-1)}$, and the factor $(A+D)^{L-l}$ 
acts on the three-magnon state to give $e^{-i(p_1+p_2+p_3)\cdot (L-l)}=e^{i(p_1+p_2+p_3)l}$, 
where in the last equality we have used the periodicity condition for the Bethe roots.
Therefore our main problem will be to find the correlation function
\be 
{\cal H}^L_{n}(\a)=\granesperado{0}{C(\l ) C(\xi+\alpha) (A+D)^n (\xi) C(\xi) B(\m _1) B(\m _2) B(\m _3)}{0} \ .
\ee
Following the procedure that we have developed in the previous section this can be done by relating ${\cal H}^L_{n+1}(\a)$ to ${\cal H}^L_{n}(\a)$. In order to do this 
let us start by introducing 
\be
{\cal H}^L_{n+1} (\l ,\a, \delta)=\lim_{\b\rightarrow \a} \granesperado{0}{C(\l ) C(\xi +\a) (A+D)(\xi+\beta) {\cal O} (\d)}{0} \ .
\ee
Now we just need to apply the commutation relations~(\ref{commCA}) and (\ref{commCD}) two times in each step to get
\begin{align}
	&{\cal H}^L_{n+1} (\l ,\a, \delta) = \lim_{\b \rightarrow \a} \Big{\{}\big[ 1+d(\xi+\b) \big] {\cal H}^L_{n} (\l ,\a, \delta) \nonumber \\
	&-\frac{i}{\l -\xi -\b} \left[ (d(\xi+\b)-1) {\cal H}^L_{n} (\l ,\a, \delta)-(d(\l)-1) {\cal H}^L_{n} (\xi +\b ,\a, \delta) \right] \nonumber \\
	&-\frac{i}{\a -\b} \left[ (d(\xi+\b) -1) {\cal H}^L_{n} (\l ,\a, \delta) -(d(\xi+\a)-1) {\cal H}^L_{n} (\l ,\b, \delta) \right] \nonumber \\
	&+\frac{i}{\a -\b} \ \frac{i}{\l -\xi -\b} \left[ (d( \xi+\b)+1) {\cal H}^L_{n} (\l ,\a, \delta) -(d(\l )+1) {\cal H}^L_{n} (\xi+\b ,\a, \delta) \right] \nonumber \\
	&-\frac{i}{\a -\b} \ \frac{i}{\l -\xi -\a} \left[ (d( \xi+\a)+1) {\cal H}^L_{n} (\l ,\b, \delta) -(d(\l )+1) {\cal H}^L_{n} (\xi+\b ,\a, \delta) \right] \Big{\}} \ . \label{Hpoles}
\end{align}
Taking the limit and applying the Bethe equation for the rapidity $\l$ we obtain
\begin{align}
	&{\cal H}^L_{n+1} (\l ,\a, \delta)=\Big( 1+d+i\partial d+\frac{\partial d-i(d-1)}{\l -\xi -\a}+\frac{d+1}{(\l-\xi-\a)^2} \Big) {\cal H}^L_{n} (\l ,\a, \delta) \notag \\
	&+\Big[ i(1-d)-\frac{d+1}{\l -\xi -\a} \Big] \dpartial{{\cal H}^L_{n} (\l ,\a, \delta)}{\a} -\frac{2}{(\l -\xi -\a)^2} {\cal H}^L_{n} (\xi+\a ,\a, \delta) \ ,
\label{Hn+1}
\end{align}
where again $d=d(\xi+\alpha)$ and $\partial d=\left. \dpartial{d}{\lambda} \right|_{\xi +\alpha}$. The next step of the calculation is a little bit more involved than in the previous cases because according to (\ref{Hn+1}) information about  
both functions ${\cal H}^L_{n} (\l ,\a, \delta)$ and ${\cal H}^L_{n+1} (\xi+\a ,\a, \delta)$ is now needed. 
This will turn the computation slightly more difficult but still manageable. For convenience in the expressions 
below we will define ${\cal H}^L_{n+1} (\a+\xi ,\a, \delta)=\hat{{\cal H}}^L_{n+1} (\a ,\a, \delta)$. This function $\hat{\cal H}$ has a nice interpretation because
\begin{align}
	\granesperado{0}{\sigma^+_k \sigma^+_{k+1}  \sigma^+_{k+n+2}}{\m _1 \m _2 \m _3} 
	&=\granesperado{0}{C(\xi ) C(\xi) (A+D)^{n} (\xi) C(\xi) (A+D)^{L-n-k-2} (\xi) }{\mu_1 \mu_2 \mu_3} \notag \\
	&=\hat{\cal H}^L_{n} e^{i(p_1+p_2+p_3)(n+k+2)} \ .
\end{align}
This proves our previous statement. When computing correlation functions having a magnon in the bra state some terms will have those magnons changed into an inhomogeneity as a consequence of the commutation relations, which makes difficult to perform computations.

We could think that, as form factors should satisfy the axioms presented in section \ref{bootstrap}, the crossing transformation will simplify these kind of computations as shown in equation~(\ref{formfactorcrossing}). However, this is not possible for our computation because the crossing condition requires to know the particle-antiparticle transformation, which is hidden in the perturbative expansion\footnote{From the point of view of the AdS/CFT correspondence, this can be seen as the degeneration of the torus that uniformized the magnon dispersion relation at weak-coupling when one of the periods becomes infinitely large, thus forbidding us the access to the crossing transformation.}.


\section[Correlation functions involving three operators]{$\granesperado{\lambda}{\sigma^+_k \sigma^+_l}{\m _1 \m _2 \m _3}$ and correlation functions involving three operators}
\sectionmark{Correlation functions involving three operators}
\label{C}

Extracting information about correlation functions becomes a challenge as the number of magnons increases. The method that we have developed along these sections can however still applied to evaluate correlations functions involving any number of magnons but it might require to trade higher number of magnons by higher number of operators, as seen in the previous section. In this section we are going to compute the correlation function $\granesperado{0}{\sigma^+_k \sigma^+_{k+1}  \sigma^+_{k+n+2}}{\m _1 \m _2 \m _3}$ and, with this information, finish the computation of $\granesperado{\lambda}{\sigma^+_k \sigma^+_l}{\m _1 \m _2 \m _3}$.

Our starting point is thus to find the recursive equation for $\hat{\cal H}$. This can be obtained setting $\l=\xi+\b$ in expression (\ref{Hpoles}) 
and taking the limit $\b \rightarrow \a$,
\begin{align}
&\hat{\cal H}^L_{n+1} (\a ,\a, \delta)=\lim _{\b\rightarrow\a}\frac{1}{\b-\a} \Big[ (1+d) 
\left. \dpartial{\hat{\cal H}^L_{n} (\l ,\a, \delta)}{\l} \right|_{\l=\a} \!\!
- (1+d) \left. \dpartial{\hat{\cal H}^L_{n} (\a ,\l, \delta)}{\l} \right|_{\l=\a} \Big] \notag \\
& + \big(1+d+2i\partial d-\frac{1}{2} \partial^2 d \big) \, \hat{\cal H}^L_{n} (\a ,\a, \delta) + \big[ 2i(1-d) +\partial d \big] 
\left. \dpartial{\hat{\cal H}^L_{n} (\l ,\a, \delta)}{\l} \right|_{\l=\a}  \notag \\
& + \frac{1+d}{2} \left. \dpartial[2]{\hat{\cal H}^L_{n} (\l ,\a, \delta)}{\l} \right|_{\l=\a}-(1+d) 
\left. \frac{\partial^2 \hat{\cal H}^L_{n} (\l _1 ,\l _2, \delta)}{\partial \l_1 \partial \l_2} \right|_{\substack{\l_{1} = \a \\ \l_{2} = \a}} \ . 
\label{c7}
\end{align}
Note that although the first term in this expression seems divergent, it vanishes because of the commutation of the $C$ operators, 
which makes the two derivatives equal. 
However, this method of calculating recursively $\hat{\cal H} (\a , \a , \d)$ is going to create more problem than it solves, 
because it will imply calculating the recurrence equation of derivative of $\hat{\cal H} (\l ,\m)$ with respect to either the first or the second argument. 
Therefore we are going to give the recursion relation of $\hat{\cal H} (\b , \a , \d)$ but without taking the limit $\b\rightarrow \a$. 
To obtain this recurrence relation we only need to substitute $\l=\xi +\b$ in equation~(\ref{Hn+1}), but {\em without} imposing $d(\l)=1$, as now it is not a solution of the BAE,
\begin{align}
& \hat{\cal H}^L_{n+1} (\b ,\a, \delta)=\Big( 1+d+i\partial d+\frac{\partial d-i(d-1)}{\b -\a}+\frac{d+1}{(\b-\a)^2} \Big) \hat{\cal H}^L_{n} (\b ,\a, \delta) \notag \\
& +\Big[ i(1-d)-\frac{d+1}{\b -\a} \Big] \dpartial{\hat{\cal H}^L_{n} (\b ,\a, \delta)}{\a} +\left[ \frac{i(d'-1)}{\b-\a} -\frac{d'+1}{(\b -\a)^2} \right] 
\lim_{\g \rightarrow \a} \hat{\cal H}^L_{n} (\g ,\a, \delta) \ ,
\label{hatHn+1}
\end{align}
where $d'=d(\xi +\b)$. Note that if we take $\b \rightarrow \a$ equation (\ref{hatHn+1}) gives (\ref{c7}). Now in the recurrence relation we need to include 
$\lim_{\b \rightarrow \a} \hat{\cal H}^L_{n} (\b ,\a, \delta)$, but this quantity is obviously known once we know $\hat{\cal H}^L_{n} (\b ,\a, \delta)$.

We also  need a recurrence equation for the derivatives. For the case of ${\cal H}_{n}$ we have
\begin{align}
	&{\cal D}^n {\cal H}^L_{m+1} (\l ,\a, \delta)=(1+d) {\cal D}^n {\cal H}^L_{m} -\frac{i}{\l -\xi -\a} \left[ (d-1) {\cal D}^n {\cal H}^L_{m} \right] \notag \\
	&+\frac{i}{n+1} (\partial^{n+1} d) {\cal H}^L_{m}+(1-d)\frac{i}{n+1} {\cal D}^{n+1} {\cal H}^L_{m} \notag\\
	&+\sum_{k=0}^n \sum_{l=0}^{k+1}{\frac{n!}{(n-k)! (k+1-l)!} \frac{1}{(\l -\xi -\a)^{l+1}} \left[ \partial^{k+1-l} (d+1) {\cal D}^{n-k} {\cal H}^L_{m} \right.} \notag \\ 
	&\left. -(d(\l )+1) {\cal D}^{n-k}_1 {\cal D}^{k+1-l}_2 \hat{\cal H}^L_{m} \right] \notag \\
	&-\sum_{k=0}^n{\sum_{l=0}^{n-k}{ \frac{n!}{(k+1)! (n-l-k)!} \frac{1}{(\l -\xi -\a)^{l+1}} \left[ \partial^{n-k-l}(d+1) {\cal D}^{k+1} {\cal H}^L_{m} \right.}} \notag \\
	&\left. -(d(\l) +1) {\cal D}^{n-k-l}_1 {\cal D}^{k+1}_2 \hat{\cal H}^L_{m} \right] \ .
\end{align}
The last two sums cancel themselves except for the terms with $k=n$. Therefore
\begin{align}
	&{\cal D}^n {\cal H}^L_{m+1} (\l ,\a, \delta)=(1+d) {\cal D}^n {\cal H}^L_{m} -\frac{i}{\l -\xi -\a}  (d-1) {\cal D}^n {\cal H}^L_{m} \notag \\
	&+\frac{i}{n+1} (\partial^{n+1} d) {\cal H}^L_{m}+(1-d)\frac{i}{n+1} {\cal D}^{n+1} {\cal H}^L_{m} \notag\\
	&+\sum_{l=0}^{n+1}{\frac{n!}{(n+1-l)!} \frac{1}{(\l -\xi -\a)^{l+1}} \left[ \partial^{n+1-l} (d+1) {\cal H}^L_{m}  - 2 \, {\cal D}^{n+1-l} \hat{\cal H}^L_{m} \right]} \notag \\
	&-\frac{1}{n+1} \frac{1}{(\l -\xi -\a)} \left[ (d+1) {\cal D}^{n+1} {\cal H}^L_{m} - 2 \, {\cal D}^{n+1} \hat{\cal H}^L_{m} \right] \ ,
\end{align}
where we have used that $d(\l)=1$. In a similar way we can obtain an expression for the derivatives of $\hat{\cal H}$,
\begin{align}
	&{\cal D}^n \hat{\cal H}^L_{m+1} (\b ,\a, \delta)=(1+d) {\cal D}^n \hat{\cal H}^L_{m} -\frac{i}{\b-\a}  (d-1) {\cal D}^n \hat{\cal H}^L_{m} \notag \\
	&+\frac{i}{n+1} (\partial^{n+1} d) \hat{\cal H}^L_{m}+(1-d)\frac{i}{n+1} {\cal D}^{n+1} \hat{\cal H}^L_{m} +\frac{i}{\b -\a} (d' -1) 
	\lim_{\g \rightarrow \a} \dpartial[n]{\hat{\cal H}^L_{n} (\g ,\a, \delta)}{\a} \notag\\
	&+\sum_{l=0}^{n+1}{\frac{n!}{(n+1-l)!} \frac{1}{(\b -\a)^{l+1}} \left[ \partial^{n+1-l} (d+1) \hat{\cal H}^L_{m}  - (d'+1)\, 
	\lim_{\g \rightarrow \a} \dpartial[{n+1-l}]{\hat{\cal H}^L_{m} (\g ,\a ,\d)}{\a} \right]} \notag \\
	&-\frac{1}{n+1} \frac{1}{(\b -\a)} \left[ (d+1) {\cal D}^{n+1} \hat{\cal H}^L_{m} - (d'+1) \, \lim_{\g \rightarrow \a} 
	\dpartial[{n+1}]{\hat{\cal H}^L_{m} (\g ,\a ,\d)}{\a} \right] \ .
\end{align}

At this point the problem is, at least formally, solved. We have found the recursion relation for $\hat{\cal H}$ and its derivatives, 
with $\granesperado{0}{C(\xi +\b) C(\xi+\a) (\xi) C(\xi)}{\mu_1 \mu_2 \mu_3} =\hat{\cal H}^L_{0} (\b , \a)$ as the initial condition. 
These functions can then be substituted in the recursion relation for $\cal H$ and thus we can obtain the desired correlation function. 
However, we are not going to present the general form for the correlation function ${\cal H}^L_n$ as function of ${\cal H}^L_0$, $\hat{\cal H}^L_0$ 
and their derivatives because, although straightforward, it becomes rather lengthy. This is because when we substitute 
the expression for the derivatives the recursion relations turn to depend on all the ${\cal H}_i$ with $0\leq i\leq n$, even after taking the limit $\a \rightarrow 0$. Instead we can present the case of correlation functions with $n$ small, 
to exhibit the nested procedure needed to write the result in terms of the initial functions ${\cal H}^L_0$ and $\hat{\cal H}^L_0$. 
In particular we are going to consider the first three functions, with $n=1$, $n=2$ and $n=3$. Thus we can safely assume that $n<L-1$ 
so that all the $d$ and $\partial^k d$ factors can be set to zero in the limit $\a \rightarrow 0$. The first of these correlation functions is given by
\be 
{\cal H}^L_{1} = \big( 1+ i c(\l) + c(\l)^2 \big) \, {\cal H}^L_0 + \big( i-c(\l) \big) \left. \dpartial{{\cal H}^L_0 (\l ,\a)}{\a} \right| _{\a =0} 
- 2 c(\l)^2 \, \hat{\cal H}^L_0 \ ,
\ee
where for convenience we have defined $c(\l) = 1/(\l-\xi)$. For simplicity, if no arguments of this functions are given, ${\cal H}^L (\l, 0) $ 
and $\hat{\cal H}^L  (0, 0)$ must be understood. The last step of the computation reduces to calculating some initial conditions, which now are
\begin{align}
	{\cal H}^L_0 (\l, \a) &=\granesperado{0}{C(\l ) C(\xi+\alpha) C(\xi) B(\m _1) B(\m _2) B(\m _3)}{0} \ , \\
	\hat{\cal H}^L_0  (\a, \b)&= {\cal H}^L_0 (\xi+\a, \b) \ .
\end{align}
These functions can be easily computed using equations~(\ref{howtoscalarproduct}). However we are not going to present the explicit expression 
for these scalar products because of its length and because we want to show the way to solve the recurrence relation rather 
than obtaining the explicit value of the correlation function.

The functional dependence of ${\cal H}_1$ on ${\cal H}_0$ is repeated for a given value of $n$ and the lower correlator. That is, 
in the limit $\a \rightarrow 0$ the recurrence relation for ${\cal H}^L_{n+1}$ is given by
\be 
{\cal H}^L_{n+1} = \big( 1+ i c(\l) + c(\l)^2 \big) {\cal H}^L_n + \big( i-c(\l) \big) {\cal D} {\cal H}^L_n - 2 c(\l)^2 \hat{\cal H}_n \, \ .
\ee
Therefore for the second correlation function we have
\be
{\cal H}^L_{2} = \big( 1+ i c(\l) + c(\l)^2 \big) {\cal H}^L_1 + \big( i-c(\l) \big) {\cal D} {\cal H}^L_1 - 2 c(\l)^2 \hat{\cal H}_1 \ .
\label{c16}
\ee
As we already know ${\cal H}^L_1$, it only remains to find the other two functions entering (\ref{c16}). This can be done using the previously obtained equations. We get
\begin{align}
{\cal D} {\cal H}^L_{1}&=c(\l)^3 {\cal H}^L_0+(1+ic(\l))  \left. \dpartial{{\cal H}^L_0 (\l ,\a)}{\a} \right| _{\a =0}+\frac{i(1+ic(\l))}{2}  \left. \dpartial[2]{{\cal H}^L_0 (\l ,\a)}{\a} \right| _{\a =0} \notag \\
&-2c(\l)^3 \hat{\cal H}^L_0 -2c(\l)^2 \left. \dpartial{\hat{\cal H}^L_0 (0 ,\a)}{\a} \right| _{\a =0} \ , \\
\hat{\cal H}^L_1 (\b , 0) &=  \Big( 1+\frac{i}{\b}+\frac{1}{\b^2} \Big) \hat{\cal H}^L_0 (\b , 0) +\Big[ i-\frac{1}{\b } \Big] 
\left. \dpartial{\hat{\cal H}^L_0 (\b ,\a)}{\a} \right|_{\a=0}-\left[ \frac{i}{\b} +\frac{1}{\b^2} \right] \hat{\cal H}^L_{0} \ , \\
\hat{\cal H}^L_1 &=  \hat{\cal H}^L_{0} +2i  \left. \dpartial{\hat{\cal H}^L_0 (0 ,\a)}{\a} \right| _{\a =0} +\frac{1}{2} 
\left. \dpartial[2]{\hat{\cal H}^L_0 (0 ,\a)}{\a} \right| _{\a =0}-\left. \frac{ \partial ^2 \hat{\cal H}^L_0 (\a ,\b)}{\partial \a \partial \b} \right| _{\substack{\a =0 \\ \b=0}} \ ,
\end{align}
which reduce again to some dependence on the initial conditions we have described before.

An identical computation can be done for ${\cal H}^L_{3}$,
\be
{\cal H}^L_{3} = \big( 1+ i c(\l) + c(\l)^2 \big) {\cal H}^L_2 + \big( i-c(\l) \big) {\cal D} {\cal H}^L_2 - 2 c(\l)^2 \hat{\cal H}_2 \ .
\ee
Now, besides ${\cal H}^L_{2}$, that has been calculated just above this lines, we need
\begin{align}
{\cal D} {\cal H}^L_{2}&=c(\l)^3 {\cal H}^L_1+(1+ic(\l))  {\cal D} {\cal H}^L_{1}+\frac{i(1+ic(\l))}{2}  {\cal D}^ 2 {\cal H}^L_{1} \notag \\
&-2c(\l)^2 \Big( c(\l) \hat{\cal H}^L_1 +{\cal D} \hat{\cal H}^L_{1} \Big) \ , \\
{\cal D}^2 {\cal H}^L_{1}&=2 c(\l)^4 {\cal H}^L_0+(1+ic(\l))  \left. \dpartial[2]{{\cal H}^L_0 (\l ,\a)}{\a} \right| _{\a =0}+\frac{i(1+ic(\l))}{3}  
\left. \dpartial[3]{{\cal H}^L_0 (\l ,\a)}{\a} \right| _{\a =0} \notag \\
&-4c(\l)^4  \hat{\cal H}^L_0 -4 c(\l)^3 \left. \dpartial{\hat{\cal H}^L_0 (0 ,\a)}{\a} \right| _{\a =0} -2c(\l)^2 \left. \dpartial[2]{\hat{\cal H}^L_0 (0 ,\a)}{\a} \right| _{\a =0} \ , \displaybreak \\
{\cal D} \hat{\cal H}^L_{1} &=\left. \dpartial{\hat{\cal H}^L_0 (0 ,\a)}{\a} \right| _{\a =0} 
+ \frac{i}{2}\left. \dpartial[2]{\hat{\cal H}^L_0 (0 ,\a)}{\a} \right| _{\a =0}+i\left. \frac{ \partial ^2 \hat{\cal H}^L_0 (\a ,\b)}{\partial \a \partial \b} 
\right| _{\substack{\a =0 \\ \b=0}} \notag \\
& + \frac{1}{3!} \left. \dpartial[3]{\hat{\cal H}^L_0 (0 ,\a)}{\a} \right| _{\a =0} 
- \frac{1}{2} \left. \frac{ \partial ^3 \hat{\cal H}^L_0 (\a ,\b)}{\partial \a \partial \b^ 2} \right| _{\substack{\a =0 \\ \b=0}} \ , \\
\hat{\cal H}^L_{2} &=\hat{\cal H}^L_{0} +4i \left. \dpartial{\hat{\cal H}^L_0 (0 ,\a)}{\a} \right| _{\a =0} -4 
\left. \frac{ \partial ^2 \hat{\cal H}^L_0 (\a ,\b)}{\partial \a \partial \b} \right| _{\substack{\a =0 \\ \b=0}}+\frac{i}{2}\left. 
\dpartial[3]{\hat{\cal H}^L_0 (0 ,\a)}{\a} \right| _{\a =0} \notag \\
& - \frac{3i}{2} \left. \frac{ \partial ^3 \hat{\cal H}^L_0 (\a ,\b)}{\partial \a \partial \b^2} \right| _{\substack{\a =0 \\ \b=0}} 
-\frac{1}{3!} \left. \frac{ \partial ^4 \hat{\cal H}^L_0 (\a ,\b)}{\partial \a \partial \b^3} \right| _{\substack{\a =0 \\ \b=0}} +\frac{1}{2!^2} 
\left. \frac{ \partial ^4 \hat{\cal H}^L_0 (\a ,\b)}{\partial \a^2 \partial \b^2} \right| _{\substack{\a =0 \\ \b=0}} \ .
\end{align}

The cases with higher values of $n$ can be obtained along similar lines. 

To conclude our analysis we will brief comment on the calculation of correlation functions $\granesperado{0}{\s^+_k \s^+_l \s^+_m}{\{ \m \}}$, 
with general values of $k$, $l $ and $m$. The initial condition we would obtain from the determinant expression of the on-shell-off-shell scalar product (\ref{scalarproduct}) would be $\granesperado{0}{\s^+_{l-1} \s^+_{l} \s^+_{l+1}}{\{ \m \}}$. With the procedure explained in this section we can separate the lattice point in which the second and the third operator act, obtaining $\granesperado{0}{\s^+_{l-1} \s^+_{l} \s^+_{m}}{\{ \m \}}$. Note that in this case the value of $n$ in $\hat{\cal H}^L_n$ will be proportional to the separation between $l$ and $m$. However it still remains to separate the first and second operator. This last step can be solved using the tools from section~\ref{0sigmasigmamumu}, as now the problem only involves commuting a set of monodromy matrices through the first $C$ operator.


\section{The long-range Bethe ansatz}

In this section we are going to apply the method that we have developed along this paper to the long-range BDS spin chain \cite{BDS}. 
This can be done quite easily because in most of our previous expressions we have kept general the homogeneous point. As we saw in section~\ref{BDSspinchain}, 
the long-range BDS spin chain can be mapped into an inhomogeneous short-range spin chain, with the inhomogeneities located at
\be 
\xi_n=\frac{i}{2} +\sqrt{2} g\cos \left( \frac{(2n-1)\pi}{2L} \right) \equiv \xi + g \k_n \ .
\ee
Therefore it is rather simple to extend all computations above to the inhomogeneous short-range version of the long-range BDS Bethe ansatz. Let us start with the computations done in section~\ref{normalizationisues} for the XXX spin chain. The normalization factor 
for the operator $B(\lambda)$ is straightforward to compute given the expressions from that section,
\be
B(\lambda)=\sum_{n=1}^L{\frac{is^-_n}{\lambda-\xi_n} \left( \prod_{l=1}^n{ \frac{\lambda-\xi_l}{\lambda-\xi_l+i}} \right) } +\dots \ .
\ee
We conclude therefore that in the inhomogeneous Bethe ansatz the difference in normalization between the ABA and the CBA depends 
on the site where the spin operator acts. An analogous result follows for the operator $C(\lambda)$.

Another example of computation that we can readily extend to the BDS Bethe ansatz is the calculation of scalar products. This is immediate because the solution 
to the inverse scattering problem in expressions~(\ref{sigmaplus})-(\ref{sigmaz}) is valid for an inhomogeneous spin chain. 
Furthermore equations~(\ref{scalarproduct}) and (\ref{howtoscalarproduct}) can be directly used without modifications. An immediate example is the calculation 
of the form factor of the single-magnon state, 
\begin{equation}
\granesperado{0}{\sigma^+_k}{\lambda}=\frac{i}{\l-\xi-g \k_k} \prod_{j=1}^k{\frac{\l - \xi-g \k_j}{\l+\xi - g \k_j}} \ ,
\end{equation}
which as in the case of the homogeneous spin chain should also be divided by the norm
\begin{equation}
\sqrt{\scal{\l}{\l}}=\sqrt{i \dpartial{d}{\l}} = i \sqrt{d(\lambda ) \sum_{m=1}^L{\frac{1}{(\l -\xi - g \k_m)(\l +\xi-g \k_m)}}} \ .
\end{equation}
The limit $g\rightarrow 0$ reduces to the result in section~\ref{normalizationisues}. In an identical way we can extend the analysis to the correlation functions obtained 
in section~\ref{evaluationcorrelationsection}. For instance, 
\begin{align}
& \granesperado{0}{\sigma^+_k \sigma^+_{k+1}}{\mu_1 \mu_2}_Z = \left[ \frac{\m_1+\xi-g\k_k}{\m_1-\xi-g \k_{k+1}} 
\frac{\m_2+\xi-g \k_{k+1}}{\m_2-\xi-g \k_k}-(\m_2 \leftrightarrow \m_1 )  \right] \nonumber \\ 
& \times \frac {1}{\big[ g(\k_{k+1}-\k_k)(\m_1 -\m_2) \big]}\prod_{j=1}^{k}{\frac{\m_1 -\xi -g \k_j}{\m_1 +\xi -g \k_j} 
\frac{\m_2 -\xi -g \k_j}{\m_2 +\xi - g \k_j}} \ .
\end{align}
The norm is now given by
\begin{align}
	\sqrt{\scal{\mu_1 \mu_2}{\mu_1 \mu_2}_Z} & = \frac{2}{(\m_2 -\m_1)^2 -4\xi^2} \sum_j {\left[ \frac{1}{(\m_1-g \k_j)^2-4\xi^2} +\frac{1}{(\m_2-g \k_j)^2-4\xi^2} \right]} \nonumber \\
	& - \sum_j{\sum_k{\frac{1}{\big[ (\m_1-g \k_j)^2-4\xi^2 \big] \big[ ( \m_2-g \k_k )^2-4\xi^2 \big]}}} \ . 
\end{align}

We should stress that an important difference when comparing with the homogeneous XXX Heisenberg spin chain in the previous sections is that 
because all the inhomogeneities are different the commutation of factors $(A+D)$ does not lead now to any of the apparent divergences we had to deal with in previous sections. Therefore we do not have 
to make use of the procedure we have developed along this chapter. For instance, the correlation function 
$\granesperado{0}{\sigma^+_k \sigma^+_{k+2}}{\mu_1 \mu_2}$ can be calculated by direct use of the commutation relations (\ref{commCB}), 
\begin{align}
& \granesperado{0}{\sigma^+_k \sigma^+_{k+2}}{\mu_1 \mu_2} = \granesperado{0}{C(\xi_k) (A+D) (\xi_{k+1}) C(\xi_{k+2}) B(\m_1) B(\m_2)}{0} p(k+2) \notag \\
& = \left[ \frac{\xi_{k}-\xi_{k+1}+i}{\xi_{k}-\xi_{k+1}} {\cal G}^{L}_0(k,k+2) + \frac{i}{\xi_{k+1}-\xi_k} {\cal G}^{L}_0(k+1,k+2) \right] e^{i\mathscr{P}(k+2)} = \\
&= \left[ {\cal G}^{L}_0(k,k+2) +\frac{i}{\xi_{k+1} - \xi_k} \left[ {\cal G}^{L}_0(k+,k+2) -{\cal G}^{L}_0(k,k+2) \right] \right] e^{i\mathscr{P}(k+2)} \ ,
\label{k2}
\end{align}
where the correlation function ${\cal G}^{L}_{0}(k,l)=\langle 0 | C(\xi_k) C(\xi_l) | \mu_1 \mu_2 \rangle$ can be computed using 
expressions~(\ref{scalarproduct}) and (\ref{howtoscalarproduct}) for the scalar product. The factor $e^{i\mathscr{P}(k+2)}$, given by
\be
e^{i\mathscr{P}(l)}=\prod_{j=1}^{l}{\frac{\m_1 -\xi -g \k_j}{\m_1 +\xi -g \k_j} \frac{\m_2 -\xi -g \k_j}{\m_2 +\xi -g \k_j}} \ ,
\ee
collects the contribution from the momenta, as it is easy to see that in the limit $g\rightarrow 0$ becomes the $e^{i(p_{\mu_1} +p_{\mu_2} ) l}$ factor. We can in fact extend rather easily expression (\ref{k2}) to the case where the spin operators are located at arbitrary sites, 
$\langle 0 | \sigma^+_k \sigma^+_{l} | \mu_1 \mu_2 \rangle$. 
As all factors $(A+D)$ have different arguments, they can be trivially commuted. Therefore
the correlation function must be invariant under exchange of the inhomogeneities, 
except for the factors coming from the correlators ${\cal G}^{L}_0(k,l)$. We find
\begin{align}
&\granesperado{0}{\sigma^+_k \sigma^+_{l}}{\mu_1 \mu_2} = \langle 0 \, |C(\xi_k) 
\prod_{j=k+1}^{l-1}{(A+D) (\xi_{j})} C(\xi_{l}) B(\m_1) B(\m_2) \, |0 \rangle \, e^{i\mathscr{P}(l)} = \notag\\
&= \left[ \, \prod_{j=k+1}^{l-1}{\frac{\xi_{k}-\xi_{j}+i}{\xi_{k}-\xi_{j}}} {\cal G}^{L}_0(k,l) 
+ \frac{-i}{\xi_{k}-\xi_{k+1}} \prod_{j=k+2}^{l-1}{\frac{\xi_{k+1}-\xi_{j}+i}{\xi_{k+1}-\xi_{j}}} {\cal G}^{L}_0(k+1,l) 
\right. \notag \\
& \left. + \left( \frac{\xi_{k+2}-\xi_{k+1}+i}{\xi_{k+2}-\xi_{k+1}} \right)
\frac{-i}{\xi_{k}-\xi_{k+2}} \prod_{j=k+3}^{l-1}{\frac{\xi_{k+2}-\xi_{j}+i}{\xi_{k+2}-\xi_{j}}} {\cal G}^{L}_0(k+2,l) + \dots \right] e^{i\mathscr{P}(l)} \ ,
\end{align}
or using the recursion relations
\begin{align}
&\granesperado{0}{\sigma^+_k \sigma^+_{l}}{\mu_1 \mu_2}
= \left[ \, \prod_{m=k+1}^{l-1}{\frac{\xi_{k}-\xi_{m}+i}{\xi_{k}-\xi_{m}}} {\cal G}^{L}_0(k,l)  \: + \right. \notag \\
& \left. +\sum_{m=k+1}^{l-1}{ \left( \prod_{n=k+1}^{m-1}{\frac{\xi_{k}-\xi_{n}+i}{\xi_{k}-\xi_{n}}} \right) \frac{-i}{\xi_{k}-\xi_{m}} 
\left( \prod_{n=k+1}^{l-1}{\frac{\xi_{m}-\xi_{n}+i}{\xi_{m}-\xi_{n}}} \right) {\cal G}^{L}_0(m,l)} \right] e^{i\mathscr{P}(l)} \ .
\end{align}
This correlation function is the all-loop generalization of the one computed in subsection~\ref{0sigmasigmamumu}. This computation was simpler than the one for the one-loop case because we did not have to remove the apparent singularities from the commutation relations, as the remaining loop corrections separate the inhomogeneities of the spin chain between themselves and from the XXX point. The one-loop results can be recovered by taking the $g\rightarrow 0$ limit, transforming this equation into equation~(\ref{GLn}) after properly dealing with the residues.

A similar discussion holds in the case of higher order correlation functions, involving a larger number of magnons, but we will not present the resulting expressions in here.

\chapter{Tailoring and hexagon form factors}
\begin{chapquote}{Albert Einstein, \textit{The Special and the General Theory-A Clear Explanation that Anyone Can understand \cite{0517029618}}.}
	I adhered scrupulously to the precept of that brilliant theoretical physicist L. Boltzmann, according to whom matters of elegance ought to be left to the tailor and to the cobbler.
\end{chapquote}

In the rest of this part we are going to focus on the study of three-point correlation functions. The first section will be devoted to the weak-coupling method for computing three-point functions as weak-coupling using the spin chain language called \emph{Tailoring} \cite{EscobedoTailoring}. The main idea behind this method is to ``cut'' the operators/spin chains in two, perform some operation to one of the halves (``flip'') and compute scalar products between two half spin chains from different operators (``sew''). A proposal for an all-loop version of the tailoring method, called \emph{hexagon proposal} \cite{BKVhexagon,2016arXiv161105577F,BKV2,Basso_Komatsu_2017}, will be presented in the second section. In the last section a rewriting of this proposal using the Zamolodchikov-Faddeev algebra \cite{Arutyunov_ZFalgebra} will be presented. The results presented in the last section will be collected in \cite{Algebraichexagon}.

\section{Tailoring method}

The tailoring method was proposed in a series of four papers \cite{EscobedoTailoring,TailoringII,TailoringIII,TailoringIV} as a method for computing the structure constants $C_{ijk}$ defined as the non-trivial part of the correlation function of three local operators
\begin{equation}
	\esperado{\O_i (x_i) \O_j (x_j) \O_k (x_k)}=\frac{\sqrt{\mathcal{N}_i \mathcal{N}_j \mathcal{N}_k} C_{ijk}}{|x_{ij}|^{\Delta_i +\Delta_j -\Delta_k} |x_{jk}|^{-\Delta_i +\Delta_j +\Delta_k} |x_{ki}|^{\Delta_i -\Delta_j +\Delta_k}} \ .
\end{equation}
The structure constants have in the planar limit a perturbative expansion in the 't Hooft coupling of the form
\begin{equation}
	N_c C_{ijk}=c_{ijk}^{(0)} +\lambda c_{ijk}^{(1)}+\lambda^2 c_{ijk}^{(2)}+\dots \ ,
\end{equation}
where $N_c$ is the number of colors. Here we are only going to discuss the zeroth-order term $c_{ijk}^{(0)}$ in the $SU(2)$ sector for non-extremal cases, that is, when all bridges $l_{ij}=\frac{L_i +L_j-L_k}{2}$ are strictly positive\footnote{These lengths are usually called \emph{bridge}, because $l_{ij}$ is the number of Wick contractions between operator $\O_i$ and $\O_j$. We must impose this positivity condition because, for non-extremal correlators, the operator mixing between single-trace operators and double-trace operators is suppressed by a color factor $\frac{1}{N_c}$ and does not need to be considered.}, as was proposed in \cite{EscobedoTailoring}. Each operator is a single trace operator made out of products of two complex scalar and is mapped to a spin chain state. As explained before, in the $SU(2)$ sector these scalars are usually denoted $Z$ and $X$, where are going to consider the first one as the vacuum and the second one as the excitation. It is important to notice that the only setup of three operators that is fully contained in the sector we are interested in is the following one: $\O_1$ is formed by $Z$ and $X$ fields, $\O_2$ is formed by $\bar{Z}$ and $\bar{X}$ fields and $\O_3$ is formed by $Z$ and $\bar{X}$ fields. The scalar products of spin chains are defined by the basic rules $\scal{Z}{Z}=\scal{\bar{Z}}{\bar{Z}}=\scal{ZX}{ZX}=1$ and $\scal{\bar{Z}}{Z}=\scal{ZX}{XZ}=0$. This will have important consequences when we proceed to compute the scalar products we are interested in.

The method to construct the zeroth-order term of the structure constant can be summarised in the following steps:
\begin{enumerate}
	\item Fixed a cyclic ordering of the three closed chains, we will break the spin chain associated to operator $\mc{O}_i$ into left and right open subchains of lenghts $l_{ij}=\frac{L_i +L_j-L_k}{2}$ and $l_{ik}=\frac{L_i -L_j+L_k}{2}$. We will do similarly to the other two. We will express the closed chain state as an entangled state of the left an right subchains. This step is called \emph{cutting}, as the basic idea is to divide a generic state with $M$ magnons into a \emph{left} subchain of length $l$ and a \emph{right} subchain of length $L-l$. The original state can be represented as an entangled state in the tensor product of both subchains,
\begin{displaymath}
	\ket{\Psi}=\sum_{k=0}^{\min\{ M,l\}} \sum_{\substack{1\leq n_1 < \dots < n_k \leq l \\ l<n_{k+1} <\dots < n_M}}{\psi (n_1 , \dots ,n_M ) \ket{n_1 , \dots , n_k} \otimes \ket{n_{k+1} -l , \dots , n_M -l} } \ .
\end{displaymath}
Here the first sum represents how to distribute the magnons between left and right chains and has $\binom{L}{M}$ terms. A Bethe state has the property that, after breaking it, the subchain states still have the same Bethe state form. This is a consequence of the Bethe states being the eigenstates of a local Hamiltonian, so magnons propagate in a local way and do not know what happens far away. Hence a Bethe states breaks in two as
\begin{equation}
	\ket{\{u_i\}}=\sum_{\alpha \cup \bar{\alpha} =\{ u_i \}} H(\alpha , \bar{\alpha}) \ket{\alpha}_l \otimes \ket{\bar{\alpha}}_r \ ,
\end{equation}
where the sum is over all $2^M$ possible ways of splitting the rapidities into two groups $\alpha$ and $\bar{\alpha}$. This is a simplification with respect to the general case as $2^M < \binom{L}{M}$ if $L\gg M$. The splitting factor $H(\alpha , \bar{\alpha})$ depend on the normalization of the states. In particular, with the choices commented on section~\ref{normalizationisues}, we have
\begin{align}
	H^{\text{c}} (\alpha , \bar{\alpha}) &=\frac{a_l^{\bar{\alpha}}}{d_l^{\bar{\alpha}}} \ \frac{f^{\alpha \bar{\alpha}} f^{\bar{\alpha} \bar{\alpha}}_< f^{\alpha \alpha}_<}{f^{\{u\} \{u\}}_<} \ , & H^{\text{a}}=f^{\alpha \bar{\alpha}} d_{L-l}^{\alpha} a_l^{\bar{\alpha}} \ ,
\end{align}
where, using the notation from the previous chapter, c and a mean the normalization from the CBA and ABA respectively and we have used the following notation to simplify the different kinds of products
\begin{align}
	F^\alpha &=\prod_{u_j\in \alpha} F(u_j) & F^{\alpha \bar{\alpha}} &=\prod_{\substack{u_i \in \alpha \\ v_j\in \bar{\alpha}}} F(u_i-v_j) \ , & F^{\alpha \alpha}_< &=\prod_{\substack{u_i , u_j\in \alpha \\ i<j}} F(u_i-u_j) \ .
\end{align}
The subindex in the $a$ and $d$ means the length of the chain in which they are defined.
	
	\item We will perform a Wick contraction of the left subchain associated with the operator $\mc{O}_i$ with the flipped version of the right subchain associated with the operator $\mc{O}_{i-1}$. The idea behind the \emph{flipping} operation is to transform a ket into a bra state in such way that the Wick contraction of two ket states gives the same answer as the scalar product of the flipped state with the other ket remaining unchanged. Of course the result should not depend on which of the two subchains we choose to flip. To be consistent we choose to always flip the right subchain.
This process is not the usual conjugation, as this one flips the order of the field and their charges while the flipping operator should only do the first operation. For example, in the $SU(2)$ fields language they act as
\begin{align*}
	\dagger &\text{ operation:} & &e^{i\phi} \ket{XZXZZ} \rightarrow \bra{XZXZZ} e^{-i\phi}  \ ,\\
	\mathcal{F} &\text{ operation:} & &e^{i\phi} \ket{XZXZZ} \rightarrow \bra{\bar{Z} \bar{Z} \bar{X} \bar{Z} \bar{X} } e^{+i\phi} \ .
\end{align*}
	If we rewrite the action in the space basis we have
\begin{align*}
	\dagger &\text{ operation: } & &\phi (n_1 , \dots ,n_M) \ket{n_1 , \dots ,n_M} \rightarrow \phi^\dagger (n_1 , \dots ,n_M) \bra{n_1 , \dots ,n_M} \ , \\
	\mathcal{F} &\text{ operation: } & &\phi (n_1 , \dots ,n_M) \ket{n_1 , \dots ,n_M} \rightarrow\\
	& & &\hspace{0.096\textwidth} \rightarrow \phi(n_1 , \dots ,n_M) \bra{L+1-n_M , \dots ,L+1-n_1} \hat{C} \ ,
\end{align*}
where $\hat{C}$ stands for charge conjugation. For a Bethe state with two magnons the flipping operation will act as
\begin{align}
	\mathcal{F} \ket{\{u_1 , u_2\}}^{\text{c}} &=\mathcal{F} \sum_{\substack{x,y=1\\x<y}}^L \left( e^{i(p_1 x + p_2 y)}+S_{21} e^{i (p_1 y +p_2 x)} \right) \ket{x,y} \notag \\
	&= e^{i(L+1)(p_1 +p_2)} S_{21} \, \mathop{\vphantom{\ket{0}}}\nolimits^{\text{c}} \!\bra{\{u^*_1 , u^*_2\}} \hat{C} \ ,
\end{align}
so applied to the split chain we are interested in we have
\begin{equation}
	(\mathbb{I}\otimes \mathcal{F}) \ket{\{u_i\}}^{\text{c}}=\sum_{\alpha \cup \bar{\alpha}=\{u_i\}} \frac{a_L^{\bar{\alpha}}}{d_L^{\bar{\alpha}}} \ \frac{g^{\bar{\alpha}-\frac{i}{2}}}{g^{\bar{\alpha}+\frac{i}{2}}} \ \frac{f^{\alpha \bar{\alpha}} f^{\bar{\alpha} \bar{\alpha}}_> f^{\alpha \alpha}_<}{f^{\{u\} \{u\}}_<} \ket{\alpha}^{\text{c}}_l \otimes \, \mathop{\vphantom{\ket{0}}}\nolimits^{\text{c}}_{r} \! \bra{\bar{\alpha}^*} \hat{C} \ ,
\end{equation}
where we have used that $S_{21}=\frac{f_{21}}{f_{12}}$. We refer to \cite{EscobedoTailoring} for the algebraic version.

	\item Now we only have to compute the scalar products (\emph{sewing}). This procedure has already been already explained in section~\ref{inversescattering}. In the general case this computation would imply computing three off-shell-off-shell scalar products. However only very particular configurations contribute when we cut and sew operators $\O_2$ and $\O_3$ because of the way these operators are constructed. In particular only the partition defined by  $\bar{\alpha_3}=\alpha_2=\emptyset$ has a non-vanishing contribution, being that of $\O_1$ the only non-trivial cut. Also the contractions between $\O_2$ and $\O_3$ are trivial, as we simply contract vacuum fields, being the contractions with $\O_1$ the non-trivial ones. 

	\item Finally we will divide by the norm of the three original spin chain. We can then write the three-point function as
\begin{equation}
	c_{123}^{(0)}=\sqrt{\frac{L_1 L_2 L_3}{\mathcal{N}_1 \mathcal{N}_2 \mathcal{N}_3}} \frac{a_{L_2}^{\{v\}}}{d_{L_2}^{\{v\}}} \ \frac{f^{\{v\}\{v\}}_>}{f^{\{v\}\{v\}}_< f^{\{u\}\{u\}}_>} \sum_{\alpha \cup \bar{\alpha}=\{u_i\}} \frac{a_{L_1+1}^{\bar{\alpha}}}{d_{L_1+1}^{\bar{\alpha}}} f^{\alpha \bar{\alpha}} f^{\bar{\alpha}\bar{\alpha}}_> f^{\alpha \alpha}_< \left( \scal{\{v^*\}}{\alpha} \scal{\bar{\alpha}^*}{\{w\}} \right)^{\text{co}} \ ,
\end{equation}
where $\{u\}$, $\{v\}$ and $\{w\}$ are the rapidities of the first, second and third operators respectively.

Luckily this generic formula can be simplified if any of the operators is a BPS operator. All these simplification are detailed in \cite{EscobedoTailoring}.
\end{enumerate}

Although we are not going to present generalizations of this method here, it has been also applied successfully to larger groups like $SU(3)$ \cite{TailoringII}, non-compact spin chains \cite{Tailoringnoncompact} and supersymmetric spin chains \cite{SUSYTailoring}.

\section{BKV hexagon}

The main idea behind the hexagon proposal \cite{BKVhexagon} is to consider the pair of pants that represent the three-string interaction/three-point correlation function as two hexagonal fundamental polygons stitched together at three of the sides. There are several similarities with the tailoring procedure we have presented in the previous section and, indeed, it is an all-loop generalization of it. This conjecture has been checked in numerous papers like \cite{Basso_2016,Eden_2016,Jiang_2016}, where an agreement has been shown with other computation methods. The only exception is the supersymmetric case, where further signs had to be included when gluing the hexagon forms factor to find the correct result \cite{Caetano_hexagon}. Interesting generalizations of the conjecture to four-point functions were proposed by an extension of the hexagon proposal \cite{2016arXiv161105577F} and by using the operator product expansion \cite{BKV2}. A generalization involving Wilson loops has been also proposed \cite{Kim2017}. The regularization of divergences appearing when gluing back the hexagons is explained in \cite{Basso_Komatsu_2017}.

When we cut the pair of pants into two hexagons we also cut each closed string into two open strings, which will carry some of the excitations of the closed string. Because an excitation can end up on either half after cutting, we should sum over all such possibilities with some weight, as we did in the cutting step of the tailoring method. With this cut we have also created three new segments (the pants' seams). Hence, when stitching back we should sum over all possible states living on them. This involves integrating over the rapidities of any number of mirror excitations and bound states of them.

This method reduces the computation of three-point functions to that of form factors of hexagon operators. These form factors in general depend on the physical rapidities of the three operators and on the mirror rapidities of the virtual particles. However this can be simplified by making use of mirror transformations which map excitations on one edge of the hexagon to a neighbouring one \cite{BKVhexagon}. The mirror transformation, which we are going to represent by a $\gamma$ superindex, is defined as the transformation that swaps the roles of space and time, therefore
\begin{align}
	E(u^\gamma ) &= i \tilde{p} (u) \ ,  & p(u^\gamma ) &= i \tilde{E} (u) \ ,
\end{align}
where the energy $\tilde{E}$ and the momentum $\tilde{p}$ are real. Note that two subsequent applications of this mirror transformations, a $2\gamma$ transformation, gives a crossing transformation. By sequential use of such transformations any generic hexagon can be related to one with all excitations in a single physical edge. We denote such a form factor, which we are going to call \emph{canonical hexagon}, as
\begin{equation}
	\mf{h}^{A_1 \dot{A}_1 , \dots , A_M \dot{A}_M} (u_1 , \dots ,u_M) \ , \label{canonicalhexagon}
\end{equation}
where $A_i \dot{A}_i$ are $SU(2|2)^2$ bifundamental indices\footnote{We will sometimes use $a$ and $\alpha$ instead of $A$ when we want to differentiate between bosonic an fermionic indices.} parametrizing the polarization of the i\textsuperscript{th} excitation $\chi^{A_i \dot{A}_i} (u)$.

Combining symmetry argument and bootstrap considerations, a conjecture for the $N$-magnon hexagon amplitude was proposed \cite{BKVhexagon},
\begin{equation}
	\mf{h}^{A_1 \dot{A}_1 , \dots , A_M \dot{A}_M} =(-1)^F \prod_{i<j} h_{ij} \granesperado{\chi_M^{\dot{A}_M} \dots \chi_1^{\dot{A}_1}}{\mathcal{S}}{\chi_1^{A_1} \dots \chi_M^{A_M}} \ , \label{hexagonconjecture}
\end{equation}
where $F$ accommodates the grading, $\chi^{A}$ is a state in the fundamental $SU(2|2)$ multiplet and $\mathcal{S}$ is Beisert $SU(2|2)$ matrix \cite{dynamicalsmatrix} with dressing phase set to one. The scalar factor $h_{ij}=h(u_i , u_j)$ can be constrained by crossing symmetry to be
\begin{equation}
	h_{12}=\frac{x^-_1 -x^-_2}{x^-_1 -x^+_2} \ \frac{1-\frac{1}{x^-_1 x^+_2}}{1-\frac{1}{x^+_1 x^+_2}} \ \frac{1}{\sigma_{12}} \ ,
\end{equation}
where $x^\pm =x\left( u\pm \frac{i}{2} \right)$ are shifted Zhukowsky variables (already introduced in section~\ref{BDSspinchain} with a different normalization), defined as $x+\frac{1}{x}=\frac{4\pi u}{\sqrt{\lambda}}$, and $\sigma_{12}$ is (half) the BES dressing phase \cite{BESphase}.

An equivalent way of thinking about the hexagon form factor is by introducing a vertex $\bra{\mf{h}}$ which can be contracted with three spin-chain states\footnote{The explicit expression for this vertex has not been explicitly computed previously, being that the aim of this chapter.}. For example, for a single magnon on the first spin chain
\begin{equation}
	\mf{h}^{A \dot{A}}=\bra{\mf{h}} \left( \ket{\chi^{A \dot{A}}}_1 \otimes \ket{0}_2 \otimes \ket{0}_3 \right) \ .
\end{equation}
We are going to use an invariant notation where each state is thought as being made out of excitations on top of the same BMN Z-vacuum, that is, we will not add the rotations and translations of having the operators at different points and R-charge conservation. We will talk about this transformation below.

\subsection{Symmetry of the hexagon form factor. Twisted translation}
\label{symmetryhexagonff}

To start working we will have to define the vacuum for the three-point functions. For the two-point function the vacuum is defined as
\begin{equation}
	\esperado{\text{Tr} [Z^L (0)] \text{Tr} [\bar{Z}^L (\infty )]} \ ,
\end{equation}
which breaks the $PSU(2,2|4)$ down to $PSU(2|2)^2$, so the excitations over the vacuum form a multiplet of the last. For three-point functions, the ``vacuum'' configuration is provided by the 1/2-BPS operators
\begin{equation}
	\O_i=\text{Tr}  [(\vec{Y}_i \cdot \Phi)^{L_i} (x_i)] \ ,
\end{equation}
where $\vec{Y}_i$'s are the $SO(6)$ polarization (null complex vectors in 6-dimensions). We can use the R-symmetry to align all the polarizations along a particular $U(1)$ direction, and conformal symmetry to put the three operators in a line. Therefore we will only have to worry about operators of the form
\begin{equation}
	\esperado{\text{Tr} [\mf{Z}^{L_1} (0)]  \text{Tr} [\mf{Z}^{L_2} (0,a,0,0)] \text{Tr} [\mf{Z}^{L_3} (\infty)] } \ ,
\end{equation}
where $\mf{Z}$ is called \emph{twisted-translated scalar}, defined by
\begin{equation}
	\mf{Z} (a)=e^{\mf{T}a} Z(0) e^{-\mf{T}a} =(z+\kappa^2 a^2 \bar{Z} +\kappa a Y -\kappa a \bar{Y}) (0,a,0,0) \ ,
\end{equation}
where $\kappa$ is a quantity with mass dimension 1, and 
\begin{equation}
	\mf{T}=-i\epsilon_{\alpha \dot{\alpha}} P^{\dot{\alpha} \alpha} +\kappa \epsilon_{\dot{a} a} R^{a \dot{a}} \ ,
\end{equation}
is the twisted translation operator. This choice of vacuum further breaks the symmetry to the diagonal part of the $PSU(2|2)^2$. In particular only the generators
\begin{align}
	\mc{L}^a_b=&L^\alpha_\beta + \dot{L}^{\dot{\alpha}}_{\dot{\beta}}\ , & \mc{Q}^\alpha_a=&Q^\alpha_a+i\kappa \epsilon^{\alpha \dot{\beta}} \epsilon_{a\dot{b}} \dot{S}^{\dot{b}}_{\dot{\beta}} \ , \notag \\
	\mc{R}^a_b=&R^\alpha_\beta + \dot{R}^{\dot{\alpha}}_{\dot{\beta}}\ , & \mc{S}^a_\alpha=&S^\alpha_a+\frac{i}{\kappa} \epsilon^{\alpha \dot{\beta}} \epsilon_{a\dot{b}} \dot{Q}^{\dot{b}}_{\dot{\beta}} \ , \label{diagonalsymmetry}
\end{align}
commute with the twisted translation $\mf{T}$. Dotted and undotted generators and indices represent the two different $PSU(2|2)$.

Non-BPS three-point functions are obtained by performing the twisted translation to the non-BPS operators constructed on the Z-vacuum at the origin,
\begin{equation}
	\mf{O}_i (a)=e^{\mf{T}a} \O_i(0) e^{-\mf{T}a} \ .
\end{equation}
Therefore the symmetry group of each state on each side of the hexagon is the usual $PSU(2|2)^2\ltimes \mathbb{R}^3$ we are familiar with \cite{dynamicalsmatrix}, and the intersection of the three symmetry groups is a single $PSU(2|2)_D$ (actually it can be centrally extended to $PSU(2|2)_D\ltimes \mathbb{R}$). 

Let us examine a little bit more this $PSU(2|2)_D$ subgroup and how it acts on the $PSU(2|2)^2$ magnons $\chi^{A_i \dot{A}_i} (u)$. The generators defined in eq.~(\ref{diagonalsymmetry}) act exactly in the same way as the usual $PSU(2|2)$ generators over the left part of the magnon, thus we can identify both quantum numbers. The only thing preventing a direct identification is that the generators act in a non-standard way over the right part of the magnon since the roles of $Q$'s and $S$'s are exchanged. Luckily, it can be checked that the quantum numbers agree with the ones with crossed rapidity $u^{-2\gamma}$ \cite{BKVhexagon}. Therefore each magnon transforms in the tensor representation of $PSU(2|2)_D$,
\begin{equation}
	\mathcal{V}_D (p, \kappa e^{-ip/2}) \otimes \mathcal{V}_D (p^{-2\gamma}, \kappa e^{-ip^{-2\gamma}/2}) \Longrightarrow \chi^{A \dot{A}} (u) \equiv \chi^A (u) \chi^{\dot{A}} (u^{-2\gamma}) \ . \label{representationpsu(2|2)D}
\end{equation}
This gives us a recipe on how to analytically continue magnon excitations under a crossing transformation $\left[ \chi^{A \dot{B}} (u) \right]^{2\gamma} =-\chi^{B \dot{A}} (u^{2\gamma})$. Another interesting way of understanding the origin of crossing is from a change of frames that modifies which operator is inserted at the origin \cite{2016arXiv161105577F}.

\subsection{Hexagon form factors for 1 and 2 particles from symmetry}

To compute simple hexagon form factors we are going to make use of the $PSU(2|2)_D$ invariance. To do that first we will centrally extend $PSU(2|2)_D$ to $PSU(2|2)_D\ltimes \mb{R}$. This central extension can be defined from the central extension $PSU(2|2)^2 \ltimes \mb{R}^3$  as
\begin{equation}
	\mathbb{P}=P-\kappa^2 K \ .
\end{equation}
This central element will appear in the anticommutators $\{\mc{Q} , \mc{Q} \} \sim \{\mc{S} , \mc{S} \} \sim \mathbb{P}$. With this central charge we can enforce the diagonal $PSU(2|2)$ symmetry by imposing that the hexagon vertex is killed by the central element,
\begin{equation}
	\granesperado{\mf{h}}{\mathbb{P}}{\psi}=0 \ ,
\end{equation}
for a generic spin chain state $\ket{\psi}$. However there is a mild non-locality in the chain of fundamentals used for describing the state. This can be addressed in various ways but we are going to choose the twisted notation of \cite{nlin/0610017}, also called \emph{spin chain frame}. In this picture our previous equation is equivalent to
\begin{equation}
	0=g\alpha (1-e^{ip}) \scal{\mf{h}}{Z^+ \psi} -\frac{g\kappa^2}{\alpha} (1-e^{-ip}) \scal{\mf{h}}{Z^- \psi} \ ,
\end{equation}
where $p$ stands for the total momentum of the state, $Z^\pm$ creates or destroys one vacuum site in the chain, and $\alpha$ is a parameter common to the left and right representations. If we assume that the $Z$ maker is diagonalized by the vertex, such that $\scal{\mf{h}}{Z^n \psi}=z^n \scal{\mf{h}}{\psi}$, this eigenvalue can be fixed to
\begin{equation}
	z^2=-\frac{\kappa^2}{\alpha^2} e^{-ip} \ .
\end{equation}

The $PSU(2|2)_D$ can be seen as the supersymmetrization of the $O(3)_{\text{Lorentz}} \times O(3)_{\text{R-charge}}$ group that preserves 3 points in space-time and 3 (generic) null vectors in R-charge space. Using this $O(3) \times O(3)$ symmetry we can fix the one particle hexagon form factors to
\begin{align}
	\mf{h}^{a \dot{a}} &=\scal{\mf{h}}{\Phi^{a \dot{a}}}=\tilde{N} \epsilon^{a \dot{a}} \ , & \mf{h}^{\alpha \dot{\alpha}} &=\scal{\mf{h}}{\mc{D}^{\alpha \dot{\alpha}}}=N \epsilon^{\alpha \dot{\alpha}} \ .
\end{align}
If we choose the normalization in such a way that $\tilde{N}=1$ and use that $\bra{\mf{h}}$ is annihilated by the right action of the supercharges, we can relate $N$ and $\tilde{N}$ imposing
\begin{equation}
	0=\Biggranesperado{\mf{h}}{\mc{Q}^\alpha_a}{\Psi^{b\dot{\beta}}} \equiv \Biggranesperado{\mf{h}}{\mc{Q}^\alpha_a}{\phi^b \psi^{\dot{\beta}}}  \ ,
\end{equation}
which gives
\begin{equation}
	N=\frac{\kappa (x^- -x^+)}{\gamma \dot{\gamma}} \ ,
\end{equation}
where $\gamma$ and $\dot{\gamma}$ are free parameters associated to relative normalization between boson and fermions in the left/right multiplet, as in \cite{nlin/0610017}. For unitary representations $|\gamma |=\sqrt{i(x^- -x^+)}$ and thus $|N|=|\kappa|=1$ by a proper choose of the phases.

The same analysis can be applied to two magnon form factors. By the $O(3) \times O(3)$ symmetry and the equations
\begin{equation}
	0=\Biggranesperado{\mf{h}}{\mc{Q}^\alpha_a}{\Phi^{b\dot{c}}_1 \Psi^{c\dot{\beta}}_2} =\Biggranesperado{\mf{h}}{\mc{S}_\alpha^a}{\Phi^{b\dot{c}}_1 \Psi^{c\dot{\beta}}_2} \ ,
\end{equation}
the form factors take the form
\begin{align}
\scal{\mf{h}}{\Phi^{a\dot{a}}_{1}\Phi^{b\dot{b}}_{2}}&= h_{12}A_{12} \epsilon^{a\dot{b}}\epsilon^{b\dot{a}} + \frac{1}{2}(h_{12}A_{12}-h_{12} B_{12}) \epsilon^{ab}\epsilon^{\dot{a}\dot{b}} \, , \\
\scal{\mf{h}}{\Phi^{a\dot{a}}_{1}\mathcal{D}^{\beta\dot{\beta}}_{2}} &= h_{12}N_{2}G_{12} \epsilon^{a\dot{a}}\epsilon^{\beta\dot{\beta}}\, , \, \, \, \scal{\mf{h}}{ \mathcal{D}^{\alpha\dot{\alpha}}_{1}\Phi^{b\dot{b}}_{2}} = h_{12}N_{1}L_{12} \epsilon^{\alpha\dot{\alpha}}\epsilon^{b\dot{b}}\, , \\
\scal{\mf{h}}{\mathcal{D}^{\alpha\dot{\alpha}}_{1}\mathcal{D}^{\beta\dot{\beta}}_{2}}&= -h_{12}N_{1}N_{2}D_{12} \epsilon^{\alpha\dot{\beta}}\epsilon^{\beta\dot{\alpha}} - \frac{1}{2}( h_{12}N_{1}N_{2}D_{12}- h_{12}N_{1}N_{2}E_{12})\epsilon^{\alpha\beta}\epsilon^{\dot{\alpha}\dot{\beta}}\, , \\
\scal{\mf{h}}{\Psi^{a\dot{\alpha}}_{1}\Psi^{b\dot{\beta}}_{2}}& = -\frac{1}{2} h_{12}N_{1}N_{2}z^{-1} C_{12} \epsilon^{ab}\epsilon^{\dot{\alpha}\dot{\beta}}\, , \,\,\, \scal{\mf{h}}{ \Psi^{a\dot{\alpha}}_{1}\Psi^{\beta\dot{b}}_{2}} = -h_{12}N_{1}H_{12} \epsilon^{a\dot{b}}\epsilon^{\beta\dot{\alpha}}\, , \\
\scal{\mf{h}}{ \Psi^{\alpha\dot{a}}_{1}\Psi^{b\dot{\beta}}_{2}}&= h_{12}N_{2}K_{12} \epsilon^{b\dot{a}}\epsilon^{\alpha\dot{\beta}}\, , \,\,\, \scal{\mf{h}}{ \Psi^{\alpha\dot{a}}_{1}\Psi^{\beta\dot{b}}_{2}} = -\frac{1}{2} zh_{12}F_{12} \epsilon^{\dot{a}\dot{b}}\epsilon^{\alpha\beta}\, ,
\end{align}
where $A_{12}, \dots ,L_{12}$ are the elements of the Beisert $S$-matrix \cite{nlin/0610017} with the dressing phase set to 1. $N$ and $z$ are defined as in the 1 magnon case, but with $p=p_1+p_2$. It is important to emphasize that the conjecture proposed (\ref{hexagonconjecture}) is in agreement with the symmetry considerations we have presented, as
\begin{equation}
	 \mf{h}^{A\dot{A},B\dot{B}}_{12}=h_{12} (-1)^{\dot{f_1}f_2} S^{AB}_{CD} (1,2) \mf{h}^{D\dot{A}}_1 \mf{h}^{C\dot{B}}_2=h_{12} (-1)^{\dot{f_1}f_2} \dot{S}^{\dot{A}\dot{B}}_{\dot{C}\dot{D}} (1,2) \mf{h}^{\dot{D}A}_1 \mf{h}^{\dot{C}B}_2 \ ,
\end{equation}
which is related with the fact that only the diagonal part of both $PSU(2|2)$ groups is involved.

We want to end this section proving that this form factor indeed fulfils the Watson equation~(\ref{watsonequation}) and the decoupling condition~(\ref{Decouplingcondition}). The Watson equation is satisfied because the hexagon vertex $\bra{\mf{h}}$ is preserved by the action of the $S$-matrix
\begin{equation}
	\Biggranesperado{\mf{h}}{(\mathbb{S}_{ii+1}-\mathbb{I})}{\ldots \chi^{A_{i}\dot{A}_{i}}_{i}\chi^{A_{i+1}\dot{A}_{i+1}}_{i+1}\ldots  } = 0 \ ,
\end{equation}
with $\mathbb{S} = S^{0}(-1)^{\dot{F}}\mathcal{S}\, \dot{\mathcal{S}}(-1)^{F}$ the $SU(2|2)^2$ $S$-matrix. Because the S-matrix is given by a left and right S-matrices, we can use Yang-Baxter and unitarity to cancel them. The scalar factor $S^{0}$ is compensated by the quotient
\begin{equation}
	\frac{h_{i,i+1}}{h_{i+1,i}}= \frac{x_{1}^{+}-x_{2}^{-}}{x_{1}^{-}-x_{2}^{+}}\frac{1-1/x^{-}_{1}x_{2}^{+}}{1-1/x^{+}_{1}x^{-}_{2}}\frac{1}{\sigma^{2}_{12}} \ .
\end{equation}
The decoupling condition~(\ref{Decouplingcondition}) is easily implemented and almost immediately satisfied as the ansatz is written in terms of the $S$-matrix
\begin{equation}
	\mc{S} \ket{\chi^{A}_{1}\chi^{B}_{2} \{\chi_{j}\}_{\text{rest}}}_{\textrm{pole}\, (12)} \propto \mc{S}_{\textrm{rest}}\prod_{\textrm{rest}}\mc{S}_{2j}\mc{S}_{1j}\ket{\{\chi_{j}\}_{\text{rest}} \otimes \mathrm{1}_{21}} \ ,
\end{equation}
where the equation holds at the level of the pole in the $(12)$ channel and $\ket{\mathrm{1}_{21}}$ is Beisert's singlet~\cite{nlin/0610017} built out of the particle-antiparticle pair $(12)$. This condition is the same as Janik's crossing equation for the $S$-matrix~\cite{Janik_2006} as derived by Beisert in~\cite{nlin/0610017}, except that the scalar factor in there should now be replaced by~$h_{12}$. Thus the decoupling condition for this ansatz boils down to the crossing equation,
\begin{equation}
	h(u_{1}^{2\gamma}, u_{2})h(u_{1}, u_{2}) = \frac{x_{1}^{-}-x_{2}^{-}}{x_{1}^{-}-x_{2}^{+}}\frac{1-1/x^{+}_{1}x_{2}^{-}}{1-1/x^{+}_{1}x_{2}^{+}}\ ,
\end{equation}
which is fulfilled by our choice of $h_{12}$ factor.

\subsection{Gluing the hexagons}

As happens in the cutting of operators in the tailoring procedure, when cutting the pair of pants into two hexagons we should keep record of the structure of the Bethe wave function. This means that, when we construct the three-point function from the hexagon form factor, for each physical operator we have to sum over all possible bipartite partitions of the set of magnons with some weight. The function $w(\alpha , \bar{\alpha})$ that weights each term can be understood in two as the product of two pieces. First, the piece that takes into account the propagation of magnons to the second hexagon and the interaction with all the magnons of the first hexagon
\begin{equation}
	w_1 (\alpha ,\bar{\alpha} )=\prod_{u_j\in\bar{\alpha}} e^{ip(u_j) l} \prod_{u_i\in \alpha} S(u_j , u_i) \ .
\end{equation}
Secondly, a part that for the three-point function is only a $(-1)$ factor for each magnon in the second hexagon
\begin{equation}
	w_2 (\alpha ,\bar{\alpha} ))=(-1)^{|\bar{\alpha}|} \ .
\end{equation}
This (originally mysterious) sign was understood in the nearly simultaneous articles \cite{2016arXiv161105577F,tessellating}. This sign takes into account the explicit dependence on the coordinates of the hexagon form factors. Hence we can write the complete weight as
\begin{equation}
	w(\alpha, \bar{\alpha} )=w_1(\alpha , \bar{\alpha} ) w_2(\alpha , \bar{\alpha} )=(-1)^{|\bar{\alpha} |} \prod_{u_j\in\bar{\alpha}} e^{ip(u_j) l} \prod_{u_i\in \alpha} S(u_j , u_i) \ .
\end{equation}

Using this weights, the asymptotic three-point function is written as
\begin{equation}
	C\propto \sum_{\substack{\alpha\cup\bar{\alpha}=\{u\} \\ \beta \cup \bar{\beta}=\{v\} \\ \gamma \cup \bar{\gamma}=\{w\}}} w(\alpha , \bar{\alpha} ) w(\beta , \bar{\beta} ) w(\gamma , \bar{\gamma} ) h(\alpha | \beta | \gamma ) h(\bar{\gamma} | \bar{\beta} | \bar{\alpha}) \ . \label{hexagonweight}
\end{equation}
This formula is only asymptotic as it does not include magnons in the mirror chanels (the ``seams''), so it can be systematically improved to incorporate finite-size corrections by adding them, with the leading correction corresponding to having a single particle passing through one of the three mirror channels (see references \cite{BKVhexagon} and \cite{Basso_Komatsu_2017} for a more complete explanation of this subject).

\section{The algebraic hexagon}

In this section we are going to try to give a motivation for the origin of the matrix elements that appear inside the hexagon form factors. To do that we are going to rewrite the proposal in a way more inspired by the ABA by using Zamolodchikov-Faddeev (ZF) operators as building blocks. First we are going to define the ZF algebra, which allows us to directly write form factors that fulfil the Watson equation (\ref{watsonequation}) by construction, together with the Fock space that we are going to use throughout this section; after that we will construct the Wick-contracting vertex taking inspiration from how the identity operator is constructed. With these tools at hand we can then check our proposal for some simple examples. Finally we will try to generalize this version of the proposal to more general hexagons and compute the weights in front of the hexagon form factors.

\subsection{The algebraic hexagon recipe}

\subsubsection{Constructing the state}

The first step will be the construction of the Fock space. For the moment we will consider only the case of canonical hexagons (\ref{canonicalhexagon}). Therefore we will put excitations only on one of the edges. In consequence we can define the Fock Space as $\mathscr{F}=\oplus_{n=0}^\infty \mathscr{H}_n$ where\footnote{Actually we should write $\otimes^6 \ket{0}$ and specify in which of the vacuums we apply our operators. However, as we are going to work mostly with the canonical hexagon, all operators will be applied to the same edge. Hence, instead of writing the whole vacuum hexagon, we are going to write only the edge in which all operators will act and drop the label on the operators to alleviate notation.}
\begin{align}
	 \mathscr{H}_0 &=\left\{ \ket{0} \, \text{such that} \, A^{A \dot{A}} (u) \ket{0}=0 \right\}\equiv \mathbb{C} \ , \\
	 \mathscr{H}_1 &=\left\{ A^\dagger_{A \dot{A}}(u) \ket{0}=\ket{\chi_{A\dot{A}} (u)} \}\right\} \ , \\
	 \mathscr{H}_N &=\left\{ \prod_{i=1}^{\substack{N \\\longrightarrow}} A^\dagger_{A_i \dot{A_i}}(u_i) \ket{0} = \ket{\chi_{A_1 \dot{A}_1} (u_1) \chi_{A_2\dot{A}_2} (u_2) \dots \chi_{A_N \dot{A}_N } (u_N)} \right\} \ .
\end{align}

Instead of this Fock space, we are going to construct an equivalent one but that will simplify our computations. The way we are going to proceed is by breaking the original generators into generators of the left and right $SU(2|2)$ representations\footnote{One way to formally do this breaking is to introduce two non-dynamical fields $\chi_A$ and $\chi_{\dot{A}}$ with a coupling with the original operators given by $\exp \left[ \chi_A (u) \otimes \chi_{\dot{A}} (u) \otimes Z_{A \dot{A}} (u) \right]$.}, $A_{A}^\dagger (u)$ and $A_{\dot{A}}^\dagger (u)$. This will imply also breaking the original vacuum hexagon into two hexagons, $(\otimes^6 \ket{0})\otimes (\otimes^6 \ket{0})$.\footnote{Of which we are only going to conserve the two edges on which the operators are going to act, $\ket{0}_L \otimes \ket{0}_R $, to alleviate notation.} We are going to define these operators in a way that they form a ZF algebra with the Beiser $SU(2|2)$ $S$-matrix as defined in~\cite{Arutyunov_ZFalgebra}
\begin{align}
	A_i^\dagger (u) A_j^\dagger (v) &=S_{ij}^{kl} (u,v) A_l^\dagger (v) A_k^\dagger (u) \ , \notag \\
	A^i (u) A^j (v) &=S^{ij}_{kl} (u,v) A^l (v) A^k (u) \ , \notag \\
	A^i (u) A_j^\dagger (v) &=S^{li}_{jk} (v,u) A_l^\dagger (v) A^k (u) +\delta_j^i \delta (u-v) \ .
\end{align}
Where the indices $i$ and $j$ are both dotted or undotted. This imposes the same Zamolodchikov-Faddeev algebra to both $SU(2|2)$ generators. We do not have to worry about the commutation between dotted and undotted operators as they are applied to different vacua.

However, the naïve construction has a problem as it has two degrees of freedom for each one we originally had. This is easily seen when we compute the commutation relation of our original operators in terms of this separated construction,
\begin{equation}
	A^{A \dot{A}} (u) A_{B\dot{B}}^\dagger (v) \propto A^{A} (u) A^{\dot{A}} (u) A_{B}^\dagger (v) A_{\dot{B}}^\dagger (v)=\dots +\delta^{A}_{B} \delta^{\dot{A}}_{\dot{B}} [\delta (u-v)]^2 \ .
\end{equation}
We get two delta functions instead of only one as an indication that we have to kill one of the degrees of freedom as they will generate $\delta (0)$ terms. A similar problem was encounter by \cite{LeClair1995} although in that case is a consequence of boundary states living in an infinite volume. Our way to solve the problem is to divide the form factors by the square root of the norm of the state, after some regularization. The way we are going to proceed is to multiply the operators by an smearing function and regularize both in the same way
\begin{equation}
	A^\dagger_{A \dot{A}} (u)= \lim_{\epsilon \rightarrow 0} \int{d\theta \, d\theta' \rho_\epsilon (u-\theta) \rho_\epsilon (u-\theta ') A_A^\dagger (\theta) A_{\dot{A}}^\dagger (\theta ') }\ ,
\end{equation}
with $\rho_\epsilon (\theta) =\frac{1}{\epsilon}$ if $|\theta |<\epsilon$.


This decomposition generates also another problem: if both algebras transform in a representation of $SU(2|2)$, the generator $A^\dagger_{A \dot{A}} (u)$ do not transform in $SU(2|2)_D$. A solution to that problem has been already commented at the end of section \ref{symmetryhexagonff}, where it was suggest to change the rapidity from the generator of dotted indices from $\theta '$ to $\theta^{\prime -2\gamma}$.

There is still one last step in the construction of the Fock space. Because our generators form a ZF algebra, they do not commute and the ordering in which we apply them is important. Our choice is going to be from left to right. To sum up,
\begin{align}
	 \tilde{\mathscr{H}}_0 &=\left\{ \ket{0}_L \otimes \ket{0}_R \right\}\equiv \mathbb{C} \ , \\
	 \tilde{\mathscr{H}}_1 &=\left\{ \lim_{\epsilon \rightarrow 0} \int{d\theta \, d\theta' \rho_\epsilon (u-\theta) \rho_\epsilon (u-\theta ') A_A^\dagger (\theta) A_{\dot{A}}^\dagger (\theta ') } \ket{0}_R=\ket{\chi_{A\dot{A}} (u)} \right\} \ , \\
	 \tilde{\mathscr{H}}_N &=\bigg\{ \lim_{\{\epsilon_i \} \rightarrow 0} \int{ \bigg( \prod_{i} d\theta_i d\theta'_i \, \rho_{\epsilon_i} (u_i -\theta_i) \rho_{\epsilon_i} (u_i-\theta '_i) \bigg) \prod_{i=1}^{\substack{N \\\longrightarrow}} A^\dagger_{A_i} (\theta_i) \ket{0}_L  \otimes \prod_{i=1}^{\substack{N \\\longrightarrow}} A^\dagger_{\dot{A_i}}\left(\theta_i^{\prime -2\gamma} \right) \ket{0}_R} \notag \\
	 & = \ket{\chi_{A_1 \dot{A}_1} (u_1) \chi_{A_2\dot{A}_2} (u_2) \dots \chi_{A_N \dot{A}_N } (u_N)} \bigg\} \ ,
\end{align}
and $\tilde{\mathscr{F}}=\oplus_{n=0}^\infty \tilde{\mathscr{H}}_n$.

We have to define also other kinds of operators, which we are going to call ``Cartan Generating Functions''. These will be a Yangian current associated to the Cartan elements of the algebra. We will assume that they behave like
\begin{align}
	H_{A \dot{A}} (u) A_{B \dot{B}}^\dagger (v) &= S^{C \dot{C},D \dot{D}}_{A \dot{A},B \dot{B}} (u,v) A_{D \dot{D}}^\dagger (v) H_{C \dot{C}} (u) \ , \\
	\left[ H_{A \dot{A}} (u) , H_{B \dot{B}} (v) \right] &=0 \ , \\
	H_{A \dot{A}} (u) \ket{0} &=e^{ip(u)l} \ket{0} \ ,
\end{align}
where $l$ will depend on the edge of the hexagon we are applying the operator  to, and $p(u)$ is the momentum associated with the magnon.

\subsubsection{Construction of the vertex}

Now that we have defined the Fock space we can start thinking about the vertex we are going to apply to a state to obtain the hexagon amplitude. To do that first we are going to construct the identity operator, as both will have similarities.

\paragraph{Identity operator}

First we can see that the identity operator can be constructed using the ZF operators we have defined in the previous section as
\begin{align}
	\mathbb{I}_{\text{undotted}}&=\mathop{\vphantom{\ket{0}}}\nolimits_{(1)} \! \! \bra{0}\exp_{BF} \left[ \int{\frac{dx}{2\pi i} A^\dagger_{A} (x) \otimes A^{A} (x) } \right] \ket{0}_{(2)} \ , \\
	\mathbb{I}_{\text{dotted}}&=\mathop{\vphantom{\ket{0}}}\nolimits_{(1)} \! \! \bra{0}\exp_{BF} \left[ \int{\frac{dx}{2\pi i} A^\dagger_{\dot{A}} (x) \otimes A^{\dot{A}} (x) } \right] \ket{0}_{(2)} \ ,
\end{align}
where $\exp_{BF}$ means the exponential defined as its formal series arranged in the following way: the terms on the left side of the tensor product are ordered from left to right, and the terms on the right side of the tensor product are ordered from right to left. It is easy to check that the operators defined in this way are $SU(2|2)_L$ and $SU(2|2)_R$ invariant, respectively. Expanding the exponential and applying the definition of the ZF algebra we can see that the first non-trivial case,
\begin{align}
	A^A (x) A^B (y) \otimes A^\dagger_B (y) A^\dagger_A (x)&= S^{AB}_{CD} (x,y) A^D (y) A^C (x) \otimes A^\dagger_B (y) A^\dagger_A (x)  \notag \\
	&= A^D (y) A^C (x) \otimes A^\dagger_C (x) A^\dagger_D (y) \ ,
\end{align}
is invariant, and so it is well defined. The same happens for dotted indices. The generalization to higher terms of the expansion is trivial. Note that the choice of the ZF algebra commented in the footnote of section \ref{bootstrap} is important here: If we choose $A_i^\dagger (u) A_j^\dagger (v) =S_{ij}^{lk} (u,v) A_l^\dagger (v) A_k^\dagger (u)$ instead (note the different ordering of upper indices in the $S$-matrix) we cannot exponentiate the identity because the invariant combination is $A^\dagger_A (x) A^\dagger_B (y) \otimes A^\dagger_A (y) A^\dagger_B (x)$, where operators in each side of the tensor product are not the same.

We still have to prove that this is the identity operator. For the cases of 0 and 1 (tensor products of) ZF operators it is trivial to prove. The first non trivial case is again the one with two operators, which we are going to prove it only for the dotted identity, as the undotted one is proven in exactly the same way
\begin{align*}
	&\frac{1}{2}\langle A^{\dot{A}} (x) A^{\dot{B}} (y) A^\dagger_{\dot{U}} (u) A^\dagger_{\dot{V}} (v) \rangle \otimes A^\dagger_{\dot{B}} (y) A^\dagger_{\dot{A}} (x) =\\
	&\frac{1}{2} \left[ S^{\dot{A}\dot{B}}_{\dot{U}\dot{V}} (u,y) \delta (y-v) \delta (x-u) + \delta^{\dot{B}}_{\dot{U}} \delta^{\dot{A}}_{\dot{V}} \delta (y-u) \delta (x-v) \right] A^\dagger_{\dot{B}} (y) A^\dagger_{\dot{A}} (x) = \\
	&\frac{1}{2}\left[ S^{\dot{A}\dot{B}}_{\dot{U}\dot{V}} (u,y) A^\dagger_{\dot{B}} (v) A^\dagger_{\dot{A}} (u) + A^\dagger_{\dot{U}} (u) A^\dagger_{\dot{V}} (v) \right]=A^\dagger_{\dot{U}} (u) A^\dagger_{\dot{V}} (v) \ .
\end{align*}
The case of higher terms in the expansion is similarly proven.

\paragraph{Constructing the vertex}

We are going to construct a vertex that contracts a dotted index and an undotted index with an $\epsilon$ symbol. Drawing inspiration from the construction of this identity operators, the vertex that gives the correct result is\footnote{Note that, as $p(u^{-2\gamma})=-p(u)$, the interpretation of this vertex as a boundary operator becomes very appealing.}
\begin{align}
	\bra{\mf{H}}=\mathop{\vphantom{\ket{0}}}\nolimits_{(L)} \! \! \bra{0}\otimes \null_{R}\bra{0} \exp_{BF}\left[ \int_{C_\infty}{\frac{dx}{2\pi i}  \epsilon_{A\dot{B}} \left( A^A (x) \otimes A^{\dot{B}} (x^{-2\gamma})\right) } \right] \ .
\end{align}
Where
\begin{equation}
	\epsilon_{A\dot{B}}= \left( \begin{array}{cc}
	\epsilon_{a \dot{b}} & 0 \\
	0 & \epsilon_{\alpha \dot{\beta}}
	\end{array} \right) \ .
\end{equation}
First we are going to prove that it indeed acts on a general product of $A_{\dot{A}}^\dagger$ operators in the form
\begin{equation}
	\bra{\mf{H}} \prod_{i=1}^{\substack{N \\\longrightarrow}}{A_{\dot{A}_i}^\dagger(u_i^{-2\gamma})} \ket{0}_{(R)}=\mathop{\vphantom{\ket{0}}}\nolimits_{(L)} \!\! \bra{0}  \prod_{i=1}^{\substack{N \\\longrightarrow}}{A^{B_i}(u_i)} \epsilon_{B_i\dot{A_i}} \ ,
\end{equation}
and the same for undotted
\begin{equation}
	\bra{\mf{H}} \prod_{i=1}^{\substack{N \\\longrightarrow}}{A^\dagger_{B_i}(u_i)} \ket{0}_{(L)}=\mathop{\vphantom{\ket{0}}}\nolimits_{(R)} \! \! \bra{0}\prod_{i=1}^{\substack{N \\\longrightarrow}}{A^{\dot{A}_i}(u_i^{-2\gamma})} \epsilon_{B_i\dot{A_i}} \ .
\end{equation}
The proof is very similar to the one for the identity operator. For the case of 0 and 1 operators the proof is trivial. The first non-trivial case is again the one with two operators, being the rest an easy generalization of this one. We are also going to do it for one combination, because the rest of the combination can be done exactly in the same way. 
\begin{align*}
	\granesperado{\mf{H}}{A^\dagger_{\dot{U}} (u^{-2\gamma}) A^\dagger_{\dot{V}} (v^{-2\gamma})}{0}_{(R)} &=\iint{\frac{dx}{2\pi i} \frac{dy}{2\pi i} \langle A^{\dot{A}} (x^{-2\gamma}) A^{\dot{B}} (y^{-2\gamma}) A^\dagger_{\dot{U}} (u^{-2\gamma}) A^\dagger_{\dot{V}} (v^{-2\gamma}) \rangle} \\
	&\times \mathop{\vphantom{\ket{0}}}\nolimits_{(L)} \! \! \bra{0} A^B(y) A^A (x) \epsilon_{A\dot{A}} \epsilon_{B\dot{B}} \ .
\end{align*}
The first factor is a scalar product that can be computed using the Zamolodchikov-Faddeev algebra. The final result is
\begin{displaymath}
	\langle A^{\dot{A}} (x) A^{\dot{B}} (y) A^\dagger_{\dot{U}} (u) A^\dagger_{\dot{V}} (v) \rangle = S^{\dot{A}\dot{B}}_{\dot{U}\dot{V}} (u,y) \delta (y-v) \delta (x-u) + \delta^{\dot{B}}_{\dot{U}} \delta^{\dot{A}}_{\dot{V}} \delta (y-u) \delta (x-v) \ .
\end{displaymath}
Where we have not added the ${-2\gamma}$ over the rapidities to alleviate notation. At this point we cannot do further unless we define a relation between the S-matrices of the two $SU(2|2)$ factors, which we are going to fix to be
\begin{equation}
	S^{CD}_{AB} (u,v) \epsilon_{C\dot{A}} \epsilon_{D\dot{B}}=S^{\dot{C} \dot{D}}_{\dot{A} \dot{B}} (u^{-2\gamma} , v^{-2\gamma}) \epsilon_{A \dot{C}} \epsilon_{B\dot{D}} \label{transformationSmatrices}\ .
\end{equation}
The reasoning behind this definition is that the contraction of a state with this vertex $\bra{\mf{H}}$ should be independent of on which set of operators (dotted or undotted) we act on first. This can also be related with the fact that we are selecting the diagonal part of the $SU(2|2)^2$ group of both generators. With this relationship at hand, the term with the $S$-matrix inverts the ordering of the dotted operators which gives the final result
\begin{align}
	&\bra{\mf{H}} A^\dagger_{\dot{U}} (u^{-2\gamma}) A^\dagger_{\dot{V}} (v^{-2\gamma}) \ket{0}_{(R)} =\frac{1}{2} \bra{0} \left( S^{\dot{A}\dot{B}}_{\dot{U}\dot{V}} (v^{-2\gamma},u^{-2\gamma}) A^{B} (v) A^{A} (u) \epsilon_{A\dot{A}} \epsilon_{B\dot{B}}+ \right. \notag \\
	& A^{B} (u) A^{A} (v) \epsilon_{A\dot{V}} \epsilon_{B\dot{U}} \Big)= \frac{1}{2} \bra{0} 2 A^{B} (u) A^{A} (v) \epsilon_{B\dot{U}} \epsilon_{A\dot{V}} \ ,
\end{align}
The transformation of undotted creation operators into dotted annihilation operators is proven in a similar way.

The invariance of this vertex with respect to the $PSU(2|2)_D$ group can be derived in a similar way we did for the identity operator, because one of the the $SU(2|2)$ component can be related to the opposite components via a $-2\gamma$ transformation, which is a particle-antiparticle transformation, replacing the operator $A^\dagger_i (x)$ by an annihilation operator of the other component does not alter the invariance properties of the vertex.

\subsubsection{Computing simple examples and checking the proposal}

To alleviate notation, from now on all the rapidities associated to an operator with dotted indices will be understood to be transformed by $-2\gamma$.

Let us first check that the $\epsilon$ symbol is our metric, as we intended. We need to check that
\begin{equation}
	\scal{\mathfrak{H}}{\chi_{A_1 \dot{A}_1}(u_1)}=\epsilon_{A_1\dot{A_1}} \ .
\end{equation}

If we substitute directly the ZF operators without the smearing we get the right answer multiplied by a delta evaluated at zero
\begin{align}
	&\bra{\mf{H}} A_{A}^\dagger (u) A_{\dot{A}}^\dagger (u) \ket{0}_L \otimes \ket{0}_R= \bra{0} A^{\dot{B}} (u) A_{\dot{A}}^\dagger (u) \epsilon_{A\dot{B}} \ket{0}= \notag \\
	&\bra{0} \left(S_{\dot{A} \dot{C}}^{\dot{D} \dot{B}} (u,u) A_{\dot{D}}^\dagger (u) A^{\dot{C}} (u) + \delta^{\dot{B}}_{\dot{A}} \delta(0^{-2\gamma}) \right) \epsilon_{A\dot{B}} \ket{0}= \delta (0)\delta^{\dot{B}}_{\dot{A}} \epsilon_{A\dot{B}}= \delta (0) \epsilon_{A\dot{A}} \ ,
\end{align}
where the first term in the third line is canceled because the right (left) vacuum is annihilated by the $A$ ($A^\dagger$) operators. If instead we choose to act with the proposed vertex on the dotted one the steps are the same mutatis mutandis, giving exactly the same answer.

If we introduce the smearing functions and divide by the square root of the norm we have to compute and regularize
\begin{align}
	&\frac{\epsilon_{A\dot{A}} \int{d\theta \, d\theta ' \rho_\epsilon (u-\theta) \rho_\epsilon (u-\theta')}}{\sqrt{\langle u | u \rangle}} = \frac{\epsilon_{A\dot{A}} \int{d\theta \rho_\epsilon (u-\theta)^2}}{\sqrt{\int{d\theta \rho_\epsilon^*(u-\theta) \rho_\epsilon(u-\theta)}\int{d\theta ' \rho_\epsilon^*(u-\theta ') \rho_\epsilon(u-\theta ')}}} \notag \\
	&= \frac{\epsilon_{A\dot{A}} \int{d\theta \rho_\epsilon (u-\theta)^2}}{\int{d\theta \rho_\epsilon^*(u-\theta) \rho_\epsilon(u-\theta)}}=\frac{1/\epsilon}{1/\epsilon}\epsilon_{A\dot{A}} =\epsilon_{A\dot{A}} \ ,
\end{align}
which eliminates the $\delta (0)$ multiplying the form factor we are interested in.

The case of two magnons is a little more involved
\begin{align}
	&\bra{\mf{H}} A_{A}^\dagger (\theta) A_{B}^\dagger (\phi) A_{\dot{A}}^\dagger (\theta ') A_{\dot{B}}^\dagger (\phi ') \ket{0}_L \otimes \ket{0}_R= \notag \\
	 &\bra{0} A^{\dot{C}} (\theta) A^{\dot{D}} (\phi) A_{\dot{A}}^\dagger (\theta ') A_{\dot{B}}^\dagger (\phi ') \ket{0} \epsilon_{A \dot{C}} \epsilon_{B \dot{D}}= \notag \\
	 &\bra{0} A^{\dot{C}} (\theta) \left[ A_{\dot{F}}^\dagger (\theta ') S_{\dot{A} \dot{G}}^{\dot{D} \dot{F}}(\theta ',\phi) A^{\dot{G}} (\phi) +\delta (\phi -\theta ' ) \delta^{\dot{D}}_{\dot{A}} \right] A_{\dot{B}}^\dagger (\phi ') \ket{0} \epsilon_{A \dot{C}} \epsilon_{B \dot{D}}= \notag \\
	 &\bra{0} \left\{ \left[ \delta^{\dot{C}}_{\dot{F}} \delta (\theta - \theta ') +A^\dagger A  \right] S_{\dot{A} \dot{G}}^{\dot{F} \dot{D}}(\theta ' , \phi) \left[ \delta^{\dot{G}}_{\dot{B}} \delta (\phi-\phi ') +A^\dagger A \right] + \right. \notag \\
	 &\left. + \delta (\phi - \theta ' ) \delta^{\dot{D}}_{\dot{A}} \left[ \delta (\theta - \phi ') \delta^{\dot{C}}_{\dot{B}} +A^\dagger A\right] \right\} \ket{0} \epsilon_{A \dot{C}} \epsilon_{B \dot{D}}= \notag \\	 
	&\left\{ [\delta (\theta - \theta ' ) \delta (\phi - \phi ')] S^{\dot{C} \dot{D}}_{\dot{A} \dot{B}} (\theta, \phi) +[\delta (\phi - \theta ' ) \delta (\theta -\phi ' )] \delta^{\dot{D}}_{\dot{A}} \delta^{\dot{C}}_{\dot{B}} \right\} \epsilon_{A \dot{C}} \epsilon_{B \dot{D}}  \ .
\end{align}
Before we perform the integration with the smearing functions we are going to take a look at this structure. We can identify two terms, a terms that is going to give us the desired result and a second term that can be considered a disconnected contribution. This last term, after the regularization, would be multiplied by $\epsilon^2 [\delta (u-v)]^2$. This factor does not contribute as no two rapidities are equal in a general on-shell state\footnote{A couples of reasons for that statement are that equal rapidities do not yield proper Bethe wavefunctions \cite{Caux_2008} and that strings (in the thermodynamical limit the solutions of the Bethe equation cluster around lines called \emph{strings}) with two equal rapidities have no weight in equilibrium problems \cite{Calabrese_2014}. Appart from that, ZF operators with equal rapidities behave like fermionic operators if the S-matrix behaves like $S_{AB}^{CD} (u,u)=-\delta^C_ A \delta^D_B$, which is our case.}. However it is important to talk about it because, by transferring excitations from a different edge, we could end with excitations with the same rapidity.

If we repeat the same regularization procedure as before, we can see that the final result is $S^{\dot{C} \dot{D}}_{\dot{A} \dot{B}} (u,v)$. It is the same that~\cite{BKVhexagon} obtained but with a different ordering of the indices of the $S$-matrix as a consequence of our different definitions of the $S$-matrix.



Let us move now to the case of three magnons. After applying the same kind of commutation relation the final expression for the three magnons form factor is, schematically
\begin{align}
	&\Big( \textcolor{red}{S_{\dot{B} \dot{C}}^{\dot{G} \dot{F}} (\phi , \chi) S_{\dot{A} \dot{F}}^{\dot{H} \dot{X}} (\theta , \chi) S_{\dot{H} \dot{G}}^{\dot{Z} \dot{Y}} (\theta , \phi ) \delta(\theta - \theta ') \delta(\phi - \phi ') \delta(\chi - \chi ')} + \notag \\
	& \textcolor{green}{S S \times \delta (\theta - \theta ') \delta(\phi - \chi ') \delta(\chi - \phi ')}+  \textcolor{green}{S S \times \delta(\theta - \phi ')\delta(\phi - \theta ') \delta (\chi - \chi ')}+ \notag \\
	& \textcolor{green}{\delta \delta \delta \times \delta(\theta - \chi ') \delta (\phi - \phi ') \delta(\chi - \theta ')} + \textcolor{Goldenrod}{S \delta \times \delta(\phi - \theta') \delta(\chi - \phi ') \delta(\theta - \chi ')} +\notag \\
	& \textcolor{Goldenrod}{S \delta \times \delta(\theta - \phi ') \delta(\phi - \chi ') \delta(\chi -\theta ')} \Big)\epsilon^{A\dot{Z}} \epsilon^{B\dot{Y}} \epsilon^{C\dot{X}} \ .\label{3magnongff}	
\end{align}
Here we can identify three different kinds of terms: the first kind (red term) is the term we want and, as in the other two cases, is the same that the authors of~\cite{BKVhexagon} obtained but with a different definition of the indices of the $S$-matrix. The second kind of terms (green terms) will be proportional to the square of a delta function of two rapidities times $\epsilon^ 2$, making these factors irrelevant unless two rapidities are the same, as happens in the case of two magnons. Terms of the third kind (dark yellow terms
) are a little different from the green terms but the regularization is very similar, giving us at the end a factor $\epsilon^3 \delta (u-v) \delta (v-w) \delta (w-u)$.


Note that we can also classify the terms we have obtained by the number of S-matrices they contain: one term with no $S$-matrix, two with one $S$-matrix, two with two S-matrices and one with three S-matrices. This sequence can be generalized to any number of magnons, and it matches the sequence of the Mahonian numbers \cite{A008302}.

\begin{figure}[t]
  \includegraphics[width=\linewidth]{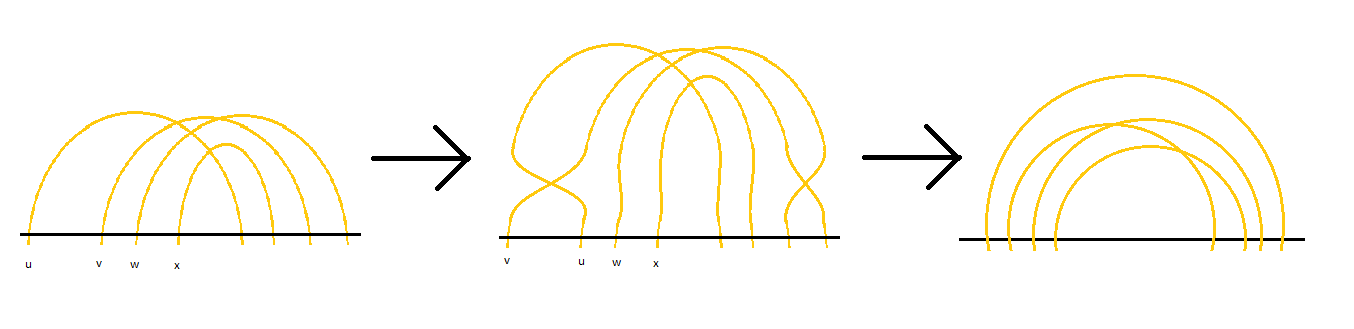}
  \caption{Reordering of the rapidities of one of the 4-magnons diagrams to explicitly show that it is non-connected. This reordering is accomplished by including two $S$-matrices (one with dotted indices and one with undotted indices) and using Yang-Baxter equation and equation~\ref{transformationSmatrices}. Other diagrams may require more than one reordering to show explicit disconnectedness.}
  \label{reordered}
\end{figure}

If we move now to the case of four magnons we will find a similar situation: one contribution which is equal to the one obtained by~\cite{BKVhexagon}, and contributions that after integration will have 
from two to four delta functions. It might seems that some of these contributions are not disconnected, but it can be shown that they actually are disconnected after a reordering of the rapidities of the diagram by including $S$-matrices, see figure~\ref{reordered}.

Therefore we have a method to compute the canonical hexagon form factor just by defining a Fock space and a ``boundary operator''.

\subsection{Generalizations of the recipe. Computations for non-canonical hexagons}

Now that we have a formal recipe for computing the canonical hexagon, in this subsection we are going to deal with the generalization of the framework. The first part of the subsection will be an explanation on how to deal with a hexagon with excitations applied to all of the edges. The second part will deal with the cutting of the pant into two hexagons.

\subsubsection{Transforming a general hexagon into the canonical hexagon}

We will first explain the recipe for computing non-canonical hexagons, that is, hexagons with excitations in all six edges.

An operator that transfers excitations from the physical sides of the hexagon (which are the third and the fifth if we count the canonical one as the first side) to the canonical side can be constructed in a similar way as we construct the identity operator, 
\begin{align}
	&\mathcal{O}^{-2\gamma}=\mathop{\vphantom{\ket{0}}}\nolimits_{\text{down,left}} \! \! \bra{0} \exp \left[ -\int{A^{B\dot{A}} (x^{2\gamma}) \otimes A^\dagger_{A\dot{B}} (x) \frac{dx}{2\pi i}} \right] \ket{0}_{\text{up}} \ , \\
	&\mathcal{O}^{-4\gamma}=\mathop{\vphantom{\ket{0}}}\nolimits_{\text{down,right}} \! \! \bra{0} \exp \left[ \int{A^{A\dot{B}} (x^{4\gamma}) \otimes A^\dagger_{A\dot{B}} (x) \frac{dx}{2\pi i}} \right] \ket{0}_{\text{up}} \ ,
\end{align}
where the choice of indices comes from the properties of our representation of $PSU(2|2)_D$, eq.~(\ref{representationpsu(2|2)D}), which imply that $\left( \chi^{A \dot{B}} (u) \right)^{2\gamma} =-\chi^{B \dot{A}} (u^{2\gamma})$ as we explained in section~\ref{symmetryhexagonff}. We refer to \cite{BKVhexagon} for a more detailed explanation.

We also have to add excitations on the three mirror sides of the hexagon. However the transformation of the indices is more involved in this case. Despite that problem, we can still add some operators to take that effect into account because, in the final formula, we have to sum over all possible excitations in the mirror edges (and over all possible number of excitations and integrate over all possible rapidities). Therefore we can forget about the transformation and directly put the sum over the already transformed indices,
\begin{align}
	\mathcal{O}^{-(2n+1)\gamma}&=\mathbb{I}+\sum_{A,\dot{A}} \int{du \, \mu (u) \, A^\dagger_{A\dot{A}} (u^{-(2n+1)\gamma})}+\frac{\left( \sum_{A,\dot{A}} \int{du \, \mu (u) \, A^\dagger_{A\dot{A}} (u^{-(2n+1)\gamma})} \right)^2}{2!} \notag \\
	&+\dots =\exp \left[\sum_{A,\dot{A}} \int{du \, \mu (u) \, A^\dagger_{A\dot{A}} (u^{-(2n+1)\gamma})} \right] \ ,
\end{align}
where $\mu$ is a measure. Obviously, when we take into account all of the transformations, we have to add them as
\begin{equation}
	\bra{\mf{H}} \mathcal{O}^{-5\gamma} \mathcal{O}^{-4\gamma} \mathcal{O}^{-3\gamma} \mathcal{O}^{-2\gamma} \mathcal{O}^{-\gamma} \ ,
\end{equation}
because, as we have already stated, the order in which we move the excitations is important.

\subsubsection{Computations with two hexagons. The Drinfeld coproduct}

In this part of the subsection we are going to see how to deal at the same time with the two hexagons we get when we cut the pair of pants and how to compute the weights for each of the partitions of the excitations. To do it we are going to start with some definitions.

The coproduct (or comultiplication) is an operation defined in a Hopf algebra $H$ such that $\Delta : H \rightarrow H \otimes H$. The physical meaning of this operation is the construction of two-particle operators from one-particle operators. One of the main properties we would like our coproduct to have is called co-associativity, defined as
\begin{equation}
	(\Delta \otimes \mathbb{I} ) \Delta= (\mathbb{I} \otimes \Delta) \Delta \ ,
\end{equation}
which physically means that the three-particle operators are well defined. This relation still leaves space for some twisting of a given coproduct by an algebra automorphism $\omega : H \rightarrow H$ as
\begin{equation}
	\Delta^{(\omega)}(a)=(\omega \otimes \omega) \Delta (\omega^{-1} (a)) \ . \label{coproducttwisting}
\end{equation}

In particular, we are interested in the so-called Drinfeld coproduct. A definiton of this coproduct in the Yangian double of $\mathfrak{gl} (m|n)$ that can be found, for example, in eq.~(3.5) of~\cite{Pakuliak_complete}\footnote{The operators they propose do not exactly form a ZF algebra because the commutation relations with the annihilation operators $E_i (z)$ are no exactly the correct ones, as they have an extra Cartan generating function. Note that their definition of Cartan generating function is different from ours, being the equivalence between notations $H_i (u)=k_{i+1}^+ (u) [k_i^+ (u)]^{-1}$. However the algebra generated by $\left( H^{-1}_i(u) E_i (u) , F_j (v) \right)$  actually form a ZF algebra. The same happens with the algebra generated by $\left( E_i (u) , F_j (v) H^{-1}_j(v)  \right)$.}
\begin{equation}
	\Delta^{(D)} (F_i (z))=F_i(z) \otimes \mathbb{I} + H_i (z) \otimes F_i (z) \ ,
\end{equation}
where $F_i (z)$ are some creation operators that commute up to an $S$-matrix, with commutation relations
\begin{equation}
	H_i (u) F_i (v) =  \frac{u-v+c_{[i+1]}}{u-v-c_{[i+1]}} F_i (v) H_i (u) \ ,
\end{equation}
as we can see, they commute up to a canonical $S$-matrix.

We can take this definition of coproduct and apply it to our original creation operators $A^\dagger_{A\dot{A}} (u)$ in the $SU(2|2)^2$ bifundamental representation\footnote{Note that this coproduct only works for the canonical S-matrix. A coproduct for the Beisert S-matrix would need more structure.}
\begin{equation}
	\Delta A^\dagger_{A \dot{B}} (u)=\mathbb{I} \otimes A^\dagger_{A \dot{B}} (u)+ A^\dagger_{A \dot{B}} (u) \otimes H_{A \dot{B}} (u) \ .
\end{equation}
If we do that we get that the coproduct of two creation operators is, schematically,
\begin{equation}
	\Delta (A^\dagger_1 A^\dagger_2)=\mathbb{I} \otimes A^\dagger_1 A^\dagger_2 + A^\dagger_1 A^\dagger_2 \otimes H_1 H_2+A^\dagger_1 \otimes H_1 A^\dagger_2 + A^\dagger_2 \otimes A^\dagger_1 H_2 \ .
\end{equation}
We recall here that our Cartan generating functions act on the vacuum hexagon in the following way $H_i (u) \ket{0}=e^{i p(u) l} \ket{0}$. Therefore if we commute $H_1 A_2 =S_{12} A_2 H_1$ and use the previous fact, we get the $w_1$ part of the weight (\ref{hexagonweight}).

One way to get the missing signs is to add by hand a minus sign to the second factor of the coproduct
\begin{equation}
	\Delta^{(D,-)} (A_{A\dot{A}}^\dagger (u))=A_{A\dot{A}}^\dagger (u) \otimes \mathbb{I} - H_{A\dot{A}} (u) \otimes A_{A\dot{A}}^\dagger (u) \ .
\end{equation}
We can check that this choice reproduces the weights(\ref{hexagonweight}) from~\cite{BKVhexagon}. However this ``twisting'' completely breaks the co-associativity property of the coproduct. It would be interesting to see if a twisting by an automorphism given by equation~(\ref{coproducttwisting}) can be used to obtain this sign and the conformal cross ratios in the hexagonalization proposal presented in \cite{2016arXiv161105577F,tessellating}.

Now we have all the tools we need to give a final construction of the complete hexagon vertex. Adding 
all the constructions from last two sections in top of our canonical hexagon, we can write
\begin{equation}
	\bra{\mf{h}}=\bra{\mf{H}} \mathcal{O}_{\otimes}^{-5\gamma} \left( \mathcal{O}^{-4\gamma}_{\text{hex. }1} \otimes \mathcal{O}^{-4\gamma}_{\text{hex. }2} \right) \mathcal{O}_{\otimes}^{-3\gamma} \left(  \mathcal{O}^{-2\gamma}_{\text{hex. }1} \otimes \mathcal{O}^{-2\gamma}_{\text{hex. }2} \right) \mathcal{O}_{\otimes}^{-\gamma} \Delta^{(D,-)}
\end{equation}
where we have to modify the operators for odd $\gamma$ translations. We still do not know how to modify the operators for $-\gamma$ and $-5\gamma$ because of the mixing between the two hexagons, but the one for $-3\gamma$ is related only with its counterpart in the other hexagon, so we can write it as
\begin{align}
	\mathcal{O}_{\otimes}^{-3\gamma}&=\exp \left[ \sum_{A,\dot{A}} \int{du \, \mu (u) \, A^\dagger_{A\dot{A}} (u^{-3\gamma})\otimes A^\dagger_{A\dot{A}} (u^{-3\gamma})} \right]
\end{align}
where each of the operators act in a different hexagon.

Then we conclude that the vertex $\bra{\mf{h}}$ provides us equivalent results as the vertex defined in \cite{BKVhexagon}, both for hexagons in the canonical configuration (as we have proved in the previous section) and general hexagons (except for the problems to define the operators $\mathcal{O}_{\otimes}^{-5\gamma}$ and $\mathcal{O}_{\otimes}^{-\gamma}$). Therefore we can consider $\bra{\mf{h}}$ as a rewriting of the original proposal as a sum of products of Zamolodchikov-Faddeev operators, so it can be considered as an ``algebraic hexagon proposal''.

\part{Conclusions and appendices}

\renewcommand{\theequation}{\thechapter.\arabic{equation}}

\chapter{Summary and conclusions}

\begin{chapquote}{C. Bowers, \textit{Calculating the Velocity of Darkness and its Possible Relevance to Lawn Maintenance}, Journal of Irreproducible Results, 1995.}
	The practical implications of these results are completely unknown, but are discussed at length anyway.
\end{chapquote}



The bulk of this thesis, formed by Part II and Part III, is divided in two halves that collect the results obtained in both sides of the AdS/CFT conjecture during my Ph.D. degree. The first half presents the computation of dispersion relations in different deformed backgrounds of $AdS_3 \times S^3$, which is interesting as its dual has to be a CFT in two dimensions, so it would have the full Virasoro symmetry. The second half presents different methods of computing of two and three-point correlation function in $\mc{N}=4$ SYM. The method for computing two-point function was obtained from the direct application of the QISM for the ABA, while the method for three-point functions is an alternative rewriting of the successful hexagon proposal.

In Part II of thesis we have studied closed string solutions rotating and pulsating in $AdS_3 \times S^3 \times T^4$ with NS-NS and R-R three-form flux. The corresponding string sigma models are equivalent to a deformed Neumann-Rosochatius integrable system. This background was chosen because it was already known to be integrable \cite{Cagnazzo-Zarembo} but the dispersion relation of spinning string had not been analysed. We have considered five different cases: strings rotating in $\mb{R}\times S^3$, strings rotating in $AdS_3$, string rotating in the full $AdS_3 \times S^3$, strings pulsating in $AdS_3 \times S^1$ and string pulsating in $\mb{R}\times S^3$. The equations of motion can be integrated and expressed in terms of Jacobi elliptic functions, reducing to trigonometric functions in the limit of pure NS-NS flux, that is, the limit $q=1$. With this solution of the equations of motion we were able to compute the classical energy as a function of the angular momenta and the windings as either a power series in the total angular momentum $\frac{J}{\sqrt{\lambda}}$ or as a power series around the pure NS-NS point $q= 1$.

The simplification in the limit of pure NS-NS flux is an appealing result present both in the case of constant radii and elliptic solutions. From our point of view it appears as a consequence of the degeneration of the elliptic curve governing the dynamics of the problem. The deep reason behind this simplification is the equivalence of the Lagrangian with a WZW model in the pure NS-NS flux limit. This limit was explored in some particular cases, for example in \cite{Moo}, using conformal field theory techniques. It would be interesting to see if these approaches can be used to get information on the neighbour of $q=1$.

Another important question is the study of the conditions of stability of our solutions and to find the spectrum of excited string states. It would also be interesting to address the question of the spectrum of small quadratic fluctuations around the solutions we have constructed, as done for the $AdS_5 \times S^5$ spinning string  \cite{Frolov_2003,Arutyunov_2003,Arutyunov_2004} and the $AdS_3 \times S^3 \times T^4$ with R-R flux \cite{Beccaria2012,Abbott2015}. These quadratic fluctuations of the Lagrangian can be computed by substituting the parametrization (\ref{parametrizationspinning}) by
\begin{align}
	X_{2j-1} +iX_{2j} &= (r_j+\tilde{r}_j+i\rho_j) e^{i\varphi_j}
\end{align}
and similarly for the AdS coordinates. The coordinates $\tilde{r}_j$ and $\rho_j$ will represent the two different kind of fluctuations we can have, and $r_j$ are fixed to the solutions we already have obtained in chapter 3.

We also studied closed string solutions rotating in $\eta$-deformed $\mb{R} \times S^3$. This second background was chosen because it was already studied for the simplest of the ansatz, presented in equation~(\ref{simpletsansatz}), so the deformed Uhlenbeck constants for the Neumann model were known. We studied the string sigma model for the more general asantz (\ref{mostgeneralansatz}), which in this background reduces to a deformation of the Neumann-Rosochatius system different from the one obtained for the flux-deformed Lagrangian. Thankfully the method used in the flux-deformed case can be applied in the same way, so we were able to compute the classical energy as a power series in the total angular momentum for the constant radii case.

In this case simplifications where found in the $\varkappa=\frac{2\eta}{1-\eta^2} \rightarrow \infty$ and $\varkappa=i$ limits, where we can find analytical expressions for the dispersion relation. It would be interesting a deeper study of these limits. Similarly to the case of the flux deformation, we could compute quadratic fluctuations in this background using the same strategy.

Another possible way to continue this work is to apply resurgence methods. Resurgence theory deal with the summation of the asymptotic series involved in perturbation theory and it is a window to non-perturbative physics. Resurgence tools have already been applied to the Principal Chiral Model and the $\eta$-deformed Principal Chiral Model for some particular groups \cite{Cherman_2014,Demulder_2016}, so it would be interesting to see if they can be applied to the backgrounds studied in this thesis: the flux-deformed $AdS_3\times S^3$ background and the $(AdS_3 \times S^3 )_\eta$ background.

The two studied deformations are very different. The flux deformation only appears in the B-field, leaving the background geometry unchanged, while the $\eta$-deformation only affects the geometry and do not have contribution to the B-field (for the $AdS_3 \times S^3$ case, as the contribution is a total derivative and, hence, trivial to the equations of motion. In contrats, the complete $\eta$-deformed $AdS_5 \times S^5$ background has a non-trivial contribution to the B-field). Therefore it would be interesting to apply our method to a deformation that affects both. An interesting example would be the Lunin-Maldacena background \cite{Lunin_2005}, which is dual to the Leigh-Strassler deformation of $\mc{N}=4$ SYM, whose geometry is a deformation of the $AdS_5 \times S^5$ space and has B-field, both of which survive when we truncate the space to $\mathbb{R}\times S^3$ or $AdS_3 \times S^3$. This background has been partialy analysed in \cite{Frolov_LMbackground}, but a computation of the Uhlenbeck constants and the dispersion relation of general spinning strings has not been carried out.

In Part III of this thesis we have studied two-point and three-point correlation functions using the spin chain interpretation of the $\mc{N}=4$ Supersymmetric Yang-Mills theory. Concerning the two-point correlation functions we have presented a systematic approach to the case of spin operators located at arbitrary sites of the spin chain. This is done both at the one-loop level, as the problem amounts to the calculation using the XXX spin chain, and at the all-loop level without including wrapping effects, as the BDS spin chain can be mapped to an inhomogeneous version of the XXX spin chain. We have found that the general case of correlation functions in a homogeneous chain is much more involved than in the BDS spin chain. This is because one needs to face apparent singular behaviour of the algebra of the elements of the monodromy matrix. The approach we used to deal with this problem was to show that the residue arising each time we commute the operators vanishes.

Our computations using the ABA are compared with results obtained using the CBA, taking special care with the normalization of the states. A good procedure to handle this is the use of Zamolodchikov-Faddeev operators instead of the usual monodromy matrix operators, which makes the agreement with the CBA immediate. The use of Zamolodchikov-Faddeev operators also allows the direct implementation of Smirnov's form factor axioms. An interesting continuation of our work in this paper would be to understand what other constraints are imposed by the remaining axioms in Smirnov's 
form factor program \cite{Smirnov}. In particular it would be very interesting to understand the behavior under crossing transformations of form factors 
evaluated using algebraic Bethe ansatz techniques. The crossing transformation corresponds to a shift in the rapidity by half the imaginary period 
of the torus that uniformizes the magnon dispersion relation in the AdS/CFT correspondence \cite{Janik_2006}. However at weak coupling one of the periods of the rapidity 
torus becomes infinitely large and thus both periodicity and the crossing transformation become invisible. In order to be able to impose periodicity most likely the 
dressing phase factor needs to be included. A natural question is therefore the extension of the method that we have developed in this paper to include the dressing 
phase factor.  The extension of Smirnov's program for relativistic integrable theories to worldsheet form factors for 
$AdS_5 \times S^5$ strings was discussed in \cite{KM1,KM2}.

Another interesting extension of our work is the analysis of the thermodynamical limit where both the number of magnons and the number of sites are large and comparable. In this limit the determinant expressions for the scalar product of Bethe states can be expressed as contour integrals. We hope our method can be combined with 
the semiclassical analysis of contour integrals in \cite{Kazama_2013,Wheeler_2013,Bettelheim_2014}.

Concerning the three-point correlation functions we have presented a review of the Tailoring method and of the hexagon framework as an introduction for our proposal. This proposal can be considered as a rewriting of the original hexagon form factor in the language similar to the ABA. This is done by first identifying our Fock space and identity operator. The information obtained from it allows to construct an expression for the hexagon vertex $\bra{\mf{h}}$ for a canonical hexagon presented in \cite{BKVhexagon} but not explicitly constructed. After this construction we have checked some simple hexagon form factor, confirming that both proposals generate the same result for one and two excitations, although at two excitations we can already see a non-connected contribution that can be safely removed. However for three excitations or more the structure becomes more involved and the non-connected contributions do not decouple so easily. A recipe is given to safely decouple them and the form factors agree with the original ``coordinate'' ones.

After confirming the agreement between both techniques, we propose generalizations of the hexagon vertex. First we show how it can be generalized to non-canonical hexagons via the introduction of operators that transport the excitations around the hexagon using mirror transformations. The second generalization we propose is the explicit computation of the weight factors that have to be added when summing over the different distribution of excitations between two hexagons when the pair of pants is cut.

One possible continuation of this work would be a better understanding of the weight factors. Here we have proposed its appearance from the presence of Cartan generating functions in the coproduct, which explains them up to a sign for the three-point functions and up to the conformal cross ratios for the hexagonalization proposal. The sign can easily be included, although this modified coproduct breaks the co-associativity property, which implies that we no longer have a Hopf algebra. It would be interesting to see if a twisting of the coproduct can reproduce the terms related with the cross ratios of the operators, explained for the original proposal in \cite{2016arXiv161105577F,tessellating}. Another interesting possibility would be the study of the symmetry of the $\bra{\mf{h}}$ vertex. Our writing of the vertex in terms of ZF operators suggest that the explicit $PSU(2|2)_D$ symmetry might be uplifted to a Yangian one.

\newpage

\null
\vfill

\begin{chapquote}{Sir Winston Churchill, just after the second battle of El Alamein}
	Now this is not the end. It is not even the beginning of the end. But it is, perhaps, the end of the beginning. 
\end{chapquote}

\vfill
\null

\appendix

\chapter{Analysis of the $\varkappa\rightarrow \infty$ and $\varkappa \rightarrow i$ Lagrangians}
\label{appendixkappa}

In this Appendix we will analyse the solutions that we have constructed in section \ref{limitingkappa} in the cases where $\varkappa \rightarrow \infty$ and $\varkappa \rightarrow i$ 
by performing the corresponding limit at the level of the Lagrangian. 
In order to deal with this problem it will be useful to think of the change of variables that brings the kinetic term in the deformed Lagrangian 
to canonical form, which is given by $r_2=\text{sn} \left( \phi , -\varkappa^2 \right)$. In the variable $\phi$ the Lagrangian is given by
\begin{equation}
L \! = \! \frac{1}{2} \!  \left[ \phi^{\prime 2} - \omega_2^2 \text{sn}^2 \left( \phi , -\varkappa^2 \right) \! - \! \frac{v_2^2}{\text{sn}^2 \left( \phi , -\varkappa^2 \right)} 
\! - \! \frac{\omega_1^2 \left( 1 + \frac{1}{\varkappa^2} \right)}{1+\varkappa^2 \text{sn}^2 \left( \phi , -\varkappa^2 \right)} -\frac{(1+\varkappa^2) v_1^2}{\text{cn}^2 \left( \phi , -\varkappa^2 \right)}\right]  .
\end{equation}
In the limit $\varkappa \rightarrow i$ the change of variables reduces to $r_2=\tanh \phi$, together with $r_1=\sech \phi$, and thus the Lagrangian becomes
\be
L_{i} = \frac{1}{2} \left[ \phi^{\prime 2} - \frac{v_2^2}{\sinh^2 \phi} + \frac{\omega_2^2}{\cosh^2 \phi} \right] \ ,
\ee
where we have shifted the Lagrangian by a constant to rewrite the term associated with $v_2^2$ with a hyperbolic secant instead of a hyperbolic cotangent. To find the limit $\varkappa \rightarrow \infty$ we need to transform the elliptic sine because its fundamental domain is defined when the elliptic modulus is between $0$ and $1$. We will write
\begin{equation}
\text{sn} (\phi, -\varkappa^2)=\frac{\text{sd} 
\left(\sqrt{1+\varkappa^2} \phi , \frac{\varkappa^2}{1+\varkappa^2} \right)}{\sqrt{1+\varkappa^2}}\approx \frac{\sinh (\varkappa \phi)}{\varkappa} \ .
\end{equation}
Therefore the change of variables is given by
$\varkappa \, r_2=\sinh \varkappa \phi=\sinh \tilde{\phi}$, which leads to 
\footnote{The extra term $(v_1^2 +\varkappa^2 v_1^2)$ accompanying $\omega_2^2$ comes from the expansion of the Jacobi cosine. 
Also, although not obvious, taking this limit implicitly assumes $r_2 \ll \mathcal{O} (\varkappa^{-1})$. That is the reason why the $1-r_2^2$ factor dividing the kinetic term 
disappears as, by direct substitution of the change of variables, it is subleading in $\varkappa^{-2}$.}
\be
L_{\infty} = \frac{1}{2\varkappa^2} \left[ \tilde{\phi}^{\prime 2}-\left( \omega_2^2 +v_1^2 +\varkappa^2 v_1^2 \right) 
\sinh^2 \tilde{\phi} - \frac{\varkappa^4 v_2^2}{\sinh^2 \tilde{\phi}} - \frac{(1+\varkappa^2 ) \omega_1^2}{\cosh^2 \tilde{\phi}} \right] \ . 
\ee
Both cases lead then to the same kind of Lagrangian, although with different coefficients in front of the terms in the potential. In what follows we will treat both of them simultaneously. 
However, even in these limiting cases the Lagrangian is not easy to handle unless some additional simplifications are performed. There simplifications will come from the choices of physical 
parameters entering the problem. We will first consider the easiest choice of parameters on the Lagrangian, which is that where only the potential with the square of the hyperbolic sine survives. Then
\begin{equation}
L = \frac {1}{2 \varkappa^2} \Big[ \phi'^2 - \alpha^2 \sinh^2 \phi \Big] \ ,
\label{Lreduced}
\end{equation}
with $\alpha$ a constant which will depend on which of the two limits we are taking. The equation of motion is then
\begin{equation}
\phi'' = - \alpha \sinh \phi \cosh \phi \ ,
\end{equation}
and can be solved in terms of the Jacobi amplitude,
\be
\phi=\pm i \text{ am}\left(\sqrt{ \alpha^{2}+c } \ \sigma,\frac{\alpha^{2}}{\alpha^{2}+c} \right) \ ,
\ee
where 
$c$ is a constant that has to be fixed using periodicity of $r_i$ (we have made use of our freedom in the choice of $\sigma$ 
to eliminate an additional integration constant). 
Note that in general depending on the sign of $\alpha^2+c$ we will have two different solutions. 

We will now focus on the limit $\varkappa \rightarrow \infty$. In this case the solutions are given by
\begin{align}
&y_2^{2} =-r_2^{2} = \frac{1}{\varkappa^2} \text{sn}^{2} \left(\sqrt{\alpha^2+c} \, \sigma ,\frac{\alpha^2}{\alpha^2+c} \right) \ , \quad \text{when} \ \alpha^2 + c >0 \ , \label{sol1} \\
&\tilde{y}_2^{2} = +r_2^{2}  = \frac{1}{\varkappa^2} \text{sc}^{2} \left(\sqrt{-(\alpha^2+c)} \, \sigma ,\frac{c}{\alpha^2 + c} \right) \ ,
\quad \text{when} \ \alpha^2 + c< 0 \ .
\end{align}
Relation (\ref{sol1}) corresponds to equation (\ref{gensolinfty}) once we set $\alpha^2+c = \varkappa^2 \omega_2^2(\tilde{I}_1-1)$. 
Note that in both cases we have to analytically continue to hyperbolic space. This is in agreement with the results obtained in~\cite{deformationsupercosets}, 
where in the limit  $\varkappa\rightarrow\infty$ the deformed sphere becomes a hyperboloid.
We must however stress that the periodicity condition for each solution is different. 
This is because the real periodicity of the sn$^2$ function is given by $2\text{K}(m)$ while its imaginary periodicity is $2i\text{K}(1-m)$, where $K(x)$ is the elliptic integral of first kind. 
Furthermore, the presence of the Jacobi sc function in the case where $\alpha^{2}+c <0$ leads to a divergence when evaluating the angular momentum,  
so from now on we will only consider the case where $\alpha^{2}+c >0$. The periodicity condition implies
\be
\frac{n}{\pi} \text{K}\left( \frac{\alpha^2}{\alpha^2+c} \right)=\sqrt{\alpha^2+c} \ .
\ee
In general this equation has no analytical solution. However, as $\alpha^{2}$ 
grows like $\varkappa^2$ we can assume that $c/\alpha^2$ is small enough to perform a series expansion in both sides of the equality. 
Then if we recall now that
\be
K[1-x] \simeq -\frac{\log (x)}{2}+2\log (2) \ ,
\ee
we find  
\begin{equation}
c \simeq \frac{n \alpha}{\pi} W \big( 16 \alpha \pi \, e^{-2 \alpha \pi /n}/n \big) \equiv \frac{n \alpha}{\pi} \bar{W} \ ,
\label{LambertW}
\end{equation}
where $W (x)$ is the Lambert W function. In fact, it is easy to check that our assumption becomes true very fast, because 
when $n=10$ and $\alpha^2=200$ we already have $c/\alpha^2 \approx 0.0014$. Now, as we have set $v_2=\omega_1=0$ 
to bring the Lagrangian to the form (\ref{Lreduced}), we have $J_1=m_2=0$ and therefore we only need to compute the angular momentum, 
\be
J_2 = \int{\frac{d\sigma}{2\pi} y_2^2 \omega_2} = \frac{\omega_2}{\varkappa^2} \ \frac{\alpha^2+c}{\alpha^2} \left[ 1- \text{E} \left( \frac{\alpha^2}{\alpha^2+c} \right) \Big/ \, \text{K} \left( \frac{\alpha^2}{\alpha^2+c} \right)  \right] \ , 
\ee
and the winding, 
\be
m_1 \! = \! \int{\frac{d\sigma}{2\pi} \, \frac{v_1 (1-\varkappa^2 y_2^2)}{y_1^2} } = v_1 \left[ \frac{1+\varkappa^2}{\varkappa^2} \ \frac{\Pi \! \left( \! -\frac{1}{\varkappa^2} , \frac{\alpha^2}{\alpha^2+c} \right)}{\text{K} \left( \frac{\alpha^2}{\alpha^2+c} \right)} -1 \right] \ ,
\ee
where we have used the periodicity condition to simplify the expressions. If we take now the large $\alpha^2$ limit, we conclude that
\begin{equation}
J_2=J=
\frac{\omega_2}{\varkappa^2} + \dots 
\end{equation}
where we have used that the first elliptic integral diverges at 1 while the second elliptic integral goes to 1. The winding can also be expanded as
\begin{align}
m_1=\frac{3 v_1}{2\varkappa^2} + \dots 
\end{align}
The only thing left is to find the dispersion relation,
\begin{align}
E^2 &=\int{\frac{d\sigma}{2\pi} \left( \frac{y_1^{\prime 2} -y_2^{\prime 2}}{1-\varkappa^2 y_2^2} +\frac{v_1^2 (1-\varkappa^2 y_2^2)}{y_1^2}-y_2^2 \omega_2^2 \right)} \nonumber \\
&=\alpha^2  \left[ 1 -\frac{(1+\varkappa^2 )\alpha^2+c}{\varkappa^2 \alpha^2} \ \frac{\Pi 
\left( -1, \frac{\alpha^2}{\alpha^2+c} \right)}{\text{K} \left( \frac{\alpha^2}{\alpha^2+c} \right)} \right] +m_1 v_1 -J\omega_2  \ ,
\end{align}
that can be easily expanded to find
\be
E^2\approx -\frac{\alpha^2}{2\varkappa^2}+m_1 v_1 -J\omega_2=-\frac{2 \varkappa^4 m_1^2}{9}+\frac{4 \varkappa^2 m_1^2}{9} -\frac{3\varkappa^2 J^2}{2} \ .
\ee
It is interesting to notice that the energy we have obtained is purely imaginary. This is again a consequence of how the space is deformed by the $\varkappa$ parameter. In particular it can be shown \cite{deformationsupercosets} that the $AdS_5$ space becomes a $dS_5$ space in the $\varkappa\rightarrow \infty$ limit. In particular the time coordinate is analytically continued (making $g_{tt}>0$), which explains the wrong sign of the square of the energy.

We will next move to the choice of parameters that brings the Lagrangian to the form
\begin{equation}
\tilde{L} = \frac {1}{2 \varkappa^2} \Big[ \phi'^2 -\frac{\alpha^2}{\cosh^2 \phi} \Big] \ .
\label{L2}
\end{equation}
Instead of writing the equations of motion for this Lagrangian and trying to integrate them it is more convenient to write the Hamiltonian associated to it, 
\begin{equation}
H=\phi'^2 +\frac{\alpha^2}{\cosh^2 \phi} \ ,
\end{equation}
and make use that is is a conserved quantity to directly integrate it.  We conclude that 
\begin{equation}
\arcsinh \phi=\sqrt{ \frac{|\alpha^2-H|}{H}} \sinh (\sqrt{H} \sigma) = \varkappa r_2 \ .
\label{s1}
\end{equation}

However we can see that this solution is not the same as the one we obtained by analysing the roots of the quartic polynomial, eq.~(\ref{omega1notzero}).  The reason for this mismatch is that, as we have previously discussed, the Lagrangian we have written implicitly ignores the $\frac{1}{1-r_2^2}$ term in the kinetic energy as it is subleading in $\varkappa$. If we restore it the modified Hamiltonian reads
\begin{equation}
H=\frac{\phi'^2}{1- \frac{1}{\varkappa^2} \sinh^2 \phi} +\frac{\alpha^2}{\cosh^2 \phi} \ ,
\end{equation}
which can be integrated to obtain
\begin{equation}
\arcsinh (\phi)=\pm \varkappa  \text{ sn} \left( \sqrt{\frac{H- \alpha^2}{\varkappa^2}} \sigma , -\frac{H \varkappa^2}{H-\alpha^2} \right) =\varkappa r_2 
\label{s2}
\end{equation}
In fact the solution (\ref{s1}) can be recovered from this one by taking the $\varkappa \rightarrow \infty$ limit after using the transformation sn$(u,m)=\frac{1}{\sqrt{m}}$ sn$\left( \sqrt{m} u , \frac{1}{m} \right)$.

From the point of view of the Uhlenbeck constant and eq.~(\ref{omega1notzero}), ignoring the $\frac{1}{1-r_2^2}$ term in the kinetic energy can be understood as explicitly taking the limit $\zeta_3 \ll \zeta_4$, giving us
\begin{equation}
    \zeta (\sigma)=\frac{\zeta_3 \tanh^2 [\tilde{n}(\sigma-\sigma_0)]}{1+\tanh^2 [\tilde{n}(\sigma-\sigma_0)]}=\zeta_3 \sinh^2 [\tilde{n}(\sigma-\sigma_0)] \ ,
\end{equation}
with $\tilde{n}=\sqrt{\omega_1^2 (1+\varkappa^2 \tilde{I}_1)}$. In both interpretations we can match the solutions obtained from the Uhlenbeck constants and the solutions obtained from the equation of motion by identifying $H=\omega_1^2 (1+\varkappa^2 \tilde{I}_1)$ and $\alpha^2=(1+\varkappa^2) \omega_1^2$.

We can now easily find the solutions to the equations of motion for the $\varkappa \rightarrow i$, $v_2=0$ limit. To do that we only have to use the transformation $r_1=\sech \phi$ in the solution~(\ref{s1}) and use that $\sech[ \arcsinh (x)]=\frac{1}{\sqrt{1+x^2}}$. With that we find
\begin{align}
	r_1^2 &=\frac{1}{1+\frac{|\alpha^2-H |}{H} \sinh^2 [\sqrt{H} (\sigma-\sigma_0)]} \notag \\
	&=\frac{1}{1-\frac{|\alpha^2-H |}{H} \cosh^2 [\sqrt{H} (\sigma-\sigma_0)]}=\frac{H}{H-|\alpha^2-H| \cosh^2 [\sqrt{H} (\sigma-\sigma_0)]} \ ,
\end{align}
where we have used a redefinition of $\sigma_0$ by a $\frac{i\pi}{2}$ shift from the first to the second line. We can see that this solution can be related with eq.~(\ref{omega2notzero}) if we perform the substitution $\alpha^2=\omega_2^2$ and $H=\tilde{I}_1 \omega_2^2$. Note that for the formula to have the correct sign we need $\tilde{I}_1 \leq 1$, which is equivalent to the condition that the roots of the elliptic curve have to be $\omega_1^2 \leq \zeta_i \leq \omega_2^2$.

To end this appendix we want to address a third simplified Lagrangian. This has the form
\begin{equation}
\hat{L} = \frac {1}{2 \varkappa^2} \Big[ \phi'^2 -\frac{\beta^2}{\sinh^2 \phi} \Big] \ .
\label{L3}
\end{equation}
So it is obvious that we can get all the solutions for this Lagrangian from the solutions of Lagrangian (\ref{L2}) after substituting $\phi_{\hat{L}}=\phi_{\tilde{L}}+\frac{i\pi}{2}$ and $\beta^2=-\alpha^2$. Let us exam one of the solutions, the limit $\varkappa=i$ and $\omega_2=0$. If we choose the solution with the hyperbolic cosine we get
\begin{equation}
	r_1=\sech \left( \phi \pm\frac{i\pi}{2} \right)=\mp i\csch (\phi)=\mp\sqrt{\frac{-H}{H+\beta^2}} \sech (\sqrt{H} \sigma ) \ ,
\end{equation}
which can be proven to be equivalent to eq.~(\ref{omega2zero}) with the identification $\beta^2=-v^2_2$ and $H=1-\tilde{I}_1$.

\chapter{General form of ${\cal F}^L_n$}
\label{A}
  
In this appendix we are going to obtain the general expression of the function ${\cal F}^L_n$ discussed in subsections \ref{noneasyway} and \ref{0sigmasigmamumu} . All along the calculation the limit $\a \rightarrow 0$ will be assumed.
Using the first recurrence relation in (\ref{recurrenceeq}) and setting both $d$ and $\dpartial{d}{\lambda}$ to zero we find
\begin{equation}
	{\cal F}^L_n={\cal F}^L_0 + i\mathcal{D} {\cal F}^L_0 + i\mathcal{D} {\cal F}^L_1 + \dots + i\mathcal{D} {\cal F}^L_{n-1} \ .
\end{equation}
If we assume that $n<L-1$, the second recurrence equation gives 
\begin{equation}
\mathcal{D} {\cal F}^L_n=\binom{n}{0} \mathcal{D} {\cal F}^L_0+\binom{n}{1} \frac{i\mathcal{D}^2}{2!} 
{\cal F}^L_0+\dots=\sum_{j=0}^n{\binom{n}{j} \frac{i^j \mathcal{D}^{j+1}}{(j+1)!} {\cal F}^L_0} \ .
\end{equation}
Therefore we need to sum the series
\be
\sum_{j=0}^{n-1}{i\mathcal{D} {\cal F}^L_j}=\sum_{j=0}^{n-1}{\sum_{k=0}^j{\binom{j}{k} \frac{i^{k+1} \mathcal{D}^{k+1}}{(k+1)!} {\cal F}^L_0}} \ .
\ee
As a first step, we can commute the two sums as $\sum_{j=0}^{n-1}{\sum_{k=0}^j{}}=\sum_{k=1}^{n-1}{\sum_{j=k}^{n-1}{}}+\sum_{j=0}^{n-1}{\delta_{k,0}}$, 
because the $j$ only appears in the limit of the sum and in the binomial coefficient, so is easy to perform first the sum over $j$. 
The second sum is easy to perform because we only have to calculate $\sum_{j=0}^{n-1}{\binom{j}{0}}=\binom{n}{1}$. 
The sum over $j$ of the other term can be evaluated using the properties of the binomial 
coefficients $\sum_{j=k}^{n-1}{\binom{j}{k}} = \binom{n-1+1}{k+1}$. 
Then the whole sum can be rewritten as
\begin{equation}
	{\cal F}^L_n={\cal F}^L_0+\sum_{k=1}^{n}{\binom{n}{k} \frac{i^{k} \mathcal{D}^{k}}{k!} {\cal F}^L_0} 
	= \sum_{k=0}^{n}{\binom{n}{k} \frac{i^{k} \mathcal{D}^{k}}{k!} {\cal F}^L_0} \ .
\end{equation}
This equation is true $\forall n < L-1$. If we want to calculate it for $n\geq L-1$ we have 
to take into account derivatives of $d$ of order greater or equal to $L$, which can be done 
independently of the calculation we have already done, because
\begin{displaymath}
	\mathcal{D}^{L+\alpha-1} {\cal F}^L_{j+1}=\frac{i}{L+\alpha} {\cal F}^L_j \dpartial[L+\alpha]{d}{\lambda} + \dots \ ,
\end{displaymath}
where the dots stand for the part that we have already taken into account. Therefore
the $d$-contribution to $\mathcal{D}{\cal F}^L_M$ will be of the form
\begin{align*}
	i\mathcal{D}{\cal F}^L_M &=\sum_{j=1}^{M-L+2}{\sum_{k=0}^{M+2-L-j}{\frac{i^{L+k-1}}{(L+k-1)!} 
	\binom{M-j}{L+k-2} \mathcal{D}^{L+k-1} {\cal F}^L_j}} \\
	&=\sum_{j=1}^{M-L+2}{\sum_{k=0}^{M+2-L-j}{\frac{i^{L+k}}{(L+k)!} \binom{M-j}{L+k-2} \dpartial[L+k]{d}{\lambda} {\cal F}^L_{j-1}}} \ ,
\end{align*}
and the derivative of $d$ can be calculated using Leibniz's rule,
\begin{displaymath}
	\left. \dpartial[L+k]{d}{\lambda} \right|_\xi =\sum_{j=0}^{L+k}{\binom{L+k}{j} 
	\dpartial[j]{(\lambda-\xi)^L}{\lambda} \, \dpartial[L+k-j]{(\lambda+\xi)^{-L}}{\lambda}} \ ,
\end{displaymath}
because we are going to evalute it at $\lambda=\xi$, the only non-zero contribution is that 
of $L$ derivatives in the first term, so that $j=L$ and
\begin{displaymath}
	\left. \dpartial[L+k]{d}{\lambda} \right|_\xi =\binom{L+k}{L} \dpartial[L]{(\lambda-\xi)^L}{\lambda} \, 
	\dpartial[k]{(\lambda+\xi)^{-L}}{\lambda}=\frac{(L+k)!}{k!} \, \frac{(L+k-1)!}{(L-1)!} \, \frac{(-1)^k}{i^{L+k}} \ .
\end{displaymath}
If we substitute that we obtain
\begin{displaymath}
i\mathcal{D}{\cal F}^L_M =\sum_{j=1}^{M-L+2}{\sum_{k=0}^{M+2-L-j}{\binom{M-j}{L+k-2} \frac{(-1)^k (L+k-1)!}{(L-1)! k!} {\cal F}^L_{j-1}}} \ .
\end{displaymath}
If we perform the sum in $k$ we have
\begin{align*}
	&\sum_{k=0}^{m+2}{(-1)^k \binom{L+m}{L+k-2} \binom{L+k-1}{k}}=\frac{(L+m)!}{(L-1)!} 
	\sum_{k=0}^{m+2}{\frac{(L+k-1)!}{(L+k-2)!} \frac{(-1)^k}{(m-k+2)! k!}} \\
	&=\frac{(L+m)!}{(L-1)! (m+2)!} \sum_{k=0}^{m+2}{\left[ (-1)^k (L-1) \binom{m+2}{k}+(-1)^k k\binom{m+2}{k} \right]} \ ,
\end{align*}
where $m=M-L-j$. Properties of the binomial coefficients say that the first sum 
is zero (unless there is a single term, that is, if $m+2=0$) and the second sum is also zero 
(except if there are two terms, so that $m+2=1$). Then the total contribution of this terms will be
\begin{align}
	& \sum_{M=L-1}^{n-1}{i\mathcal{D}{\cal F}^L_M}=\sum_{M=L-1}^{n-1}{\sum_{j=1}^{M-L+2}}{\frac{(M-j)!}{(L-1)! (M-L-j+2)!}} \nonumber \\
	& \times \big[ (L-1)\delta_{M-L-j+2,0} - (M-L-j+2) \delta_{M-L-j+2,1}) \big] {\cal F}^L_{j-1} \ , \end{align}
which telescopes, so that
\be
\sum_{M=L-1}^{n-1}{i\mathcal{D}{\cal F}^L_M} = {\cal F}^L_{n-L} \ . 
\ee
Therefore, the most general form of the correlation function ${\cal F}^L_n$ is 
\begin{equation}
{\cal F}^L_n=\sum_{k=0}^{n}{\binom{n}{k} \frac{i^{k} \mathcal{D}^{k}}{k!} {\cal F}^L_0}+\theta(n-L) {\cal F}^L_{n-L} \ . 
\label{FLn}
\end{equation}
where $\theta (x)$ is the Heaviside step function, with $\theta (x)=0$ if $x<0$ and $\theta (x)=1$ if $x \geq 0$.

\bibliography{tesis}{}
\bibliographystyle{ieeetr}

\end{document}